\documentclass[12pt,epsf]{article}
\usepackage{amssymb,amscd}
\usepackage{graphics}
\usepackage{epsfig}

\DeclareFontFamily{U}{rsf}{}
\DeclareFontShape{U}{rsf}{m}{n}{
  <5> <6> rsfs5 <7> <8> <9> rsfs7 <10-> rsfs10}{}
\DeclareMathAlphabet\Scr{U}{rsf}{m}{n}

\catcode`\@=11
\def\@citex[#1]#2{%
\if@filesw \immediate \write \@auxout {\string \citation {#2}}\fi
\@tempcntb\m@ne \let\@h@ld\relax \def\@citea{}%
\@cite{%
  \@for \@citeb:=#2\do {%
    \@ifundefined {b@\@citeb}%
      {\@h@ld\@citea\@tempcntb\m@ne{\bf ?}%
      \@warning {Citation `\@citeb ' on page \thepage \space undefined}}%
      {\@tempcnta\@tempcntb \advance\@tempcnta\@ne%
      \@tempcntb\number\csname b@\@citeb \endcsname \relax%
      \ifnum\@tempcnta=\@tempcntb 
        \ifx\@h@ld\relax%
          \edef \@h@ld{\@citea\csname b@\@citeb\endcsname}%
        \else%
          \edef\@h@ld{\ifmmode{-}\else--\fi\csname b@\@citeb\endcsname}%
        \fi%
      \else
        \@h@ld\@citea\csname b@\@citeb \endcsname%
        \let\@h@ld\relax%
      \fi}%
    \def\@citea{,\penalty\@highpenalty\,}%
  }\@h@ld
}{#1}}

\def\@citeb#1#2{{[#1]\if@tempswa , #2\fi}}
\def\@citeu#1#2{{$^{#1}$\if@tempswa , #2\fi }}
\def\@citep#1#2{{#1\if@tempswa , #2\fi}}

\def\bcites{         
        \catcode`\@=11
        \let\@cite=\@citeb
        \catcode`\@=12
}

\def\upcites{         
        \catcode`\@=11
        \let\@cite=\@citeu
        \catcode`\@=12
}

\def\plaincites{      
        \catcode`\@=11
        \let\@cite=\@citep
        \catcode`\@=12
}

\def\section{\@startsection {section}{1}{\z@}{3.ex plus 1ex minus
 .2ex}{2.ex plus .2ex}{\Large\bf}}
\def\subsection{\@startsection{subsection}{2}{\z@}{2.75ex plus 1ex minus
 .2ex}{1.5ex plus .2ex}{\large\bf}}        

\def\appendix{{\newpage\section*{Appendix}}\let\appendix\section%
        {\setcounter{section}{0}
        \gdef\thesection{\Alph{section}}}\section}

\def\abstract{\if@twocolumn
\section*{Abstract}
\else 
\begin{center}
{\bf Abstract\vspace{-.5em}\vspace{0pt}}
\end{center}
\quotation
\fi}

\catcode`\@=12

\newcommand{\beq}{\begin{equation}}
\newcommand{\eeq}{\end{equation}}
\newcommand{\beqa}{\begin{eqnarray}}
\newcommand{\eeqa}{\end{eqnarray}}
\newcommand{\dd}{{\rm d}}

\newcommand{\Z}{{\mathbb Z}}

\newcommand{\R}{{\mathbb R}}
\newcommand{\RR}{{\mathbb R}}
\newcommand{\C}{{\mathbb C}}
\newcommand{\CC}{{\mathbb C}}
\newcommand{\PP}{{\mathbb P}}
\newcommand{\e}{\,{\rm e}}
\newcommand{\CP}{{\CC\PP}}
\newcommand{\RP}{{\RR\PP}}

\newcommand{\ket}{\rangle}
\newcommand{\no}{\nonumber}

\newcommand{\be}{\begin{equation}}
\newcommand{\ee}{\end{equation}}
\newcommand{\bea}{\begin{eqnarray}}
\newcommand{\eea}{\end{eqnarray}}

\def\to{\rightarrow}

\def\ket#1{\left| #1 \right\rangle}

\def\Tr{{\rm Tr}}

\catcode`\@=11

\catcode`\@=12

\bcites

\catcode`\@=11

\@addtoreset{equation}{section}
\@addtoreset{footnote}{section}
\@addtoreset{footnote}{subsection}
\catcode`\@=12

\newcommand{\opsi}{\overline{\psi}}

\newcommand{\btheta}{\overline{\theta}}

\newcommand{\bepsilon}{\overline{\epsilon}}
\newcommand{\bPhi}{\overline{\Phi}}

\newcommand{\bphi}{\overline{\phi}}
\newcommand{\bpsi}{\overline{\psi}}

\newcommand{\bi}{\overline{\imath}}
\newcommand{\bj}{\overline{\jmath}}
\newcommand{\bz}{\overline{z}}
\newcommand{\bchi}{\overline{\chi}}

\newcommand{\bareta}{\overline{\eta}}

\newcommand{\bD}{\overline{D}}

\newcommand{\tpsi}{\widetilde{\psi}}

\newcommand{\tr}{{\rm tr}}
\newcommand{\nn}{\nonumber}
\newcommand{\bartial}{\overline{\partial}}

\newcommand{\wtJ}{\widetilde{J}}

\newcommand{\bLambda}{\bar{\Lambda}}

\newcommand{\V}{\widehat{V}}

\newcommand{\bG}{\overline{G}}
\newcommand{\tG}{\widetilde{G}}
\newcommand{\btG}{\overline{\tG}}

\newcommand{\bQ}{\overline{Q}}
\newcommand{\half}{{1\over 2}}
\newcommand{\wtq}{\widetilde{q}}
\newcommand{\s}{\sigma}
\newcommand{\bk}{\overline{k}}

\newcommand{\kket}[1]{\vert #1\rangle\!\rangle}
\newcommand{\bket}[1]{\vert \Scr{B},#1\rangle\!\rangle}

\newcommand{\cket}[1]{\vert \Scr{C},#1\rangle\!\rangle}

\begin{document}

\begin{titlepage}

\begin{center}

\today\hfill
hep-th/0303135\\
\hfill CERN-TH/2003-061 \\

\vskip 1.5 cm
{\large \bf Orientifolds and Mirror Symmetry}
\vskip 1 cm 
{Ilka Brunner${}^*$ and Kentaro Hori${}^{\dag}$}\\
\vskip 0.5cm
{\it ${}^*$Theory Division, CERN\\
CH-1211 Geneva 23, Switzerland}\\[0.3cm]
{\it ${}^{\dag}$University of Toronto, Ontario, Canada}

\end{center}

\vskip 0.5 cm
\begin{abstract}
We study parity symmetries and crosscap states
in classes of ${\mathcal N}=2$ supersymmetric quantum field theories in
$1+1$ dimensions, including non-linear sigma models,
gauged WZW models,
Landau-Ginzburg models, and linear sigma models.
The parity anomaly and its cancellation
play important roles in many of them.
The case of the ${\mathcal N}=2$
minimal model are studied in complete detail, from all 
three realizations
--- gauged WZW model, abstract RCFT, and LG models.
We also identify mirror pairs of orientifolds, extending the correspondence
between symplectic geometry and algebraic geometry by including
unorientable worldsheets.
Through the analysis in various models and comparison in the overlapping
regimes, we obtain a global picture of orientifolds and D-branes.

\end{abstract}

\end{titlepage}

\pagestyle{empty}

{\footnotesize

\tableofcontents
}

\newpage

\pagenumbering{arabic}
\pagestyle{plain}

\section{Introduction}

String compactifications with ${\mathcal N}=1$ supersymmetry
in $3+1$ dimensions are theoretically very interesting and are believed to be
important for real world physics.
There are several approaches to the constructions of models,
starting with Heterotic strings on Calabi--Yau manifolds \cite{CHSW}.
The approach that has been attracting more recent attention is to consider
Type II strings involving D-branes, which fill
out the $(3+1)$-dimensional world.
In such constructions, orientifolds \cite{ASI,BSII,ASII,HoI,HoII},  are indispensable
elements in order to have consistent theories with finite Newton's
constant and supersymmetry.
Despite this importance, orientifolds are less studied
compared to D-branes which have been investigated extensively
in recent years.
In particular, it is not well understood
what kinds of orientifolds are possible
in which kinds of models.

In this paper, we systematically study parity symmetries
of $(2,2)$ theories in $1+1$ dimensions
commuting with one half of the worldsheet $(2,2)$ supersymmetry,
which are relevant for the construction of supersymmetric
orientifolds.
We particularly study general properties of parity symmetries and the
associated crosscap states, such as the Witten index twisted by parity
symmetry and the dependence of certain
$\RP^2$ diagrams on the parameters of the theory.
Our emphasis is on supersymmetry rather than superconformal invariance,
and we do not limit ourselves to conformal field theories.
This attitude allows us to treat a broader class of models and has proved to
be useful in various other contexts.
The general story is examined and illustrated in several important
classes of theories including the non-linear sigma
models on  K\"ahler manifolds, gauged Wess--Zumino--Witten (WZW) models,
Landau--Ginzburg (LG) models, and linear sigma models.
All these models are related in one way or another
and understanding relations between parities in these
models will be very important.

As the primary example, we perform a complete study of
 parity symmetries and crosscap states in the
${\mathcal N}=2$ minimal model.
The minimal model \cite{ZF,Qiu1}
is the simplest non-trivial theory with $(2,2)$ superconformal
invariance \cite{Ademollo};
it has been playing a central role in the study of
supersymmetric string compactification.
In particular, it can be used as the building block of the Gepner model
of critical supersymmetric string theory
in $3+1$ dimensions \cite{Gepner1,Gepner2}.
The model is realized in three different ways:

(i) as an abstract RCFT using modular matrices, $S$, $T$
\cite{Gepner1,Yang,Qiu2} and $P$,

(ii) as the $SU(2)$ mod $U(1)$ supersymmetric gauged WZW model
\cite{KS,Nmatrix},

(iii) as the IR limit of the
LG model with superpotential $W=\Phi^{k+2}$
\cite{M,VW,minW}.

\noindent
D-branes in the minimal model are studied in these realizations
in \cite{RS,BDLR},\cite{MMS},\cite{HIV,HKKPTVVZ} respectively.
Exact results on the crosscap states are obtained in
the realization (i), following the general RCFT
procedure \cite{BPS0,PSS,bantay,HSS,FSHSS,SaAn,BH1}.
However, the information obtained in this way is about the theory
with a particular GSO projection, in which the
${\mathcal N}=2$ supersymmetry is not manifest.
The essential task required here is to
 entangle the GSO projection and obtain the information on the parities
 and crosscap states of the {\it full} ${\mathcal N}=2$
theory before the GSO projection.
This is done as one of the important achievements of the paper.
The results are in complete agreement with the results
from approach (ii) and (iii) whenever available.

We also study parity symmetries of linear sigma models.
These are simple gauge theories that flow under
renormalization group to the models of interest \cite{phases}.
In many important cases,
they are defined on the whole moduli space of theories,
which interpolates the Gepner models and large volume Calabi--Yau sigma model,
and provide a good understanding of the singularity
of the worldsheet theory.
Moreover they can be used to derive mirror symmetry \cite{HV}.
Thus, by understanding parity symmetries of linear sigma models,
one can first of all
argue on the existence or absence of orientifolds on the moduli
space, one can provide a relation between the Gepner model orientifold
(that is obtained as the application of the orientifolds
 of ${\mathcal N}=2$ minimal models) and the orientifolds of large
volume sigma model, and one can find the mirror correspondence
of orientifold models. 
D-branes are studied in the context of linear sigma models in
\cite{HIV,GJS,Mayr,Hlin,Takayanagi,Hellerman,GJ,HKLM,Kennaway,Distler}.

As for any other symmetry,
one needs to check if the parity symmetry of the classical system
is maintained in the quantum theory.
In many of the examples studied in this paper, we do encounter anomalies
of classical parity symmetries.
They are anomalous because the path-integral measure is not invariant:
in certain topologically non-trivial backgrounds,
there is an odd number of fermion zero mode pairs
that are exchanged under some of the parities.
The anomaly can be cancelled by combining it with another anomalous symmetry.
One possibility is to use $(-1)^{F_{L}}$ that flips the sign of the
left-moving fermions. This works when the theory is conformal
and is indeed applied in the ${\mathcal N}=2$ minimal model
in this paper.
Another way is to turn on a $B$-field. We recall that
the $B$-field term $\int_{\Sigma} \phi^*B$ flips its
sign under the orientation
reversal of $\Sigma$, and for this reason it can generate or cancel
phase factors in the parity transformation of the path-integral measure.

We also present a number of new observations in this paper.
For example, in specifying a parity of non-linear sigma models,
in addition to the action $\tau$ on the target space $X$,
one must specify its action on a complex line bundle
on $X$ whose first Chern class is
$(\tau^*[B]+[B])/2\pi$. 
We also show (with the help of M. Kapranov and Y.-G. Oh)
that the deformation theory of holomorphic Calabi-Yau
orientifolds
is {\it not} obstructed, namely, the classical moduli space of
holomorphic orientifolds is smooth. This is in contrast to the
case of holomorphic D-branes whose deformation {\it is} obstructed 
in general \cite{BDLR,Katz}.

This paper is organized as follows.

In Section~\ref{sec:PC},
we describe general features of
parity symmetries and crosscap states
of theories with ${\mathcal N}=(2,2)$ supersymmetry.
In particular, we consider parities commuting with  half of the
$(2,2)$ supersymmetry. As in the case with boundary conditions
\cite{OOY,HIV},
we will find essentially two types of parities and call them
A-parities and B-parities, following \cite{OOY}.
We show that the overlap of the crosscap and the supersymmetric
ground states obey  certain differential equations with respect to
the parameters of the theory.
We also find the relation of these overlaps and the parity-twisted
Witten indices, which will be called bilinear identities.

This general story is illustrated in 
Section~\ref{sec:NLSM}, in the examples of non-linear sigma models
with  K\"ahler target spaces. A-parities are associated with anti-symplectic
isometries, while B-parities correspond to holomorphic isometries.
We determine the conditions on the complex structure and complexified
K\"ahler parameters for the parity symmetry.
We also compute the parity-twisted Witten index using supersymmetric
localization applied to path-integrals and interpret the result from
the canonical formalism.
For A-type parities and branes, the index is
the self-intersection number of the orientifold plane (O-plane)
for closed string, while it is the intersection number
of the O-plane and D-brane for open string.
The overlaps with the RR ground states are period integrals, and the
bilinear identity is nothing but
the classical Riemann bilinear identity.
For B-type objects, the index is the $\Z_2$-signature  for closed string
while it is the holomorphic Lefschetz number for the open string.
The path-integral computation reproduces the
$\Z_2$-signature theorem and the Lefschetz fixed-point theorem.

In Sections~\ref{sec:Omin} and \ref{sec:OMMII}, we consider
${\mathcal N}=2$ minimal model. We introduce the model
as the gauged WZW model
(realization (ii)) which can be regarded as the sigma model
on the unit disk.
We find A-parities that act on the disk as complex conjugation,
folding along diameters, and B-parities that act as rotation around the
center. We compute the parity-twisted partition functions
(Klein bottle  amplitudes).
We next consider a non-chiral GSO projection that leads
to the realization (i), and determine the crosscap states
following the general procedure
\cite{PSS,PSSI,PSSII,bantay,HSS,FSHSS,SaAn,BH1}. (A part of the computation
given here was also done in \cite{Hikida}, and some earlier results 
in the context of
Gepner models have already been obtained in \cite{BlWi,ABPSS}) 
In the final subsection of Section~\ref{sec:Omin},
we entangle the GSO projection and
determine the crosscap states of the original ${\mathcal N}=2$ minimal model.
We compute the Klein bottle amplitudes, including parity-twisted Witten index
for the closed string, and the result matches with the one from
the gauged WZW computation.
Section~\ref{sec:OMMII} is devoted to the study of parity actions on
the D-branes and stretched open strings.
The D-branes we consider are A-branes (straight segments in the disk)
 and B-branes (concentric disks).
The geometric picture allows us to read off  how the A- and B-parities act
on them, which is confirmed by the M\"obius strip  amplitudes.
We also compute the parity-twisted open string Witten index,
after entangling the GSO projection for the boundary states.

In Section~\ref{sec:LG}, we consider Landau--Ginzburg models.
We find that A-parities are antiholomorphic maps of the LG fields
 such that the superpotential is complex conjugated, and B-parities are
holomorphic maps that reverse the sign of the superpotential.
We show that the overlaps of the crosscap states and the RR ground states
are given by a weighted period integral on the suitably modified 
orientifold planes, and the parity-twisted Witten indices are
intersection numbers of suitably modified branes and O-planes.
This general result is applied to the particular example
of the LG model of a single field $\Phi$ with superpotential
$W=\Phi^{k+2}$, which flows in the IR limit to
the ${\mathcal N}=2$ minimal model (realization (iii)).
We compute the closed and open string parity-twisted Witten index as well as
the overlaps with the RR ground states. The results are in complete agreement
with the results from Sections~\ref{sec:Omin} and ~\ref{sec:OMMII}.

In Sections~\ref{sec:LSM} and ~\ref{sec:CY},
we study parity symmetries of linear sigma models.
We determine the conditions on the parameters for the theory
to be invariant under A-type and B-type parities. These conditions match
the ones derived from the non-linear sigma model
in the large volume limit,
and the ones coming from the LG model
at the Gepner point.
We also determine the corresponding parity in
the mirror Landau--Ginzburg model.
In particular, we find that the information on the parity actions on
the line bundle ${\mathcal L}_{\tau^*B+B}$ mentioned above
 has a natural counterpart in the mirror LG model in terms of the
type of the orientifold planes.
The results are applied to several specific examples
where we find highly non-trivial agreement of the mirror models.
In Section~\ref{sec:CY},
we discuss orientifolds of the system including
compact Calabi--Yau sigma models in its moduli space.
We classify the possible orientifolds of the quintic hypersurface in
$\CP^4$, at least those present for the Fermat type quintic.
We find six of them, three A-type and three B-type.
Using the linear sigma model, we identify the mirror parities in the
mirror quintic.
We also discuss issues concerning the
spacetime physics of Type II orientifolds on Calabi--Yau manifolds.
We count the number of light chiral multiplets and vector multiplets
from the closed string.
We also discuss spacetime superpotential.
Especially, we argue that the moduli space of
holomorphic orientifolds is smooth.
This last section is a preparation for a more complete analysis
of the full orientifold models of string compactification,
which is now possible to do as an application of the present paper.

\section{Parities and Crosscaps in ${\mathcal N}=2$ Theories}\label{sec:PC}

In this section, we describe general features of
parity symmetry of theories with ${\mathcal N}=(2,2)$ supersymmetry
(not necessarily with conformal invariance).
In particular, we consider parities that preserve half of the
$(2,2)$ supersymmetry. As in the case with boundaries,
we will find essentially two types of parities, A-type and B-type.
We will also study and describe the properties of the corresponding
crosscap states.
Especially, we show that the overlap of the crosscap and the supersymmetric
ground states obey  certain differential equations with respect to
the parameters of the theory.
We also find the relation of these overlaps and the parity-twisted
Witten indices.

\subsection{A-parity and B-parity}

\subsubsection{Parity of the (2,2) superspace}

Let us first classify parities of the $(2,2)$ superspace in $1+1$ dimensions.
The superspace has two bosonic coordinates
$x^0,x^1$ (or $x^{\pm}=x^0\pm x^1$) and four fermionic coordinates
$\theta^{\pm},\btheta^{\pm}=(\theta^{\pm})^{\dag}$.
By definition, parity reverses the orientation of the space coordinates
$x^1\to -x^1+$ constant, and therefore exchanges the chirality.
Let us consider the ones
maintaining the holomorphy of ${\mathcal N}=2$ supersymmetry:
\beqa
&&\Omega_A:\,(x^{\pm},\theta^+,\theta^-,\btheta^+,\btheta^-)\longmapsto
(x^{\mp},-\btheta^-,-\btheta^+,-\theta^-,-\theta^+),
\nn\\
&&\Omega_B:\,(x^{\pm},\theta^+,\theta^-,\btheta^+,\btheta^-)\longmapsto
(x^{\mp},\theta^-,\theta^+,\btheta^-,\btheta^+).
\nn
\eeqa
We shall call the former {\it A-parity},
and the latter {\it B-parity}.
The supersymmetry generators
$Q_{\pm}=\partial/\partial\theta^{\pm}
+i\btheta^{\pm}\partial/\partial x^{\pm}$,
$\bQ_{\pm}=-\partial/\partial\btheta^{\pm}
-i\theta^{\pm}\partial/\partial x^{\pm}$
are then transformed
as
\beqa
&&A: \quad
Q_{\pm}\longrightarrow \bQ_{\mp},\quad
\bQ_{\pm}\longrightarrow Q_{\mp},
\label{PASUSY}\\
&&B: \quad
Q_{\pm}\longrightarrow Q_{\mp},\quad
\bQ_{\pm}\longrightarrow \bQ_{\mp}.
\label{PBSUSY}
\eeqa
The vector and axial R-rotations, $U(1)_V$ and $U(1)_A$,
 are also transformed:
A-parity reverses $U(1)_V$ and preserves $U(1)_A$ while
B-parity preserves $U(1)_V$ and reverses $U(1)_A$.
The differential operators
$D_{\pm}=\partial/\partial\theta^{\pm}
-i\btheta^{\pm}\partial/\partial x^{\pm}$,
$\bD_{\pm}=-\partial/\partial\btheta^{\pm}
+i\theta^{\pm}\partial/\partial x^{\pm}$
are transformed as
$A:$ $D_{\pm}\leftrightarrow \overline{D}_{\mp}$ and
$B:$ $D_+\leftrightarrow D_-$, $\bD_+\leftrightarrow \bD_-$.
Accordingly, chiral and twisted chiral superfields
are mapped by A-parity and B-parity as
\beqa
&&A:\quad \begin{array}{rcl}
\mbox{chiral}& \longleftrightarrow &\mbox{antichiral}\\
\mbox{twisted chiral}& \longleftrightarrow & \mbox{twisted chiral},
\end{array}
\nn
\\
&&B:\quad \begin{array}{rcl}
\mbox{chiral}&\longleftrightarrow &\mbox{chiral}\\
\mbox{twisted chiral}&\longleftrightarrow &\mbox{twisted antichiral}.
\end{array}
\nn
\eeqa
One can modify the above parities by the $U(1)$ R-rotations,
$A_{\alpha,\beta}:$
$\theta^{\pm}\mapsto -\e^{i\alpha\mp i\beta}\btheta^{\mp}$,
and
$B_{\alpha,\beta}:$
$\theta^{\pm}\mapsto \e^{-i\alpha\pm i\beta}\theta^{\mp}$.
They transform the supercharges as
\beqa
&&A_{\alpha,\beta}: \quad
Q_{\pm}\rightarrow \e^{-i\alpha\mp i\beta}\bQ_{\mp},\,\,
\bQ_{\pm}\rightarrow \e^{i\alpha \pm i\beta} Q_{\mp},
\label{Aab}\\
&&B_{\alpha,\beta}: \quad
Q_{\pm}\rightarrow \e^{-i\alpha\mp i\beta}Q_{\mp},\,\,
\bQ_{\pm}\rightarrow \e^{i\alpha \pm i\beta} \bQ_{\mp}.
\label{Bab}
\eeqa

The fixed-point set of $\Omega_A$ and $\Omega_B$ is
$x^1=0,$ $\theta^++\btheta^-=0,$ $\btheta^++\theta^-=0$
and
$x^1=0,$ $\theta^+=\theta^-,$ $\btheta^+=\btheta^-$,
respectively.
These are nothing but the A-boundary and B-boundary that
are relevant to the superfield description of
boundary ${\mathcal N}=2$ theories \cite{Martinec,Hlin}.
One can also consider parity actions on boundary superspace
whose bosonic subspace is the strip
$0\leq x^1\leq \pi$ preserved by
$x^1\leftrightarrow \pi -x^1$.
One can consider A-boundaries $\theta^++\btheta^-=\btheta^++\theta^-=0$
or B-boundaries $\theta^+-\theta^-=\btheta^+-\btheta^-=0$
at $x^1=0$ and $\pi$.
Under both A-parity and B-parity, an
A(B)-boundary at $x^1=0$ is mapped to an A(B)-boundary at $x^1=\pi$
and vice versa.
Chiral superfields on an A-boundary at $x^1=0$
are mapped by A-parity (B-parity) to
chiral (antichiral) superfields on an A-boundary at $x^1=\pi$.

\subsubsection{A-parity and B-parity in $(2,2)$ theories}

\newcommand{\wtx}{\widetilde{x}}

In any quantum field theory in $1+1$ dimensions,
a parity symmetry takes the form
$\tau\circ \Omega$, where $\Omega$ is the space inversion
$x=(x^0,x^1)\to \wtx=(x^0,-x^1)$
and $\tau$ is an internal action of the fields.
In general, only the combination $\tau\circ \Omega$ is a symmetry, not 
$\tau$ and $\Omega$ individually.
The parity symmetry
is realized as an operator $P$ on the Hilbert space of states
that commutes with the Hamiltonian but inverts the momentum.
In a $(2,2)$ supersymmetric theory,
an A-parity and a B-parity take the form $P_A=\Scr{T}_A\circ \Omega_A$
and $P_B=\Scr{T}_B\circ\Omega_B$ respectively,
where $\Scr{T}_{A,B}$ are internal actions of the superfields.
They are realized as operators on the Hilbert space that transform
the supercharges as
 (\ref{PASUSY}) and (\ref{PBSUSY}) respectively.

In particular, they transform the supercurrents
$G_{\pm}^{\mu}$, $\bG_{\pm}^{\mu}$ ($\mu=0,1$) as
\beqa
&&P_A:
G_{\pm}^{\mu}(x)\rightarrow (-1)^{\mu}\bG_{\mp}^{\mu}(\wtx),
\,\,\,
\bG_{\pm}^{\mu}(x)\rightarrow (-1)^{\mu} G_{\mp}^{\mu}(\wtx),
\label{actOAG}
\\
&&P_B:
G_{\pm}^{\mu}(x)\rightarrow (-1)^{\mu}G_{\mp}^{\mu}(\wtx),\,\,\,
\bG_{\pm}^{\mu}(x)\rightarrow (-1)^{\mu} \bG_{\mp}^{\mu}(\wtx).
\label{actOBG}
\eeqa
If the system has vector and/or axial R-symmetry, and if the parity
respects them, the R-currents are transformed as
\beqa
&&P_A:
J_V^{\mu}(x)\to -(-1)^{\mu}J_V^{\mu}(\wtx),
\,\,\,
J_A^{\mu}(x)\to (-1)^{\mu}J_A^{\mu}(\wtx),
\label{aAJ}\\
&&P_B:
J_V^{\mu}(x)\to (-1)^{\mu}J_V^{\mu}(\wtx),
\,\,\,
J_A^{\mu}(x)\to -(-1)^{\mu}J_A^{\mu}(\wtx).
\label{aBJ}
\eeqa
For each A-parity $P_A$
we obtain an $A_{\alpha,\beta}$-parity
by combining it with the R-symmetry:
$P_{A_{\alpha,\beta}}=\e^{-i\alpha F_V-i\beta F_A}P_A$.
Similarly, a $B_{\alpha,\beta}$-parity can be obtained:
$P_{B_{\alpha,\beta}}=\e^{-i\alpha F_V-i\beta F_A}P_B$.
(We define the transformation of operators by a symmetry $U$ by
${\mathcal O}\to U^{-1}{\mathcal O}U$.)

An A-parity in one theory is mapped to a B-parity of the mirror,
since mirror symmetry exchanges $Q_-$ and $\bQ_-$.
Also, mirror symmetry exchanges
$A_{\alpha,\beta}$ and $B_{\beta,\alpha}$.

\subsubsection{Parity actions on chiral superfields}

For example, let us consider the theory of a single chiral superfield
$\Phi(x,\theta)=\Phi(x^{\pm},\theta^{\pm},\btheta^{\pm})$ with the Lagrangian
$$
L=\int\dd^4\theta\,\, \bPhi\Phi.
$$
Since the measure $\dd^4\theta=\dd\theta^+\dd\theta^-\dd\btheta^-\dd\btheta^+$
is invariant under both $\Omega_A$ and $\Omega_B$, the Lagrangian
is invariant under
\beqa
&&A:\,\Phi(x,\theta)\longrightarrow \overline{\Omega_A^*\Phi(x,\theta)}
=\overline{\Phi(\Omega_A(x,\theta))},
\label{Aacchi}\\
&&B:\,\Phi(x,\theta)\longrightarrow\Omega_B^*\Phi(x,\theta)
=\Phi(\Omega_B(x,\theta)).
\label{Bacchi}
\eeqa
The right hand sides are both chiral superfields:
$\Omega_B^*\Phi$ is chiral as we have seen, while
$\Omega_A^*\Phi$ is antichiral and therefore
its hermitian conjugate $\overline{\Omega_A^*\Phi}$ is chiral.
In this way they determine consistent transformations of the field,
$\Phi\to P_A^{-1}\Phi P_A$ and
$\Phi\to P_B^{-1}\Phi P_B$.
They realize an A-parity and a B-parity,
$P_A^{-1}Q_{\pm}P_A=\bQ_{\mp}$ and $P_B^{-1}Q_{\pm}P_B=Q_{\mp}$.
For $P_B$ this is because (\ref{PBSUSY}) says that
$\Omega_B^*[Q_{\pm},\Phi]=[Q_{\mp},\Omega_B^*\Phi]$,
which means that $P_B^{-1}[Q_{\pm},\Phi]P_B=[Q_{\mp},P_B^{-1}\Phi P_B]$.
For $P_A$, (\ref{PASUSY}) says
$\Omega_A^*[\bQ_{\pm},\Phi]=[Q_{\mp},\Omega_A^*\Phi]$
and its hermitian conjugate 
equation is
$P_A^{-1}[Q_{\pm},\Phi]P_A=[\bQ_{\mp},P_A^{-1}\Phi P_A]$.
In terms of the component fields,
$\Phi=\phi+\theta^+\psi_++\theta^-\psi_-+\theta^+\theta^-F+\cdots$,
the actions are as follows:
\beqa
&&
P_A:\left\{
\begin{array}{c}
\phi(x)\longrightarrow \overline{\phi}(\wtx)\\[0.1cm]
\psi_{\pm}(x)\longrightarrow \overline{\psi}_{\mp}(\wtx)\\[0.1cm]
F(x)\longrightarrow\overline{F}(\wtx)
\end{array}
\right.\\
&&
P_B:
\left\{
\begin{array}{c}
\phi(x)\longrightarrow \phi(\wtx)\\[0.1cm]
\psi_{\pm}(x)\longrightarrow \psi_{\mp}(\wtx)\\[0.1cm]
F(x)\longrightarrow -F(\wtx)
\end{array}
\right.
\eeqa

The parity transformations 
(\ref{Aacchi}) and (\ref{Bacchi}) are essentially
those  used
in more interesting systems described in terms of
chiral superfields, such as non-linear sigma models (Section~\ref{sec:NLSM}),
Landau--Ginzburg models (Section~\ref{sec:LG})
and linear sigma models (Sections~\ref{sec:LSM} and \ref{sec:CY}).
In the last example, we will also encounter parity actions on
twisted chiral superfields.

\subsubsection{Unbroken supersymmetry and the Witten index}

Half of the $(2,2)$ supersymmetry is invariant under
A-parity and B-parity.
The invariant combinations are respectively
\beqa
&&Q_A=\bQ_++Q_-,\,\,\,Q_A^{\dag}=Q_++\bQ_-,\\
&&Q_B=\bQ_++\bQ_-,\,\,\,Q_B^{\dag}=Q_++Q_-.
\eeqa
These are the same as the supercharges that are preserved by
A-branes and B-branes \cite{OOY,HIV}.
$Q=Q_A$ or $Q_B$ obey
\beq
\{Q,Q^{\dag}\}=2H,\quad Q^2=0,
\label{SUSY1}
\eeq
which are the relations of ${\mathcal N}=2$
supersymmetric quantum mechanics.
The symmetry generated by
$Q_A, Q_A^{\dag}$ and $Q_B, Q_B^{\dag}$
shall be called ${\mathcal N}=2_A$ and ${\mathcal N}=2_B$
supersymmetries respectively.

One may consider the Witten index with a twist by $P=P_A$ or $P_B$
\beq
I_{P}=\Tr\!\!\mathop{}_{{\mathcal H}_{\rm RR}}\!\!
(P (-1)^F \e^{-\beta H}).
\label{defIP}
\eeq
As a consequence of supersymmetry (\ref{SUSY1}),
it receives a contribution only from the ground states, and
is invariant under supersymmetric deformations of the theory.
In particular,
it is independent of $\beta$ and of 
the radius of the circle on which the system is quantized.

One can also consider the parity of an open string stretched between D-branes.
Under both A-parity and B-parity, A(B)-branes are mapped to
A(B)-branes. Furthermore an open string stretched between A(B)-branes
preserves the ${\mathcal N}=2$ supersymmetry generated by $Q_A$ and
$Q_A^{\dag}$ ($Q_B$ and $Q_B^{\dag}$),
which are invariant under an A(B)-parity.
For an A(B)-brane $a$ and its image $Pa$ under A(B)-parity
$P=P_A$($P_B$), one can also consider the Witten index
\beq
I_P(a,Pa)=\Tr\!\!\mathop{}_{{\mathcal H}_{a,Pa}}\!\!
(P (-1)^F \e^{-\beta H}),
\label{defIPa}
\eeq
where ${\mathcal H}_{a,Pa}$ is the space of states of the
$a$-$Pa$ string.
It receives contributions only from the supersymmetric
ground states and is a topological invariant of the open string system.

The modified versions $A_{\alpha,\beta}$ and
$B_{\alpha,\beta}$ may or may not preserve  half of the
supersymmetry.
$A_{\alpha,\beta}$-parity (resp. $B_{\alpha,\beta}$-parity)
preserves an ${\mathcal N}=2$ supersymmetry if and only if
$\beta\in \pi \Z$ (resp. $\alpha\in\pi\Z$).
The invariant combinations are
\beqa
&&Q_{A_{\alpha,\beta}}=\bQ_++\e^{i\alpha+i\beta}Q_-,
\,\,\,Q_{A_{\alpha,\beta}}^{\dag}=Q_++\e^{-i\alpha-i\beta}\bQ_-,
\quad \beta\in\pi\Z,
\nn\\
&&Q_{B_{\alpha,\beta}}=\bQ_++\e^{i\alpha+i\beta}\bQ_-,
\,\,\,Q_{B_{\alpha,\beta}}^{\dag}=Q_++\e^{-i\alpha-i\beta}Q_-,
\quad \alpha\in\pi\Z.
\nn
\eeqa
Thus for such values of $\beta$ (resp. $\alpha$), the
twisted Witten indices are deformation invariants of the theory.

\subsection{Crosscap states}

\begin{figure}[tb]
\centerline{\includegraphics{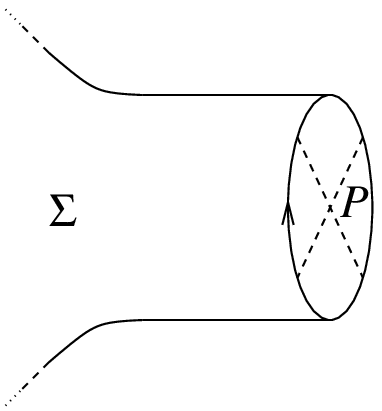}}
\caption{Crosscap $\langle\Sigma|\Scr{C}_P\rangle$}
\label{cc}
\end{figure}

For each parity symmetry $P=\tau\circ\Omega$, there is a so-called
`crosscap state'
$|\Scr{C}_P\rangle$
which can be used to express partition functions on
unorientable surfaces \cite{Callanetal,PolCai}.
Let $\Sigma$ be an orientable or unorientable
 surface with an oriented boundary circle around which $\Sigma$
is flat, as in Fig.~\ref{cc}.
We choose
a coordinate system $(\sigma^1,\sigma^2)$, $\sigma^1\equiv \sigma^1+2\pi$,
$\sigma^2\geq 0$, where the boundary circle is at
 $\sigma^2=0$ and is parametrized by $\sigma^1$.
We glue $\Sigma$ and its copy
$\overline{\Sigma}$
along the boundary circles to make a double
$\Sigma\# \overline{\Sigma}$ which has an involution
$\Omega$ that extends $(\sigma^1,\sigma^2)
\mapsto (\sigma^1+\pi,-\sigma^2)$.
Consider a path integral over the fields on this double
$\Sigma\# \overline{\Sigma}$
obeying the condition
${\mathcal O}=\tau \Omega^*{\mathcal O}$.
The crosscap state is defined by the property that
the path-integral is expressed as $\langle\Sigma|\Scr{C}_P\rangle$,
where $\langle \Sigma|$ is the state at the boundary circle
resulting from the path-integral over the fields on $\Sigma$.
The fields are periodic along the circle if and only if
$P$ is involutive, $P^2={\rm id}$.
If not, $|\Scr{C}_P\rangle$ belongs to the sector
in which the fields obey the twisted boundary condition
${\mathcal O}(\sigma^1,\sigma^2)
=P^{-2}{\mathcal O}(\sigma^1+2\pi,\sigma^2)P^2$.
In such a case, the pairing $\langle\Sigma|\Scr{C}_P\rangle$
makes sense only if $\langle\Sigma|$ belongs to the sector with the same
periodicity (which can be realized, say, by
inserting a twist operator in the interior of $\Sigma$).

We study its properties when $P$ is an A-parity or a B-parity,
or their variants.

\subsubsection{Current conditions}

\newcommand{\wtz}{\widetilde{z}}

The transformation rule of the currents
(\ref{actOAG})-(\ref{aAJ}) or (\ref{actOBG})-(\ref{aBJ})
yields current conditions on
the crosscap states.
We write them down using
the `tree-channel' coordinates $(\sigma^1,\sigma^2)$
which are obtained from the `loop-channel'
Minkowski coordinates $(x^0,x^1)$
via Wick rotation and
90${}^{\circ}$-rotation.\footnote{$(\sigma^1,\sigma^2)$ here
is $(ix^0,-x^1)$ there.
The tree-channel supercurrents are related to those of
the loop channel as
$G_{\pm}^{\rm loop}=\e^{\pm \pi i/4}G_{\pm}^{\rm tree}$.
The factors of $i$ that appear in (\ref{condCA})
or (\ref{condCB}) have their origin in the phase factor
$\e^{\pm \pi i/4}$ here.
}
The crosscap state $|\Scr{C}_{P_A}\rangle$
for an A-parity $P_A$ obeys the following condition
for $z=(\sigma^1,\sigma^2)\to \wtz=(\sigma^1+\pi,-\sigma^2)$:
\beqa
&&\bG_+^{\mu}(z)-i(-1)^{\mu}G_-^{\mu}(\wtz)
=G_+^{\mu}(z)-i(-1)^{\mu}\bG_-^{\mu}(\wtz)\nn\\
&&
=G_-^{\mu}(z)+i(-1)^{\mu}\bG_+^{\mu}(\wtz)
=\bG_-^{\mu}(z)+i(-1)^{\mu}G_+^{\mu}(\wtz)=0,
\label{condCA}
\\
&&J_V^{\mu}(z)-(-1)^{\mu}J_V^{\mu}(\wtz)=
J_A^{\mu}(z)+(-1)^{\mu}J_A^{\mu}(\wtz)=0.
\nn
\eeqa
where $\mu=1,2$.
The crosscap state $|\Scr{C}_{P_B}\rangle$
for a B-parity $P_B$ obeys
\beqa
&&\bG_+^{\mu}(z)-i(-1)^{\mu}\bG_-^{\mu}(\wtz)
=G_+(z)-i(-1)^{\mu}G_-(\wtz)\nn\\
&&=\bG_-^{\mu}(z)+i(-1)^{\mu}\bG_+(\wtz)
=G_-(z)+i(-1)^{\mu}G_+(\wtz)=0,
\label{condCB}
\\
&&J_V^{\mu}(z)+(-1)^{\mu}J_V^{\mu}(\wtz)=
J_A^{\mu}(z)-(-1)^{\mu}J_A^{\mu}(\wtz)=0.
\nn
\eeqa
The supercurrents are periodic along the circle
since $P_A^2$ and $P_B^2$ act trivially on the
supercurrent.
Namely, {\it the crosscap states 
for A-parity and B-parity belong to sectors in which
the supercurrents are periodic,}
such as Ramond-Ramond sector.
The R-charges $q_V=\int J_V^2(\s^1)\dd\s^1$,
$q_A=\int J_A^2(\s^1)\dd \s^1$
of the crosscap states are also constrained;
{\it The crosscap state for A-parity (B-parity)
has vanishing axial (vector) R-charge
$q_A=0$ ($q_V=0$).}

\begin{figure}[tb]
\centerline{\includegraphics{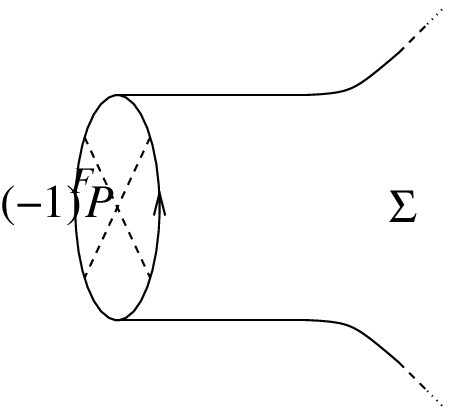}}
\caption{Crosscap $\langle\Scr{C}_P|\Sigma \rangle$}
\label{rcc}
\end{figure}

Let us next consider the
`bra-crosscap' which is defined as the dagger
of the `ket-crosscap'
$$
\langle \Scr{C}_P|:=|\Scr{C}_P\rangle^{\dag}.
$$
For $P=P_A$ or $P_B$, the condition obeyed by this state is obtained by
taking the dagger of (\ref{condCA}) or
(\ref{condCB}).
Note that each factor of $i$ receives a minus sign under dagger.
Thus, the bra-crosscap
$\langle \Scr{C}_P|$ fulfills the same condition
as the ket-crosscap $|\Scr{C}_{(-1)^FP}\rangle$
for the parity $(-1)^FP$.
In other words, the
fields are subject to the condition
${\mathcal O}=(-1)^{|\mathcal O|}\tau \Omega^*{\mathcal O}$ at
the state $\langle \Scr{C}_P|$ where
$|{\mathcal O}|$ is the mod 2 fermion number of
${\mathcal O}$. \footnote{To be more precise, $|{\mathcal O}|$ is
twice the mod $\Z$ spin of ${\mathcal O}$. However, we only consider theories
in which the spin-statistics correlation holds and the two
definitions of $|{\mathcal O}|$ agree.}
See Fig.~\ref{rcc}.

The crosscap states
$|\Scr{C}_{P_{A_{\alpha,\beta}}}\rangle$ and
$|\Scr{C}_{P_{B_{\alpha,\beta}}}\rangle$
for $A_{\alpha,\beta}$ and
$B_{\alpha,\beta}$-parities
obey the same R-current conditions as above but
the supercurrent conditions are modified as
\beqa
&&A_{\alpha,\beta}:\quad
\bG_{\pm}^{\mu}(z)\mp i(-1)^{\mu}\e^{i\alpha\pm i\beta}
G_{\mp}^{\mu}(\wtz)
=G_{\pm}^{\mu}(z)\mp i(-1)^{\mu}\e^{-i\alpha\mp i\beta}
\bG_{\mp}^{\mu}(\wtz)=0,\quad
\label{coCA}\\
&&B_{\alpha,\beta}:\quad
\bG_{\pm}^{\mu}(z)\mp i(-1)^{\mu}\e^{i\alpha\pm i\beta}
\bG_{\mp}^{\mu}(\wtz)
=G_{\pm}(z)\mp i(-1)^{\mu}\e^{-i\alpha\mp i\beta}G_{\mp}(\wtz)=0.\quad
\label{coCB}
\eeqa
The supercurrents fulfill the boundary condition
$G_{\pm}^{\mu}(\s^1)=\e^{\mp 2i\beta}G_{\pm}^{\mu}(\s^1+2\pi)$
(for $|\Scr{C}_{P_{A_{\alpha,\beta}}}\rangle$)
and
 $G_{\pm}^{\mu}(\s^1)=\e^{-2i\alpha}G_{\pm}^{\mu}(\s^1+2\pi)$
(for $|\Scr{C}_{P_{B_{\alpha,\beta}}}\rangle$).
Note that they are periodic if and only if
the parity preserves an ${\mathcal N}=2$ supersymmetry.
We often call those parities $\widetilde{A}$-parity or
$\widetilde{B}$-parity if the crosscap states belong to 
sectors in which the supercurrents are
anti-periodic, such as the Neveu-Schwarz-Neveu-Schwarz sector.
($A_{\alpha,\beta}$ is an $\widetilde{A}$-parity iff
$\beta\in \pi(\Z+\half)$, and $B_{\alpha,\beta}$ is a
$\widetilde{B}$-parity iff
$\alpha\in \pi(\Z+\half)$.)
Taking the dagger, one realizes that
the bra-crosscap state
$\langle \Scr{C}_P|$ obeys the same condition
as the ket-crosscap state
for the parity $(-1)^FP$ (see Fig.~\ref{rcc}).

\subsubsection{Partition functions and crosscaps}

Let us consider the pairing of the crosscap states
$\langle\Scr{C}_{P_1}|q_t^H|\Scr{C}_{P_2}\rangle$
for two parities $P_i=\tau_i\circ\Omega$.
This can be identified as the partition function on
the Klein bottle
 $(x,y)\equiv (x+2,y)\equiv (-x,y+1)$ with a suitable metric,
where the fields fulfill the following boundary conditions
$$
{\mathcal O}(x,y)=\tau_2{\mathcal O}(2-x,y+1)
=(-1)^{|\mathcal O|}\tau_1{\mathcal O}(-x,y+1).
$$
Here $(-1)^{|\mathcal O|}$ is the mod 2 fermion number of
${\mathcal O}$ whose appearance here is explained above.
It follows that
$
{\mathcal O}(x,y)=(-1)^{|\mathcal O|}\tau_1
{\mathcal O}(2-(x+2),y+1)
=(-1)^{|\mathcal O|}\tau_1\tau_2^{-1}{\mathcal O}(x+2,y).
$
Namely, the fields obey the boundary condition
\beq
{\mathcal O}(x,y)=U^{-1}{\mathcal O}(x+2,y)U,\quad
\mbox{where $U=(-1)^FP_1P_2^{-1}$.}
\label{twU}
\eeq
Thus, the pairing can be identified as the
twisted partition function
\beq
\langle\Scr{C}_{P_1}|q_t^H|\Scr{C}_{P_2}\rangle
=\Tr\!\!\mathop{}_{{\mathcal H}_{(-1)^FP_1P_2^{-1}}}\!\!
(-1)^FP_2 q_l^H,
\eeq
where ${\mathcal H}_{(-1)^FP_1P_2^{-1}}$
is the space of states with the twisted boundary condition
(\ref{twU}).
Using this, one can express various twisted partition functions 
with the help of the crosscap states.
For example,
the partition function in the NSNS sector
can be written as
\beq
\Tr\!\!\mathop{}_{{\mathcal H}_{\rm NSNS}}\!\!Pq^H
=\langle\Scr{C}_{(-1)^FP}|q_t^H|\Scr{C}_{(-1)^FP}\rangle.
\label{idNSNS}
\eeq
Also, the twisted Witten index can be expressed as
\beq
I_P=\Tr\!\!\mathop{}_{{\mathcal H}_{\rm RR}}\!\!(-1)^FPq^H
=\langle\Scr{C}_{(-1)^FP}|q_t^H|\Scr{C}_{P}\rangle.
\label{idRR}
\eeq

The twisted Witten index for a supersymmetric
open string can be obtained
in terms of crosscap and boundary states
\beq
I_P(a,Pa)=\Tr\!\!\mathop{}_{{\mathcal H}_{a,Pa}}\!\!
(P (-1)^F \e^{-\beta H})
=\langle\Scr{B}_a|q_t^H|\Scr{C}_P\rangle.
\eeq
Here it is important that the boundary state
$\langle\Scr{B}_a|$ is chosen in such a way 
that it preserves the same supersymmetry
as $|\Scr{C}_P\rangle$ does.

\subsection{Overlap with supersymmetric ground states}

For D-branes,
the overlaps of the boundary states and the RR ground states
are their important characteristics
---
they obey certain differential equations with respect to
the parameters of the theory, and also carry information on the 
RR charge and tension \cite{HIV}.
Here we study the analogs for orientifolds.
Let $P$ be an A-parity or a B-parity.
One may also consider a variant that preserves
an ${\mathcal N}=2$ supersymmetry
(namely $A_{\alpha,\beta}$ with $\beta\in\pi\Z$ or
$B_{\alpha,\beta}$ with $\alpha\in\pi\Z$).
In such cases, the crosscap states $|\Scr{C}_P\rangle$
and $\langle \Scr{C}_{(-1)^FP}|$
are in the sector in which the
supercurrents are periodic, that is, a sector with $(2,2)$
supersymmetry.
Therefore, one can consider the overlaps
with the supersymmetric ground states $|i\rangle$
in that sector:
\begin{equation}
\begin{array}{l}
\\
\Pi_{i}^P=\langle \Scr{C}_{(-1)^FP}|i\rangle,
\\[0.2cm]
\widetilde{\Pi}_i^P=\langle i|\Scr{C}_P\rangle.
\\[0.2cm]
\end{array}
\label{pairi}
\end{equation}
We study the properties of such overlaps.

\subsubsection{Dependence on parameters}

Let us study the dependence of the overlaps on the parameters of the theory.
Let $P$ be an A-parity (or an
$A_{\alpha,\beta}$-parity with $\beta\in\pi\Z$)
in a theory that admits a B-twist.
As the ground states $|i\rangle$,
we use those corresponding to $cc$ ring elements $\phi_i$.
One important point is that parity symmetry imposes constraints on
the allowed deformations of the theory.
Thus the parameter space is generally reduced.
For A-parities, the constraints are holomorphic for
twisted chiral parameters and antiholomorphic for the chiral parameters.
This will be explained in several examples in later sections.

Let us first consider twisted F-term deformations.
Since the constraints are holomorphic,
the allowed twisted chiral parameters are complex.
It is easy to see, using the standard techniques as in \cite{HIV}
that the overlaps are invariant under the allowed twisted F-term deformations
\begin{equation}
{\partial \Pi^P_i\over\partial t_{ac}}=0,~~
{\partial \Pi^P_i\over\partial \overline{t}_{ac}}=0.
\label{acinv}
\end{equation}

Let us next study the F-term deformations.
Since the constraints are antiholomorphic,
the allowed chiral parameters are real. To be more precise,
the allowed moduli space is a
middle dimensional real subspace of the (complex) moduli space of all
chiral parameters $t^i$.
Let
$t^i=x^i+iy^i$ be the decomposition into the tangent direction
$x^i$ and orthogonal directions $y^i$.
\begin{figure}[tb]
\centerline{\includegraphics{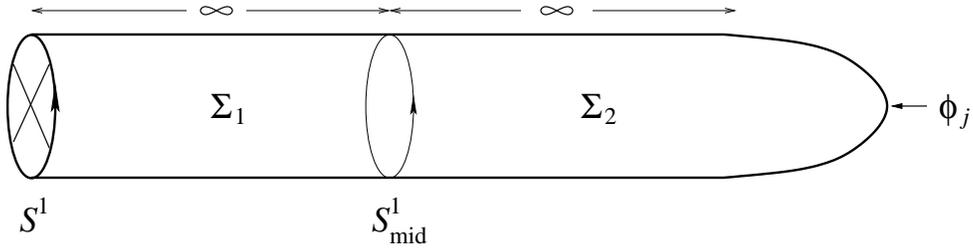}}
\caption{The semi-infinite $\RP^2$}
\label{nand}
\end{figure}
It can then be shown that the overlaps obey the following differential
equations:
\begin{equation}
\begin{array}{l}
(\nabla_{x^i}\Pi^P)_j=(D_{x^i}\delta^k_j+\beta C_{y^ij}^k)\Pi^P_k=0,
\\[0.2cm]
(\nabla_{x^i}\Pi^P)_{\bj}=(D_{x^i}\delta^{\bk}_{\bj}
+\beta C_{y^i \bj}^{\bk})\Pi^P_{\bk}=0,
\end{array}
\label{parallel}
\end{equation}
where $\beta$ is the circumference of the boundary circle $S^1$.
Here $D_{x^i}$ is the covariant derivative of the vacuum bundle \cite{CV}
in the direction of $x^i$,
and
$$
C_{y^ij}^k=iC_{ij}^k-iC_{\bi j}^k,
$$
where $C_{ij}^k$
are the structure constants of the chiral ring
and $C_{\bi j}^k=g^{k\overline{l}}g_{j\overline{m}}
C_{\bi \overline{l}}^{\overline{m}}$.
The relations (\ref{acinv})-(\ref{parallel})
can be shown by the standard gymnastics
in $tt^*$ equation, using the worldsheet  in Figure~\ref{nand},
just as in the derivation of
the similar equation for overlaps with boundary states.
The essential point is that
the contour integral of the supercurrent
bounces back at the boundary of the cigar, with
$\bG_{\pm}$ turned into
$\pm i G_{\mp}$ via the supercurrent condition at the crosscap
shown in Eq. (\ref{condCA}).
In the derivation of (\ref{parallel}), we consider
$t^i$ and $\overline{t}^{\bi}$ variations
in the combination of
$\partial/\partial x^i
=\partial/\partial t^i+\partial/\partial \overline{t}^{\bi}$.
From the $t^i$-variation we obtain the term
$-i\beta C_{ij}^k\phi_k$ and from the $\overline{t}^{\bi}$-variation
we obtain the term $+i\beta C_{\bi j}^k\phi_k$.
The sum is $-\beta C_{y^ij}^k\phi_k$ which is the origin of
the second term in (\ref{parallel}).
Essentially the same relation holds for the other overlaps
$\widetilde{\Pi}^P_i=\langle i|P\rangle$:
they do not depend on the twisted F-term deformations, and
satisfy the follwing equation
for the F-term deformations
\begin{equation}
\begin{array}{l}
(\nabla_{x^i}\widetilde{\Pi}^P)_j=(D_{x^i}\delta^k_j-\beta C_{y^ij}^k)
\widetilde{\Pi}^P_k=0,
\\[0.2cm]
(\nabla_{x^i}\widetilde{\Pi}^P)_{\bj}=(D_{x^i}\delta^{\bk}_{\bj}
-\beta C_{y^i\bj}^{\bk})\widetilde{\Pi}^P_{\bk}=0.
\end{array}
\label{parallel2}
\end{equation}

\subsubsection{Bilinear identities}

\begin{figure}[tb]
\centerline{\includegraphics{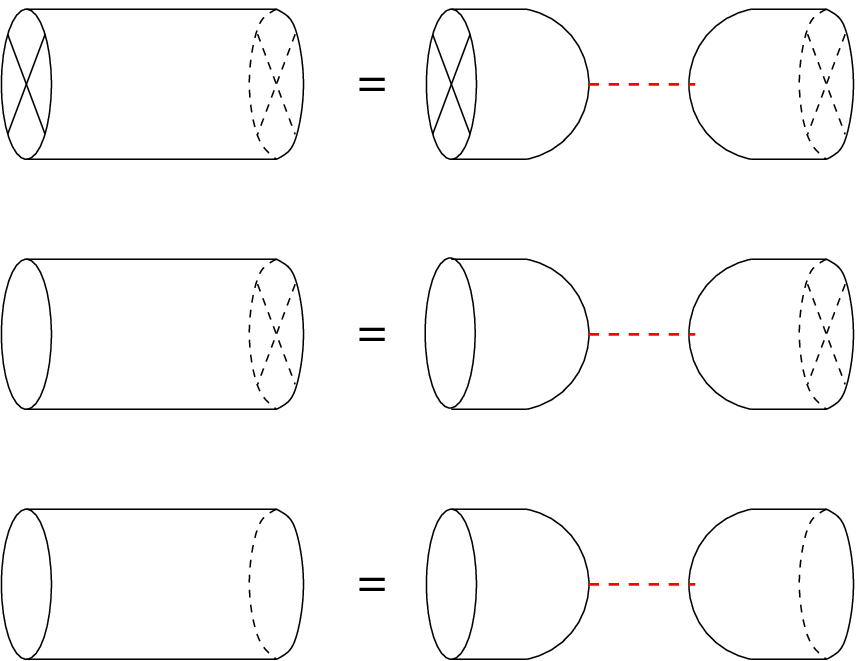}}
\caption{The Bilinear Identities (\ref{biIP}), (\ref{biIPa})
and (\ref{biIab}).}
\label{bili}
\end{figure}

Let $P$ be a supersymmetric parity
(an A-parity or a B-parity or their variant preserving
an ${\mathcal N}=2$ supersymmetry).
We have seen that the twisted Witten index is expressed as
the pairing
$I_P=\langle\Scr{C}_{(-1)^FP}|\e^{-TH}|\Scr{C}_P\rangle$.
Using a complete basis $|N\rangle$ of the closed string states,
this can be rewritten as
$\sum_N\langle \Scr{C}_{(-1)^FP}|N\rangle \e^{-TE_N}\langle N
|\Scr{C}_P\rangle$.
Note that $|\Scr{C}_P\rangle$ belongs to a sector
in which there is a $(2,2)$ supersymmetry
and hence the intermediate energies are non-negative, $E_N\geq 0$,
with $E_N=0$ corresponding to the supersymmetric ground states.
Now we use the fact that the Witten index is independent of the deformation
parameters, in particular $T$.
The limit $T\to\infty$ projects out the positive energy states
and we are left with
$I_P=\sum_{i,\bj}\langle \Scr{C}_{(-1)^FP}|i\rangle
g^{i\bj}\langle \bj |\Scr{C}_P\rangle$,
or
\beq
I_P=\Pi^P_ig^{i\bj}\widetilde{\Pi}_{\bj}^P,
\label{biIP}
\eeq
where $g^{i\bj}$ is the inverse of $g_{\bi j}=\langle \bi |j\rangle$.
Similarly, for the twisted Witten index for the $a$--$Pa$ open string
we have
\beq
I_P(a,Pa)=\Pi^a_ig^{i\bj}\widetilde{\Pi}_{\bj}^P.
\label{biIPa}
\eeq
The following must also hold
\beq
I_P(Pa,a)=\Pi^P_ig^{i\bj}\widetilde{\Pi}_{\bj}^{a}.
\label{biIPa2}
\eeq
These generalize the more standard expression for
the open string Witten index
\beq
I(a,b)=\Pi_i^ag^{i\bj}\widetilde{\Pi}_{\bj}^b.
\label{biIab}
\eeq
These `bilinear identities' are summarized in Fig.~\ref{bili}.

\subsubsection{Other identities}

\begin{figure}[htb]
\centerline{\includegraphics{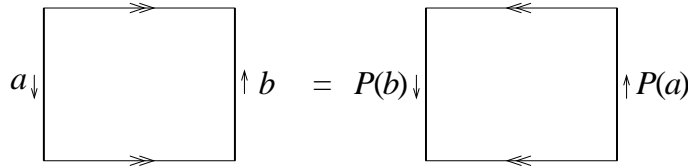}}
\caption{Applying parity to the cylinder}
\label{id1}
\end{figure}

Applying the parity symmetry to partition and correlation functions,
one can derive several identities of different type.
The first identity is obtained by applying the parity symmetry to the
path-integral on the cylinder. As is evident from Fig.~\ref{id1},
we find the relation between the open string Witten indices
\beq
I(a,b)=I(Pb,Pa).
\label{id1eq}
\eeq
By applying the parity to semi-infinite cigars,
we obtain
\beq
\Pi^a_i=\widetilde{\Pi}^{Pa}_{P(i)},
\qquad
\widetilde{\Pi}^a_{\bi}=\Pi^{Pa}_{P(\bi)},
\label{id2eq}
\eeq
where $P(i)$ is the label for the ground state obtained by applying
$P$ to the ground state labelled by $i$.
(Similarly for $P(\bj)$.)
Note that (\ref{id1eq}) also follows from these relations applied to
the bilinear identity (\ref{biIab}), with the help of unitarity of the
parity operator $P$.
Application of parity to the semi-infinite $\RP^2$ yields
\beq
\Pi^P_{i}=\widetilde{\Pi}^P_{P(i)},
\qquad
\widetilde{\Pi}^P_{P(\bi)}=\Pi^P_{P(\bi)}.
\label{id3eq}
\eeq
It follows from (\ref{id2eq}) and (\ref{id3eq}) and from the unitarity
of $P$ that
\beq
\Pi^a_ig^{i\bj}\widetilde{\Pi}^P_{\bj}
=\Pi^P_ig^{i\bj}\widetilde{\Pi}^{Pa}_{\bj}.
\label{rell}
\eeq
This relation ensures the consistency of the two bilinear identities
for boundary and crosscap states,
(\ref{biIPa}) and
(\ref{biIPa2}).

\section{Geometric Picture}\label{sec:NLSM}

In this section, the general theory of parity
symmetry and crosscap states developed in the previous section
is applied to and illustrated by
the examples of non-linear sigma models on
K\"ahler manifolds.
The classical action of the sigma model on a K\"ahler manifold
$(X,g)$ is
$$
L=\int K(\Phi,\bPhi)\,\dd^4\theta.
$$
$K(z,\bz)$ is a K\"ahler potential in a coordinate patch
on which the metric is expressed as $g_{i\bj}=\partial_i\partial_{\bj}K$.
We start by studying the parity invariance of this classical action.

\subsection{Antiholomorphic and holomorphic involutions}

We recall the basic parity actions on a chiral superfield,
(\ref{Aacchi}), (\ref{Bacchi}),
\beqa
&&A:\, \Phi\longrightarrow \overline{\Omega_A^*\Phi},
\nn
\\
&&B:\,\Phi\longrightarrow \Omega_B^*\Phi,
\nn
\eeqa
which are A- and B-parities respectively.
Similarly, for a holomorphic coordinate transformation
$f^i(z)=f^i(z^1,...,z^n)$ ($i=1,...,n$),
the transformations of the form
$\Phi^i\to \overline{f^i(\Omega_A^*\Phi)}$
and $\Phi^i\to f^i(\Omega_B^*\Phi)$ are A- and a B-parities.
We would like to find such actions that leave the Lagrangian invariant.
Since the measure $\dd^4\theta$ is invariant under both
$\Omega_A$ and $\Omega_B$, what we need to find is a transformation
that leaves the K\"ahler potential invariant, up to a K\"ahler transformation
$K(z,\bz)\to K(z,\bz)+g(z)+\overline{g(z)}$.

Let $f:X\to X$ be a holomorphic and isometric diffeomorphism.
Using complex coordinates, it can be represented as
$f:z^i\to f^i(z)$ where $f^i(z)$ are
holomorphic functions of $z=(z^1,...,z^n)$
obeying
$K(f(z),\overline{f(z)})=K(z,\bz)$, up to a K\"ahler transformation.
Thus, the sigma model action is invariant under
a B-parity 
$$\Phi^i\to f^i(\Omega_B^*\Phi).$$
Let $f:X\to X$ be an antiholomorphic and isometric diffeomorphism.
It can be represented as
$f:z^i\to \overline{h^{\bi}(z)}$ where $h^{\bi}(z)$
are holomorphic functions of $z$ obeying
$K(\overline{h(z)},h(z))=K(z,\bz)$,
up to a K\"ahler transformation.
Thus, the sigma model action is invariant under
an A-parity
$$\Phi^i\to \overline{h^{\bi}(\Omega_A^*\Phi)}.$$
Thus, classically {\it the sigma model has an A-parity symmetry
for each antiholomorphic
isometry and a B-parity symmetry for each holomorphic isometry.}
Note that both of the above can be regarded as the action
$f\circ \Omega$ on the component fields, $\phi^i,\psi_{\pm}^i$.

One can state (anti-)holomorphicity of a map in terms of the K\"ahler form
$\omega={i\over 2}g_{i\bj}\dd z^i\wedge \dd \bz^{\bj}$.
An isometry $f:X\to X$ is holomorphic if and only if
it preserves the K\"ahler form,
$$
f^*\omega=\omega.
$$
It is antiholomorphic if and only if it
reverses the K\"ahler form, 
$$
f^*\omega=-\omega.
$$
Thus,
 holomorphic and antiholomorphic isometries can be regarded respectively
as symplectic and anti-symplectic maps with respect to the K\"ahler form
$\omega$.

\subsection{Parity anomaly}\label{paano}

So far our considerations have been exclusively in the classical system.
In the quantum theory, we always have to check the potential anomaly.
It turns out that the B-parity is always anomaly-free
but the A-parity is potentially in danger.
Let $\phi:\Sigma\to X$ be a map of the worldsheet into the target space.
%
For this bosonic background the path-integral measure of the fermions
$\Psi=(\psi_{\pm},\bpsi_{\pm})$ changes under A-parity
(that acts on the bosonic part as $\phi\to \phi'=f\circ \phi\circ\Omega$
where $\Omega$ is the worldsheet orientation reversal) as
$$
{\mathcal D}_{\phi}\Psi
\longrightarrow
(-1)^{\int_{\Sigma}\phi^*c_1(X)}
{\mathcal D}_{\phi'}\Psi.
$$
This is seen by looking at the action on
the fermion zero modes.
The detail will be discussed in Section~\ref{subsub:panomaly}
where the parity anomaly of supersymmetric
gauged WZW models is considered (in which `A' and `B' are exchanged).
The sign $(-1)^{\int_{\Sigma}\phi^*c_1(X)}$ is the anomaly.
It is always trivial when $X$ is spin so that $c_1(X)$ is even.
(Recall that the second Stiefel-Whitney class $w_2(X)$, 
the obstruction against spin structures, is the mod 2 reduction of
$c_1(X)$.)
In particular, the A-parity is anomaly-free if $X$ is Calabi--Yau.

The anomaly can be cancelled by
combining the parity action with $(-1)^{F_R}$
which flips the sign of $\psi_-$ and $\bpsi_-$.
This is the option that will be used in the gauged WZW model.
There is an alternative way using $B$-fields which we discuss now.

\subsubsection{Anomaly cancellation by $B$-field}

The $B$-field term of the sigma model $\int_{\Sigma}\phi^*B$
flips by sign
under the worldsheet orientation reversal $\Omega$.
The only way to make the term invariant is to combine it with
a diffeomorphism $f:X\to X$ such that $f^*[B]=-[B]$, where
$[B]\in H^2(X,\R)$ is the cohomology class represented by $B$.
However, since the $B$-field enters into 
the path-integral weight in the form
$\e^{i\int_{\Sigma}\phi^*B}$, one may have a shift of
$[B]$ by $2\pi$ times an integral class.
Thus the condition of invariance is
$f^*[B]=-[B]$ mod $2\pi H^2(X,\Z)$.
This is the whole story for a B-parity which is always anomaly-free.
In the case of an A-parity, one may use this freedom of choosing
the $B$-field to cancel the anomaly $(-1)^{\int_{\Sigma}\phi^*c_1(X)}$.
This works out if the $B$-field is chosen such that
$f^*[B]=-[B]+\pi c_1(X)$ mod $2\pi H^2(X,\Z)$.
Recalling the action on the K\"ahler form of holomorphic and
antiholomorphic isometries,
we find the following:
{\it An A-parity is a symmetry if it acts on the complexified K\"ahler
form as}
$$
A:\quad
f^*[\omega-iB]=-[\omega-iB]+\pi i c_1(X)\quad\mbox{mod $2\pi iH^2(X,\Z)$},
$$
{\it whereas a B-parity is a symmetry if it acts
as the complex conjugation}
$$
B:\quad
f^*[\omega-iB]=\overline{[\omega-iB]}\quad\mbox{mod $2\pi iH^2(X,\Z)$}.
$$
Let us choose an integral basis $\{\omega_a\}$ of $H^2(X,\R)$
and express the complexified K\"ahler class as
$[\omega-iB]=\sum_{a}\omega_at^a$.
If the action of $f$ on $H^2(X)$ is
$f^*\omega_a=\sum_bf_a^{\,b}\omega_b$,
the condition of unbroken symmetry is written as
\beqa
&&A:\quad t^bf_b^{\,a}=-t^a+\pi i c_1(X)^a+2\pi i n^a,\quad
n^a\in\Z,
\label{Asymc}
\\
&&B:\quad t^bf_b^{\,a}=\overline{t^a}+2\pi i m^a,\quad
m^a\in\Z,
\label{Bsymc}
\eeqa
where $c_1(X)=\sum_a\omega_ac_1(X)^a$.
We find holomorphic constraints on $t^a$ for A-parities but
antiholomorphic constraints for B-parities.
For B-parities, we loose a half of the complexified
K\"ahler moduli.

\noindent
{\it Remark}.
If $\dim H^2(X)=1$ (as is the case for $\CP^n$, Grassmannian, and
submanifolds therein of dimensions $>2$),
an antiholomorphic map $f$ acts on $H^2(X)$ by a sign flip. Then,
the equation (\ref{Asymc}) is satisfied if and only
if $c_1(X)$ is even, i.e. $X$ is spin. Thus, for a non-spin manifold $X$
with $b^2(X)=1$
(such as $\CP^{\it even}$), the A-parity anomaly cannot be canceled by
a $B$-field.
However, if $\dim H^2(X)>1$, there are cases in which 
this works.
For example consider $X=\CP^{2m}\times \CP^{2m}$ and
the map $(z_1,z_2)\mapsto (\overline{z_2},\overline{z_1})$, where
$z\mapsto \bz$ is an antiholomorphic map of $\CP^{2m}$.
Since $c_1(\CP^{2m})$ is $(2m+1)$ times the integral generator,
the equation (\ref{Asymc}) reads 
$$
\left(\begin{array}{c}
-t^2\\
-t^1
\end{array}
\right)
=-
\left(\begin{array}{c}
t^1\\
t^2
\end{array}
\right)
+\pi i \left(\begin{array}{c}
2m+1\\
2m+1
\end{array}
\right)
+2\pi i \left(\begin{array}{c}
n^1\\
n^2
\end{array}
\right).
$$
This has a solution $(t^1,t^2)=(t,t-\pi i)$.

\subsection{Witten index}

We compute the parity-twisted Witten indices
in the non-linear sigma model. Throughout this subsection we consider
involutive parities.
We start with the case without $B$-field.

\subsubsection{General formula}

We first present the index formula that applies
to a general Riemannian manifold $X$ for which the sigma model has
${\mathcal N}=(1,1)$ supersymmetry.
We consider a parity of the form
$P=\tau\circ \Omega$ where $\tau:X\to X$ is an involution of $X$,
which acts on the fields as
$\phi^I(x)\to \tau^I(\phi(\wtx))$,
$\psi_{\pm}^I(x)\to \tau_{*J}^I \psi_{\mp}^J(\wtx)$, in the notation using
real coordinates of $X$.
This preserves the diagonal ${\mathcal N}=1$ supersymmetry.

The closed string twisted Witten index is the partition function
on the Klein bottle $(x_1,x_2)\equiv (x_1+L_1,x_2)\equiv
(-x_1,x_2+L_2)$ with  periodic boundary conditions along $x_1$,
$x_1\to x_1+L_1$, but with the twisted boundary condition along $x_2$:
\beqa
&&\phi^I(x_1,x_2)=\tau^I(\phi(-x_1,x_2+L_2)),
\nn\\
&&
\psi_{\pm}^I(x_1,x_2)=\tau_{*J}^I \psi_{\mp}^J(-x_1,x_2+L_2).
\nn
\eeqa
Because of the ${\mathcal N}=1$ supersymmetry, the computation
localizes on the zero modes, which are the constant maps to
the submanifold
$X^{\tau}\subset X$ of $\tau$-fixed points.
The relevant computation is performed in \cite{SS} (following \cite{AG,FWin})
and the result is
\beq
I_{\tau\Omega}
=\int_{X^{\tau}}{L(T(X^{\tau}))\over
L(N(X^{\tau}))}e(N(X^{\tau})),
\label{ItO}
\eeq
where $T(X^{\tau})$ and $N(X^{\tau})$ are
the tangent and normal bundles of $X^{\tau}$ (in $X$),
which has an orientation determined by the type of parity
action.
 $L(V)$ and $e(V)$ are the Hirzebruch L-genus and the Euler class.
An outline of the derivation is recorded in Appendix~\ref{app:index}.
Note that we allow $X^{\tau}$ to have many components ---
the above formula is understood as the sum of the integrals over all
the components of $X^{\tau}$.
The same remark applies to the formulae below.

We next consider an open string with one end on a D-brane wrapped on
$W\subset X$ and supporting a vector bundle $E$ and the other end
on the parity image $(\tau W,\tau E)$.
The image bundle $\tau E$ is topologically
$\tau^*\overline{E}$, where the
complex conjugation is involved
because the left and the right boundaries of the string worldsheet
have opposite orientation.
The ${\mathcal N}=1$ supersymmetry
survives the D-brane boundary condition
as well as $\tau\Omega$, and one can consider the $\tau\Omega$-twisted
Witten index.
It is represented as the partition function on the
M\"obius strip $(x_1,x_2)\equiv (L_1-x_1,x_2+L_2)$, $0\leq x_1\leq L_1$,
with the standard Dirichlet/Neumann boundary conditions on
$x_1=0$ and $x_1=L_1$, and the periodicity
along $x_2$ as above (where $-x_1$ there replaced by $L_1-x_1$).
This also localizes on the zero modes, which are constant maps to
$W\cap X^{\tau}$.
By a computation similar to \cite{SS} we find the following expression
\beqa
\lefteqn{I_{\tau\Omega}((W,E),(\tau W,\tau E))}
\nn\\&&
=\int_{W\cap X^{\tau}}2^{\dim_r\!X^{\tau}-{1\over 2}\dim_r\!X}
{\rm ch}(\overline{E})
\sqrt{\widehat{A}(T(W))\over\widehat{A}(N(W))}
\sqrt{L({1\over 4}T(X^{\tau}))\over
L({1\over 4}N(X^{\tau}))}e(N(W)\cap N(X^{\tau})),
\nn\\
\label{ItOE}
\eeqa
where $\dim_r X^{\tau}$ and $\dim_r X$ are real dimensions of
$X^{\tau}$ and $X$.
See Appendix~\ref{app:index} for the derivation.
$\widehat{A}(V)$ is the A-roof genus.
We note that the formula could be changed by an  overall sign,
$I\to -I$,
depending on the type of the parity action.

Let us now specialize to the parity symmetries
that preserve an ${\mathcal N}=2$ supersymmetry.
$X$ is thus assumed to be a K\"ahler manifold.

\subsubsection{A-parity}

\newcommand{\rank}{{\rm rank}}

Let us consider an A-parity $\tau\Omega$ that is associated with
an antiholomorphic and isometric involution $\tau:X\to X$.
In such a case, the fixed-point set $X^{\tau}$
is a middle dimensional Lagrangian submanifold of $X$.
Then, $\rank N(X^{\tau})=\rank T(X^{\tau})$ and the Euler class
$e(N(X^{\tau}))$ in the index formula (\ref{ItO})
saturates the dimension of $X^{\tau}$.
The index is thus the integral of just the Euler class of the normal bundle.
The Euler class is the obstruction against trivialization and
counts the number of zeroes of a generic section of the bundle.
On the other hand, a section of the normal bundle
simply corresponds to a deformation of $X^{\tau}$ inside
$X$, and its zero corresponds to the intersection of
$X^{\tau}$ and its deformation. Thus, the index is nothing but
the self-intersection number
\beq
I_{\tau\Omega}=\int_{X^{\tau}}e(N(X^{\tau}))
=\#(X^{\tau}\cap X^{\tau}).
\label{indpapa}
\eeq
Let us next consider an A-brane wrapped on a Lagrangian submanifold
$L\subset X$.
Since $L$ and $X^{\tau}$ are both middle dimensional, we have
$\rank(N(L)\cap N(X^{\tau}))=\dim(L\cap X^{\tau})$.
The Euler class in the formula (\ref{ItOE}) again saturates, and we find
\beq
I_{\tau \Omega}(L,\tau L)=\int_{L\cap X^{\tau}}
e(N(L)\cap N(X^{\tau}))=\#(L\cap X^{\tau}).
\label{indbapa}
\eeq

\subsubsection{B-parity}

We next consider the B-parity associated with a holomorphic involution
$\tau:X\to X$. The general formula from the path-integral
is compared with the consideration from the canonical formalism.
This reproduces the various signature and fixed-point theorems.

\subsubsection*{\it Closed string}

The supersymmetric ground states of the sigma model
are in one to one correspondence with the Harmonic forms or de Rham
cohomology classes
where the correspondence is given by
$\psi_-^i\sim\dd z^i$,
$\bpsi_+^{\bj}\sim \dd \bz^{\bj}$,
$\bpsi_-^{\bj}\sim g^{i\bj}i_{\partial/\partial z^i}$
and $\psi_+^i\sim g^{i\bj}i_{\partial/\partial \bz^{\bj}}$.
The parity action $\Omega:\psi_-^i\leftrightarrow \psi_+^i,
\bpsi_-^{\bj}\leftrightarrow \bpsi_+^{\bj}$ therefore corresponds to
the Hodge $*$-operator that sends $H^{p,q}(X)$ to
$H^{n-q,n-p}(X)$. Thus, the twisted Witten index, which receives
contribution only from the ground states is identified as the
{\it signature} of $X$;
\beq
I_{\Omega}=\sum_{p+q=n}(-1)^n
\tr\!\!\mathop{}_{H^{p,q}(X)}\!\!(*)
={\rm Sign}(X).
\label{signa}
\eeq
Since the fixed-point set $X^{\tau}$ is $X$ itself the formula (\ref{ItO})
tells
\beq
{\rm Sign}(X)=\int_XL(T(X)),
\eeq
which is nothing but the Hirzebruch signature formula.
If $\tau$ is a non-trivial (holomorphic) involution,
the twisted Witten index is identified as
the $\Z_2$-signature,
\beq
I_{\tau\Omega}=\sum_{p+q=n}(-1)^n
\tr\!\!\mathop{}_{H^{p,q}(X)}\!\!(\tau^*\circ *)
={\rm Sign}(\tau,X).
\label{signat}
\eeq
The formula (\ref{ItO}) is nothing but the
$G$-signature formula for the case of $G=\Z_2$.

\subsubsection*{\it Open string}

Let us next consider the twisted Witten index for open string
stretched between B-branes.
Here we restrict our attention to B-branes wrapped totally on
the target space, $W=X$, and supporting a holomorphic vector bundle
$E$ over $X$. We note the standard subtlety in the
normal coordinate expansion in the evaluation of the index, and here we take
the one natural for the holomorphic category, resulting in
a replacement of the A-roof
genus in the formula by the Todd class.
Then the index formula is
\beqa
I_{\tau\Omega}(E,\tau^*\overline{E})&=&
\int_{X^{\tau}}2^{\dim_{r}\!X^{\tau}-{1\over 2}\dim_r\!X}
{\rm ch}(\overline{E})\sqrt{{\rm td}(X)}
\sqrt{L({1\over 4}T(X^{\tau}))\over
L({1\over 4}N(X^{\tau}))}
\nn\\
&=&\int_{X^{\tau}}{\rm ch}(2\overline{E})
{{\rm td}(X^{\tau})\over
{\rm ch}(\bigwedge\overline{N\,\,}_{\!\!\!\!X^{\tau}})}.
\label{indfo}
\eeqa
The second equality is an algebraic identity, where
$\overline{N\,\,}_{\!\!\!\!X^{\tau}}$ is the
antiholomorphic part of the complexified normal bundle
(see Appendix~\ref{app:index} for a proof).

Let us now study the index in the canonical formalism.
In the zero mode approximation (which gives the exact answer for the index),
the open string states are antiholomorphic forms on $X$
with values in $\overline{E}\otimes \tau^* \overline{E}$, where the
identification is based on $\bpsi_-^{\bi}+\bpsi_+^{\bi}\sim \dd \bz^{\bi}$
and $\psi_-^i+\psi_+^i\sim g^{i\bj}i_{\partial/\partial \bz^{\bj}}$.
The supercharges are identified as the Dolbeault operator $\bartial$
and the supersymmetric ground states are the Dolbeault cohomology classes
$$
{\mathcal H}_{\rm SUSY}^{\rm zero\, mode}=\bigoplus_{p=1}^n
H^{0,p}(X,\overline{E}\otimes \tau^*\overline{E}).
$$
The parity $\tau\Omega$ acts naturally on antiholomorphic forms
as $\dd \bz^{\bi}\cdots\to \tau^*(\dd \bz^{\bi}\cdots)$ since
$\Omega$ simply exchanges $\bpsi_-^{\bi}\leftrightarrow \bpsi_+^{\bi}$,
leaving $\bpsi_-^{\bi}+\bpsi_+^{\bi}$ fixed.
The action $\tau_{{}_{\rm CP}}$ of the parity on the Chan-Paton factor
$\overline{E}\otimes \tau^*\overline{E}$ can be of various types
\cite{PS,GimPol,DouMoo},
although it is basically the exchange of the left and the right factors.
$\tau^*$ combined with such an
action $\tau_{{}_{\rm CP}}$ on the Chan-Paton bundle defines a map $\tau$ of
$H^{0,p}(X,\overline{E}\otimes \tau^*\overline{E})$ into itself.
This is the action of parity on the ground states
in the zero mode approximation.
Thus, the index is identified as
\beq
I_{\tau\Omega}(E,\tau^*\overline{E})
=\sum_{p=1}^n(-1)^p\,
\tr\!\!\mathop{}_{H^{0,p}(X,\overline{E}\otimes \tau^*\overline{E})}\!\!
(\,\tau\,)
=:L(\tau,\Scr{E}^{\vee}\otimes \tau^*\Scr{E}^{\vee}).
\label{indfoC}
\eeq
This number is known as the {\it holomorphic 
Lefschetz number}.

We obtained two representations of the Witten index,
one (\ref{indfo}) from the path-integral and
another (\ref{indfoC}) from the canonical formalism.
The two must agree.
Here we quote the
Lefschetz fixed-point theorem \cite{ASegal} which expresses
the holomorphic Lefschetz number by topological data.
Let $g:V\to V$ be a holomorphic bundle isomorphism covering a
holomorphic automorphism $g:X\to X$. The theorem states
\beq
L(g,\Scr{V}):=\sum_{p=0}^n(-1)^p
\tr\!\!\mathop{}_{H^{0,p}(X,V)}\!\!
(\,g\,)
=\int_{X^{g}}{\rm ch}_{g}(V|_{X^{g}})
{{\rm td}(X^{g})\over
{\rm ch}_{g}(\wedge_{-1}\overline{N\,\,}_{\!\!\!\!X^{g}})}.
\label{fpt}
\eeq
Here ${\rm ch}_{g}$ is the $g$-twisted Chern character
and $\wedge_{-1}\overline{N\,\,}_{\!\!\!\!X^{g}}
:=\wedge^{\rm even}\overline{N\,\,}_{\!\!\!\!X^{g}}
-\wedge^{\rm odd}\overline{N\,\,}_{\!\!\!\!X^{g}}$.
In the present case, since $\tau$ is involutive
and hence is just a $(-1)$ on the normal bundle,
we find ${\rm ch}_{\tau}(\wedge_{-1}\overline{N\,\,}_{\!\!\!\!X^{\tau}})
={\rm ch}(\wedge\overline{N\,\,}_{\!\!\!\!X^{\tau}})$.
Now, with this fixed-point theorem (\ref{fpt}),
the canonical formula (\ref{indfoC}) 
looks very much close
to the path-integral formula (\ref{indfo}).
Indeed, they agree 
as we now see in several
cases (up to an important sign difference in certain cases).
In other words, the path-integral result
reproduces the fixed-point theorem.
\\
{\bf Case I:} $E=\Scr{O}$, $\tau_{{}_{\rm CP}}=$ simple exchange.\\
In this case the action on the Chan-Paton factor
$\overline{\Scr{O}}\otimes \tau^*\overline{\Scr{O}}\cong \Scr{O}$
is trivial. Thus, the twisted index can be identified as
\beq
I_{\tau\Omega}(\Scr{O},\Scr{O})
=\sum_{p=1}^n(-1)^p\,\tr\!\!\mathop{}_{H^{0,p}(X)}\!\!(\tau^*)
=L(\tau,\Scr{O}),
\eeq
the original holomorphic Lefschetz number of the map $\tau$.
It is evident that the two index formulae
agree.
\\
{\bf Case II:} $E$ general,
$\tau_{{}_{\rm CP}}=$ simple exchange.\\
On the fixed-point locus $X^{\tau}$, the Chan-Paton factor is
$\overline{E}\otimes \overline{E}$.
$\tau_{{}_{\rm CP}}$ acts trivially
 on the symmetric part ${\rm Sym}^2\overline{E}$ but 
as $(-1)$-multiplication on the anti-symmetric part $\wedge^2\overline{E}$.
Thus, we find
$$
{\rm ch}_{\tau}(\overline{E}\otimes \tau^*\overline{E}|_{X^{\tau}})
={\rm ch}({\rm Sym}^2\overline{E})
-{\rm ch}(\wedge^2\overline{E})
={\rm ch}(2\overline{E}).
$$
The last equality is a simple algebraic identity.
Thus, the two index formulae
agree in these cases as well.\\
{\bf Case III:} $E=F\oplus F$, $\tau_{{}_{\rm CP}}=$ exchange with
symplectic action.\\
One may also consider combining the exchange with
an internal action $\gamma_{ij}$,
$$
|i,j\rangle\mapsto\sum_{i',j'}\gamma_{ii'}|j',i'\rangle\gamma^{-1}_{j'j}.
$$
Here we consider the case where $E=F\oplus F$,
with $F$ a rank $r$ bundle, and $\gamma_{ij}$ is of the form
$$
\Bigl(\gamma_{ij}\Bigr)=\left(
\begin{array}{cc}
{\bf 0}&\!\!-{\bf 1}_r\\
\,\,\!{\bf 1}_r\!\!&{\bf 0}
\end{array}\right)
$$
We focus on the fixed-point locus $X^{\tau}$ on which the Chan-Paton bundle
is $\overline{E}\otimes \overline{E}$.
The Chan-Paton factors of the forms
$|i_{(1)},j_{(1)}\rangle\pm |j_{(2)},i_{(2)}\rangle$,
$|i_{(1)},j_{(2)}\rangle\mp |j_{(1)},i_{(2)}\rangle$,
$|i_{(2)},j_{(1)}\rangle\mp |j_{(2)},i_{(1)}\rangle$
have eigenvalue $\pm 1$ under $\tau_{{}_{\rm CP}}$, where $i_{(1)}$
and $i_{(2)}$ are the index for the first and the second factor of
$\overline{E}=\overline{F}\oplus \overline{F}$.
These vectors are the basis of bundles isomorphic to
$\overline{F}\otimes\overline{F}$,
$\overline{F}\otimes\overline{F}$,
$\wedge^2\overline{F}$,
${\rm Sym}^2\overline{F}$,
$\wedge^2\overline{F}$,
${\rm Sym}^2\overline{F}$, respectively.
On the other hand, it is easy to see that
\beqa
{\rm Sym}^2\overline{E}&\cong&(\overline{F}\otimes\overline{F})\oplus
({\rm Sym}^2\overline{F})\oplus({\rm Sym}^2\overline{F}),
\nn\\
\wedge^2\overline{E}&\cong&(\overline{F}\otimes\overline{F})\oplus
(\wedge^2\overline{F})\oplus(\wedge^2\overline{F}).
\nn
\eeqa
Thus, we find
$$
{\rm ch}_{\tau}(\overline{E}\otimes \tau^*\overline{E}|_{X^{\tau}})
={\rm ch}(\wedge^2\overline{E})
-{\rm ch}({\rm Sym}^2\overline{E})
=-{\rm ch}(2\overline{E}).
$$
The two index formulae differ  by just an overall sign.
We see that we must have minus sign $-1$ in the path-integral formula
(\ref{indfo}) in this case.
This corresponds to the sign flip of the crosscap state,
which is the standard feature for $Sp$-type orientifolds.

\subsubsection{Inclusion of the $B$-field}\label{subsubsec:Bfield}

Let us now turn on a $B$-field.
As we have seen, we are allowed to have discrete values of $B$-field.
In some cases for A-parity,
a non-zero $B$-field is enforced to cancel parity anomaly.
See the general conditions of parity symmetry (\ref{Asymc}) and (\ref{Bsymc}).

The inclusion of a $B$-field does not affect the closed string Witten index.
This is because the index computation localizes on constant maps,
for which the $B$-field has no effect.
Thus, the formulae (\ref{ItO}), (\ref{indpapa})
and (\ref{signa}) or (\ref{signat}) remain valid.

The $B$-field does affect the open string indices.
The effect is to shift the first Chern class
 of the gauge bundle
on the brane by $-B/2\pi$. Thus, the Chan-Paton bundle is effectively
replaced as $E\to E\otimes {\mathcal L}^{-1}_B$,
where `${\mathcal L}_B$' is the `line bundle
whose first Chern-class is $B/2\pi$'.
This affects the parity transformation of the Chan-Paton bundle,
which would be
$$
\tau\Omega: (W,E)\longrightarrow (\tau W, \tau^* \overline{E})
$$
if $B$ were zero.
For simplicity let us consider how $\Omega={\rm id}\circ\Omega$ acts on
the open string stretched from
the brane $(X,E)$ to the brane $(X,F)$.
The effective
Chan-Paton factor is $E\otimes {\mathcal L}_B^{-1}$ backward in time
on the left boundary and $F\otimes {\mathcal L}_B^{-1}$ forward in time
on the right boundary.
After the parity action which swaps the left and the right boundary,
the Chan-Paton factor is $F\otimes {\mathcal L}_B^{-1}$ forward on the left
and $E\otimes {\mathcal L}_B^{-1}$ backward on the right, or equivalently
$\overline{F\otimes {\mathcal L}_B^{-1}}$ backward on the left
and $\overline{E\otimes {\mathcal L}_B^{-1}}$ forward on the right.
Since
$$
\overline{E\otimes {\mathcal L}_B^{-1}}
\cong \overline{E}\otimes {\mathcal L}_B
=(\overline{E}\otimes {\mathcal L}_B^{\otimes 2})\otimes {\mathcal L}_B^{-1},
$$
we see that the parity $\Omega$ transforms the bundle
$E$ to the bundle $\overline{E}\otimes {\mathcal L}_B^{\otimes 2}$.
More generally the transformation rule becomes
\beq
\tau\Omega: (W,E)\longrightarrow
(\tau W, \tau^* (\overline{E}\otimes {\mathcal L}_B)\otimes {\mathcal L}_B)
\eeq
It differs from $\tau^*\overline{E}$
by the factor $\tau^*{\mathcal L}_B\otimes {\mathcal L}_B$
whose `first Chern class' is
\beq
c_1(\tau^*{\mathcal L}_B\otimes {\mathcal L}_B)
=\left.{1\over 2\pi}(\tau^*[B]+[B])\right|_{\tau W}.
\label{c1fo}
\eeq
The parity symmetry condition (\ref{Asymc}) and (\ref{Bsymc})
includes
$$
{1\over 2\pi}(\tau^*[B]+[B])=\left\{
\begin{array}{ll}
{1\over 2}c_1(M)~\mbox{mod $H^2(X,\Z)$}&\mbox{for A-parity}
\\
0~\mbox{mod $H^2(X,\Z)$}&\mbox{for B-parity}
\end{array}
\right.
$$
Thus the class (\ref{c1fo}) is evidently
integral, except in the cases of A-parity
on non-spin manifolds. In the latter case, however,
if we assume $W$ to be an A-brane wrapped on Lagrangian submanifold
supporting a flat bundle, the 
$B$-field has to be zero on $W$ \cite{HIV}.\footnote{There are
non-Lagrangian A-branes with non-flat connection or non-zero $B$-field
\cite{OOY,KapuOr}.
We do not include this in our paper. It would be interesting to see
the consistency of parity with such A-branes in
of non-spin manifolds.}
Since $\tau$ is anti-symplectic, $\tau W$ is also Lagrangian and
$B=\tau^*B=0$ on $\tau W$.
Thus, the class (\ref{c1fo}) is zero and hence integral also in this case.
In any case,
$\tau^*{\mathcal L}_B\otimes {\mathcal L}_B$
is a well-defined complex line bundle
${\mathcal L}_{\tau^*B+B}$ on $\tau W$.
Then, the Chan-Paton factor for the stretched open string
is the bundle
${\mathcal L}_{\tau^*B+B}\otimes\overline{E}\otimes \tau^*\overline{E}$.
To specify the system, {\it one needs to specify a hermitian connection
and a parity action on ${\mathcal L}_{\tau^*B+B}$ as well.}
In the discussion below, we suppose that a choice has been made.

Now it is straightforward to generalize the
index formulae obtained above to the present situation.
The computation again localizes on the fixed-point set $X^{\tau}$.
On this set, the parity action on the bundle
${\mathcal L}_{\tau^*B+B}={\mathcal L}_{2B}$ is just a bundle map
which is $+1$ or $-1$ at each connected  component of $X^{\tau}$.
Let
\beq
\varepsilon_B:X^{\tau}\to\{\pm 1\}
\label{signfcn}
\eeq
 be the locally constant function
determined by this sign.
Then, the general index formula is given by
\beqa
\lefteqn{I^B_{\tau\Omega}((W,E),(\tau W,\tau E))}
\nn\\&&
=\int_{W\cap X^{\tau}}2^{\dim_r\!X^{\tau}-{1\over 2}\dim_r\!X}
{\rm ch}(\overline{E})\varepsilon_B\e^{B/2\pi}
\sqrt{\widehat{A}(T(W))\over\widehat{A}(N(W))}
\sqrt{L({1\over 4}T(X^{\tau}))\over
L({1\over 4}N(X^{\tau}))}e(N(W)\cap N(X^{\tau})),
\nn
\eeqa
where $\tau E$ is now
$\tau^*\overline{E}\otimes {\mathcal L}_{\tau^*B+B}$.

Let us consider an A-parity associated with an antiholomorphic isometry
$\tau:X\to X$ and a Lagrangian A-brane $L$ with trivial $U(1)$
connection.
The only effect of the $B$-field is a possible non-trivial parity action
on the trivial bundle ${\mathcal L}_{\tau^*B+B}$.
This modifies the index formula by the sign $\varepsilon_B$.
Let $X^{\tau}=\sum_iX^{\tau}_i$ be the decomposition into
the connected components.
The $B$-field modifies it to
$X^{\tau}_B=\sum_i\varepsilon_B(i)X^{\tau}_i$ in the index formula
\beq
I_{\tau\Omega}^B(L,\tau L)
=\#(L\cap X_B^{\tau})
\eeq

Let us now consider a B-parity associated with a holomorphic isometry
$\tau:X\to X$ and a B-brane wrapped on $X$
and supporting a holomorphic vector bundle $E$.
The general index formula, with the replacement $\hat{A}\to{\rm td}$,
is given by
\beq
I^B_{\tau\Omega}(E,\tau E)
=\int_{X^{\tau}}{\rm ch}(2\overline{E})\varepsilon_B\e^{B/\pi}
{{\rm td}(X^{\tau})\over
{\rm ch}(\wedge\overline{N\,\,}_{\!\!\!\!X^{\tau}})}.
\eeq
One the other hand, the canonical formalism identifies the twisted index as
the holomorphic Lefschetz number
\beqa
I^B_{\tau\Omega}(E,\tau E)
&=&L(\tau, \Scr{L}_{\tau^*B+B}\otimes \Scr{E}^{\vee}\otimes
\tau^*\Scr{E}^{\vee})
\nn\\
&=&\int_{X^{\tau}}{\rm ch}_{\tau}({\mathcal L}_{\tau^*B+B}\otimes\overline{E}
\otimes\tau^*\overline{E}|_{X^{\tau}})
{{\rm td}(X^{\tau})\over
{\rm ch}(\wedge\overline{N\,\,}_{\!\!\!\!X^{\tau}})}.
\eeqa
Here $\Scr{L}_{\tau^*B+B}$ is the holomorphic line bundle associated with
the hermitian connection of ${\mathcal L}_{\tau^*B+B}$ which we assumed to
have specified.
By definition of $\varepsilon_B$, we have
$$
{\rm ch}_{\tau}({\mathcal L}_{\tau^*B+B}|_{X^{\tau}})
=\varepsilon_B{\rm ch}({\mathcal L}_{2B})
=\varepsilon_B\e^{B/\pi}
$$
Thus, the two index formula agree with each other
(except the overall sign for the
symplectic orientifolds).

\subsection{Overlaps with supersymmetric ground states}

We now compute the overlaps of the crosscap states and the
RR ground states. For A-parity, we consider Calabi--Yau sigma model
which is B-twistable.
For B-parity, we consider general sigma model which can always be A-twisted.
In both cases, we examine or use the bilinear identities
(\ref{biIP}) (\ref{biIPa}), (\ref{biIPa2}).

\subsubsection{A-parity in Calabi--Yau sigma model}

The overlap $\Pi^{\tau\Omega}_i$ for an A-parity in Calabi--Yau sigma model
can be computed exactly. The point is that they are independent of the twisted
chiral parameters. In particular, they are constant
along the moduli space of complexified K\"ahler class and the computation in
the large volume limit is exact.
In this limit, the theory reduces to the quantum mechanics, and the
overlaps are simply the integration of the ground state wavefunctions
$\omega_i$
over the orientifold plane,
\beq
\Pi^{\tau\Omega}_i=\widetilde{\Pi}^{\tau\Omega}_i
=\int_{X^{\tau}}\varepsilon_B\omega_i.
\eeq
$\varepsilon_B$ is
the sign function on $X^{\tau}$ associated
with the $B$-field.
The bilinear identity can be shown to hold with this factor.
We recall the overlaps for the A-branes
$$
\Pi^L_i=\widetilde{\Pi}^L_i=\int_L\omega_i.
$$
Then, the bilinear identity is nothing but Riemann's bilinear identity.
For example,
$$
\#(L\cap X^{\tau}_B)
=\left(\int_L\omega_i\,\right)\eta^{ij}
\left(\int_{X^{\tau}}\varepsilon_B\omega_j\right).
$$
Note that the overlap is non-vanishing only if $\omega_i\in H^n(X)$.
This is consistent with the selection rule ---
the crosscap state for an A-parity has vanishing axial R-charge
and thus have non-zero overlaps only with ground states of zero axial R-charge
(which are middle dimensional forms).
The most important of them is the overlap with the ground states of minimum
vector R-charge. This is simply the period integral over the
orientifold plane
\beq
\Pi^{\tau\Omega}_0=\int_{X^{\tau}_B}{\mit\Omega},
\eeq
where ${\mit\Omega}$ is the holomorphic volume form of
the Calabi--Yau manifold $X$.
This has an interpretation of the tension of
the orientifold plane in superstring theory.

\subsubsection{B-parity in general model}

The overlaps of the crosscap states for
a B-parity $\tau\Omega$ depends on the K\"ahler class parameter
and the exact result is hard to obtain.
However, an approximate formulae valid at large volume
is obtained by requiring the
bilinear identities and by the differential equation.
We recall that the overlaps of the B-brane $(X,E)$ and the RR ground states
are
\beqa
&&\Pi^E_i
=\int_X{\rm ch}(\overline{E})\e^{B+i\omega}\sqrt{{\rm td}(X)}\,\omega_i
+\cdots,
\nn\\
&&
\widetilde{\Pi}^E_i
=\int_X{\rm ch}(E)\e^{-B-i\omega}\sqrt{{\rm td}(X)}\,\omega_i
+\cdots,
\nn
\eeqa
where $\omega$ is the K\"ahler class and
$+\cdots$ are corrections that vanish in the 
large volume limit.
Requiring the bilinear identity, we find that the
overlap of the crosscap states and the RR ground states are
\beqa
&&\Pi^{\tau\Omega}_i
=2^{\dim_rX^{\tau}-{1\over 2}\dim_r X}
\int_{X^{\tau}}\varepsilon_B\e^{i\omega}\sqrt{L({1\over 4}T(X^{\tau}))
\over L({1\over 4}N(X^{\tau}))}\omega_i+\cdots,
\\
&&\widetilde{\Pi}^{\tau\Omega}_i
=2^{\dim_rX^{\tau}-{1\over 2}\dim_r X}
\int_{X^{\tau}}\varepsilon_B\e^{-i\omega}\sqrt{L({1\over 4}T(X^{\tau}))
\over L({1\over 4}N(X^{\tau}))}\omega_i+\cdots.
\eeqa

\subsection{D-Branes from parity}\label{dbrfp}

One can sometimes associate a D-brane to a parity symmetry.
For example, let us consider a bosonic
sigma model on the real line described by
a scalar field $X(x^0,x^1)$, and its parity symmetries
$P_{\pm}:X(x^0,x^1)\to \pm X(x^0,-x^1)$.
Consider a smooth field configuration invariant under $P_+$:
$X(x^0,-x^1)=X(x^0,x^1)$.
It can be regarded as the extension of a configuration on the
left half-plane $x^1\leq 0$ which obey  Neumann boundary conditions
at the boundary, $\partial_1X|_{x^1=0}=0$.
Likewise, a $P_-$ invariant configuration can be regarded
as the extension of a configuration on the left half-plane
obeying Dirichlet boundary conditions $X|_{x^1=0}=0$.
In other words, D1-brane is associated with the parity $P_+$
and D0-brane at $X=0$ is associated with $P_-$.
It should be noted, however, that
it is not always possible to associate a D-brane to a parity.
For instance, $P_+':X(x^0,x^1)\to X(x^0,-x^1)+\Delta X$ is also a parity
symmetry, but it is impossible to have a
$P_+'$-invariant configuration if $\Delta X\ne 0$.

This can be generalized to systems with $(2,2)$ supersymmetry.
Suppose a $(2,2)$ theory has an A-parity $P_A$.
Since it acts on the supercurrents as (\ref{actOAG}),
a field configuration invariant under $P_A$ obeys in particular
$$
G^1_+(x^0,x^1)+\bG^1_-(x^0,-x^1)=0.
$$
Such a configuration can be considered as an extension of a configuration on
the left half-plane obeying a boundary condition at $x^1=0$
such that
$$
(G^1_++\bG_-^1)|_{x^1=0}=0.
$$
The latter is the condition on D-branes to preserve an A-type supersymmetry.
Thus, {\it
a D-brane associated with an A-parity is an A-brane.}
Likewise, {\it
a D-brane associated with a B-parity is a B-brane.}

For example, let us consider a supersymmetric
sigma model on the complex plane $\C$,
described by a chiral superfield $\Phi$.
The transformation
$
\Phi\longrightarrow \overline{\Omega_A^*\Phi}
$
is an A-parity. The condition of invariance under this parity is
$$
\Phi=\overline{\Omega_A^*\Phi}.
$$
Such a configuration can be regarded as the smooth extension of the
configuration on the left (super) half-plane
which obeys the boundary condition
$$
~~~\Phi=\overline{\Phi} \qquad\mbox{at A-boundary}.
$$
This is nothing but the ${\mathcal N}=2_A$ preserving
boundary condition corresponding to
the D1-brane at the real line $\phi=\bphi$.

Similarly, the transformations
$
\Phi\longrightarrow \pm \Omega_B^*\Phi
$
are B-parities. Invariant configurations
$$
\Phi=\pm \Omega_B^*\Phi
$$
can be regarded as
the smooth extensions of the
configuration on the left (super) half-plane
which obey the boundary condition
$$
~~~
\left\{
\begin{array}{l}
D_+\Phi=D_-\Phi\\
\Phi=0
\end{array}
\right.
\qquad\mbox{at B-boundary}.
$$
They are nothing but the ${\mathcal N}=2_B$ preserving
boundary condition corresponding to
D2-brane filling $\C$ or D0-brane at $\phi=0$ 
respectively.

These examples generalize to any parity symmetry of the form
$\tau\Omega$ where $\tau:X\to X$ is an antiholomorphic or
holomorphic isometry of the target K\"ahler manifold $X$.
If $\tau$ has fixed points, the corresponding boundary
condition  is the one for the D-brane wrapped on the fixed-point set
$X^{\tau}$.
If $\tau$ has no fixed point, there is no invariant configurations
and therefore no associated boundary condition.

It is important that $\tau\Omega$ is
a symmetry of the theory.
If that is anomalous, the corresponding D-brane boundary condition
is expected to suffer from some pathology.
Let us consider the example $\tau:(z_1,z_2)
\mapsto (\bz_2,\bz_1)$ for $X=\CP^n\times \CP^n$,
where the two $\CP^n$
have the same size and no $B$-field $t_1=t_2=r$.
The fixed-point set is
$$
X^{\tau}=\Bigl\{\,(z_1,z_2)\in \CP^n\times \CP^n\,|\,z_2=\bz_1\,\Bigr\}
\cong \CP^n.
$$
If $n$ is odd, $\tau\Omega$ is a symmetry as long as $t_1=t_2$,
and the D-brane boundary condition for $X^{\tau}$ is expected to be
good.
If $n$ is even, however, we have seen that
$\tau\Omega$ is anomalous if $t_1=t_2$.
Then, the D-brane boundary condition for $X^{\tau}$ is expected to be
bad. In fact, the quantization of open string in such examples are
studied from the point of view of
symplectic geometry \cite{FOOO}, and it was found that
definition of (a finite dimensional model of)
the open string Hilbert space suffers from a problem
if $n$ is even.\footnote{Roughly speaking, the definition involves
a study of the moduli space of holomorphic discs, and the moduli space
is required to be orientable.
If $n$ is even the moduli space is unorientable.
We thank Y.-G. Oh for explanation of this point.}
However, if we turn on a $B$-field of period $\pi$ on one of the $\CP^n$ in
$X=\CP^n\times \CP^n$,
the parity anomaly is cancelled and the D-brane at $X^{\tau}$ will not suffer
from pathology. 
Pathology of $X^{\tau}$ for the case $n$ even, $t_1=t_2$, as well as
the remedy by taking $t_1=t_2+\pi i$
 will be explicitly seen in the mirror description
in Section~\ref{subsub:AsmBlg}.

\section{Orientifolds of ${\cal N}=2$ Minimal Models I}
\label{sec:Omin}

In this section, we study parity symmetries of the ${\mathcal N}=2$
minimal model. The discussion crucially depends on whether or not,
and how, GSO projection is imposed.
We introduce the model as a supersymmetric gauged WZW model where the GSO 
projection is (of course) not imposed.
We find the A and B parities of the system and compute the twisted Witten indices
for the closed string.
We then move on to the model with non-chiral GSO projection.
The procedure proposed by Pradisi, Sagnotti and Stanev (PSS)
\cite{PSS} (and established by later works
\cite{bantay,HSS,FSHSS,SaAn,BH1})
gives a prescription to construct the crosscap 
states of the GSO projected model,
and we will compute the Klein bottle amplitudes.
Then, we determine the crosscap states of the model before the GSO
projection, using the PSS procedure combined with
the supercurrent condition.
This enables us to compute the Witten indices
as well as the overlap of the crosscap states and supersymmetric ground states.

Sections \ref{sub:minimal} through \ref{sub:DOF}
 deal with the system before GSO projection, Section~\ref{sub:RCFT}
discusses the GSO projected model, and Section~\ref{sub:beGSO}
provides the relation between them.
In Section \ref{sec:OMMII}, we extend the study by including
D-branes and open strings.

\subsection{The minimal model}\label{sub:minimal}

The ${\cal N}=2$ minimal model is realized as
the $SU(2)$ WZW model and a Dirac fermion system,
which are coupled through a $U(1)$ gauge field.
Let $G=SU(2)$ and $H=U(1)\subset SO(3)$.
Let us consider a Dirac fermion
arranged into the hermitian $2\times 2$ matrix
$$
\Psi_{\pm}=\left(\begin{array}{cc}
0&\bpsi_{\pm}\\
\psi_{\pm}&0
\end{array}
\right).
$$
It preserves its form under 
the $H$-action $\Psi_{\pm}\to h^{-1}\Psi_{\pm} h$:
the component $\psi_{\pm}$ transforms as a charge $1$ field.
We consider the system with the action
$$
S(A,g,\Psi)=kS(A,g)+{i\over 2\pi}\int_{\Sigma}
\dd^2x\,\tr\left(
\Psi_-D_+\Psi_-+\Psi_+D_-\Psi_+\right),
$$
where $kS(A,g)$ is the WZW action \cite{Wnab} and
$D_{\mu}\Psi:=\partial_{\mu}\Psi+[A_{\mu},\Psi]$.
The action is invariant under the $H$-valued gauge transformation
$A\to h^{-1}Ah+h^{-1}\dd h$,
$g\to h^{-1}gh$, $\Psi_{\pm}\to h^{-1}\Psi_{\pm}h$.
Under the $H$-valued constant chiral gauge transformation
\beq
g\to h_1^{-1}gh_2,
\,\,\,
\Psi_-\to h_1^{-1}\Psi_-h_1,
\,\,\,
\Psi_+\to h_2^{-1}\Psi_+h_2,
\label{scgt}
\eeq
the path-integral measure changes by the factor \cite{PW}
$$
\exp\left[\,(k+2){i\over 2\pi}\int
\tr\Bigl(F_A\log(h_1h_2^{-1})\Bigr)\,\right].
$$
The origin of `$k$' in the exponent is  the change in the action
$kS(A,g)$
while `$+2$' comes from the chiral anomaly of the charged
fermion.
Supersymmetric gauged WZW models have been studied in
\cite{SchnitzerS,Nmatrix}
(see also \cite{GawKup,Schnitzer} for bosonic models).

\subsubsection{$(2,2)$ Superconformal symmetry}

The important property of the system is
$(2,2)$ supersymmetry.
Using
$$
\eta_{\pm}
={1\over\sqrt{k}}\psi_{\pm}\sigma_-
={1\over\sqrt{k}}\left(\begin{array}{cc}
0&0\\
\psi_{\pm}&0
\end{array}
\right),
\,\,\,\,\,\,\,
\bareta_{\pm}
={1\over\sqrt{k}}\bpsi_{\pm}\sigma_+
={1\over \sqrt{k}}\left(\begin{array}{cc}
0&\bpsi_{\pm}\\
0&0
\end{array}
\right),
$$
the supersymmetry transformations are written as
\beqa
&\delta g=-i\sqrt{2}\left(\,
\epsilon_+\eta_-g-\bepsilon_+\bareta_-g-\bepsilon_-g\eta_+
+\epsilon_-g\bareta_+\,\right),\nn\\
&\delta\eta_-=\sqrt{2}\bepsilon_+\left[D_-gg^{-1}\right]_-,
\,\,\,\,
\delta\eta_+=-\sqrt{2}\epsilon_-\left[g^{-1}D_+g\right]_-,
\label{SUSY}\\
&\delta\bareta_-=-\sqrt{2}\epsilon_+\left[D_-gg^{-1}\right]_+,
\,\,\,\,
\delta\bareta_+=\sqrt{2}\bepsilon_-\left[g^{-1}D_+g\right]_+,
\nn
\eeqa
where $[X]_{\pm}$ is the projection to the upper-right/lower-left
entry of $X$.
The naive chiral R-symmetry
$\psi_{\mp}\to\e^{\mp i\alpha_{\mp}}\psi_{\mp}$
has an anomaly, but it can be cancelled  using the
`anomalous' chiral gauge transformations (\ref{scgt}).
The following combination is anomaly-free:
\beq
U(1)_R:\left\{
\begin{array}{l}
g\to h_{\alpha_-}^{-1}g,\\
\eta_-\to\e^{-i\alpha_-}h_{\alpha_-}^{-1}\eta_-h_{\alpha_-},\\
\bareta_-\to\e^{i\alpha_-}
h_{\alpha_-}^{-1}\bareta_-h_{\alpha_-},
\end{array}
\right.
\,\,\,
U(1)_L:
\left\{
\begin{array}{l}
g\to gh_{\alpha_+}^{-1},\\
\eta_+\to \e^{i\alpha_+}h_{\alpha_+}\eta_+h_{\alpha_+}^{-1},\\
\bareta_+\to \e^{-i\alpha_+}h_{\alpha_+}\bareta_+h_{\alpha_+}^{-1},
\end{array}
\right.
\label{RLR}
\eeq
where $h_{\alpha}:=\exp(i\alpha\sigma_3/(k+2))$.
The R-symmetries are trivial at $\alpha_{\pm}=2\pi m_{\pm} (k+2)$
with $m_{\pm}\in\Z$ since $h_{2\pi m_{\pm}(k+2)}=1$.
Note also that $(\alpha_-,\alpha_+)=(2\pi m, -2\pi m)$
with $m\in\Z$
is gauge-equivalent to the trivial transformation.

The supercurrents and the R-currents are found by the Noether procedure.
In fact,
the left and right moving currents decouple as
$G_-^0=G_-^1=G$, $G_+^0=-G_+^1=\tG$, and
$J_R^0=J_R^1=J$, $J_L^0=-J_L^1=\wtJ$,
reflecting the
$(2,2)$ super{\it conformal} symmetry that the system actually has.
The expressions of these currents are
\beqa
&&\bG=i\sqrt{2}k\,\tr(\bareta_-D_-gg^{-1})
=\sqrt{2\over k}\bpsi_-J^G(\sigma_+),
\label{Gexpr1}\\
&&G=i\sqrt{2}k\,\tr(\eta_-D_-gg^{-1})
=\sqrt{2\over k}\psi_-J^G(\sigma_-),
\\
&&\btG=-i\sqrt{2}k\,\tr(\eta_+g^{-1}D_+g)
=\sqrt{2\over k}\psi_+\wtJ^G(\sigma_-),
\\
&&\tG=-i\sqrt{2}k\,\tr(\bareta_+g^{-1}D_+g)
=\sqrt{2\over k}\bpsi_+\wtJ^G(\sigma_+),
\label{Gexpr4}\\
&&J=k\,\tr(\bareta_-\eta_-+{i\over k+2}\sigma_3(
D_-gg^{-1}+i[\bareta_-,\eta_-]))
=\bpsi_-\psi_-+{1\over k+2}J^H,
\label{Rright}\\
&&\wtJ=-k\,\tr(\bareta_+\eta_++{i\over k+2}\sigma_3(
D_-gg^{-1}+i[\bareta_+,\eta_-]))
=-\bpsi_+\psi_+-{1\over k+2}\wtJ^H.
\label{Rleft}
\eeqa
Here
 $J^G(X)$ and $\wtJ^G(X)$ are the currents of the 
$G$-WZW sector
$$
J^G(X)=ik\,\tr(D_-gg^{-1}X),
\,\,\,\,
\wtJ^G(X)=-ik\,\tr(g^{-1}D_+g X),
$$
while $J^H$ and $\wtJ^H$ are ($2$ times)
the right and left components of the gauge current:
\beq
J^H=2(J^G(\sigma_3/2)+\psi_-\bpsi_-),
\,\,\,\,\,\,
\wtJ^H=2(\wtJ^G(\sigma_3/2)+\psi_+\bpsi_+).
\label{JH}
\eeq
By definition, R-symmetries transform the supercurrents as
$G\to \e^{-i\alpha_-}G, \bG\to\e^{i\alpha_-}\bG$,
$\tG\to\e^{-i\alpha_+}\tG,\btG\to\e^{i\alpha_+}\btG$.
Those that transform them 
only by sign form the subgroup $\Z_{2(k+2)}\times \Z_2$
generated by $(\alpha_-,\alpha_+)=(\pi,0)$
and by $(\alpha_-,\alpha_+)=(\pi,-\pi)$.
Up to gauge transformations,
the former is equivalent to the axial rotation of order $2(k+2)$:
\beq
a=\e^{-\pi iJ_0}: A\to A,\,\,\,
g\to h_{\pi\over 2}^{-1}gh_{\pi\over 2}^{-1},\,\,\,
\Psi_-\to -h_{\pi\over 2}^{-1}\Psi_-h_{\pi\over 2},\,\,\,
\Psi_+\to h_{\pi\over 2}\Psi_+h_{\pi\over 2}^{-1},
\label{axsym}
\eeq
and the latter is
equivalent to the fermion number
\beq
(-1)^F=\e^{-\pi i(J_0-\widetilde{J}_0)}:(A,g,\Psi)\to (A,g,-\Psi).
\label{mod2F}
\eeq
We will later consider the model where $(-1)^F$ is gauged --- the model
with non-chiral GSO projection.

\subsubsection{Geometric picture}

The model has a geometrical interpretation.
We parametrize the group element by
$$
g=\e^{i(\phi+t)\sigma_3/3}\e^{i\theta\sigma_1}\e^{i(\phi-t)\sigma_3/2}
=\left(\begin{array}{cc}
\e^{i\phi}\cos\theta&i\e^{it}\sin\theta\\
i\e^{-it}\sin\theta&\e^{-i\phi}\cos\theta
\end{array}\right),
$$
and the fermions by
\beqa
&&\bchi_+=-\sqrt{2\over k}\,\psi_+\e^{it}\sin\theta,
\,\,\,
\chi_+=-\sqrt{2\over k}\,\bpsi_+\e^{-it}\sin\theta,
\nn\\
&&\bchi_-=\sqrt{2\over k}\,\bpsi_-\e^{-it}\sin\theta,
\,\,\,
\chi_-=\sqrt{2\over k}\,\psi_-\e^{it}\sin\theta,
\nn
\eeqa
After integrating out the gauge field, we obtain the following
action which involves $(\theta,\phi)$ or $z=\e^{-i\phi}\cos\theta$
but does not contain $t$:
$$
S={1\over 2\pi}\int_{\Sigma}
\dd^2x\,\Bigl\{\,-g_{z\bz}\partial^{\mu}z\partial_{\mu}\bz
+2ig_{z\bz}(\bchi_-{\cal D}_{\!+}^{}\chi_-+\bchi_+{\cal D}_{\!-}^{}\chi_+)
+R_{z\bz z\bz}\chi_+\chi_-\bchi_-\bchi_+
\,\Bigr\}.
$$
Here $g_{z\bz}={k\over 2(1-|z|^2)}$
and ${\cal D}_{\mu}:=\partial_{\mu}+\Gamma^z_{zz}\partial_{\mu}z$.
This is the action for the supersymmetric sigma model
whose target space is the disc $|z|\leq 1$ with metric
\beq
\dd s^2=k\Bigl[\,(\dd\theta)^2+\cot^2\theta(\dd\phi)^2\,\Bigr]
=k{|\dd z|^2\over 1-|z|^2}.
\label{disc}
\eeq
The supersymmetry variation (\ref{SUSY}) transforms
the complex coordinate as
$\delta z=\epsilon_+\chi_--\epsilon_-\chi_+$, which shows that
$z$ and $\chi_{\pm}$ form a chiral multiplet.
The R-symmetry acts on $z$ as
$$
z\to \e^{i(\alpha_-+\alpha_+)/(k+2)}z.
$$

\subsection{Parity symmetry}\label{sub:pari}

Let us study the parity symmetry of the system.
The coordinate
transformation $(x^0,x^1)\to (x^0,-x^1)$ that reverses the
orientation of the worldsheet is denoted by $\Omega$. 
We have seen in \cite{BH1} that the bosonic gauged WZW model has two types of
parity invariance: $\Omega$ combined with
$(A,g)\to (A,g^{-1})$ or
$(A,g)\to (g_*^{-1}Ag_*,g_*^{-1}g^{-1}g_*)$, where
$$
g_*:=i\sigma_1=\left(\begin{array}{cc}
0&i\\
i&0
\end{array}\right),\,\,\,\,
\mbox{modulo $H$-action,}
$$
extending the parity symmetries of the ordinary WZW model 
\cite{PoSt,BCW,Schel,Br} to the gauged case.
Analogs of these in the supersymmetric model
are $\Omega$ combined with
\beq
{\cal I}_A:(A,g,\Psi)\to (A,g^{-1},\Psi)
\eeq
and
\beq
{\cal I}_B:(A,g,\Psi)\to
(g_*^{-1}Ag_*,g_*^{-1}g^{-1}g_*,g_*^{-1}\Psi g_*).
\eeq
Note that $\Omega$ reverses the worldsheet chirality,
and hence it exchanges the left and  right components of the fermion $\Psi$:
$(\Omega\Psi)_{\pm}(x^0,x^1)=\Psi_{\mp}(x^0,-x^1)$.
(It also maps the gauge field as
$(\Omega A)_{\pm}(x^0,x^1)=A_{\mp}(x^0,-x^1)$, just as in the bosonic case.)
In particular,
${\cal I}_A\Omega$ and ${\cal I}_B\Omega$
act on the components $\psi_{\pm}, \bpsi_{\pm}$ as
\beqa
&&{\cal I}_A\Omega:\psi_{\pm}(x^0,x^1)\to \psi_{\mp}(x^0,-x^1),
\,\,\,
\bpsi_{\pm}(x^0,x^1)\to \bpsi_{\mp}(x^0,-x^1),
\label{Apsit}\\
&&{\cal I}_B\Omega:
\psi_{\pm}(x^0,x^1)\to \bpsi_{\mp}(x^0,-x^1),
\,\,\,
\bpsi_{\pm}(x^0,x^1)\to \psi_{\mp}(x^0,-x^1).
\label{Bpsit}
\eeqa
The classical action $S(A,g,\Psi)$ is invariant under both
${\cal I}_A\Omega$ and ${\cal I}_B\Omega$.
Thus, these are the candidates for parity symmetry of the
supersymmetric model.

\subsubsection{Parity anomaly}\label{subsub:panomaly}

Let us examine whether there is an anomaly from the fermionic sector.
We recall that there are fermionic zero modes in a background $A$
in which the first Chern class
$$
c_1={i\over 2\pi}\int_{\Sigma}\tr\left(F_A
(\mbox{${-1\over 2}$}\sigma_3)\right)
$$
is non-zero.
If it is positive,  
there are generically $c_1$ zero modes for both $\psi_-$ and $\bpsi_+$
and none for $\bpsi_-$ and $\psi_+$.
Thus the path-integral measure contains a factor
\beq
{\cal D}_A^{(0)}\Psi=\prod_{i=1}^{c_1}\dd\psi_{-}^{(0)i}
\dd\bpsi_{+}^{(0)i},
\label{zeromme}
\eeq
where $\psi_-^{(0)i}$ and $\bpsi_+^{(0)i}$ are the zero modes,
which are complex conjugates of each other.
If $c_1$ is negative, the zero modes originate from $\bpsi_-$ and $\psi_+$.
Observe that $c_1$ flips its sign under the worldsheet
orientation reversal $\Omega$.
It also flips under the $g_*$-conjugation
since $g_*\sigma_3 g_*^{-1}=-\sigma_3$.
Thus, the first Chern class flips its sign under ${\cal I}_A\Omega$ but
remains invariant under ${\cal I}_B\Omega$:
\beqa
&&
{\cal I}_A\Omega:c_1\to -c_1,
\label{Ac1t}\\
&&
{\cal I}_B\Omega:c_1\to c_1.
\label{Bc1t}
\eeqa
Let us first examine ${\cal I}_B\Omega$,
which preserves the topology of the $U(1)$ gauge bundle by (\ref{Bc1t}).
By (\ref{Bpsit}),
${\cal I}_B\Omega$ sends the $(\psi_-,\bpsi_+)$ zero modes
in the background $A$ to
the $(\bpsi_+,\psi_-)$ zero modes
in the background $-\Omega^*A$.
Because of the Fermi statistics
$\dd\bpsi_{+}^{(0)i}
\dd\psi_{-}^{(0)i}=-\dd\psi_{-}^{(0)i}
\dd\bpsi_{+}^{(0)i}$,
the measure (\ref{zeromme}) is transformed as
\beq
{\cal D}_A^{(0)}\Psi\longrightarrow
(-1)^{c_1}
{\cal D}_{-\Omega^*A}^{(0)}\Psi.
\label{anoma}
\eeq
There is no extra sign from
the measure of the non-zero modes.
Since the field configurations $A$ and $-\Omega^*A$ are smoothly
connected,
there is no way to define the measure so that it is invariant under
${\cal I}_B\Omega$.
Thus, the parity ${\cal I}_B\Omega$ suffers from an anomaly.
Let us next consider ${\cal I}_A\Omega$.
By (\ref{Ac1t}), it changes the topology of the gauge bundle, except for
the trivial one $c_1=0$.
In the latter case there is no net fermionic zero mode and therefore
it is ${\cal I}_A\Omega$-invariant. 
For $c_1\ne 0$, one can choose the phase of the measure,
first for $c_1>0$ and then for $c_1<0$, so that
 it is invariant under ${\cal I}_A\Omega$,which sends $c_1$ to $-c_1$.
Thus, the parity ${\cal I}_A\Omega$ is anomaly-free.

\subsubsection{Cancellation of the anomaly}

We have seen that
\beq
P_A:={\cal I}_A\Omega
\eeq
is anomaly-free
but ${\cal I}_B\Omega$ suffers from a $\Z_2$ anomaly
%
originating from the sign
$\dd\bpsi_+^{(0)}\dd\psi_-^{(0)}
=-\dd\psi_-^{(0)}\dd\bpsi_+^{(0)}$.
It is obviously cancelled by combination with the
transformation
$$
(-1)^{F_R}:(A,g,\Psi_-,\Psi_+)\to(A,g,-\Psi_-,\Psi_+).
$$
In other words, $(-1)^{F_R}$ has the same anomaly (\ref{anoma}) as
${\cal I}_B\Omega$.
Thus,
\beq
P_B:=(-1)^{F_R}{\cal I}_B\Omega
\eeq
is an anomaly-free parity symmetry of the system.

\subsubsection{Action on the supercurrents}

We determine how the parity symmetries transform the  supercurrents of the
system.
Let us start with $P_A:(A,g,\Psi)\to\Omega(A,g^{-1},\Psi)$,
which acts in the following way
$\psi_{\pm}(x)\to \psi_{\mp}(\wtx)$,
$D_-gg^{-1}(x)\to D_+g^{-1}g(\wtx)=-g^{-1}D_+g(\wtx)$,
and
$g^{-1}D_+g(x)\to gD_-g^{-1}(\wtx)=-D_-gg^{-1}(\wtx)$.
It follows that $P_A$ transforms the currents as
\beqa
&&G(x)\to \btG(\wtx),\,\,\,
\bG(x)\to \tG(\wtx),\,\,\,
J(x)\to -\wtJ(\wtx)
\nn\\
&& \tG(x)\to \bG(\wtx),\,\,\,
\btG(x)\to G(\wtx),\,\,\,
\wtJ(x)\to -J(\wtx)
\nn
\eeqa
and therefore $P_A$ is an A-parity.
One may also consider the combination
\beq
P_A^{\alpha,\beta}:=\e^{-i\alpha F_V-i\beta F_A}P_A.
\eeq
This is an ${\rm A}_{\alpha,\beta}$-parity of the system.
It preserves an ${\mathcal N}=2$ supersymmetry if $\beta\in \pi\Z$.
Note that $a^{\ell}P_A=P_A^{{\pi\ell\over 2},-{\pi\ell\over 2}}$
and $(-1)^Fa^{\ell}P_A=P_A^{{\pi\ell\over 2},-{\pi(\ell-2)\over 2}}$
preserve an ${\mathcal N}=2$ supersymmetry if $\ell$ is even,
while they are $\widetilde{A}$-parities if $\ell$ is odd.



We next consider $P_B=
(-1)^{F_R}\circ {\rm ad} g_*^{-1}\circ P_A$.
Note that ${\rm ad}g_*^{-1}$ exchanges $G\leftrightarrow\bG$,
$\tG\leftrightarrow \btG$, $J\leftrightarrow -J$, and
$\wtJ\leftrightarrow -\wtJ$. Thus $P_B$ transforms the currents as
\beqa
&&G(x)\to -\tG(\wtx),\,\,\,
\bG(x)\to -\btG(\wtx),\,\,\,
J(x)\to \wtJ(\wtx)
\nn\\
&& \tG(x)\to G(\wtx),\,\,\,
\btG(x)\to \bG(\wtx),\,\,\,
\wtJ(x)\to J(\wtx).
\nn
\eeqa
Thus, $P_B$ is a $\widetilde{\rm B}$-parity.
It can also be modified by the R-symmetry:
\beq
P_B^{\alpha,\beta}
:=\e^{-i\alpha F_V-i\beta F_A}P_B.
\eeq
This is a ${\rm B}_{\alpha+{\pi\over 2},\beta-{\pi\over 2}}$-parity
of the system. It preserves an ${\mathcal N}=2$ supersymmetry if
$\alpha+{\pi\over 2}\in\pi\Z$.
Note that $a^{\ell}P_B=P_B^{{\pi\ell\over 2},-{\pi\ell\over 2}}$
and $(-1)^Fa^{\ell}P_B=P_B^{{\pi\ell\over 2},-{\pi(\ell-2)\over 2}}$
preserve an ${\mathcal N}=2$ supersymmetry if $\ell$ is odd,
while they are $\widetilde{B}$-parities if $\ell$ is even.




\subsubsection{The square of the parity}

In order to determine the sector to which the crosscap states
belong, we compute
the square of the above parities.

The basic A-parity $P_A={\cal I}_A\Omega$ is clearly
involutive, $P_A^2:(A,g,\Psi)\to \Omega(A,g^{-1},\Psi)
\to \Omega^2(A,(g^{-1})^{-1},\Psi)=(A,g,\Psi)$.
To find the square of the modified parities $P_A^{\alpha,\beta}$,
we note that conjugation by $P_A$ flips the sign of
$F_V=J_0+\widetilde{J}_0$, while
$F_A=-J_0+\widetilde{J}_0$ is invariant,
thus we see that
$$
(P_A^{\alpha,\beta})^2=\e^{-2i\beta F_A}.
$$
Since $\e^{-2i\beta F_A}$ is gauge-equivalent to
the fermion phase rotation $\psi_{\pm}\to\e^{-2i\beta}\psi_{\pm}$,
$P_A^{\alpha,\beta}$ is involutive as long as $\beta\in\pi \Z$, that is,
if and only if it preserves an ${\mathcal N}=2$ supersymmetry.
Namely, $|\Scr{C}_{P_A^{\alpha,\beta}}\rangle$ belongs to
 the RR-sector if and only if
$P_A^{\alpha,\beta}$ is supersymmetric.
Both $a^{\ell}P_A$ and $(-1)^Fa^{\ell}P_A$
square to $(-1)^{\ell F}$, and  their crosscaps therefore
belong to the RR (resp.~NSNS) sector if $\ell$ is even (resp.~odd),

Let us compute the square of $P_B=(-1)^{F_R}{\mathcal I}_B\Omega$.
The transformation ${\mathcal I}_B\Omega$ is involutive since
$g_*^2=-1$ is the central element of $SU(2)$.
On the other hand, $(-1)^{F_R}$ is transformed to $(-1)^{F_L}$
under conjugation by ${\mathcal I}_B\Omega$.
Thus, we find $P_B^2=(-1)^F$.
To find the square of $P_B^{\alpha,\beta}$, we note that
conjugation by $P_B$ keeps $F_V$ invariant but
flips the sign of $F_A$.
Thus, we see that
$$
(P_B^{\alpha,\beta})^2=\e^{-2i\alpha F_V}(-1)^F.
$$
$\e^{-2i\alpha F_V}$ is the fermion phase rotation
$\psi_{\pm}\to\e^{\mp 2i\alpha}\psi_{\pm}$ combined with the
axial rotation by $h_{-2\alpha}$.
This cancels $(-1)^F$ if and only if $2\alpha\in \pi(k+2)\Z$ as well as
$2\alpha\in\pi(2\Z+1)$.
This is possible only if $k$ is odd, in which case
one may take $\alpha=\pi{k+2\over 2}$.
Thus, if $k$ is even, none of the Parities $P_B^{\alpha,\beta}$ is involutive.
If $k$ is odd, $P_B^{\pi(k+2)/2,\beta}$ is involutive.
Both of $a^{\ell}P_B$ and $(-1)^Fa^{\ell}P_B$
square to $\e^{-\pi i \ell F_V}(-1)^F=(-1)^{(\ell+1)F}a^{2\ell}$
and hence the crosscap states belong to the sector twisted by this symmetry.
If $k$ is odd, $a^{k+2}P_B$ and $(-1)^Fa^{k+2}P_B$
are involutive.

\subsubsection{Geometric picture}
\label{subsub:geoac}

The above parities yield transformations of the
disc $z\mapsto P^{-1}zP$, which are
isometries of the metric (\ref{disc}).
The A-parities act on the coordinate $z$ as
\beq
P_A^{\alpha,\beta}:z\to \e^{i {2\alpha\over k+2}}\bz,
\label{PAdisk}
\eeq
which is the reflection with respect to
the line $\e^{i{\alpha\over k+2}}\R$.
The action of the B-parities are
\beq
P_B^{\alpha,\beta}:z\to \e^{i {2\alpha\over k+2}}z,
\label{PBdisk}
\eeq
which are rotations of the disc. The involutive parity
$P_B^{\pi(k+2)/2,\beta}$ acts on the disc as a rotation by $\pi$,
or an inversion $z\to -z$.

\subsection{Description in the operator formalism}\label{sub:DOF}

In this subsection,
we see how the parity symmetries act on the states.

\newcommand{\Pk}{{\rm P}_k}
\newcommand{\ta}{\widetilde{a}}
\newcommand{\wts}{\widetilde{s}}
\newcommand{\wtn}{\widetilde{n}}
\newcommand{\wtr}{\widetilde{r}}
\newcommand{\tMk}{\widetilde{\rm M}_k}

\subsubsection{The spectrum}

The space of states of the theory are the
gauge invariant states in the parent theory, $SU(2)$ WZW model plus
the free Dirac fermion system, modulo the large gauge transformations.
We formulate the system on a circle of radius $2\pi$.
The space of states of the WZW model is
$$
{\mathcal H}^{G,k}=\bigoplus_{j\in \Pk}\V_j\otimes\V_j,
$$
where $\V_j$ is the spin $j$ representation of the
$SU(2)$ current algebra at level $k$
and  $\Pk=\{0,{1\over 2},1,...,{k\over 2}\}$ is the set of integrable spins
at level $k$.
The space of states of the Dirac fermion system
is composed of the Fock space, which depends on the choice of
periodicity along the circle.
For the boundary condition
$\psi_-(\s+2\pi)=\e^{2\pi i a}\psi_-(\s)$
and $\psi_+(\s+2\pi)=\e^{-2\pi i \ta}\psi_+(\s)$,
the space of states is
$$
{\mathcal H}^f_{a,\ta}={\mathcal F}_a\otimes {\mathcal F}_{\ta}.
$$
Here ${\mathcal F}_a$ is the Fock space
for the fermion oscillator algebra
$\{\psi_{r_1},\psi_{r_2}\}=\{\bpsi_{r_1'},\bpsi_{r'_2}\}=0$,
$\{\psi_r,\bpsi_{r'}\}=\delta_{r+r',0}$ where,
$r\in \Z+a$ and $r'\in\Z-a$.
It is built on the vacuum state $|0\rangle_a$,
which is annihilated by $\psi_r$ ($r\geq 0$)
and $\opsi_{r'}$ ($r'>0$).
The fermion number operator is given by
\beq
J_0^f=\sum_{r\in \Z+a}:\!\bpsi_{-r}\psi_r\!:+a-[a]-\half.
\label{J0f}
\eeq
Here, $a=0$ and $a={1\over 2}$ are the R  and
NS  sectors respectively.
The space of states of the total parent theory is the tensor product
$$
{\mathcal H}^{\rm parent}_{a,\ta}
={\mathcal H}^{G,k}\otimes {\mathcal H}^f_{a,\ta}
=
\bigoplus_{j\in {\rm P}_k}
\bigoplus_{
s\in 2\Z+2a-1
\atop
\wts\in 2\Z+2\ta-1}
\V_j\otimes {\mathcal F}_a|_{s}
\otimes
\V_j\otimes {\mathcal F}_{\ta}|_{\wts},
$$
where ${\mathcal F}_{a}|_s$ is the subspace of 
${\mathcal F}_{a}$ in which $J_0^f=s/2$ ($s$ takes values
in $2(\Z+(a-1/2))$ by (\ref{J0f})).
The gauge invariant states are those obeying
the condition
\beqa
&&J_n^H=\wtJ_n^H=0,\quad n\geq 1,
\nn\\
&&J_0^H+\wtJ_0^H+2(a+\ta)=0,
\nn
\eeqa
where $J_n^H=J_n^3-2J_n^f$ and $\wtJ^H_n=\wtJ_n^3-2\wtJ_n^f$
according to (\ref{JH}). $J_n^H$ and $\wtJ_n^H$ generate the $U(1)$
current algebra at level $k+2$.
Let us denote by $B_{j,n,s}$
the subspace of $\V_j\otimes {\mathcal F}_a|_{s}$ in which
$J^H_n=0$ ($n\geq 1$) and $J^H_0=-n$, so that
$$
\V_j^{G,k}\otimes {\mathcal F}_a|_{s}=\bigoplus_{n\in 2\Z+2j+s}B_{j,n,s}
\otimes \V_{-n}^{H,k+2}.
$$
Then, the space of gauge invariant states is
$$
\left({\mathcal H}^{\rm parent}_{a,\ta}\right)^{H-{\rm inv}}
=\bigoplus_{j\in \Pk}
\bigoplus_{
s\in 2\Z+2a-1
\atop
\wts\in 2\Z+2\ta-1}
\bigoplus_{n\in 2\Z+2j+s}B_{j,n,s}\otimes
B_{j,-n+2(a+\ta),\wts}.
$$
The topology of the gauge group $\pi_1(H)=\pi_1(U(1))=\Z$
allows  large gauge
transformations, which act on the labels as
$(j,n,s,\tilde{s})\to ({k\over 2}-j,n\pm(k+2),s\pm 2,\tilde{s}\mp 2)
\to (j,n\pm 2(k+2),s\pm 4,\tilde{s}\mp 4)\to\cdots$.
This induces the so-called ``field identification'' \cite{FIGep,MS,FIH}.
Namely, the space of states of our gauge system is given by
\beq
{\cal H}_{a,\ta}=\bigoplus_{(j,n,s,\tilde{s})\in \tMk(a,\ta)}
B_{j,n,s}\otimes B_{j,-n+2(a+\ta),\wts}
\eeq
where $\tMk(a,\ta)$ is the infinite set
$$
\tMk(a,\ta)=\left\{(j,n,s,\tilde{s}) \Biggl|
\begin{array}{l}
j\in \Pk,\quad
2j-s+n\,\,{\rm even}\\
s\in 2\Z+2a-1,\,\,
\wts\in 2\Z+2\ta-1
\end{array}
\right\}\Bigl/\pi_1(H).
$$
The R-currents (\ref{Rright}) and (\ref{Rleft}) are
expressed as
$$
J_n={1\over k+2}J_n^H+J_n^f,
\,\,\,\,
\wtJ_n=-{1\over k+2}\wtJ_n^H-\wtJ_n^f,
$$
and hence the R-charges ($(J_0,\wtJ_0)$-eigenvalues) are given by
\beq
(q,\wtq)=\left(-{n\over k+2}+{s\over 2},
{\wtn\over k+2}-{\wts\over 2}\right)
\,\,\,\,
\mbox{on $B_{j,n,s}\otimes B_{j,\wtn,\wts}$}.
\eeq
The supercharges (\ref{Gexpr1})-(\ref{Gexpr4})
are expressed, after the standard shift $k\to k+2$, as
\beqa
&&\bG_{r'}=\sqrt{2\over k+2}\sum_{n\in\Z}
\bpsi_{r'-n}J_n^+,
\quad
G_r=\sqrt{2\over k+2}\sum_{n\in\Z}
\psi_{r-n}J_n^-,
\quad (r,-r'\in \Z+a),
\nn\\
&&\btG_{\wtr}=\sqrt{2\over k+2}\sum_{n\in\Z}
\tpsi_{\wtr-n}\wtJ_n^-,
\quad
\tG_{\wtr'}=\sqrt{2\over k+2}\sum_{n\in\Z}
\overline{\tpsi}_{\wtr'-n}\wtJ_n^+,
\quad (\wtr,-\wtr'\in \Z+\ta).
\nn
\eeqa
{\bf Remark.}
We regard the $(a,\ta)$ sector to be
the sector in which the fields obey the boundary condition
\beqa
&&\Phi(\s)=U_{a,\ta}^{-1}\Phi(\s+2\pi)U_{a,\ta},
\nn\\
&&\quad U_{a,\ta}:=\e^{-2\pi i(aJ_0+\ta\wtJ_0)}
=\e^{-2\pi i (a-{1\over 2})J_0-2\pi i(\ta+{1\over 2})\wtJ_0}(-1)^F.
\nn
\eeqa
Note that $(-1)^F:=\e^{\pi i (-J_0+\wtJ_0)}$ is the mod 2 fermion number
(\ref{mod2F}).
In particular the standard NSNS sector is
$(a,\ta)=({1\over 2},-{1\over 2})$, and it indeed includes
the $SL(2,\C)$-invariant
ground state $[|0;0\rangle\otimes |0\rangle_{1\over 2}]\otimes
[|0;0\rangle\otimes |0\rangle_{-{1\over 2}}]\in B_{0,0,0}\otimes B_{0,0,0}
\subset {\mathcal H}_{{1\over 2},-{1\over 2}}$.
The $(a,\ta)$ sector is the spectral flow \cite{SchSei}
from this
by $(a-{1\over 2},\ta+{1\over 2})$.
Because of the periodicity of the R-symmetries, we have the
identifications
$(a,\ta)\equiv (a+(k+2),\ta)\equiv
(a,\ta+(k+2))\equiv (a+1,\ta-1)$.

\subsubsection*{\it Chiral primaries}

The local operators of the system can be mapped one-to-one to
the states in the NSNS sector ($a=-\ta={1\over 2}$).
Chiral primary fields correspond to those obeying the conditions
$\bG_{-\half}=\btG_{-\half}=0$ and $G_{\half}=\tG_{\half}=0$.
One can show that they are given by
$$
|j\rangle^{}_{\it cc}=\Bigl[|j,j\rangle\otimes |0\rangle^{}_{\rm NS}\Bigr]
\otimes
\Bigl[|j,-j\rangle\otimes |0\rangle^{}_{\rm NS}\Bigr],
\,\,\,\,
j=0,\half,1,\ldots,{k\over 2},
$$
which belongs to $B_{j,-2j,0}\otimes B_{j,2j,0}$. 
The corresponding chiral primary ${\mathcal O}_j$
has R-charges $(q,\wtq)=({2j\over k+2},{2j\over k+2})$.
There are also antichiral primaries $\overline{\mathcal O}_j$
corresponding to
$|j\rangle^{}_{\it aa}=[|j,-j\rangle\otimes |0\rangle^{}_{\rm NS}]
\otimes
[|j,j\rangle\otimes |0\rangle^{}_{\rm NS}]$ in
$B_{j,2j,0}\otimes B_{j,-2j,0}$
with charge $(q,\wtq)=(-{2j\over k+2},-{2j\over k+2})$.
There are no twisted (anti)chiral primaries, except for the identity operator.

\subsubsection*{\it Supersymmetric ground states}

The supersymmetry of the sector with $a,\ta\in \Z$ is
generated by $G_0,\bG_0,\tG_0,\btG_0$. We would like to find
the supersymmetric ground states, namely the states
annihilated by all of $G_0,\bG_0,\tG_0,\btG_0$.

We start with the RR sector ($a=\ta=0$).
The supersymmetric ground states are
$$
|j\rangle^{}_{{}_{\rm RR}}
=\Bigl[|j;j\rangle\otimes |0\rangle^{}_{\rm R}\Bigr]
\otimes
\Bigl[|j;-j\rangle\otimes \bpsi_0|0\rangle^{}_{\rm R}\Bigr],
\,\,\,\,
j=0,\half,1,\ldots,{k\over 2},
$$
where $|0\rangle_{\rm R}$ is the vacuum state
$|0\rangle_0\in {\mathcal F}_0$, which is
annihilated by $\psi_0$ and has $J_0^f=-1/2$.
The state $|j\rangle^{}_{{}_{\rm RR}}$
belongs to $B_{j,-(2j+1),-1}\otimes B_{j,2j+1,1}$ and has R-charges
$(q,\wtq)=({2j+1\over k+2}-\half,{2j+1\over k+2}-\half)$.
Another representation of the same state (up to a phase) is
$$
|j\rangle'_{{}_{\rm RR}}=
\Bigl[|{k\over 2}-j;-{k\over 2}+j\rangle\otimes \bpsi_0
|0\rangle^{}_{\rm R}\Bigr]
\otimes \Bigl[|{k\over 2}-j;{k\over 2}-j\rangle
\otimes |0\rangle^{}_{\rm R}\Bigr],
$$
which belongs to
$B_{{k\over 2}-j,(k+2)-(2j+1),1}\otimes B_{{k\over 2}-j,-(k+2)+(2j+1),-1}$.
The state $|j\rangle^{}_{{}_{\rm RR}}\propto |j\rangle'_{{}_{\rm RR}}$
 corresponds to the chiral primary field
${\mathcal O}_j$.

We next consider the sectors with
twisted boundary conditions $(a,\ta)\not\equiv (0,0)$.
(There are $k+1$ such sectors labelled by $a+\ta\in\{1,2,\ldots,k+1\}$.)
For each such sector, there is a unique supersymmetric ground state
$$
|{\rm G}\rangle_{a,\ta}
=\Bigl[|j_{*};j_{*}\rangle\otimes |0\rangle^{}_{\rm R}\Bigr]
\otimes
\Bigl[|j_{*};j_{*}\rangle\otimes |0\rangle^{}_{\rm R}\Bigr]
\in
B_{j_{*},-2j_{*}-1,-1}\otimes
B_{j_{*},-2j_{*}-1,-1},
$$
where $j_*\in \Pk$ is defined by 
$2j_{*}+1\equiv -a-\ta$ mod $(k+2)$.
It has R-charge
$(q,\wtq)=({2j_*+1\over k+2}-{1\over 2},-{2j_*+1\over k+2}+{1\over 2})$.

The above results are consistent with the equivariant Witten indices.
Let us consider the partition function on the torus
$(x,y)\equiv (x+1,y)\equiv (x,y+1)$
with the (twisted) boundary condition $\Phi(x,y)=\Phi(x+1,y)
=U_{a,\ta}^{-1}\Phi(x,y+1)U_{a,\ta}$.
There is a supersymmetry as long as $a,\ta\in\Z$,
and the partition function is regarded as the Witten index.
If we consider $x$ as the space and $y$ as the time coordinate,
this can be regarded as the trace over the RR sector of the 
operator $U_{a,\ta}(-1)^F\e^{-\beta H}$.
If, on the other hand, we consider $x$ as the time and $y$ as the space
coordinate, the partition function is identified as the
trace over the $(a,\ta)$-sector of the operator
$(-1)^F\e^{-\beta' H}$.
We thus find the identity
$$
\Tr\!\!\mathop{}_{{\mathcal H}_{\rm RR}}\!\!
\e^{2\pi i (aJ_0+\ta\wtJ_0)}(-1)^F\e^{-\beta H}
=\Tr\!\!\mathop{}_{{\mathcal H}_{a,\ta}}\!\!
(-1)^F\e^{-\beta' H}.
$$
The left hand side is the equivariant Witten index and can be computed,
using our knowledge of the supersymmetric ground states
in the RR sector, as
\beqa
{\rm LHS}&=&
\sum_{j\in\Pk}\e^{2\pi i (a-{1\over 2})q_j+2\pi i (\ta+{1\over 2})\wtq_j}
=\sum_{2j=0,1,...,k}\e^{2\pi i(a+\ta)({2j+1\over k+2}-{1\over 2})}
\nn\\
&=&
-\e^{-\pi i(a+\ta)}=\pm 1
\qquad \mbox{if $a+\ta \not\equiv 0$ mod $(k+2)$}.
\nn
\eeqa
On the other hand, the right hand side is the ordinary Witten index
for the system twisted by $U_{a,\ta}$, which computes the number of bosonic
supersymmetric ground states minus the number of fermionic ones.
That it is equal to LHS=$\pm 1$ if $a+\ta \not\equiv 0$
 is consistent with the above conclusion
 that there is a unique supersymmetric ground states in
${\mathcal H}_{a,\ta}$ with $a+\ta \not\equiv 0$.

\subsubsection{The parity action}

Let us now see how  parity acts on the states.
We will also compute the twisted partition function for the
NSNS and RR sectors.

\newcommand{\ch}{{\rm ch}}
\newcommand{\bepi}{\mbox{$\beta\over\pi$}}

\subsubsection*{\it A-Parity}

We start with the basic parity symmetry $P_A={\cal I}_A\Omega$.
Since the space of states is basically
the subspace of the tensor product
${\mathcal H}^{G,k}\otimes {\mathcal H}_{a,\ta}^f$,
we separate the discussion into bosonic and fermionic sectors.
The action on the bosonic part is determined in \cite{BH1}:
\beq
u_b\otimes \widetilde{v_b}\in \V_j\otimes\V_j
\,\,\longmapsto\,\,
(-1)^{2j} v_b\otimes \widetilde{u_b}\in \V_j\otimes\V_j.
\eeq
For the fermionic sector it exchanges the periodicity parameter $a$, $\ta$
of the left and right movers, ${\mathcal H}^f_{a,\ta}\to
{\mathcal H}_{\ta,a}^f$, mapping
the oscillators as
\beq
\psi_r,\,\,\bpsi_{r'},\,\,
\tpsi_{\wtr},\,\,\overline{\tpsi}_{\wtr'}
\longrightarrow
\tpsi_r,\,\,\overline{\tpsi}_{r'},\,\,
\psi_{\wtr},\,\,\bpsi_{\wtr'},
\eeq
where $r,-r'\in\Z+a$ and $\wtr,-\wtr'\in\Z+\ta$.
It follows  that the ground state
$|0\rangle_{a,\ta}=|0\rangle_a\otimes |0\rangle_{\ta}$
(annihilated by $\psi_{r\geq 0}$, $\opsi_{r'>0}$,
$\tpsi_{\wtr\geq 0}$ and $\overline{\tpsi}_{\wtr'>0}$)
is mapped
to the ground state $|0\rangle_{\ta,a}$
(annihilated by $\psi_{\wtr\geq 0}$, $\opsi_{\wtr'>0}$,
$\tpsi_{r\geq 0}$ and $\overline{\tpsi}_{r'>0}$)
up to a phase,
$|0\rangle_{a,\ta}\mapsto \epsilon_{a,\ta}|0\rangle_{\ta,a}$.
More general states are mapped as
\beq
{\mathcal O}_1\widetilde{{\mathcal O}_2}
|0\rangle_{a,\ta}\,\,
\longmapsto\,\,
\epsilon_{a,\ta}
\widetilde{{\mathcal O}_1}{\mathcal O}_2
|0\rangle_{\ta,a}
=
\epsilon_{a,\ta}(-1)^{|{\mathcal O}_1||{\mathcal O}_2|}
{\mathcal O}_2\widetilde{{\mathcal O}_1}
|0\rangle_{\ta,a}.
\eeq
Here, ${\mathcal O}_i$ are polynomials of the fermion oscillators
$\psi_{\bullet}$, $\opsi_{\bullet}$,
and $\widetilde{{\mathcal O}_i}$ are
the ones where $\psi_{\bullet}$, $\opsi_{\bullet}$ 
are replaced by $\tpsi_{\bullet}$, $\overline{\tpsi}_{\bullet}$.
$|{\mathcal O}_i|$ is the fermion number of ${\mathcal O}_i$.
Thus, we find that the states of the combined system are mapped as
follows
\beqa
P_A&:&u_a\otimes \widetilde{v_b}\otimes
{\mathcal O}_1\widetilde{{\mathcal O}_2}|0\rangle_{a,\ta}
\in B_{j,n,s}\otimes B_{j,-n+2(a+\ta),\wts}
\nn\\
&&\,\,\longmapsto\,\,
\epsilon_{a,\ta}(-1)^{2j+|{\mathcal O}_1||{\mathcal O}_2|}
v_b\otimes\widetilde{u_b}\otimes
{\mathcal O}_2\widetilde{{\mathcal O}_1}
|0\rangle_{\ta,a}
\in B_{j,-n+2(a+\ta),\wts}\otimes B_{j,n,s},\quad
\label{PAaction}
\eeqa
where
$$
|{\mathcal O}_1|={s\over 2}-(a-[a]-{1\over 2}),
\quad
|{\mathcal O}_2|={\wts\over 2}-(\ta-[\ta]-{1\over 2}).
$$
One can check that this action is compatible with the field identification.
Let us show this for the action on the RR ground states,
$|j\rangle^{}_{{}_{\rm RR}}=|j;j\rangle\otimes |j;-j\rangle\otimes
\overline{\tpsi}_0|0\rangle_{0,0}$
$\propto$
$|j\rangle'_{{}_{\rm RR}}
=|{k\over 2}-j;-({k\over 2}-j)\rangle\otimes
|{k\over 2}-j;{k\over 2}-j\rangle\otimes
\opsi_0|0\rangle_{0,0}$
Using (\ref{PAaction}), we find that they are mapped as
$$
|j\rangle^{}_{{}_{\rm RR}}\mapsto 
\epsilon_{0,0}(-1)^{2j}|{k\over 2}-j\rangle'_{{}_{\rm RR}},
\quad
|j\rangle'_{{}_{\rm RR}}\mapsto \epsilon_{0,0}(-1)^{k-2j}
|{k\over 2}-j\rangle^{}_{{}_{\rm RR}}.
$$
This is consistent with the field identification
$|j\rangle^{}_{{}_{\rm RR}}\propto |j\rangle'_{{}_{\rm RR}}$, provided
that
\beq
|{k\over 4}\rangle^{}_{{}_{\rm RR}}=\pm |{k\over 4}\rangle'_{{}_{\rm RR}},
\label{sign}
\eeq
which is non-vacuous only if $k$ is even.
This also shows that $P_A$ is involutive only if $\epsilon_{0,0}$
is $1$ or $-1$.

Let us compute the twisted partition function
for the NSNS and RR sectors.
The subspaces $B_{j,n,s}\otimes B_{j,-n,\wts}$
 that contribute to it are such that
it is equivalent to $B_{j,-n,\wts}\otimes B_{j,n,s}$
up to field identification.
This is so for
$B_{j,0,s}\otimes B_{j,0,s}$
and $B_{{k\over 4},-{k+2\over 2},s}\otimes B_{{k\over 4},{k+2\over 2},s+2}$
($k$ even).
It is then straightforward to compute the partition functions.
We present the result for the more general parity
$P_A^{\alpha,\beta}=\e^{-i\alpha F_V-i\beta F_A}P_A$.
For the NSNS sector it is
\beqa
\Tr\!\!\mathop{}_{{\mathcal H}_{\rm NSNS}}\!\!
(P_A^{\alpha,\beta}q^H)
&=&\epsilon_{{}_{\rm NSNS}}\sum_{2j,s\,{\rm even}}
(-1)^{s\over 2}\e^{i\beta s}\ch_{j,0,s}(2\tau)
\nn\\
&=&\epsilon_{{}_{\rm NSNS}}
\sum_{2j\,{\rm even}}(\chi_{j,0,0}
-\chi_{j,0,2})(2\tau,\bepi),
\eeqa
and for the RR sector
\beqa
\lefteqn{\Tr\!\!\mathop{}_{{\mathcal H}_{\rm RR}}\!\!
(P_A^{\alpha,\beta}q^H)}\nn\\
&=&-\epsilon_{{}_{\rm RR}}\sum_{2j,s\,{\rm odd}}
(-1)^{s+1\over 2}\e^{i\beta s}\ch_{j,o,s}(2\tau)
\pm \epsilon_{{}_{\rm RR}}(-1)^{k\over 2}\sum_{s\,{\rm odd}}
(-1)^{s+1\over 2}\e^{2i\beta {s+1\over 2}}
\ch_{{k\over 2},-{k+2\over 2},s}(2\tau)
\nn\\
&=&
-\epsilon_{{}_{\rm RR}}\sum_{2j\,{\rm odd}}(\chi_{j,0,-1}
-\chi_{j,0,1})(2\tau,\bepi)
\pm\epsilon_{{}_{\rm RR}}(-1)^{k\over 2}
(\chi_{{k\over 2},-{k+2\over 2},-1}
-\chi_{{k\over 2},-{k+2\over 2},1})(2\tau,\bepi).
\nn\\
\label{partPARR}
\eeqa
In the above expressions, $\ch_{j,n,s}$ is the character
\beq
\ch_{j,n,s}(\tau)
=\Tr\!\!\mathop{}_{B_{j,n,s}}\!\!
q^{L_0-{c\over 24}},
\label{defcha}
\eeq
and $\chi_{j,n,s}$, for $s\in \Z/4\Z$, is
$$
\chi_{j,n,s}(\tau,u)
=\sum_{p\in\Z}\Tr\!\!\mathop{}_{B_{j,n,s+4p}}\!\!
q^{L_0-{c\over 24}}\e^{2\pi i J_0 u}
=\sum_{p\in\Z}\e^{2\pi i u(-{n\over k+2}+{s+4p\over 2})}
\ch_{j,n,s+4p}(\tau),
$$
The sign $\pm$ of the second term on
the right hand side of (\ref{partPARR}) is
the same as the one that appears in the
Field Identification (\ref{sign}).

The special cases are the ones with $(-1)^{\nu F}a^{\ell}P_A$-twists
($\nu=0,1$):
\beqa
&&
\Tr\!\!\mathop{}_{{\mathcal H}_{\rm NSNS}}\!\!
((-1)^{\nu F}a^{\ell}P_Aq^H)
=\epsilon_{{}_{\rm NSNS}}\sum_{2j\,{\rm even}}
(\chi_{j,0,0}-(-1)^{\ell}\chi_{j,0,2})(2\tau),
\label{pfNSNS1}\\
&&
\Tr\!\!\mathop{}_{{\mathcal H}_{\rm RR}}\!\!
((-1)^{\nu F}a^{\ell}P_Aq^H)
=-\e^{\pi i\ell\over 2}(-1)^{\nu}\epsilon_{{}_{\rm RR}}
\sum_{2j\,{\rm odd}}(\chi_{j,0,-1}
-(-1)^{\ell}\chi_{j,0,1})(2\tau)
\nn\\
&&\qquad\qquad\qquad\qquad\qquad
\pm\epsilon_{{}_{\rm RR}}\delta^{(2)}_k(-1)^{k\over 2}
(\chi_{{k\over 4},-{k+2\over 2},-1}
-(-1)^{\ell}\chi_{{k\over 4},-{k+2\over 2},1})(2\tau).
\label{pfRR1}
\eeqa
We recall that $(-1)^{\nu F}a^{\ell}P_A$
with even $\ell$ preserves an ${\mathcal N}=2$ supersymmetry and the
twisted partition function in the RR-sector can be regarded as
the Witten index.
Indeed, since $\chi_{j,n,-1}-\chi_{j,n,1}=\pm\delta_{n,\mp(2j+1)}$,
we find
\beq
I_{a^{\rm even}P_A}=
\pm\epsilon_{{}_{\rm RR}}(-1)^{k\over 2}\delta^{(2)}_k.
\label{Wtind1}
\eeq
For odd $\ell$, $(-1)^{\nu F}a^{\ell}P_A$ breaks all supersymmetry
and the partition function is indeed a non-trivial function
of $\tau$.

\subsubsection*{\it B-Parity}

\newcommand{\alpi}{\mbox{$\alpha\over\pi$}}

We next consider B-parity $P_B=(-1)^{F_R}{\cal I}_B\Omega$, which is
the same as $P_A$ followed by $(-1)^{J_0^f}{\rm ad}g_*^{-1}$.
The action of ${\rm ad}g_*^{-1}$ on the bosonic sector
is the usual one, $u_b\otimes \widetilde{v_b}\mapsto
g_*u_b\otimes \widetilde{g_*v_b}$,
where $g_*$ acts on the ground states of $\V_j$ as
$g_*|j;m\rangle=i^{2j}|j;-m\rangle$ \cite{BH1}.
Let us next see how ${\rm ad}g_*^{-1}$ acts on the fermionic sector.
Since it exchanges $\psi_{\pm}$ and $\opsi_{\pm}$,
it flips the sign of the periodicity parameter $(a,\ta)$,
${\mathcal H}_{a,\ta}^f\to{\mathcal H}_{-a,-\ta}^f$, and maps
the oscillators as
\beq
{\rm ad}g_*^{-1}\,:\,\psi_r,\,\,\bpsi_{r'},\,\,
\tpsi_{\wtr},\,\,\overline{\tpsi}_{\wtr'}
\longrightarrow
\bpsi_r,\,\,\psi_{r'},\,\,
\overline{\tpsi}_{\wtr},\,\,\tpsi_{\wtr'}.
\eeq
It follows from this that the ground state
$|0\rangle_{a,\ta}$ (annihilated by $\psi_{r\geq 0}$,
$\opsi_{r'>0}$, $\tpsi_{\wtr\geq 0}$ and $\overline{\tpsi}_{\wtr'>0}$)
is mapped to the ground state
$|0\rangle_{-a,-\ta}'$ (annihilated by $\psi_{r'\geq 0}$,
$\opsi_{r>0}$, $\tpsi_{\wtr'\geq 0}$ and $\overline{\tpsi}_{\wtr>0}$)
up to a phase, $|0\rangle_{a,\ta}\mapsto\eta_{a,\ta}|0\rangle'_{-a,-\ta}$,
Note that one may set $|0\rangle_{-a,-\ta}'=|0\rangle_{-a,-\ta}$,
as long as $a,\ta\not\in\Z$.
If $a$ or $\ta$ is an integer, they are not proportional to
each other. For example $|0\rangle_{0,0}$ is
annihilated by $\psi_0$, $\tpsi_0$, while
$|0\rangle_{0,0}'$ is annihilated by $\opsi_0$, $\overline{\tpsi}_0$,
and therefore one may set
$|0\rangle_{0,0}'=\opsi_0\overline{\tpsi_0}|0\rangle_{0,0}$.
More general states are mapped as
$$
{\rm ad}g_*^{-1}\,:\,
{\mathcal O}_1\widetilde{{\mathcal O}_2}|0\rangle_{a,\ta}
\,\,\longmapsto\,\,
\eta_{a,\ta}\overline{\mathcal O}_1\widetilde{\overline{\mathcal O}_2}
|0\rangle'_{a,\ta}.
$$
Combining with the action of $P_A$ and $(-1)^{F_R}=\e^{\pi i J^f_0}$,
we find that $P_B=\e^{\pi i J^f_0}{\rm ad}g_*^{-1}P_A$ acts on the states as
\beqa
P_B&:&
u_a\otimes \widetilde{v_b}\otimes
{\mathcal O}_1\widetilde{{\mathcal O}_2}|0\rangle_{a,\ta}
\in B_{j,n,s}\otimes B_{j,-n+2(a+\ta),\wts}
\nn\\
&&\!\!\!\!\!\!\longmapsto\,
\varepsilon_{a,\ta}(-1)^{2j+|{\mathcal O}_1||{\mathcal O}_2|}
\e^{-\pi i{\wts\over 2}}
g_*(v_b)\otimes \widetilde{g_*(u_b)}\otimes
\overline{\mathcal O}_2\widetilde{\overline{\mathcal O}_1}
|0\rangle_{-\ta,-a}'\in
B_{j,n-2(a+\ta),-\wts}\otimes B_{j,-n,-s},
\nn\\
\label{PBaction}
\eeqa
where $\varepsilon_{a,\ta}=\epsilon_{a,\ta}\eta_{\ta,a}$.
Let us see if it is compatible with the field identification.
We examine it in the action on the RR ground states.
Using (\ref{PBaction}), we find the following $P_B$ action
\beqa
&&|j\rangle^{}_{{}_{\rm RR}}=|j;j\rangle\otimes |j;-j\rangle\otimes
\overline{\tpsi}_0|0\rangle_{0,0}
\,\longmapsto\,
\varepsilon_{0,0}\e^{-{\pi i\over 2}}|j;j\rangle\otimes |j;-j\rangle\otimes
\psi_0|0\rangle_{0,0}'=
\nn\\
&&\quad\qquad\quad\qquad\quad\qquad
=\varepsilon_{0,0}\e^{-{\pi i\over 2}}|j;j\rangle\otimes |j;-j\rangle\otimes
\psi_0\opsi_0\overline{\tpsi}_0|0\rangle_{0,0}
=\varepsilon_{0,0}\e^{-{\pi i\over 2}}|j\rangle^{}_{{}_{\rm RR}}
\nn\\[0.2cm]
&&|j\rangle'_{{}_{\rm RR}}
=|j';-j'\rangle\otimes
|j';j'\rangle\otimes
\opsi_0|0\rangle_{0,0}
\,\longmapsto\,
\varepsilon_{0,0}\e^{\pi i\over 2}|j';-j'\rangle\otimes
|j';j'\rangle\otimes
\tpsi_0|0\rangle_{0,0}'=
\nn\\
&&\quad\qquad\quad\qquad\quad\qquad
=\varepsilon_{0,0}\e^{\pi i\over 2}|j';-j'\rangle\otimes
|j';j'\rangle\otimes
\tpsi_0\opsi_0\overline{\tpsi}_0|0\rangle_{0,0}
=-\varepsilon_{0,0}\e^{\pi i\over 2}|j\rangle'_{{}_{\rm RR}},
\nn
\eeqa
where $j':={k\over 2}-j$.
We see that the action is compatible with the field identification
$|j\rangle^{}_{{}_{\rm RR}}\propto |j\rangle'_{{}_{\rm RR}}$.
Note that it would have been incompatible without the
$(-1)^{F_R}=\e^{\pi i J_0^f}$ factor. This corresponds to the
anomaly of ${\mathcal I}_B\Omega={\rm ad}g_*^{-1}P_A$ and its cancellation by
$(-1)^{F_R}$.

Let us now compute the twisted partition function in the NSNS and RR
sectors. A non-zero contribution comes from the subspaces
$B_{j,n,s}\otimes B_{j,-n,-s}$.
The result is
\beqa
\Tr\!\!\mathop{}_{{\mathcal H}_{\rm NSNS}}\!\!
(P_B^{\alpha,\beta}q^H)
&=&\varepsilon_{{}_{\rm NSNS}}\sum_{2j+n,\,s\,{\rm even}}
\e^{-2i\alpha (-{n\over k+2}+{s\over 2})}\ch_{j,n,s}(2\tau)
\nn\\
&=&\varepsilon_{{}_{\rm NSNS}}\sum_{2j+n\,{\rm even}\atop
s=0,2}\chi_{j,n,s}(2\tau,-\alpi),
\\
\Tr\!\!\mathop{}_{{\mathcal H}_{\rm RR}}\!\!
(P_B^{\alpha,\beta}q^H)
&=&\varepsilon_{{}_{\rm RR}}
\sum_{2j+n,\,s\,{\rm odd}}
\e^{-2i\alpha (-{n\over k+2}+{s\over 2})}\ch_{j,n,s}(2\tau)
\nn\\
&=&\varepsilon_{{}_{\rm RR}}
\sum_{2j+n\,{\rm odd}\atop
s=\pm 1}\chi_{j,n,s}(2\tau,-\alpi),
\eeqa
where $\varepsilon_{{}_{\rm RR}}=\varepsilon_{0,0}\e^{-{\pi i\over 2}}$.

Let us consider the special cases with $(-1)^{\nu F}a^{\ell}P_B$-twists
($\nu=0,1$):
\beqa
&&
\Tr\!\!\mathop{}_{{\mathcal H}_{\rm NSNS}}\!\!
((-1)^{\nu F}a^{\ell}P_Bq^H)
=\varepsilon_{{}_{\rm NSNS}}\sum_{2j+n\,{\rm even}\atop s=0,2}
\e^{\pi i\ell({n\over k+2}-{s\over 2})}
\chi_{j,n,s}(2\tau)
\label{pfNSNS2}
\\
&&
\Tr\!\!\mathop{}_{{\mathcal H}_{\rm RR}}\!\!
((-1)^{\nu F}a^{\ell}P_Bq^H)
=\varepsilon_{{}_{\rm RR}}
\sum_{2j+n\,{\rm odd}\atop s=-1,1}
\e^{\pi i\ell({n\over k+2}-{s\over 2})}
\chi_{j,n,s}(2\tau)
\label{pfRR2}
\eeqa
$(-1)^{\nu F}a^{\ell}P_B$
with odd $\ell$ preserves an ${\mathcal N}=2$ supersymmetry and the
twisted partition function in the RR-sector can be regarded as
the Witten index.
Indeed, it is just a number
\beqa
I_{a^{\ell}P_B}&=&\varepsilon_{{}_{\rm RR}}
\sum_{j\in\Pk}\e^{\pi i\ell({2j+1\over k+2}-{1\over 2})}
=\varepsilon_{{}_{\rm RR}}\e^{-\pi i{\ell\over 2}}
(z+z^2+\cdots+z^{k+1})|_{z=\e^{\pi i\ell/(k+2)}}
\nn
\\
&=&
\varepsilon_{{}_{\rm RR}}\e^{-\pi i{\ell\over 2}}{z-z^{k+2}\over 1-z}
=
i\varepsilon_{{}_{\rm RR}}\e^{-{\pi i\ell\over 2}}
\cot\left[{\pi \ell\over 2(k+2)}\right],
\quad\mbox{$\ell$ odd}.
\label{IBparity}
\eeqa
For even $\ell$, $(-1)^{\nu F}a^{\ell}P_B$ breaks all supersymmetry
and the partition function is indeed a non-trivial function
of $\tau$.

\subsection{RCFT point of view}\label{sub:RCFT}

We next study the system in which a certain GSO projection is imposed.
The system can be regarded as
a rational conformal field theory, the crosscaps can be studied using the
standard procedure of Pradisi--Sagnotti--Stanev \cite{PSS}
which is reviewed (along with more recent developments such as 
\cite{HSS,FSHSS})
and extended in \cite{BH1}.

\newcommand{\Mk}{{\rm M}_k}

\subsubsection*{\it GSO projection}

We perform the GSO projection with respect to the operator
$$
(-1)^F=\e^{-\pi i (J_0-\widetilde{J}_0)}.
$$
This is a non-chiral projection and the
projected theory consists of NSNS as well as RR sectors,
in each of which only the states with $(-1)^F=1$ are kept.
RR and NSNS sectors correspond to the
twist parameters $(a,\widetilde{a})=(0,0)$ and
$({1\over 2},-{1\over 2})$ respectively, and
the GSO operator $(-1)^F$ is $\e^{-\pi i(s+\widetilde{s})/2}$
on the subspace $B_{j,n,s}\otimes B_{j,-n,\widetilde{s}}$.
We therefore keep only the subspaces
$$
B_{j,n,s}\otimes B_{j,-n,\widetilde{s}},
\qquad
\mbox{with $s,\widetilde{s}\in \Z$,~ $s+\widetilde{s}=0$ mod $4$,~
and $2j+n-s$ even.}
$$
Note that
$(-1)^F$ acts only on the Dirac fermions
$\Psi=(\psi_{\pm},\bpsi_{\pm})$ and the GSO projection of the latter system
is equivalent to the rational $U(1)$ at level $2$,
the the circle sigma model of radius $R=\sqrt{2}$.
Thus the GSO projection of the full minimal model can be regarded as
the $SU(2)_k\times U(1)_2$ mod $U(1)$ gauged WZW model.
From this point of view, it is natural to group the spaces as
$$
\Scr{H}_{j,n,s}=\bigoplus_{p\in\Z}B_{j,n,s+4p},
$$
where $s$ is now regarded as a mod $4$ integer.
The character $\chi_{jns}(\tau,u)$ that appears in (\ref{defcha})
is simply the trace on this space.
The Hilbert space of states of the GSO projected theory is expressed as
\beq
{\cal H}^{\rm GSO} = \bigoplus_{(j,n,s)\in \Mk}
\Scr{H}_{j,n,s} \otimes \Scr{H}_{j,-n,-s},
\eeq
where $\Mk$ is the set of $(j,n,s)\in \Pk\times\Z\times \Z_4$
modulo $\pi_1(H)\cong\Z$, or more explicitly
$$
\Mk={\left\{\,\,(j,n,s)\in \Pk\times\Z_{2(k+2)}\times \Z_4\,\,\Bigl|\,\,
2j+n+s\,\,\mbox{even}\,\,
\right\}\over (j,n,s)\sim({k\over 2}-j,n+(k+2),s+2)}.
$$
In this subsection $n$ and $s$ are thus mod $2(k+2)$ and mod $4$ integers
that label $U(1)_{k+2}$ and $U(1)_2$ RCFT.
We usually assume them to be in the standard range 
$-k-1,\dots, k+2$ and $-1,0,1,2$, and
addition
modulo $2k+4$ (or modulo $4$ for $s$) is denoted by the symbol $\hat{+}$.
We often put hat $\widehat{n}$, $\widehat{s}$ to
the $U(1)_{k+2}$ and $U(1)_2$ labels $n$, $s$, to stress that they are
brought into the respective standard ranges.

\subsubsection{RCFT aspects of the theory}

\subsubsection*{\it Modular matrices}

The $S$- and $T$-matrices of the coset model have the factorized form
\beq
S_{(j,n,s)(j',n',s')} = 2 \ S_{jj'} \ S^*_{nn'} \ S_{ss'},~~~
T_{(j,n,s),(j',n',s')}=T_{jj'}T^*_{nn'} T_{ss'},
\label{factorPF}
\eeq
where it is understood that the matrices with
pure $j$ labels are those of the $SU(2)_k$
WZW model, matrices with pure $n$ or pure $s$ labels are
those of $U(1)_{k+2}$ or $U(1)_2$. 
Using this factorization property, we find
\beq
N_{(j,n,s)(j',n',s')}^{\,\,\,(j'',n'',s'')}
=N_{jj'}^{\,\,j''}\delta_{n+n',n''}^{(2k+4)}\delta_{s+s',s''}^{(4)}
+N_{jj'}^{\,\,\frac{k}{2}-j''}\delta_{n+n',n''+k+2}^{(2k+4)}\delta_{s+s',s''+2}^{(4)}.
\label{paraN}
\eeq
We also need to have expressions for
$P=\sqrt{T} S T^2 S \sqrt{T}$ and
$Y_{ab}^c=\sum_dS_{ad}P_{bd}P_{cd}^*/S_{0d}$ \cite{BSII,PSS}.
$P$ relates the open and closed
string channel of the M\"obius strip and
$Y$ appears in the loop channel of the M\"obius strip and Klein bottle.
For the computation, it is useful to consider
$$
Q=ST^2S, \quad 
\tilde{Y}_{ab}^c = \sum_d \frac{S_{ab} Q_{bd}Q^*_{cd}}{S_{0d}}
=\sqrt{\frac{T_c}{T_b}} \ Y_{ab}^c.
$$
Thanks to the factorization of $S$ and $T$, we find
\beqa
&&
Q_{(j,n,s)(j',n',s')} = Q_{jj'} Q_{nn'}^* Q_{ss'}+ 
Q_{\frac{k}{2}-j,j'} Q_{n\hat{+}(k+2),n'}^* Q_{s\hat{+}2,s'}
\\
&&
\tilde{Y}_{(j,n,s)(j',n',s')}^{(j'',n'',s'')} = \tilde{Y}_{jj'}^{j''} \
\overline{\tilde{Y}}_{nn'}^{n''} \tilde{Y}_{ss'}^{s''}+ 
\tilde{Y}_{jj'}^{\frac{k}{2}-j''} \
\overline{\tilde{Y}}_{nn'}^{n''+k+2} \tilde{Y}_{ss'}^{s''+2}.
\eeqa
 From this, one can compute $P=\sqrt{T}Q\sqrt{T}$ and
$Y_{ab}^{c}=\sqrt{T_b/T_c}\tilde{Y}_{ab}^c$, using the
following expressions for $\sqrt{T}$ in the coset model
\beq\label{roott}
\sqrt{T_{j,n,s}} = \sigma_{j,n,s}\sqrt{ T_j T^*_n T_s},
\eeq
where $\sigma$ is a sign factor defined by this equation and
explicitly computed in Appendix~\ref{app:weights}.

\subsubsection*{\it Discrete symmetries}

The group of simple currents is given by the primaries
$(0,n,s)$. For odd $k$ the symmetry group is $\Z_{4k+8}$ and
is generated by $(0,1,1)$. For even $k$ it is $\Z_{2k+4} \times \Z_2$,
generated by $(0,1,1)$ and $(0,0,2)$.
The monodromy charge of the field $(j,n,s)$ under
the simple currents is 
\beq
Q_{n,s}(j',n',s') = \frac{nn'}{2(k+2)} - \frac{ss'}{4} \ \ {\rm mod} \ 1 \ .
\eeq
Accordingly, there is a symmetry action on states such that
\beq
g_{n,s}=\e^{\pi i \left( \frac{nn'}{k+2} -\frac{ss'}{2}
\right)}\quad\mbox{on $\Scr{H}_{j',n',s'}\otimes\Scr{H}_{j',-n',-s'}$}
\eeq
$g_{1,1}$ corresponds to the generator
$a=\e^{-\pi i J_0}$ of the axial rotations in the gauged WZW model;
$g_{0,2}$ is the element $(-1)^{\widehat{F}}$ that distinguishes the RR and
NSNS sectors.

\subsubsection*{\it Orbifold}

We consider the orbifold by the subgroup $\Z_{k+2} \times \Z_2$
generated by the currents $g_{2,0}$ and $g_{0,2}$,
whose space of states is 
\beq
{\cal H}^M = \bigoplus_{(j,n,s)\in \Mk}
\Scr{H}_{j,n,s} \otimes \Scr{H}_{j,n,s}.
\label{orbi}
\eeq
This can be regarded as the mirror of the original model.
The mirror map $\Psi: {\cal H} \to {\cal H}^M$ 
acts on states as
$\Psi=V_M \otimes 1: 
\ket{j,n,s} \otimes \ket{j,-n,-s} \to  \ket{j,-n,-s} \otimes \ket{j,-n,-s}$.

\subsubsection{A-type parities}

We now turn to the construction of the standard PSS parities and crosscaps,
which we shall call A-type crosscaps.
For each simple current $(0,n,s)$ there is a crosscap state
given by
\beq
\ket{\Scr{C}_{n,s}} = \sum_{(j',n',s')\in \Mk}
\frac{P_{(0,n,s)(j',n',s')}}{\sqrt{S_{(0,0,0)(j',n',s')}}} 
\ \cket{j',n',s'}.
\eeq
Explicit expressions for the A-type crosscap states can be found
in Appendix~\ref{app:cc}.
Note that the crosscap states with
$n$, $s$ even contain only Ishibashi states in the NSNS-sector, whereas
those with $n$, $s$ odd contain only Ishibashi states
in the RR-sector.
One can compute the Klein bottle amplitudes using the $Y$-tensor.
The result is
\beqa
\lefteqn{\langle \Scr{C}_{\bar{n},\bar{s}}|\e^{-{\pi i\over 2\tau}H}
|\Scr{C}_{n,s}\rangle
=}
\nn\\&&
\sigma_{0,\bar{n},\bar{s}}\sigma_{0,n,s}\delta^{(2)}_{\bar{n}+n}
 \delta^{(2)}_{\bar{s} +s}
\Biggl\{\,
\sum_{2j+\frac{\bar{n}-n}{2}+ \frac{\bar{s}-s}{2} \, {\rm even}}  
(-1)^{2j}(\chi_{j,\frac{\bar{n}-n}{2}, \frac{\bar{s}-s}{2}}
+ (-1)^s \chi_{j,\frac{\bar{n}-n}{2}, \frac{\bar{s}-s}{2}+2})(2\tau)
\nn\\
&&\qquad\qquad
 + \delta_k^{(2)} \e^{\frac{\pi i}{2} (-\bar{n}+\bar{s})}
( \chi_{\frac{k}{4}, \frac{\bar{n}-n}{2}+\frac{k+2}{2}, 
\frac{\bar{s}-s}{2}+1} + (-1)^s 
\chi_{\frac{k}{4}, \frac{\bar{n}-n}{2}+\frac{k+2}{2}, \frac{\bar{s}-s}{2}-1}
) (2\tau) \Biggr\}.
\nn\\
\label{atypekb}
\eeqa
For $n=\bar{n}$ and $s=\bar{s}$, this expression simplifies to
\beqa
\lefteqn{\langle \Scr{C}_{n,s}|\e^{-{\pi i\over 2\tau}H}
|\Scr{C}_{n,s}\rangle}
\nn\\
&=& \sum_{j \in \Z }  
(\chi_{j,0,0}+ (-1)^s\chi_{j,0,2}) (2\tau) 
 + \delta_k^{(2)}\e^{\frac{\pi i}{2} (-n+s)}
(\chi_{\frac{k}{4}, \frac{k+2}{2}, 1}+ (-1)^s
\chi_{\frac{k}{4}, \frac{k+2}{2}, -1})(2\tau).
\nn
\eeqa
The crosscaps correspond to involutive parity symmetries
$P_{n,s}$ which are related among themselves as
$$
P_{n,s}=g_{n,s}P_{0,0}.
$$
The above expression for
the Klein bottle function
$\Tr P_{n,s}q^H=\langle \Scr{C}_{n,s}|q_t^H
|\Scr{C}_{n,s}\rangle$
is consistent with this relation.

\subsubsection{B-type parities}

Another class of crosscap states can be constructed
from A-type crosscap states in
the $G=\Z_{k+2} \times \Z_2$ orbifold model, 
with an application of the mirror map.
We shall call them B-type crosscaps.
We refer to \cite{BH1} for notation and conventions.
For an RCFT ${\cal C}$ with
the charge conjugation modular invariant partition function
there are $|G|$  A-type
crosscap states in the orbifold theory for each $G$-orbit of simple currents.
These crosscap states are given by
\beq\label{orbicross}
\ket{\Scr{C}_{P^{\theta}_{g'}}}^{{\cal C}/G}
= \frac{e^{i\omega_{g'}}}{\sqrt{|G|}} \sum_{g\in G}
e^{-\pi i (\theta(g)-\hat{Q}_{g'}(g))} \ket{\Scr{C}_{P_{gg'}}}^{{\cal C}},
\eeq
where $\hat{Q}_g(i):= h_g+h_i-h_{g(i)}$
and $g'$ is a fixed representative
of a simple current orbit.
$\theta$ is a solution to the constraint equation
\beq
\theta(g_1g_2)= \theta(g_1) + \theta(g_2) -\hat{Q}_{g_2}(g_1)
+2q(g_1, g_2) \quad \quad {\rm mod} \, 2
\eeq
where $q$ is a symmetric bilinear form of $G$
that determines the orbifold theory.
($q$ is a form obeying $q(g,g)=-h_g$ and 
$Q_{g_1}(g_2) = 2q(g_1,g_2)$ mod $1$.)
The crosscap state corresponds to a parity symmetry
$P^{\theta}_{g'}$, which squares to
\beq
(P_{g'}^\theta)^2 = \e^{2\pi i (\theta(g)-Q_{g'}(g))} \quad \quad
\mbox{on the $g$-twisted Hilbert space ${\cal H}_g$}.
\eeq

We apply this construction to
the orbifold of ${\mathcal C}=SU(2)_k\times U(1)_2/U(1)_{k+2}$
by $G=\Z_{k+2}\times\Z_2$ with the Hilbert space (\ref{orbi}).
This is the orbifold with respect to
the bilinear form given by
\beq
q(g_{n,s},g_{n',s'}) = \frac{nn'}{4(k+2)} - \frac{ss'}{8}.
\eeq
($g_{n,s}$ is in $G$ if $n$ and $s$ are both even.)
We first need expressions modulo $2$ for $\hat{Q}_g(h)$.
The conformal weight of a simple current
$(0,n,s)$ with $(0,n,s) \neq (0,\pm 1, \mp 1)$ is given by
\beq
h_{(0,n,s)} = -\frac{n^2}{4(k+2)} + \frac{s^2}{8} + \frac{|n|-|s|}{2}
\quad {\rm for} \quad (0,n,s) \neq (0,\pm 1, \mp 1).
\eeq
Using this we find
\beqa
\hat{Q}_{g_{n,s}}(g_{n',s'}) &=& h_{(0,n,s)} + h_{(0,n',s')}
- h_{(0,n\hat{+} n', s\hat{+} s')} \\ \no
&=& -\frac{n^2}{4(k+2)} + \frac{s^2}{8}
-\frac{(n')^2}{4(k+2)} + \frac{(s')^2}{8}
+\frac{(n\hat{+} n')^2}{4(k+2)} - \frac{(s\hat{+} s')^2}{8} \\ \no
&& ~~~~+\frac{1}{2} (|n| - |s| + |n'| - |s'| - |n\hat{+} n'| + |s \hat{+}s'|) \\ \no
&=&
\frac{nn'}{2(k+2)} - \frac{ss'}{4} - \frac{n\hat{+} n'}{2}
+ \frac{n+n'}{2} + \frac{s\hat{+} s'}{2} - \frac{s+s'}{2} \\ \no
&& ~~~~+  \frac{1}{2} (|n| - |s| + |n'| - |s'| - |n\hat{+} n'| + |s \hat{+}s'|)
\eeqa
In the last step, we have used  $n,n',s,s'$ even.
We thus conclude that
\beq
\hat{Q}_{g_{n,s}}(g_{n',s'}) = \frac{nn'}{2(k+2)} - \frac{ss'}{4}
\quad {\rm mod} \, 2,
\eeq
in particular $\hat{Q} = 2q$ mod $2$. Therefore, we obtain a homogeneous
equation for $\theta$, $\theta(gh) = \theta(g)+ \theta(h)$, whose
solutions are given by
\beq
\theta_{rq} (g_{n,s}) = -\frac{rn}{k+2} + \frac{qs}{2} \ .
\eeq
The set of simple currents $(0,n,s)$ splits up into two orbits
under the orbifold group $\Z_{k+2} \times \Z_2$. The first orbit is the one
of $(0,0,0)$, which contains only currents $(0,n,s)$ with $n,s$ even; the
other orbit is the one of $(0,1,1)$, which contains only currents with
$n,s$ odd. Accordingly, there are two types of crosscap states. 
Following the general procedure, one first constructs
A-type crosscaps in the orbifold and then applies the
mirror map. These steps are performed in the appendix, and here we
merely list the results. 
B-type crosscap states are labelled by an element $(r,q)\in \Z_{k+2}\times
\Z_2$ and an orbit label $p$ which can take the values $0$ and $1$.
They are given by
\beq
\ket{\Scr{C}_{rqp}}= (2(k+2))^{\frac{1}{4}} \sum_j 
\sigma_{j,-2r-p,-2q-p}
\frac{P_{j\frac{k}{2}}}{\sqrt{S_{0j}}} 
(-1)^{\frac{\widehat{2r+p}-p}{2}+q} 
\cket{j,2r+p,2q+p}_B,
\label{Crqp}
\eeq
where $P_{j\frac{k}{2}}$ is the $P$-matrix of the $SU(2)_k$-theory.
These states are elements of the sector twisted by $g_{4r+p,2p}$.
Hence, the square of the parity action $P_{rqp}$ is given as
\beq\label{Psquare}
P_{rqp}^2 = g_{4r+2p, 2p}.
\eeq

One can compute the Klein bottle amplitudes using the average formula
(\ref{orbicross}) and the results
for A-type Klein bottles (\ref{atypekb}).
The result is 
\beq\label{rcftbkb}
\langle \Scr{C}_{rqp}
|e^{-\frac{\pi i}{2\tau}H}\ket{\Scr{C}_{rqp}}  =
\sum_{(j,n,s)\in \Mk}
e^{\pi i \frac{\widehat{(2r+p)}n}{k+2}} \ 
\e^{-\pi i \frac{\widehat{(2q+p)}s}{2}}
\chi_{j,n,s}(2\tau) 
\eeq
We also note that
$$
\langle \Scr{C}_{r,q,p}
|e^{-\frac{\pi i}{2\tau}H}\ket{\Scr{C}_{r,q+1,p}} =0,
$$
because $\ket{\Scr{C}_{r,q,p}}$
and $\ket{\Scr{C}_{r,q+1,p}}$ belong to orthogonal subspaces.
The expression (\ref{rcftbkb}) for the Klein bottle function
$\Tr P_{rqp}q^H=\langle \Scr{C}_{rqp}
|q_t^H\ket{\Scr{C}_{rqp}} $ implies the following relations
among the parities,
\beq
P_{rqp}=g_{2r+p,2q+p}P_{000},
\label{procla}
\eeq
at least in the action on closed string states.
This is also consistent with the square formula
(\ref{Psquare}).

\subsection{Crosscaps in the theory before GSO projection}
\label{sub:beGSO}

In the previous subsection, we  obtained the crosscaps
for the theory in which the non-chiral GSO projection is imposed.
In this subsection, we use this result to reconstruct the
crosscaps in the theory before the GSO projection.
This enables us to compute the overlaps with the supersymmetric
ground states, as well as to reproduce the Witten index with a twist
by supersymmetric parities.

\newcommand{\whF}{\widehat{F}}

GSO projection is in a sense an {\it orbifold} by the symmetry
$(-1)^F$ where NSNS and RR sectors are regarded as the untwisted and
twisted sectors respectively.
Thus, one can find the relation of the crosscaps
before and after the GSO projection by following the argument
used in finding the relation of crosscaps before and after orbifolding
\cite{BH1}.
In this subsection, we shall refer to the theory before GSO projection
simply as `the theory' or `the ${\mathcal N}=2$ theory'
and the theory after GSO projection as
`the (GSO) projected theory' or `the RCFT'.
We shall also put a superscript `GSO' to the space of states, the crosscaps,
etc, of the projected theory.
Let $P$ be a parity of the theory, and consider the parity of
the projected theory induced from $P$. The twisted partition
function of the latter is
$$
\Tr\!\!\mathop{}_{{\mathcal H}^{\rm GSO}}\!\!
Pq^H
=\half\Tr\!\!\mathop{}_{{\mathcal H}_{\rm NSNS}}\!\!
\left((1+(-1)^F)Pq^H\right)
+\half\Tr\!\!\mathop{}_{{\mathcal H}_{\rm RR}}\!\!
\left((1+(-1)^F)Pq^H\right).
$$
Using (\ref{idNSNS}) and (\ref{idRR}), the four terms of the
right hand side can be expressed as
$$
\half\langle\Scr{C}_{(-1)^FP}|q_t^H|\Scr{C}_{(-1)^FP}\rangle
+\half\langle\Scr{C}_{P}|q_t^H|\Scr{C}_{P}\rangle
+\half\langle\Scr{C}_{P}|q_t^H|\Scr{C}_{(-1)^FP}\rangle
+\half\langle\Scr{C}_{(-1)^FP}|q_t^H|\Scr{C}_{P}\rangle.
$$
This shows that the crosscap of the projected theory
is given by
$$
|\Scr{C}_P\rangle^{\rm GSO}
=\e^{i\theta}\left[
{1\over\sqrt{2}}|\Scr{C}_{(-1)^FP}\rangle
+{1\over\sqrt{2}}|\Scr{C}_{P}\rangle\right].
$$
One may also consider the parity $(-1)^{\whF}P$ of the projected theory,
where $(-1)^{\whF}$ is $1$ on NSNS sector and $-1$ on RR sector.
Repeating the same procedure, we find
$$
|\Scr{C}_{(-1)^{\whF}P}\rangle^{\rm GSO}
=\e^{i\theta'}\left[
{1\over\sqrt{2}}|\Scr{C}_{(-1)^FP}\rangle
-{1\over\sqrt{2}}|\Scr{C}_{P}\rangle\right].
$$
We would like to invert these equations
to express $|\Scr{C}_{P}\rangle$ and $|\Scr{C}_{(-1)^FP}\rangle$
as linear combinations of $|\Scr{C}_P\rangle^{\rm GSO}$
and $|\Scr{C}_{(-1)^{\whF}P}\rangle^{\rm GSO}$, which we know from
the RCFT computation of the previous section.
However, in order to find the right combination
it is important to know the phases
$\e^{i\theta}$ and $\e^{i\theta'}$, but there is no canonical way
to fix them.
Moreover, the phases of the RCFT crosscaps
$|\Scr{C}_P\rangle^{\rm GSO}$
and $|\Scr{C}_{(-1)^{\whF}P}\rangle^{\rm GSO}$
are highly non-canonical.

\subsubsection{The right combination}\label{subsub:right}

In fact, one can overcome this difficulty by making use of one independent
constraint --- the supercurrent condition.
If $P$ is an $A_{\alpha,\beta}$-parity or
a $B_{\alpha,\beta}$-parity of the ${\mathcal N}=2$ theory,
the crosscap $|\Scr{C}_P\rangle$
obeys a certain supercurrent condition, which is satisfied
only for a particular linear combination of 
$|\Scr{C}_P\rangle^{\rm GSO}$
and $|\Scr{C}_{(-1)^{\whF}P}\rangle^{\rm GSO}$.
The other crosscap, $|\Scr{C}_{(-1)^FP}\rangle$, obeys a different
condition which is satisfied by another linear combination.
In this way one can find the right linear combinations, up to an overall
phase.

In what follows we carry out this program.
It turns out that A-type (resp.~B-type) parities in the RCFT correspond
to $A_{\alpha,\beta}$-parities (resp.~$B_{\alpha,\beta}$-parities)
of the ${\mathcal N}=2$ theory.
We thus separate the discussions into the two types.

\subsubsection*{\it A-type}

The A-type crosscaps in RCFT are the PSS crosscaps
$|\Scr{C}_{n,s}\rangle$
labelled by simple currents, $(n,s)$ with $n+s$ even.
The symmetry $(-1)^{\whF}$ is nothing but the global symmetry
labelled by $(n,s)=(0,2)$. Thus, $|\Scr{C}_{n,s}\rangle$
and $|\Scr{C}_{n,s+2}\rangle$ corresponds to
$|\Scr{C}_{P_{n,s}}\rangle^{\rm GSO}$
and $|\Scr{C}_{(-1)^{\whF}P_{n,s}}\rangle^{\rm GSO}$ for a suitable
$P_{n,s}$. The task is to identify $P_{n,s}$ and find the right combination
to express $|\Scr{C}_{P_{n,s}}\rangle$
and $|\Scr{C}_{(-1)^{F}P_{n,s}}\rangle$.
Since $|\Scr{C}_{n,s}\rangle$ belongs to NSNS-sector (resp.~RR-sector)
for even $s$ (resp. odd $s$),
$P_{n,s}$ squares to $(-1)^F$ if $s$ is even and it is involutive
if $s$ is odd.

The crosscap state of an $A_{\alpha,\beta}$-parity must
obey the supercurrent condition
(\ref{coCA}). In terms of the Fourier modes, it is
\beq
G_r+i\e^{-i\alpha}(-1)^{r+{\beta\over \pi}}\overline{\tG}_{-r}=
\tG_r-i\e^{-i\alpha}(-1)^{r+{\beta\over \pi}}\bG_{-r}=0,
\quad\,
\mbox{$r\in \Z-{\beta\over \pi}$.}
\eeq
Let us put
\beq
|\Scr{C}_{n,s}(\pm)\rangle:=
{1\over \sqrt{2}}|\Scr{C}_{n,s}\rangle
\mp {1\over \sqrt{2}}{\sqrt{T_{0,n,s}}\over\sqrt{T_{0,n,s+2}}}
|\Scr{C}_{n,s+2}\rangle.
\eeq
One can show that they obey the
conditions of the types described in the following
table
(see Appendix~\ref{app:SC} for the proof):
$$
\begin{array}{|c|c|c|}
\hline
&\mbox{$s$ odd}&\mbox{$s$ even}\\
\hline
|\Scr{C}_{n,s}(+)\rangle&A_{0,0}&A_{{\pi\over 2},{\pi\over 2}}\\
\hline
|\Scr{C}_{n,s}(-)\rangle&A_{\pi,0}&A_{{\pi\over 2},-{\pi\over 2}}
\\
\hline
\end{array}
$$
We know that the $A_{0,0}$, $A_{\pi,0}$,
$A_{{\pi\over 2},-{\pi\over 2}}$, $A_{{\pi\over 2},{\pi\over 2}}$-parities
of the theory
are $a^{\it even}P_A$, $(-1)^Fa^{\it even}P_A$,
$a^{\it odd}P_A$, $(-1)^Fa^{\it odd}P_A$ respectively.
We also know that the axial symmetry $a$
induces
the global symmetry of the projected theory
labelled by $(n,s)=(1,\pm 1)$.
These are enough to show that
\beqa
&&
|\Scr{C}_{a^{2m}P_A}\rangle=|\Scr{C}_{2m-1,2m-1}(+)\rangle,
\label{CApa1}
\\[0.2cm]
&&
|\Scr{C}_{(-1)^Fa^{2m}P_A}\rangle
=|\Scr{C}_{2m-1,2m-1}(-)\rangle,
\label{CApa2}
\\[0.2cm]
&&
|\Scr{C}_{a^{2m+1}P_A}\rangle=|\Scr{C}_{2m,2m}(-)\rangle,
\label{CApa3}
\\[0.2cm]
&&
|\Scr{C}_{(-1)^Fa^{2m+1}P_A}\rangle=|\Scr{C}_{2m,2m}(+)\rangle.
\label{CApa4}
\eeqa
The overall phases are not fixed by this argument.
Here we have chosen the ones that will be justified
by later computations.

\subsubsection*{\it B-type}

B-type crosscaps in RCFT are
$|\Scr{C}_{r,q,p}^B\rangle$ given in (\ref{Crqp})
labelled by $(r,q)\in\Z_{k+2}\times \Z_2$ and $p\in\{0,1\}$.
Combining the parity with $(-1)^{\whF}$ corresponds to the shift $q\to q+1$.
Thus, we need to find the right combinations of
$|\Scr{C}_{r,q,p}^B\rangle$
and $|\Scr{C}^B_{r,q+1,p}\rangle$.
Since $|\Scr{C}^B_{r,q,p}\rangle$ belongs to
the sector twisted by the symmetry labelled by $(n,s)=(4r+2p,2p)$,
it corresponds to the parity that
squares to $a^{4r}(-1)^F$ if $p=0$ and $a^{4r+2}$ if $p=1$.
This shows that $P_B^{r,q,p}$ is induced from
$a^{2r+p}P_B$ or $(-1)^F a^{2r+p}P_B$ (or those
combined with axial R-symmetry).

The supercurrent
condition on the crosscap state of a $B_{\alpha,\beta}$-parity
is (\ref{coCB}), or in terms of the Fourier modes
\beq
G_r+i\e^{i\beta}(-1)^{r-{\alpha\over \pi}}\tG_{-r}=
\overline{\tG}_r-i\e^{i\beta}(-1)^{r-{\alpha\over \pi}}\bG_{-r}=0,
\quad\,
\mbox{$r\in \Z+{\alpha\over \pi}$.}
\eeq
Let us put
\beq
|\Scr{C}^B_{r,p}(\pm)\rangle:=
{1\over\sqrt{2}}|\Scr{C}^B_{r,0, p}\rangle\pm 
{1\over\sqrt{2}}
{\sqrt{T_{2+p}^{(2)}}\over \sqrt{T^{(2)}_{p}}}
|\Scr{C}^B_{r,1,p}\rangle,
\label{BCexp}
\eeq
where $\sqrt{T^{(2)}_{n}}=\e^{\pi i \hat{n}^2/8}$ is the square-root
of the T-matrix of rational $U(1)$ at level $2$.
One can show that $|\Scr{C}^B_{r,p}(\pm)\rangle$ obey the
conditions of the types described in the following
table
(see Appendix~\ref{app:SC} for the proof):
$$
\begin{array}{|c|c|c|}
\hline
&p=1&p=0\\
\hline
|\Scr{C}^B_{r,p}(+)\rangle&B_{0,0}&B_{{\pi\over 2},{\pi\over 2}}\\
\hline
|\Scr{C}^B_{r,p}(-)\rangle&B_{0,\pi}&B_{{\pi\over 2},-{\pi\over 2}}
\\
\hline
\end{array}
$$
Since $B_{0,0}$, $B_{0,\pi}$, $B_{{\pi\over 2},-{\pi\over 2}}$,
$B_{{\pi\over 2},{\pi\over 2}}$-parities are
$a^{\it odd}P_B$, $(-1)^Fa^{\it odd}P_B$,
$a^{\it even}P_B$, $(-1)^Fa^{\it even}P_B$, we can conclude that
\beqa
&&
|\Scr{C}_{a^{2m+1}P_B}\rangle=(-1)^m|\Scr{C}^B_{m,1}(+)\rangle,
\label{CBpa1}
\\[0.2cm]
&&
|\Scr{C}_{(-1)^Fa^{2m+1}P_B}\rangle=|\Scr{C}^B_{m,1}(-)\rangle,
\label{CBpa2}
\\[0.2cm]
&&
|\Scr{C}_{a^{2m}P_B}\rangle=(-1)^m|\Scr{C}^B_{m,0}(-)\rangle,
\label{CBpa3}
\\[0.2cm]
&&
|\Scr{C}_{(-1)^Fa^{2m}P_B}\rangle=|\Scr{C}^B_{m,0}(+)\rangle.
\label{CBpa4}
\eeqa
Again, the overall phases cannot be fixed by this argument.
The choice we made will be justified by later computations.

\subsubsection{Overlaps with supersymmetric ground states}

Using the above results, one can compute the overlaps of the
supersymmetric crosscaps and supersymmetric ground states.

\subsubsection*{\it A-parities}

To compute the overlaps of the crosscap for
the basic A-parity $P_A$ and the RR ground states
$|j\rangle^{}_{{}_{\rm RR}}$ one has to read off the coefficient of
$\cket{j,-2j-1,-1}$ in the expansion of
$|\Scr{C}_{-1,-1}(+)\rangle$.
Since $\sqrt{T_{0,-1,-1}/T_{0,-1,1}}=-1$ by supersymmetry,
the latter is $(|\Scr{C}_{-1,-1}\rangle+|\Scr{C}_{-1,1}\rangle)/\sqrt{2}$.
The result is
\beq
{}^{}_{{}_{\rm RR}}\langle j|\Scr{C}_{P_A}\rangle
=\sqrt{2\over (k+2)\sin\bigl({\pi(2j+1)\over k+2}\bigr)}
\e^{\pi i(2j+1)\over 2(k+2)}
\left\{
-i\delta_{2j+k}^{(2)}\sin\left(\mbox{${\pi(2j+1)\over 2(k+2)}$}\right)
+\delta^{(2)}_{2j}\cos\left(\mbox{${\pi(2j+1)\over 2(k+2)}$}\right)
\right\}.
\label{ovPj}
\eeq
Also, we find
\beq
\langle \Scr{C}_{(-1)^FP_A}|j\rangle^{}_{{}_{\rm RR}}
=\sqrt{2\over (k+2)\sin\bigl({\pi(2j+1)\over k+2}\bigr)}
\e^{-\pi i(2j+1)\over 2(k+2)}
\left\{
\delta_{2j+k}^{(2)}\sin\left(\mbox{${\pi(2j+1)\over 2(k+2)}$}\right)
+i\delta^{(2)}_{2j}\cos\left(\mbox{${\pi(2j+1)\over 2(k+2)}$}\right)
\right\}.
\label{ovFPj}
\eeq
For other A-parities the overlaps can be easily obtained by
using $|\Scr{C}_{a^{2m}P_A}\rangle=\pm a^m|\Scr{C}_{P_A}\rangle$,
\beqa
&&{}^{}_{{}_{\rm RR}}\langle j|\Scr{C}_{a^{2m}P_A}\rangle
=\pm\e^{-2\pi i m ({2j+1\over k+2}-{1\over 2})}
{}^{}_{{}_{\rm RR}}\langle j|\Scr{C}_{P_A}\rangle,
\nn\\
&&\langle \Scr{C}_{(-1)^Fa^{2m}P_A}|j\rangle^{}_{{}_{\rm RR}}
=\pm\e^{2\pi i m ({2j+1\over k+2}-{1\over 2})}
\langle \Scr{C}_{(-1)^FP_A}|j\rangle^{}_{{}_{\rm RR}}.
\nn
\eeqa

\subsubsection*{\it B-parities}

Since the B-parity $a^{2m+1}P_B$
squares to $a^{2(2m+1)}$, the crosscap state
belongs to the sector with the twist parameter $(a,\ta)=(2m+1,0)$.
Such a sector has a $(2,2)$ supersymmetry and has
a unique supersymmetric ground state $|{\rm G}\rangle_{2m+1,0}$
which is an element of
$B_{j_*,-2j_*-1,-1}\otimes B_{j_*,-2j_*-1,-1}$
where $j_*\in \Pk$ is defined by
$2j_*+1\equiv -(a+\ta)=-(2m+1)$ mod $(k+2)$.
Namely,
$$
j_*=\left\{
\begin{array}{ll}
{k\over 2}-m&\mbox{if\, $m=0,1,...,[{k\over 2}]$,}
\\
k+1-m&\mbox{if\, $m=[{k\over 2}]+1,...,k+1$.}
\end{array}\right.
$$
In the two cases, $(j_*,-2j_*-1,-1)$ are equivalent to
$(m,2m+1,1)$ and $(k+1-m,2m+1,-1)$ respectively.
We are interested in the overlaps of this ground state
$|{\rm G}\rangle_{a,\ta}$ and the crosscap states
$$
\left.
\begin{array}{l}
|\Scr{C}_{a^{2m+1}P_B}\rangle(-1)^m\\
|\Scr{C}_{(-1)^Fa^{2m+1}P_B}\rangle
\end{array}
\right\}
={1\over\sqrt{2}}|\Scr{C}_{m,0,1}^B\rangle
\pm {1\over\sqrt{2}}|\Scr{C}_{m,1,1}^B\rangle.
$$
The overlaps are obtained by reading
the coefficient of
$\cket{m,2m+1,1}$ or $\cket{k+1-m,2m+1,-1}$ depending on
$m\equiv 0,1,...,[{k\over 2}]$
or $m\equiv [{k\over 2}]+1,...,k+1$ (mod $(k+2)$).
The result is
\beq
{}_{2m+1,0}\langle {\rm G}|\Scr{C}_{a^{2m+1}P_B}\rangle
=\left\{
\begin{array}{ll}
(-1)^m\sqrt{\cot\left({\pi(2m+1)\over 2(k+2)}\right)}
&m\equiv 0,1,...,[{k\over 2}],\\
(-1)^{m+k+1}\sqrt{-\cot\left({\pi(2m+1)\over 2(k+2)}\right)}
&m\equiv [{k\over 2}]+1,...,k+1,
\end{array}
\right.
\eeq
and
\beq
{}_{2m+1,0}\langle{\rm G}|\Scr{C}_{(-1)^Fa^{2m+1}P_B}\rangle
=\left\{
\begin{array}{ll}
\sqrt{\cot\left({\pi(2m+1)\over 2(k+2)}\right)}
&m\equiv 0,1,...,[{k\over 2}],\\
(-1)^{k}\sqrt{-\cot\left({\pi(2m+1)\over 2(k+2)}\right)}
&m\equiv [{k\over 2}]+1,...,k+1.
\end{array}
\right.
\eeq
This is sufficient to show that
the twisted Witten index is
\beq
I_{a^{2m+1}P_B}
=\langle\Scr{C}_{(-1)^Fa^{2m+1}P_B}|q_t^H|\Scr{C}_{a^{2m+1}P_B}\rangle
=(-1)^m\cot\left({\pi(2m+1)\over 2(k+2)}\right),
\label{IBparity2}
\eeq
which reproduces the loop channel
result (\ref{IBparity}), provided $\varepsilon_{{}_{\rm RR}}=1$.

\subsubsection{Partition function and Witten index}

The expressions for the crosscap states obtained above
can now be used to compute the parity-twisted partition functions,
or equivalently Klein bottle amplitudes,
$\Tr_{{}_{\rm NSNS}}(Pq^H)
=\langle\Scr{C}_{(-1)^FP}|q_t^H|\Scr{C}_{(-1)^FP}\rangle$,
$\Tr_{{}_{\rm RR}}(Pq^H)
=\langle\Scr{C}_{(-1)^FP}|q_t^H|\Scr{C}_{P}\rangle$.

\subsubsection*{\it A-type}

We first evaluate those amplitudes for A and $\widetilde{A}$-parities
$P=a^{\ell}P_A, (-1)^Fa^{\ell}P_A$.
Since each crosscap is a sum of two RCFT crosscaps,
the partition function is a sum of four terms.
The summands are
\beqa
&&
\langle \Scr{C}_{\ell-1,\ell-1}|q_t^H|\Scr{C}_{\ell-1,\ell-1}\rangle
=\sum_{j\in\Z}(\chi_{j00}-(-1)^{\ell}\chi_{j02})
+\delta^{(2)}_k(\chi_{{k\over 4},{k+2\over 2},1}-(-1)^{\ell}
\chi_{{k\over 4},{k+2\over 2},-1})
\nn\\
&&
\langle\Scr{C}_{2m-1,2m-1}|q_t^H|\Scr{C}_{2m-1,2m+1}\rangle
=0,
\nn\\
&&
\langle\Scr{C}_{2m,2m}|q_t^H|\Scr{C}_{2m,2m+2}\rangle
=-{\sigma_{0,2m,2m}\over\sigma_{0,2m,2m+2}}
\sum_{j\in\Z+{1\over 2}}(\chi_{j,0,1}+\chi_{j,0,-1}),
\nn
\eeqa
where the argument of the characters are all $2\tau$.
Using these formulae and also the relation
$\sqrt{T_{0,2m,2m}\over T_{0,2m,2m+2}}
{\sigma_{0,2m,2m}\over\sigma_{0,2m,2m+2}}
=\sqrt{T^{(2)}_{2m}\over T^{(2)}_{2m+2}}=-i(-1)^m$,
we find
\beqa
&&\Tr_{{}_{\rm NSNS}}(-1)^{\nu F}a^{\ell}P_Aq^H
=\sum_{j\in\Z}(\chi_{j00}-(-1)^{\ell}\chi_{j02}),
\\
&&\Tr_{{}_{\rm RR}}(-1)^{\nu F}a^{\ell}P_Aq^H
=-\e^{\pi i\ell\over 2}(-1)^{\nu}\sum_{j\in\Z+{1\over 2}}
(\chi_{j,0,1}-(-1)^{\ell}\chi_{j,0,-1})
\nn\\
&&\qquad\qquad\qquad\qquad\qquad
+\delta^{(2)}_k(\chi_{{k\over 4},{k+2\over 2},1}-(-1)^{\ell}
\chi_{{k\over 4},{k+2\over 2},-1}).
\eeqa
Note that this reproduces the results
(\ref{pfNSNS1}) and (\ref{pfRR1}) obtained in the gauged WZW model,
where the constants undetermined there are now fixed as
\beqa
&&\epsilon_{{}_{\rm NSNS}}=1,
\nn\\
&&\epsilon_{{}_{\rm RR}}=1,
\nn\\
&&\pm (-1)^{k\over 2}=1.
\nn
\eeqa
In particular, the Witten index is fixed as
\beq
I_{a^{2m}P_A}=\left\{
\begin{array}{ll}
1&\mbox{$k$ even}\\
0&\mbox{$k$ odd}.
\end{array}\right.
\label{WindMMA}
\eeq

\subsubsection*{\it B-type}

We now consider the partition functions for
B- and $\widetilde{B}$-parities.
The computation is simpler here
since the pairings of crosscaps of different $q$ (for the same $r,p$)
vanish, $\langle\Scr{C}^B_{r,q,p}|\Scr{C}^B_{r,q+1,p}\rangle=0$.
Using the formula (\ref{rcftbkb}),
we find
\beqa
\Tr_{{}_{\rm NSNS}}(-1)^{\nu F}a^{2m+p}P_Bq^H
&=&
{1\over 2}\langle \Scr{C}_{m,0,p}|q_t^H|\Scr{C}_{m,0,p}\rangle
+{1\over 2}\langle \Scr{C}_{m,1,p}|q_t^H|\Scr{C}_{m,1,p}\rangle
\nn\\
&=&
\sum\e^{\pi i {(2m+p)n\over k+2}}{\e^{-\pi i {ps\over 2}}
+\e^{\pi i {(p+2)s\over 2}}\over 2}
\chi_{jns}(2\tau)
\nn\\
&=&
\sum_{s\,{\rm even}}\e^{\pi i (2m+p)({n\over k+2}-{s\over 2})}
\chi_{jns}(2\tau),
\\
\Tr_{{}_{\rm RR}}(-1)^{\nu F}a^{2m+p}P_Bq^H
&=&
{(-1)^{m}\over 2}\langle \Scr{C}_{m,0,p}|q_t^H|\Scr{C}_{m,0,p}\rangle
-{(-1)^m\over 2}\langle \Scr{C}_{m,1,p}|q_t^H|\Scr{C}_{m,1,p}\rangle
\nn\\
&=&
(-1)^m\sum\e^{\pi i {(2m+p)n\over k+2}}{\e^{-\pi i {ps\over 2}}
-\e^{\pi i {(p+2)s\over 2}}\over 2}
\chi_{jns}(2\tau)
\nn\\
&=&
\sum_{s\,{\rm odd}}\e^{\pi i (2m+p)({n\over k+2}-{s\over 2})}
\chi_{jns}(2\tau).
\eeqa
This reproduces the results (\ref{pfNSNS2}) and (\ref{pfRR2}) we obtained in 
the gauged WZW model.
In particular, the undetermined coefficients there are determined now as
\beqa
&&\varepsilon_{{}_{\rm NSNS}}=1,
\nn\\
&&\varepsilon_{{}_{\rm RR}}=1.
\nn
\eeqa
The Witten index $I_{a^{2m+1}P_B}$ is given by (\ref{IBparity2}).

\section{Orientifold of ${\mathcal N}=2$ Minimal Models II --- Open Strings}
\label{sec:OMMII}

This is a continuation of the previous section.
We include D-branes into the discussion.

\subsection{Facts on D-branes in ${\mathcal N}=2$ minimal models}

We start with reviewing some known facts about D-branes in the minimal model
\cite{MMS}.

\subsubsection{A geometrical picture}

It is convenient to provide a basic geometrical picture of the branes.
As we have seen, the $SU(2)$ mod $U(1)$ supersymmetric gauged
WZW model can be interpreted as the sigma model on the disk $|z|\leq 1$
with the metric $\dd s^2=k|\dd z|^2/(1-|z|^2)$, with non-trivial dilaton.
As in ordinary sigma models, A-branes are wrapped on Lagrangian submanifolds
(1-dimension) and B-branes are at complex submanifolds
(0 or 2-dimensions).

\subsubsection*{\it A-branes}

A-branes are denoted as $\Scr{B}_{j,n,s}$ where $(j,n,s)\in\Mk$.
There are $2k+4$ special points at the boundary
of the disk. They fall into two classes, the even and the odd points.
The A-branes are D1 branes which stretch between these special
points. The branes with $s$ even extend between the even points, while
those with $s$ odd extend between the odd points.
More precisely, the brane $\Scr{B}_{jns}$ stretches between
boundary points
\beq
z_i=\e^{{\pi i\over k+2}(n-2j-1)}
\quad {\rm and} \quad
z_f=\e^{{\pi i\over k+2} (n+2j+1)}
\label{inifi}
\eeq
For $s=0,-1$ the orientation of the brane is from $z_i$ to $z_f$,
and for $s=1,2$ the other way around meaning that $s\to s+2$
is an orientation flip.
The corresponding boundary condition on the right boundary of the string
preserves the combination $\bQ_+-(-1)^sQ_-$
of the supersymmetry.
The boundary condition on the left boundary
preserves the opposite
combination $\bQ_++(-1)^sQ_-$ for the standard reason.
The axial rotation $a$ is a rotation of the disc by
angle $\pi/(k+2)$ and thus rotates the brane as
$n\to n+1$.

\subsubsection*{\it B-branes}

There are unoriented B-branes denoted as $\Scr{B}^B_{[j,s]}$, where
$j\in \Pk$,
$j<[{k\over 4}]$ and
$s\in \Z_2$.
They are located at concentric smaller disks,
whose radius depends on the label $j$. The $j=0$ states represent
D0 branes at the center of the disk whereas the higher $j$
states correspond to D2-branes.
For even $k$, there are also oriented
B-branes $\Scr{B}^{B}_{{k\over 4},s}$
where $s\in \Z_4$.
They are D2 branes wrapping the whole disk. $s\to s+2$ corresponds to
an orientation flip.
As before, the corresponding boundary condition
on the right boundary
preserves $Q_+-(-1)^sQ_-$, while on the left boundary
the opposite
combination $Q_++(-1)^sQ_-$ is preserved.
The axial rotation $a$ does the exchanges
$\Scr{B}^B_{[j,0]}\leftrightarrow \Scr{B}^B_{[j,1]}$
and also $\Scr{B}^{B}_{{k\over 4},s}\leftrightarrow
\Scr{B}^{B}_{{k\over 4},s\pm 1}$.\footnote{Note
that `orientation of branes' is defined in the GSO projected theory, while
the `axial rotation $a$' is defined before the GSO.
This and the above paragraphs contain a certain abuse of language.
This is the reason for the arbitrariness in, say, the ``action of
$a$ on the branes.''
There is of course a well-defined description in both
before and after GSO, separately (as given in more detail
in the following discussions).}

\subsubsection{Boundary states and one-loop amplitudes}

We next provide the boundary states and cylinder amplitudes,
both in the GSO projected theory.

\subsubsection*{\it A-branes}

In the RCFT, the A-branes $\Scr{B}_{j,n,s}$ are described by
the Cardy states \cite{Cardy}
$$
|\Scr{B}_{j,n,s}\rangle
=\sum_{(j'n's')\in \Mk}
{S_{(j,n,s)(j',n',s')}\over
\sqrt{S_{(0,0,0)(j',n',s')}}}\kket{j',n',s'}.
$$
Under the global symmetry $g_{n,s}$, the branes are transformed as
$$
g_{n,s}:\Scr{B}_{j,n',s'}
\to \Scr{B}_{j,n'+n,s'+s}.
$$
The cylinder amplitudes are
\beq
\langle \Scr{B}_{j,n,s} | \e^{-{\pi i\over \tau}H}\ket{\Scr{B}_{j',n',s'}}
= \sum_{j''\in\Pk} N_{jj'}^{j''}
\ \chi_{j'',n-n',s-s'}(\tau)
\label{cylAbrane}
\eeq
This shows that the open string Hilbert space is the sum of
$\Scr{H}_{j'',n-n',s-s'}$ where $j''$ runs over $\Pk$ such that
$N_{jj'}^{j''}=1$.

\subsubsection*{\it B-branes}

B-type boundary states are obtained by taking the $\Z_{k+2}
\times \Z_2$ orbit of A-type boundary states, followed by an
application of the mirror map. They are labelled by $(j,n,s)$ modulo
the action of the group, $(j,n,s)\sim (j,n+2,s)\sim (j,n,s+2)$.
For each $j$ there are only two orbits distinguished by $s$ even
or $s$ odd. $n$ is then even or odd as required by the selection rule.
We also note that $(j,n,s)\sim ({k\over 2}-j,n+k,s)$. Thus,
the states are labelled by $[j,s]$ where $j\leq
[{k\over 2}]$ and $s\in \Z_2$.
The boundary states are
\beqa
|\Scr{B}^B_{[j,s]}\rangle 
& = & \frac{1}{\sqrt{2k+4}} \sum_{n',r}
(V_M \otimes 1) \ |\Scr{B}_{j,n+2n', s+2r}\rangle \\ \no
&=&
(k+2)^{\frac{1}{4}} \sum_{j'} \delta^{(2)}_{2j'}
\frac{S_{j j'}}{\sqrt{S_{0j'}}}
\left( \kket{j',0,0}_B + (-1)^s \kket{j',0,2}_B \right)
\eeqa
Since there is no RR-part, the brane is unoriented.
If $k$ is even, the orbifold action has fixed points at $j={k\over 4}$.
The above boundary states for $j={k\over 4}$ should be further
resolved \cite{MMS} (see \cite{FS} for a general discussion
and \cite{PSSI,PSSII} for the original discussion in the context
of WZW models with non-diagonal modular invariant) as
\beq\label{shortorbit}
\ket{\Scr{B}_{\frac{k}{4},S}} = \frac{1}{2}\left(
\ket{\Scr{B}^B_{[{k\over 4},S]}} + \sqrt{k+2} \e^{-i\frac{\pi S^2}{2}}
\sum_{s=\pm1} \e^{-i\frac{\pi Ss}{2}} \kket{\frac{k}{4}, \frac{k+2}{2},s}_B
\right)
\eeq
Here, $S$ can take the values $-1,0,1,2$. $S\to S+2$ flips
the sign of the RR part and thus corresponds to the orientation flip.

The action of the global symmetry
$g_{n,s}$ on the branes can be read off from the boundary states.
\beq
\begin{array}{ll}
g_{1,1}:\Scr{B}^B_{[j,s]}\to \Scr{B}^B_{[j,s+1]},
&
g_{0,2}:\Scr{B}^B_{[j,s]}\to\Scr{B}^B_{[j,s]},
\\
g_{1,1}:\Scr{B}^B_{{k\over 4},S}\to
\Scr{B}^B_{{k\over 4},S-(-1)^S},
&
g_{0,2}:\Scr{B}^B_{{k\over 4},S}\to
\Scr{B}^B_{{k\over 4},S+2}.
\end{array}
\label{actPBonBb}
\eeq
In particular, the unoriented brane $\Scr{B}^B_{[j,s]}$ is invariant under
the subgroup $\Z_{k+2}\times \Z_2$ generated by
$g_{2,0}$ and $g_{0,2}$, while
the oriented brane $\Scr{B}^B_{{k\over 4},S}$ is invariant under
the subgroup $\Z_{k+2}$ generated by $g_{2,2}$.

We will later compute parity-twisted partition functions.
Since the B-parities are not always involutive
(\ref{Psquare}), we need to have the boundary state on the circles
twisted by $P_{rqp}^2=g_{4r+2p,2p}$.
All B-branes are invariant under this, and one can indeed
think about the boundary states on the circle twisted by this symmetry.
For the construction, we use the fact
\cite{BH1} that boundary states of the orbifold ${\mathcal C}/G$
on the circle twisted by a quantum symmetry $g_\rho$
associated with the character
$g\to \e^{2\pi i \rho(g)}$
are given by
\beq
\ket{\Scr{B}_{[i]}}_{g_\rho}^{{\cal C}/G} =\frac{e^{i\lambda}}{\sqrt{|G|}}
\sum_{g\in  G} \e^{-2\pi i \rho(g)} \ket{\Scr{B}_{g(i)}}^{{\cal C}}.
\eeq
Since the boundary states associated to long orbits remain invariant under the
action of the symmetry group $\Z_{k+2} \times \Z_2$, one can
consider boundary states on circles with $\Z_{k+2}\times \Z_2$ twisted
boundary conditions. To construct those states, note that the group element
$g_{2,0}^r$ is mapped to a quantum symmetry of the orbifold model
associated to the character $g_{2n,2q} \to \e^{-2\pi i \frac{rn}{k+2}}$.
Similarly, $g_{0,2}^r$ is mapped to the character $g_{2n,2q}\to
e^{\pi i rq}$. Hence, the boundary states on the twisted circles
are
\beqa
\ket{\Scr{B}^B_{[j,s]}}_{g_{2n',2s'}}&=&
\frac{e^{i\lambda}}{\sqrt{2(k+2)}}
\sum_{\bar{n},\bar{s}:\, {\rm even}} \e^{\pi i \frac{\bar{n}n'}{k+2}} \ 
e^{-\pi i \frac{\bar{s}s'}{2}} \ (V_M \otimes 1) \
\ket{\Scr{B}_{j,n\hat{+}\bar{n},s\hat{+}\bar{s}}} \\ \no
&=& (2k+4)^{\frac{1}{4}} \e^{i\lambda} \e^{-\frac{\pi i nn'}{k+2}} 
\e^{\frac{\pi i ss'}{2}}
\sum_{j'} \frac{S_{jj'}}{\sqrt{S_{0j'}}} \left(
\kket{j',n',s'}_B + (-1)^s \kket{j',n',s'+2}_B \right)
\eeqa
The boundary state is an element of the twisted Hilbert space with
twist $g_{2n',2s'}$ as indicated by the subscript. Note that
$|\Scr{B}^B_{[j,s]}\rangle_{g_{2n',2s'}} = 
|\Scr{B}^B_{[j,s]}\rangle_{g_{2(n'+k+2),2(s'+2)}}$. We now choose the $n',s'$
dependent phase $e^{i\lambda} = \e^{\pi i \frac{nn'}{k+2} }
\e^{-\frac{\pi i ss'}{2}}$, which makes the boundary state real. This
choice breaks the explicit invariance of the expression under shifts
of $n'$ by $k+2$ and of $s'$ by $2$, effectively reducing the range
of $n',s'$ to $n'\in \{ -\frac{k}{2},\dots, \frac{k+2}{2} \}$ 
and $s'\in \{ 0,1 \}$. With this choice, the equations $ \frac{2n'}{2} = n'$
and $\frac{2s'}{2} = s'$ hold exactly, not only modulo $k+2$ or $2$.
The short orbit state (\ref{shortorbit}) is only invariant under the
elements of the subgroup generated by $g_{2,2}$,
therefore, we can only construct twisted
boundary states with the corresponding twists. They are
\beq
\ket{\Scr{B}_{\frac{k}{4},S}}_{g_{2n',2n'}} = \frac{1}{2}\left(
\ket{\Scr{B}^B_{[k/4,S]}}_{g_{2n',2n'}} + 
\sqrt{k+2} \e^{-i\frac{\pi S^2}{2}}
\sum_{s=\pm 1} \e^{-i\frac{\pi Ss}{2}} \kket{\frac{k}{4}, \frac{k+2}{2}+n',s+n'}_B
\right)
\label{twshortorbit}
\eeq

\subsubsection*{\it The one-loop amplitudes}

The cylinder amplitude 
makes sense for any two boundary states on the same twisted circle.
Hence, we can consider the following amplitudes between long orbit
branes
\beq
\langle \Scr{B}^B_{[j,s]}|e^{-\frac{\pi i}{\tau}H}
|\Scr{B}^B_{[j's']} \rangle_{g_{2\bar{n},2\bar{s}}}
=\sum_{2j''+n''+s''\,{\rm even}} N_{jj'}^{j''} \,\delta^{(2)}_{s'-s+s''}
\e^{\frac{\pi i \bar{n}n''}{k+2}-\frac{\pi i \bar{s}s''}{2}} 
\chi_{j'',n'',s''}(\tau).
\label{bcyl}
\eeq
For short orbit branes, we obtain
\beqa
\lefteqn{\langle \Scr{B}^B_{\frac{k}{4},S}|e^{-\frac{\pi i}{\tau}H}
|\Scr{B}^B_{\frac{k}{4},S'}\rangle_{g_{2\bar{n},2\bar{n}}} }
\nn\\
&=& \sum_{(j,n,s)\in \Mk}
\delta_{2j}^{(2)}
\delta^{(2)}_{S-S'-s}
{1+(-1)^{\frac{2j+n-s}{2}} \e^{\frac{i\pi}{2}(S^2+S-S^{'2}-S')}\over 2}
\e^{\frac{\pi i \bar{n}n}{k+2}-\frac{\pi i \bar{n}s}{2}}
\chi_{j,n,s}(\tau).
\label{bcyl4}
\eeqa
Since the long orbit branes only have non-vanishing overlap with the
orbit part of the short orbit boundary state, one can easily obtain
the cylinder involving a long and a short orbit brane 
from (\ref{bcyl}) by setting $j'=k/4$ and dividing by two. 

\subsection{Parity Actions on D-branes and open strings}

Let us now compute the M\"obius strip amplitudes in the GSO projected theory
in order to find how the parity acts
on the D-branes and on the open string states.

\subsubsection{Geometrical picture}

Before doing the CFT computation, consider  the actions in
the geometrical picture.
In Section~\ref{subsub:geoac},
we have seen that the A-parities act as the complex conjugation
(\ref{PAdisk}),
folding the disk along the diameters,
while B-parities act as the rotations (\ref{PBdisk}), including
the identity $z\to z$ as well as the inversion $z\to -z$.
This can already tell, roughly, how these parities act on the D-branes.

Let us first look how A-branes are transformed.
The A-type parity $a^{\ell}P_A$ acts on the disk as the reflection
$z\to \e^{{\pi i\over k+2}\ell}\bz$, mapping the initial/final points
$z_i(j,n)$ and $z_f(j,n)$ of the brane $\Scr{B}_{j,n,s}$ (\ref{inifi})
as $z_i(j,n)\to z_f(j,\ell-n)$ and $z_f(j,n)\to z_i(j,\ell-n)$.
Thus, the label $(j,n)$ is mapped to $(j,\ell-n)$ and the orientation is
flipped.
If we denote the oriented segment from $z_i(j,n)$ to $z_f(j,n)$
by $\overrightarrow{L}_{j,n}$ and the one with reversed orientation
by $\overleftarrow{L}_{j,n}$, the parity maps them as
\beq
a^{\ell}P_A:\overrightarrow{L}_{j,n}
\to \overleftarrow{L}_{j,\ell-n}.
\label{geomactA}
\eeq
The B-type parity $a^{\ell}P_B$ acts on the disk as
the rotation $z\to\e^{{\pi i\over k+2}\ell}z$, mapping the initial/final
points as $z_{i,f}(j,n)\to z_{i,f}(j,n+\ell)$.
Thus, the transformation rule is
\beq
a^{\ell}P_B:\overrightarrow{L}_{j,n}
\to \overrightarrow{L}_{j,n+\ell}.
\eeq

B-branes are D0-brane at the center or D2-branes 
located at the concentric disks whose radii are determined by $j$.
Since concentric disks are invariant under both
reflections $z\to \e^{{\pi i\over k+2}\ell}\bz$
and rotations $z\to\e^{{\pi i\over k+2}\ell}z$,
both A-type and B-type parities preserve the $j$-label of the B-branes.
To find the action on the $s$ (or $S$) label,
we need a geometrical interpretation of $s$ (or $S$),
which is currently missing.
This is found, however, by computing and reading the M\"obius strip amplitudes,
which we now do.

\subsubsection{M\"obius strips in RCFT}

\subsubsection*{\it A-type}

A-type parities corresponds to PSS crosscaps
and A-branes correspond to Cardy states.
Thus, the actions of A-type parities on A-branes follow the general rule
in RCFT:
\beq
P_{\bar{n},\bar{s}}:\Scr{B}_{j,n,s}
\to \Scr{B}_{j,\bar{n}-n,\bar{s}-s}.
\label{RulE}
\eeq
To compare this with the geometric action (\ref{geomactA}),
we recall that the parity $a^{\ell}P_A$ becomes the PSS-parity
$P_{\ell-1,\ell\mp 1}$ after GSO projection (\ref{CApa1})-(\ref{CApa4}),
where the ambiguity in the $s$-index is due to the symmetry
$(-1)^{\widehat{F}}$ absent before GSO projection.
Thus, (\ref{geomactA}) appears to suggest the action
$
P_{\bar{n},\bullet}:
\Scr{B}_{j,n,\bullet}\to\Scr{B}_{j,\bar{n}+1-n,\bullet},
$
where the $s$-indices are hidden by $\bullet$ due to the ambiguity mentioned.
The $j$-index is invariant as in
(\ref{RulE}) but the transformation of the $n$-index is different from
the RCFT rule (\ref{RulE}).
A possible resolution of this problem is that
the correspondence between the geometry and the boundary state
depends on whether the boundary is on the left or on the right
of the string. If
 the boundary state for the brane located at $L_{j,n}$ is
$|\Scr{B}_{j,n,\bullet}\rangle$ on the right-boundary but is
$\langle \Scr{B}_{j,n-1,\bullet}|$ on the right-boundary,
then the two transformation rules (\ref{geomactA})
and (\ref{RulE}) are consistent with each other.
This is exactly the case in the Landau--Ginzburg
description of the model where there is a similar geometric picture of
the branes \cite{HIV,HKKPTVVZ}, as we will see in the next section.

To find the parity action on the states,
we compute the M\"obius strip amplitudes, which is represented
in the tree-channel by the overlap of the crosscap states and the
boundary states.
A computation as shown in Appendix~\ref{app:cc}
leads to the following
result 
\beq\label{amobius}
\langle \Scr{C}_{\bar{n},\bar{s}}| q_t^H |\Scr{B}_{j,n,s}\rangle
=
\sum_{j'\in \Pk}
N_{jj}^{j'}\delta_{2n+n'-\bar{n}}^{(2k+4)}\delta^{(4)}_{2s+s'-\bar{s}}
\epsilon_{\bar{n},\bar{s}}^{j,n,s}(j',n',s')
\widehat{\chi}_{j',n',s'}(\tau),
\eeq
where $\epsilon_{\bar{n},\bar{s}}^{j,n,s}$ is the sign factor
$$
\epsilon_{\bar{n},\bar{s}}^{j,n,s}(j',n',s')
=(-1)^{2j+j'
+\bar{s}\cdot {s-{\widehat{\bar{s}}-\widehat{s'}\over 2}\over 2}
+(\bar{n}+k){n-{\widehat{\bar{n}}-\widehat{n'}\over 2}\over k+2}
}\sigma_{0,\bar{n},\bar{s}}
\sigma_{j'n's'}.
$$
The factor
$N_{jj}^{j'}\delta_{2n+n'-\bar{n}}^{(2k+4)}\delta^{(4)}_{2s+s'-\bar{s}}$
indeed selects the character $\chi_{j',n',s'}$ that appears in
the $\Scr{B}_{j,\bar{n}-n,\bar{s}-s}$-$\Scr{B}_{j,n,s}$ open string.
From the above result one can read that
the parity $P_{\bar{n},\bar{s}}$ acts on
the open string Hilbert space as
\beq
P_{\bar{n},\bar{s}}=\epsilon_{\bar{n},\bar{s}}^{j,n,s}(j',n',s')
\e^{\pi i (L_0-h_{j,n,s})}
\qquad \mbox{on the subspace $\Scr{H}_{j',n',s'}$}.
\eeq

\subsubsection*{\it B-type}

For B-type parity, there is no general rule on the action on B-branes,
but one can read it from M\"obius strip amplitudes.

The M\"obius strip with an unoriented B-brane at the
boundary is given by
\beqa
\lefteqn{\langle \Scr{C}_{rqp}|q_t^H|\Scr{B}^B_{[j,s]}
\rangle_{g_{4r+2p,2p}}}
\nn\\
&=& (-1)^{sq+p}\!\!\!\!\!\!\!\!
\sum_{(j',n',s'):\, {\rm even}}\!\!\!\!
\delta^{(2)}_{s',p}
N_{jj}^{j'} \e^{\frac{\pi i \widehat{(2r+p)}n'}{2(k+2)}
-\frac{\pi i \widehat{(2q+p)}s'}{4}}
(-1)^{\frac{2j'+n'+s'}{2}} \sigma_{j',n',s'} 
\widehat{\chi}_{j',n',s'}(\tau)
\nn\\
\label{rcftbm1} 
\eeqa
A comparison of the M\"obius and cylinder amplitude shows
that the boundary label $j$ is mapped to itself by all B-type parities,
as expected from the geometrical consideration.
The selection factor $\delta^{(2)}_{s',p}$
shows that the $p=0$ parities leave also the label $s$ invariant,
whereas those with $p=1$ exchange even and odd boundary labels.
Furthermore, the structure
of the M\"obius strip implies that the B-parity
$P_{rqp}$
can be obtained 
from $P_{000}$ by combination with the symmetry elements
$g_{2r+p,2q+p}$.
This extends the claim (\ref{procla}) from
the closed string sector to the open string sector.

The M\"obius strip with an oriented B-brane at the boundary is given
by
\beqa
\lefteqn{\langle \Scr{C}_{rqp}|q_t^H|\Scr{B}^B_{\frac{k}{4},S}
\rangle_{g_{4r+2p,2p}}}
\nn\\
&&=(-1)^{Sq+r+q+p}\!\!\!\!\!\!
\sum_{(j,n,s)\in \Mk} \!\!\delta_{2j}^{(2)}
\delta^{(2)}_{s,p}
{1+(-1)^{r+q} 
(-1)^{\frac{2j+n-s}{2}} \over 2}
e^{\frac{\pi i (2r+p)n}{2(k+2)}-\frac{\pi i (2q+p)s}{4}}
\sigma_{j,n,s} \hat\chi_{j,n,s}(\tau)
\nn\\
\eeqa
Again, we see that the Cardy label $j=k/4$ remains invariant as
expected from geometry. Also, 
parities with $p=0$ map $S$-even branes to $S$-even branes and odd
branes to odd branes, while $p=1$ parities exchange them, as before.
In the present case, however, the label $S$ is defined mod $4$.
To obtain the refined information,
note that the $\Scr{B}_{{k\over 4},S'}$-$\Scr{B}_{{k\over 4},S}$
open string partition function (\ref{bcyl4}) has the selection factor
$(1+(-1)^{2j+n-s\over 2}\e^{{\pi i\over 2}((S')^2+S'-S^2-S)})/2$.
Comparing with this, we find that the label $S'$
of the image brane is identified as
\beq\label{signs}
(-1)^{r+q}
= \e^{\frac{\pi i }{2} (S^{'2}+S' - S^{2} -S)}.
\eeq
This determines the parity action on $S$ mod $4$:
\beqa
p=0,\ r+q\ {\rm even}&:& \Scr{B}^B_{\frac{k}{4}, S} \to 
                        \Scr{B}^B_{\frac{k}{4}, S} \no\\ \no
p=0,\ r+q \ {\rm odd}&:&  \Scr{B}^B_{\frac{k}{4}, S} \to 
                        \Scr{B}^B_{\frac{k}{4}, S+2} \\ \no
p=1,\  r+q \ {\rm even}&:& \Scr{B}^B_{\frac{k}{4}, 0}
                         \leftrightarrow \Scr{B}^B_{{k\over 4},-1},\quad
                       \Scr{B}^B_{\frac{k}{4}, 2}
                         \leftrightarrow \Scr{B}^B_{{k\over 4},1}\\\no
p=1,\ r+q \ {\rm odd}&:& \Scr{B}^B_{\frac{k}{4}, 0}
                         \leftrightarrow \Scr{B}^B_{{k\over 4},1},\quad
                       \Scr{B}^B_{\frac{k}{4}, 2}
                         \leftrightarrow \Scr{B}^B_{{k\over 4},-1}
\eeqa
In comparison with the action of $g_{n,s}$ on the branes
(\ref{actPBonBb}), we see that this action is consistent with
the claim $P_{rqp}=g_{2r+p,2q+p}P_{000}$.

\subsection{Resolving GSO}

Let us now entangle the GSO projection and derive the
M\"obius strip amplitudes in the original ${\mathcal N}=2$
minimal model.
This in particular enables us to compute the open string Witten indices.

\subsubsection{Oriented branes vs unoriented branes}

The first step is to find the relation of
 the boundary states of the model before and after the GSO projection.
We use the following prescription.
Let us consider a theory involving fermions
with a mod-2 fermion number $(-1)^F$
that can be used to define the non-chiral GSO projection. We are interested
in how the open string amplitudes of the GSO projected theory can be defined
in terms of the underlying theory.
Let $\{O_a\}$ be oriented D-branes and $\{U_j\}$ be unoriented branes
of the GSO projected theory.
For each oriented brane $O_a$ there is another brane
$O_{r(a)}$ which
corresponds to the same boundary condition $[a]$ but has an opposite sign
in the GSO projection on the open string sector:
\beqa
&&\Tr\!\!\mathop{}_{{\mathcal H}_{a,b}^{\rm GSO}}\!\!q^H
=\Tr\!\!\mathop{}_{{\mathcal H}_{r(a),r(b)}^{\rm GSO}}\!\!q^H
=\Tr\!\!\mathop{}_{{\mathcal H}_{[a],[b]}}\!\!{1+(-1)^F\over 2}q^H,\\
&&\Tr\!\!\mathop{}_{{\mathcal H}_{a,r(b)}^{\rm GSO}}\!\!q^H
=\Tr\!\!\mathop{}_{{\mathcal H}_{r(a),b}^{\rm GSO}}\!\!q^H
=\Tr\!\!\mathop{}_{{\mathcal H}_{[a],[b]}}\!\!{1-(-1)^F\over 2}q^H.
\eeqa
$O_{r(a)}$ is the orientation reversal of $O_a$ in this sense.
The partition functions involving unoriented branes are
\beqa
&&\Tr\!\!\mathop{}_{{\mathcal H}_{i,j}^{\rm GSO}}\!\!q^H
=\Tr\!\!\mathop{}_{{\mathcal H}_{i,j}}\!\!q^H,\\
&&\Tr\!\!\mathop{}_{{\mathcal H}_{a,i}^{\rm GSO}}\!\!q^H
={1\over \sqrt{2}}
\Tr\!\!\mathop{}_{{\mathcal H}_{[a],i}}\!\!q^H.
\eeqa
The factor of $1/\sqrt{2}$ may appear odd. However, in the spectrum between
oriented and unoriented branes, there is always an odd number of real
(or Majorana) fermion zero modes whose partition functions are odd powers of
$\sqrt{2}$.
Thus, it is only with the factor of $1/\sqrt{2}$
that the open string partition functions of
the GSO projected theory have integer coefficients.

The above definition leads to the following expressions for
the boundary states of the GSO projected theory
\beqa
&&|\Scr{B}_{r^s(a)}\rangle^{\rm GSO}
={1\over \sqrt{2}}|\Scr{B}_{[a]}\rangle^{}_{{}_{\rm NSNS}}
+{(-1)^s\over \sqrt{2}}|\Scr{B}_{[a]}\rangle^{}_{{}_{\rm RR}},
\\
&&|\Scr{B}_i\rangle^{\rm GSO}
=|\Scr{B}_i\rangle^{}_{{}_{\rm NSNS}}.
\eeqa

\subsubsection*{\it Example: free Dirac fermion}

For illustration, we consider the free Dirac fermion $\psi_{\pm},\bpsi_{\pm}$,
and the following two boundary conditions
\beqa
&&A: \psi_-=\bpsi_+, \,\,\,\bpsi_-=\psi_+;
\nn\\
&&B: \psi_-=\psi_+,\,\,\, \bpsi_-=\bpsi_+.
\nn
\eeqa
For both AA and BB open strings, the space of states is the Fock space
of the complex Clifford algebra generated by $\psi_n,\bpsi_n=\psi_{-n}^{\dag}$
($n\in \Z$) obeying the relation
$\{\psi_n,\bpsi_m\}=\delta_{n+m,0}$,
$\{\psi_n,\psi_m\}=0$. In particular, the partition functions
$\Tr_{{}_{AA}}\e^{i\alpha F_A}q^H$ and
$\Tr_{{}_{BB}}\e^{i\alpha F_V}q^H$ have integer coefficients in the expansion
by $\e^{i\alpha Q_N}q^{E_N}$, where
$F_A=\bpsi_-\psi_-+\bpsi_+\psi_+$ and
$F_V=-\bpsi_-\psi_-+\bpsi_+\psi_+$ are the fermion numbers
conserved in the respective open string system.
Let us now consider the AB-string.
There is one real fermion zero mode, the zero mode of
${\rm Re}(\psi_-)$ or equivalently of ${\rm Re}(\psi_+)$,
whose partition function
is \cite{WK-theory}
$$
\sqrt{2}.
$$
The non-zero modes are positive-integer
as well as positive-half-integer moded
complex fermions, with the zero point energy
$-{1\over 24}+{1\over 16}={1\over 48}$.
Thus the total partition function is
$$
\Tr_{{}_{AB}}q^H=\sqrt{2}q^{1\over 48}
\prod_{n=1}^{\infty}(1+q^n)(1+q^{n-{1\over 2}}).
$$
Let us now consider GSO projection by the operator
$$
(-1)^F=\e^{\pi i F_A}.
$$
The projected theory is
the sigma model on the circle of radius $R=\sqrt{2}$,
$X\equiv X+2\pi R$, where the
correspondence is $\bpsi_{\pm}\psi_{\pm}
=(\partial_tX\pm\partial_{\sigma}X)/\sqrt{2}$.
The boundary condition A implies
$\bpsi_-\psi_-=-\bpsi_+\psi_+$ or equivalently the Dirichlet
boundary condition $\partial_tX=0$ (for a D0-brane), while
the boundary condition B implies
$\bpsi_-\psi_-=\bpsi_+\psi_+$ or equivalently the Neumann
boundary condition $\partial_{\sigma}X=0$ (for a D1-brane).
A more careful inspection of the boundary states shows that
the location of the D0-brane is at $X=0$ or $X=\pi R$, and 
the Wilson line for the D1-brane is zero.
Namely,
\beqa
&&
|D_{\pi s R}\rangle={1\over \sqrt{2}}|\Scr{B}_A\rangle^{}_{{}_{\rm NSNS}}
+{(-1)^s\over\sqrt{2}}|\Scr{B}_A\rangle^{}_{{}_{\rm RR}},
\label{DBA}\\
&&
|N_0\rangle =|\Scr{B}_B\rangle^{}_{{}_{\rm NSNS}}.
\label{NBB}
\eeqa
In fact, the DN-string has odd-integer moded bosonic field
and the partition function is
\beqa
\Tr_{{}_{DN}}q^H
&=&q^{1\over 48}\prod_{n=1}^{\infty}(1-q^{n-{1\over 2}})^{-1}
\nn\\
&=&q^{1\over 48}\prod_{n=1}^{\infty}(1+q^n)(1+q^{n-{1\over 2}})
={1\over \sqrt{2}}\Tr_{{}_{AB}}q^H.
\nn
\eeqa
This is indeed in agreement with the relation
$\langle D |q_t^H|N\rangle={1\over \sqrt{2}}
{}_{{}_{\rm NSNS}}\langle\Scr{B}_{A}|q_t^H|
\Scr{B}_B\rangle_{{}_{\rm NSNS}}$ that follows from
(\ref{DBA}) and (\ref{NBB}).

\subsubsection{A-type}

All the A-branes $\Scr{B}_{j,n,s}$ are oriented where
the orientation is flipped by the shift $s\to s+2$.
Thus, the boundary states of the projected theory
is composed of the NSNS and RR sector states as
\beq
|\Scr{B}_{j,n,s}\rangle
={1\over \sqrt{2}}|\Scr{B}_{j,n}\rangle^{}_{{}_{\rm NSNS}}
+{1\over \sqrt{2}}|\Scr{B}_{j,n,(s)}\rangle^{}_{{}_{\rm RR}}.
\eeq
In other words, the boundary states before the GSO projection are given by
\beqa
&&|\Scr{B}_{j,n}\rangle^{}_{{}_{\rm NSNS}}
={1\over \sqrt{2}}|\Scr{B}_{j,n,s}\rangle
+{1\over \sqrt{2}}|\Scr{B}_{j,n,s+2}\rangle,
\nn\\
&&|\Scr{B}_{j,n,(s)}\rangle^{}_{{}_{\rm RR}}
={1\over \sqrt{2}}|\Scr{B}_{j,n,s}\rangle
-{1\over \sqrt{2}}|\Scr{B}_{j,n,s+2}\rangle.
\nn
\eeqa
We have retained the orientation dependence for the RR-boundary states,
so that
the orientation flip corresponds to the sign flip
$|\Scr{B}_{j,n,(s+2)}\rangle^{}_{{}_{\rm RR}}=-
|\Scr{B}_{j,n,(s)}\rangle^{}_{{}_{\rm RR}}$.

The spectrum of the open string before the GSO projection
is found by computing the ordinary partition function
$\Tr_{(j,n),(j',n')}q^H={}_{{}_{\rm NSNS}}\langle
\Scr{B}_{j,n}|q_t^H|\Scr{B}_{j',n'}\rangle_{{}_{\rm NSNS}}$.
Using the above identification of $|\Scr{B}_{j,n}\rangle_{{}_{\rm NSNS}}$
and the formula (\ref{cylAbrane}), we find
\beq
{\mathcal H}_{(j,n),(j',n')}
=\bigoplus_{2j''+n-n'+s''\,{\rm even}}N_{jj'}^{j''}\,
\Scr{H}_{j'',n-n',s''}^{{\mathcal N}=2}
\eeq
where
$\Scr{H}_{j,n,0}^{{\mathcal N}=2}$ and
$\Scr{H}_{j,n,1}^{{\mathcal N}=2}$ are the representations of the
Neveu-Schwarz and Ramond ${\mathcal N}=2$ super-Virasoro algebra
which is defined as
\beq
\Scr{H}_{j,n,[s]}^{{\mathcal N}=2}
=\Scr{H}_{j,n,s}\oplus \Scr{H}_{j,n,s+2}.
\label{NSRrepr}
\eeq
where $j\in \Pk, n\in \Z_{k+2}$, and $[s]\in \Z_2$ is the mod 2 reduction of
$s\in\Z_4$.

Let us now compute the open string Witten index.
$\Scr{B}_{j,n,(s)}$ at the left boundary and $\Scr{B}_{j',n',(s')}$
at the right boundary preserve the same supersymmetry if and only if
$s-s'$ is odd. Then, the index is given by
\beqa \no
I(\Scr{B}_{j,n,(s)},\Scr{B}_{j',n',(s')}) &=&
{}^{}_{{}_{\rm RR}}\!\langle \Scr{B}_{j,n,(s)}|
q_t^H| \Scr{B}_{j',n',(s')}\rangle^{}_{{}_{\rm RR}}\\ \no
&=& \langle \Scr{B}_{j,n,s}|q_t^H| \Scr{B}_{j',n',s'}\rangle
- \langle \Scr{B}_{j,n,s} |q_t^H | \Scr{B}_{j',n',s'+2} \rangle \\ \no
&=& \sum_{j'':ev} N_{jj'}^{j''} \delta^{(2)}_{s-s'+1}
\left( \chi_{j'', n-n',s-s'}
- \chi_{j'',n-n',s-s'+2} \right)(\tau) \\ \no
&=& \sum_{j''} N_{jj'}^{j''} (\delta_{s-s',1}^{(4)}-\delta_{s-s',-1}^{(4)})
(\delta_{n-n',2j''+1}^{(2k+4)}-\delta_{n-n',-2j''-1}^{(2k+4)})
\nn\\
&=& (-1)^{\frac{s-s'+1}{2}} N_{jj'}^{\frac{n'-n-1}{2}}
 \label{cylindex}
\eeqa
Here, $N$ is the periodically continued fusion rule coefficient \cite{BDLR},
$N_{jj'}^{j''}=N_{jj'}^{j''+(k+2)}=-N_{jj'}^{-j''-1}$,
$N_{jj'}^{\frac{1}{2}}= N_{jj'}^{\frac{k+1}{2}} =0$.
This continuation is the same as the analytic continuation
using the Verlinde formula, which expresses the fusion rule coefficient
$N$ in terms of elements of the modular $S$-matrix.

Let us next compute the open string index twisted by the parity symmetry $P_A$.
We recall that $P_A$ commutes with
the supercharge $\bQ_++Q_-$ which is the combination 
preserved by the $\Scr{B}_{j,n,(s)}$-$\Scr{B}_{j',n',(s')}$-string
with $s$ even and $s'$ odd.
Recall also that $P_A$ corresponds to $P_{-1,-1}$
in the GSO projected theory under which the
$\Scr{B}_{j,n,s}$-$\Scr{B}_{j,-n-1,-s-1}$-string is invariant.
Thus, we consider the twisted Witten index for the
$\Scr{B}_{j,n,(s)}$-$\Scr{B}_{j,-n-1,(-s-1)}$-string
in the original minimal model:
\beqa
I_{P_A}(\Scr{B}_{j,n,(s)},\Scr{B}_{j,-n-1,(-s-1)})
&=&{}^{}_{{}_{\rm RR}}\!
\langle \Scr{B}_{j,n,(s)}|q_t^H|\Scr{C}_{P_A}\rangle
\nn\\
&=&\langle \Scr{B}_{j,n,s}|q_t^H|\Scr{C}_{-1,-1}\rangle
+\langle \Scr{B}_{j,n,s}|q_t^H|\Scr{C}_{-1,1}\rangle.
\nn
\eeqa
Using the formula (\ref{thecorrectone}),
we find that the summands are
\beqa
\lefteqn{
\langle \Scr{B}_{j,n,s}|q_t^H|\Scr{C}_{-1,\mp 1}\rangle}
\nn\\
&&=(-1)^{s\over 2}\sum_{2j'\,{\rm even}\atop
n'\,{\rm odd}}
\left(
\delta^{(2k+4)}_{n,{-1- \widehat{n'}\over 2}}-(-1)^{k}
\delta^{(2k+4)}_{n,{-1- \widehat{n'}\over 2}+k+2}
\right)
(-1)^{2j+j'}N_{jj}^{j'}{\sigma_{0,-1,\mp 1}\over
\sigma_{j',n',\mp 1}}\widehat{\chi}_{j',-n',\pm 1}.
\nn
\eeqa
Note that $\sigma_{0,-1,\mp 1}=\pm 1$ and
$$
{\widehat{\chi}_{j',-n',1}\over \sigma_{j',n',-1}}
-{\widehat{\chi}_{j',-n',-1}\over \sigma_{j',n',1}}
=\delta_{n',-2j'-1}-\delta_{n',2j'+1}.
$$
This shows that
\beqa
\lefteqn{I_{P_A}(\Scr{B}_{j,n,(s)},\Scr{B}_{j,-n-1,(-s-1)})}
\nn\\
&=&(-1)^{s\over 2}\sum_{j'\in \Pk}
(-1)^{2j+j'}N_{jj}^{j'}
\left(\delta_{n,j'}^{(2k+4)}
-(-1)^{k}\delta_{n, j'+k+2}^{(2k+4)}
-\delta_{n,-j'-1}^{(2k+4)}
+(-1)^{k}\delta_{n,-j'-1+k+2}^{(2k+4)}\right)
\nn\\
&=&
(-1)^{s\over 2}\sum_{j'\in \Pk}N_{jj}^{j'}
\left(\delta_{j',n}^{(2k+4)}
-\delta_{j',n+(k+2)}^{(2k+4)}
+\delta_{j',-n-1}^{(2k+4)}
-\delta_{j',(k+2)-n-1}^{(2k+4)}
\right).
\label{IpABB}
\eeqa
Similar to the untwisted open string Witten index, this index can
be expressed as a periodically continued fusion rule coefficient:
\beq
I_{P_A}(\Scr{B}_{j,n,(s)},\Scr{B}_{j,-n-1,(-s-1)})
=(-1)^{s\over 2}\widetilde{N}_{jj}^{n},
\label{IpABBa}
\eeq
where $\widetilde{N}_{jj}^{j'}$ is the $SU(2)$ fusion rule coefficient
with the periodic continuation
$\widetilde{N}_{jj}^{j'}=-\widetilde{N}_{jj}^{j'+k+2}
=\widetilde{N}_{jj}^{-j'-1}$.
Note that this periodic continuation is
different than the one that appears in the untwisted open string
Witten index (\ref{cylindex}).
While the fusion rule coefficients
appear  in the open string channel of the cylinder diagram, 
the quantity that governs 
the M\"obius strip amplitude is the $Y$-tensor, and the
periodicity is in fact the one inherited  from the $Y$-tensor.
To see this, recall that
$$
N_{jj}^{j'} = (-1)^{j'+2j} Y_{j0}^{j'} = (-1)^{j'+2j}
\sum_{j''} \frac{4}{k+2} \frac{\sin\pi \frac{(2j'+1)(2j''+1)}{2(k+2)}
\sin\pi \frac{(2j+1)(2j''+1)}{k+2}\sin\pi \frac{(2j''+1)}{2(k+2)}}
{\sin\pi\frac{2j''+1}{k+2}},
$$
whose analytic continuation indeed agrees with the one derived above.

We will compute the same index in the Landau--Ginzburg description of the
model in the next section.

\subsubsection{B-type}

We now consider B-branes and B-parities.
We shall omit the superscript ``$B$'' for the branes
in this subsubsection.
The boundary states of the unoriented branes $\Scr{B}_{[j,s]}$
and the oriented branes
 $\Scr{B}_{{k\over 4},S}$
are written in terms of those before GSO projection as
\beqa
&&
|\Scr{B}_{[j,s]}\rangle=|\Scr{B}_{[j,s]}\rangle^{}_{{}_{\rm NSNS}},
\label{idbst1}
\\
&&
|\Scr{B}_{{k\over 4},S}\rangle
={1\over \sqrt{2}}|\Scr{B}_{{k\over 4},[S]}\rangle^{}_{{}_{\rm NSNS}}
+{\e^{-{\pi i\over 2}(S^2+S)}\over \sqrt{2}}
|\Scr{B}_{{k\over 4},[S]}\rangle^{}_{{}_{\rm RR}}.
\label{idbst2}
\eeqa
$[S]$ is the mod 2 reduction of the mod 4 integer $S$,
and the phase factor $\e^{-{\pi i\over 2}(S^2+S)}$
represents the sign flip  of  the RR-part
under the orientation reversal
$S\to S+2$.

Let us find the open string spectrum before the GSO projection.
As in the A-type case,
this can be read off from the ordinary partition function
$\Tr_{ab}q^H={}_{{}_{\rm NSNS}}\langle a|q_t^H|b\rangle_{{}_{\rm NSNS}}$,
which is computable using the relation to the boundary states of
the GSO projected theory, (\ref{idbst1})-(\ref{idbst2}),
and the formulae (\ref{bcyl})-(\ref{bcyl4}).
We find that the spectrum of
$\Scr{B}_{[j,s]}$-$\Scr{B}_{[j',s']}$-strings and
$\Scr{B}_{{k\over 4},[S]}$-$\Scr{B}_{{k\over 4},[S']}$-strings
is
\beqa
&&
{\mathcal H}_{[j,s],[j',s']}
=\bigoplus_{2j''+n''+s-s'\, {\rm even}}
N_{jj'}^{j''}\,\Scr{H}_{j'',n'',s-s'}^{{\mathcal N}=2},
\\
&&
{\mathcal H}_{({k\over 4},[S]), ({k\over 4},[S'])}
=\bigoplus_{2j, n+S-S'\,\,{\rm even}
\atop (j,n)\equiv
({k\over 2}-j,n+k+2)}
\Scr{H}_{j,n,[S-S']}^{{\mathcal N}=2},
\eeqa
where $\Scr{H}_{j,n,0/1}^{{\mathcal N}=2}$ are
the ${\mathcal N}=2$ NS/R modules (\ref{NSRrepr}).
We would like to identify the action of the symmetries
 $a^{2m}$ and $(-1)^Fa^{2m}$ on these open string Hilbert spaces.
We shall find it using the following guideline:\\
$\bullet$ $a^{2m}$ must commute with the supersymmetry,\\
$\bullet$ $(-1)^Fa^{2m}$ must anticommute with the supersymmetry,\\
$\bullet$ They must induce the symmetry
$g_{2m,2m}$ or $g_{2m,2m+2}$ after GSO projection.\\
Let us first look at the
$\Scr{B}_{[j,s]}$-$\Scr{B}_{[j',s']}$-string.
Looking at the Cylinder amplitudes 
(\ref{bcyl}) one observes that $g_{2\bar{n},2\bar{s}}$
acts as the phase $\e^{\pi i ({\bar{n}n''\over k+2}-{\bar{s}s''\over 2})}$
on the subspace $\Scr{H}_{j''n''s''}$.
This shows that $g_{2m,0}$ commutes with the supersymmetry
and $g_{2m,2}$ anti-commutes with the supersymmetry.
This fixes the identification as
\beq
\begin{array}{l}
a^{2m}\equiv g_{2m,0},\\
(-1)^Fa^{2m}\equiv g_{2m,2}.
\end{array}
\eeq
The boundary state identification $|\Scr{B}_{[j,s]}\rangle
=|\Scr{B}_{[j,s]}\rangle_{{}_{\rm NSNS}}
:=|\Scr{B}_{[j,s]}\rangle_{(-1)^F}$ is now generalized by the twists
as
\beqa
&&
|\Scr{B}_{[j,s]}\rangle_{g_{2m,0}}=
|\Scr{B}_{[j,s]}\rangle_{(-1)^Fa^{2m}},
\label{idbst3}\\
&&
|\Scr{B}_{[j,s]}\rangle_{g_{2m,2}}=
|\Scr{B}_{[j,s]}\rangle_{a^{2m}}.
\label{idbst4}
\eeqa
Next, consider the
$\Scr{B}_{{k\over 4},[S]}$-$\Scr{B}_{{k\over 4},[S']}$-string.
Looking at the formula
 (\ref{bcyl4}), we find that $g_{2\bar{n},2\bar{n}}$ acts as
$\e^{-\pi i \bar{n}J_0}$ on the subspace of
${\mathcal H}_{({k\over 4},[S]), ({k\over 4},[S'])}$
that survives the GSO projection.
The operator $\e^{-\pi i\bar{n}J_0}$ commutes (anti-commutes)
with the supersymmetry for even $\bar{n}$ (odd $\bar{n}$).
Thus, we identify
\beq
g_{2m,2m}=\e^{-\pi i mJ_0}
=(-1)^{m F}a^{2m}.
\eeq
In particular, the twisted version of
the boundary state identification (\ref{idbst2}) is
\beq
|\Scr{B}_{{k\over 4},S}\rangle_{g_{2m,2m}}
={1\over \sqrt{2}}|\Scr{B}_{{k\over 4},[S]}\rangle_{(-1)^{(m+1)F}a^{2m}}
+{\e^{-{\pi i\over 2}(S^2+S)}\over \sqrt{2}}
|\Scr{B}_{{k\over 4},[S]}\rangle_{(-1)^{mF}a^{2m}}.
\label{idbst5}
\eeq
In particular, we find
\beqa
|\Scr{B}_{{k\over 4},[S]}\rangle_{a^{2m}}
&=&{\e^{-i\theta_S}\over\sqrt{2}}
\left[|\Scr{B}_{{k\over 4},S}\rangle_{g_{2m,2m}}
+(-1)^{m-1}|\Scr{B}_{{k\over 4},S+2}\rangle_{g_{2m,2m}}\right]
\nn\\
&=&\left\{
\begin{array}{ll}
{1\over \sqrt{2}}
|\Scr{B}_{[{k\over 4},S]}\rangle_{g_{2m,2m}}&\mbox{$m$ odd}
\\[0.1cm]
\e^{-i\theta_S}\sqrt{k+2\over 2}\e^{-\pi i S^2\over 2}\sum_{s=\pm 1}
\e^{-\pi i Ss\over 2}
\kket{{k\over 4},{k+2\over 2}+m,s+m}_B
&\mbox{$m$ even}
\end{array}
\right.
\nn\\
\label{idbst6}
\eeqa
where $\e^{i\theta_S}=1$ for odd $m$ and
$\e^{i\theta_S}=\e^{-{\pi i\over 2}(S^2+S)}$ for even $m$.

\subsubsection*{\it Open string Witten index}

As an application of
(\ref{idbst3}), (\ref{idbst4}) and (\ref{idbst5}),
 let us compute the open string Witten index from the tree-channel.

We start with the open string stretched between unoriented B-branes.
$\Scr{B}_{[j,s]}$ on the left boundary and  $\Scr{B}_{[j',s']}$ on the right
boundary preserve the same supersymmetry if $s-s'$ is odd.
Thus we consider the Witten index for odd $s-s'$,
twisted by an axial rotation symmetry:
\beqa
\Tr_{{}_{[j,s],[j',s']}}a^{2m}(-1)^Fq^H
&=&{}_{a^{2m}}\!\langle\Scr{B}_{[j,s]}|q_t^H|
\Scr{B}_{[j',s']}\rangle_{a^{2m}}
\nn\\
&=&{}_{g_{2m,2}}\!\langle\Scr{B}_{[j,s]}|q_t^H|
\Scr{B}_{[j',s']}\rangle_{g_{2m,2}}
\nn\\
&=&
\sum_{j''\in \Pk}N_{jj'}^{j''}
\Bigl(\e^{\pi i({m(2j''+1)\over k+2}-{1\over 2})}
-\e^{\pi i(-{m(2j''+1)\over k+2}-{1\over 2})}\Bigr)
\eeqa
The ordinary index (the one with $m=0$) vanishes.
This means that there are equal number of bosonic and fermionic supersymmetric
ground states ---
for each $j''$ with $N_{jj'}^{j''}\ne 0$ there is one ground state
from $\Scr{H}_{j'',2j''+1,1}^{{\mathcal N=2}}$
and another from $\Scr{H}_{j'',-2j''-1,1}^{{\mathcal N=2}}$
which contribute to the index with opposite signs.
The index escapes
from vanishing if twisted by an operator $a^{2m}$
that acts  differently on those ground states.

Let us next consider the index for the string stretched between oriented
B-branes. The boundary conditions $\Scr{B}_{{k\over 4},[S]}$ on the left and
$\Scr{B}_{{k\over 4},[S']}$ on the right preserve the same supersymmetry.
Thus, the open string Witten index is defined for odd $S-S'$ 
and is given by
\beqa
\Tr_{{}_{({k\over 4},[S]),({k\over 4},[S'])}}
a^{2m}(-1)^Fq^H
&=&
{}_{a^{2m}}\langle \Scr{B}_{{k\over 4},[S]}|q_t^H|
\Scr{B}_{{k\over 4},[S']}\rangle_{a^{2m}}
\nn\\
&=&
\left\{\begin{array}{ll}
{1\over 2}\langle \Scr{B}_{[{k\over 4},S]}|
\Scr{B}_{[{k\over 4},S']}\rangle_{g_{2m,2m}}
&(\mbox{$m$ odd})\\
\e^{i(\theta_S-\theta_{S'})}\!\left[\langle \Scr{B}_{{k\over 4},S}|
\Scr{B}_{{k\over 4},S'}\rangle
\!-\!\langle \Scr{B}_{{k\over 4},S}|
\Scr{B}_{{k\over 4},S'+2}\rangle\right]_{\!\!g_{2m,2m}}\!\!\!\!\!\!\!\!
\!\!\!\!
&(\mbox{$m$ even})
\end{array}\right.
\nn\\
&=&
\sum_{j\in \Pk\cap\Z}
\e^{\pi i m({2j+1\over k+2}-{1\over 2})}
\eeqa
This is consistent with the fact that
there are ${k+2\over 2}$ ground states
from $\Scr{H}_{j,2j+1,1}^{{\mathcal N}=2}$, $j\in \Pk\cap \Z$.
They are all regarded bosonic and $a^{2m}$ acts on them by 
phase $\e^{\pi i m({2j+1\over k+2}-{1\over 2})}$.
For even $m$, the result can also be evaluated as
$$
\left\{\begin{array}{ll}
0 & (\mbox{$m$ even}, m\neq 0)\\
\frac{k+2}{2} & (m=0)\\
\end{array}\right.
$$
The vanishing for twists with $m$ even, $m\neq0$, 
is most easily understood in terms
of the closed string channel: by (\ref{idbst6}) we see that
the propagating closed string states are in the representation
$(\frac{k}{4}, \frac{k+2}{2}+m,s+m)$. There are no supersymmetric
ground state in this sector, unless $m=0$.

In both cases, the index behaves in the way it should.
This may be regarded as a strong consistency test of the twisted versions of
the boundary state identification,
(\ref{idbst4}) and (\ref{idbst5}).\footnote{It would also be interesting
to study the open string stretched between oriented brane and
unoriented brane. Just as in the case of free fermion,
we obtain $\sqrt{2}$ in the partition function.
This signals the presence
of an odd number of real fermion zero modes in the open string system.
We do not, however, study it further in this paper.}
Now let us use them to
compute the M\"obius strip amplitudes
of the theory before GSO projection.

\subsubsection*{\it Parity actions on boundary conditions}

From the actions of parities $P_{rqp}$ on the B-branes
in the GSO projected theory, we see that the parities
in the original model transforms the boundary conditions as
\beqa
(-1)^{\nu F}a^{2m+1}P_B&:&\left\{
\begin{array}{l}
\Scr{B}_{[j,s]}\to \Scr{B}_{[j,s+1]}
\\
\Scr{B}_{{k\over 4},[S]}\to \Scr{B}_{{k\over 4},[S+1]}
\end{array}\right.
\nn\\
(-1)^{\nu F}a^{2m}P_B&:&\left\{
\begin{array}{l}
\Scr{B}_{[j,s]}\to \Scr{B}_{[j,s]}
\\
\Scr{B}_{{k\over 4},[S]}\to \Scr{B}_{{k\over 4},[S]}
\end{array}\right.
\nn
\eeqa
In particular, the B-parities $a^{2m+1}P_B$
transform the $\Scr{B}_{[j,s]}$-$\Scr{B}_{[j,s+1]}$-string to itself and
$\Scr{B}_{{k\over 4},[S]}$-$\Scr{B}_{{k\over 4},[S+1]}$-string to itself.

\subsubsection*{\it parity-twisted open string Witten index}

The B-parities $a^{2m+1}P_B$ commute with the supercharge
$\bQ_++\bQ_-$ which is the supersymmetry preserved by the
$\Scr{B}_{[j,0]}$-$\Scr{B}_{[j,1]}$-string
as well as by the 
$\Scr{B}_{{k\over 4},[0]}$-$\Scr{B}_{{k\over 2},[1]}$-string.
Thus, we consider parity-twisted Witten index in such open string sectors.

We first consider the $\Scr{B}_{[j,0]}$-$\Scr{B}_{[j,1]}$-string.
The index is expressed as
$$
I_{a^{2m+1}P_B}(\Scr{B}_{[j,0]},\Scr{B}_{[j,1]})
:=\Tr_{{}_{[j,0],[j,1]}}
(-1)^Fa^{2m+1}P_Bq^H
=\langle \Scr{B}_{[j,0]}|q_t^H|\Scr{C}_{a^{2m+1}P_B}\rangle,
$$
where the boundary state should be the one on the circle twisted by
the square of the parity $(a^{2m+1}P_B)^2=a^{4m+2}$,
which is ${}_{g_{4m+2,2}}\langle\Scr{B}_{[j,0]}|$
by the identification (\ref{idbst4}).
The crosscap state is the one given in Eq.~(\ref{CBpa1}), and hence
$$
I_{a^{2m+1}P_B}(\Scr{B}_{[j,0]},\Scr{B}_{[j,1]})
={}_{g_{4m+2,2}}\langle\Scr{B}_{[j,0]}|q_t^H|\Scr{C}_{m,1}(+)\rangle
\times (-1)^m.
$$
Since $\sqrt{T^{(2)}_{-1}\over T^{(2)}_{1}}=1$, we find
$|\Scr{C}_{m,1}(+)\rangle=
[|\Scr{C}_{m01}\rangle+|\Scr{C}_{m11}\rangle]/\sqrt{2}$.
The sum of the two pairings is the complex conjugate of the following
\beqa
\lefteqn{
\langle\Scr{C}_{m01}|\Scr{B}_{[j,0]}\rangle_{g_{4m+2,2}}
+\langle\Scr{C}_{m11}|\Scr{B}_{[j,0]}\rangle_{g_{4m+2,2}}
}
\nn\\
&&=
\sum_{j'\in \Pk\cap \Z
\atop
n'\,{\rm odd},\,s'=\pm 1}
N_{jj}^{j'}\e^{\pi i(\widehat{2m+1})n'\over 2(k+2)}
(\e^{-{\pi i s'\over 4}}+\e^{\pi i s'\over 4})
(-1)^{{2j'+n'+s'\over 2}+1}\sigma_{j'n's'}\widehat{\chi}_{j'n's'}
\nn\\
&&=\sqrt{2}\sum_{j'\in\Pk\cap\Z\atop n'\,{\rm odd}}
N_{jj}^{j'}\e^{\pi i(\widehat{2m+1})n'\over 2(k+2)}
(-1)^{{2j'+n'+s'\over 2}+1}
(\sigma_{j'n'1}\widehat{\chi}_{j'n'1}-
\sigma_{j'n',-1}\widehat{\chi}_{j'n',-1})
\nn\\
&&=
\sqrt{2}\sum_{j'\in\Pk\cap\Z}N_{jj}^{j'}
\Bigl(\e^{\pi i(\widehat{2m+1})(2j'+1)\over 2(k+2)}
+\e^{-{\pi i(\widehat{2m+1})(2j'+1)\over 2(k+2)}}\Bigr).
\nn
\eeqa
Thus, we find
\beq
I_{a^{2m+1}P_B}(\Scr{B}_{[j,0]},\Scr{B}_{[j,1]})
=(-1)^m\sum_{j'\in\Pk\cap\Z}
N_{jj}^{j'}\left(\e^{\pi i {(\widehat{2m+1})(2j'+1)\over 2(k+2)}}
+\e^{-\pi i {(\widehat{2m+1})(2j'+1)\over 2(k+2)}}\right)
\eeq
This shows that the B-parity $a^{2m+1}P_B$ acts on
the supersymmetric ground states in
$\Scr{H}_{j',\pm(2j'+1),1}^{{\mathcal N}=2}$
by the phase $\pm(-1)^m
\e^{\pm\pi i{(\widehat{2m+1})(2j'+1)\over 2(k+2)}}$.

We next consider the 
$\Scr{B}_{{k\over 4},[0]}$-$\Scr{B}_{{k\over 2},[1]}$-string.
The index is represented in the tree-channel as
the pairing $\langle \Scr{B}_{{k\over 4},[0]}|q_t^H|
\Scr{C}_{a^{2m}P_B}\rangle$. The boundary state is the one on
the circle twisted by $a^{4m+2}=a^{2(2m+1)}$, which is 
${}_{g_{4m+2,4m+2}}\langle\Scr{B}_{[{k\over 4},0]}|\times{1\over\sqrt{2}}$ by
the identification (\ref{idbst6}):
$$
I_{a^{2m+1}P_B}(\Scr{B}_{{k\over 4},[0]},\Scr{B}_{{k\over 4},[1]})
={}_{g_{4m+2,2}}\langle
\Scr{B}_{[{k\over 4},0]}|\Scr{C}_{m,1}(+)\rangle\times {(-1)^m\over\sqrt{2}}.
$$
The computation thus reduces to the special case $j={k\over 4}$ of
the unoriented branes, the difference being
a division by $\sqrt{2}$.
Since $N_{{k\over 4}{k\over 4}}^{j'}=1$ for any $j'\in \Pk\cap\Z$
we find
\beqa
I_{a^{2m+1}P_B}(\Scr{B}_{{k\over 4},[0]},\Scr{B}_{{k\over 4},[1]})
&=&{(-1)^m\over\sqrt{2}}
\sum_{j'\in\Pk\cap\Z}
\Bigl(\e^{\pi i(\widehat{2m+1})(2j'+1)\over 2(k+2)}
+\e^{-{\pi i(\widehat{2m+1})(2j'+1)\over 2(k+2)}}\Bigr)
\nn\\
&=&(-1)^m
\sum_{j'\in\Pk\cap\Z}{1+\e^{-\pi i {\widehat{2m+1}\over 2}}\over \sqrt{2}}
\e^{\pi i(\widehat{2m+1})(2j'+1)\over 2(k+2)},
\nn
\eeqa
where, in the second step, we made the change of variable
$j'\to {k\over 2}-j'$ for the second term of the summand.
Note at this point that
$(1+\e^{-\pi i {\widehat{2m+1}\over 2}})/\sqrt{2}=\e^{\pm \pi i /4}$.
This shows that the index is given by
\beq
I_{a^{2m+1}P_B}(\Scr{B}_{{k\over 4},[0]},\Scr{B}_{{k\over 4},[1]})
=\alpha_m\sum_{j\in \Pk\cap\Z}
\e^{\pi i{\widehat{2m+1}\over 2}({2j'+1\over k+2}-{1\over 2})},
\eeq
where $\alpha_m$ is a phase $\pm 1, \pm i$ depending only on $m$.
This shows that the B-parity $a^{2m+1}P_B$ acts on the supersymmetric
ground state in $\Scr{H}_{j,2j+1,1}^{{\mathcal N}=2}$
as the phase multiplication by
$\alpha_m\e^{\pi i{\widehat{2m+1}\over 2}({2j'+1\over k+2}-{1\over 2})}$.

\section{Landau--Ginzburg Orientifolds}\label{sec:LG}

The ${\mathcal N}=2$ minimal model
is realized as the infra-red fixed point
of the ${\mathcal N}=2$ Landau--Ginzburg model of
a single chiral superfield $\Phi$ with superpotential
\beq
W=\Phi^{k+2}.
\label{Wmin}
\eeq
In this section, we study parity invariance and orientifolds of
Landau--Ginzburg models and apply the result to
the particular example (\ref{Wmin}).
We start by determining the condition
for a parity to preserve
A-type or B-type supersymmetry.
We next 
find the integral expression of the overlap
of the crosscap states and the supersymmetric ground states.
Also, the twisted Witten indices are interpreted as the
`intersection numbers' of O-planes and O-planes
or O-planes and D-branes.
We then specialize to the model with the superpotential
(\ref{Wmin}). The results on the overlaps of crosscaps and
supersymmetric ground states and Witten indices
agree with the ones of the ${\mathcal N}=2$ minimal model
obtained in Section~\ref{sub:beGSO}.

\subsection{A-parity and B-parity}

Let us consider a Landau--Ginzburg model
of chiral superfields $\Phi=(\Phi^i)$
with the Lagrangian
\begin{equation}
{\mathcal L}
=\int \dd^4\theta\,
K(\Phi,\bPhi)
+\int \dd\theta^-\dd\theta^+\,W(\Phi)|_{\btheta^{\pm}=0}
+\int \dd\btheta^+\dd\btheta^-\,
\overline{W(\Phi)}|_{\theta^{\pm}=0}.
\label{actionLG}
\end{equation}
$K(\phi,\bphi)$ is the K\"ahler potential for a non-degenerate
K\"ahler metric
$g_{i\bj}=\partial^2 K/\partial\phi^i\partial\bphi^{\bj}$,
and the superpotential $W(\phi)$ is a holomorphic function of
$(\phi^1,...,\phi^n)$.
In terms of the component fields $(\phi^i,\psi_{\pm}^i)$ of $\Phi^i$,
the Lagrangian has the kinetic and four-Fermi terms
for the non-linear sigma model and also 
a potential term $-g^{i\bj}\partial_iW\partial_{\bj}\overline{W}$
as well as the fermion mass term (or `Yukawa coupling') 
$-(D_i\partial_jW\psi_+^i\psi_-^j+c.c.)$.
We may regard $\phi^i$ as local complex coordinates of
some K\"ahler manifold $X$, but we
assume that the first Chern class of $X$ is zero, so that the B-twist
is possible. For the existence of a non-trivial holomorphic function $W$,
$X$ has to be non-compact.

The supercharges are expressed as 
\beqa
&&Q_+=\int\dd x^1\left(
g_{i\bj}(\partial_0+\partial_1)\bphi^{\bj}\psi_{+}^i
- i\bpsi_{-}^{\bi}\partial_{\bi}\overline{W}\right)
\nn\\
&&Q_-=\int\dd x^1\left(
g_{i\bj}(\partial_0-\partial_1)\bphi^{\bj}\psi_{-}^i
+i\bpsi_{+}^{\bi}\partial_{\bi}\overline{W}\right)
\nn\\
&&\bQ_+=\int\dd x^1\left(
g_{i\bj}\bpsi_{+}^{\bj}
(\partial_0+\partial_1)\phi^{i}
+i\psi_{-}^{i}\partial_{i}W\right)
\nn\\
&&\bQ_-=\int\dd x^1\left(
g_{i\bj}\bpsi_{-}^{\bj}
(\partial_0-\partial_1)\phi^{i}
- i\psi_{+}^{i}\partial_{i}W\right).
\nn
\eeqa
We would like to find a parity symmetry that
transforms the supercharges as (\ref{PASUSY}) or (\ref{PBSUSY})
--- A-parity or B-parity.

Let us first consider the A-parity that exchanges $Q_+\leftrightarrow \bQ_-$.
By looking at the expression for the supercharges, it is clear that an
A-parity should map the holomorphic coordinates to antiholomorphic
coordinates.
Furthermore, the superpotential should be conjugated.
Namely, an A-parity is given by $\tau_A\Omega$, where
$\tau_A$ {\it is an antiholomorphic involution of the
K\"ahler manifold $X$ such that}
\begin{equation}
W(\tau_A\phi)=\overline{W(\phi)}+{\rm constant}.
\label{WA}
\end{equation}
This is anomaly-free since we assume that the first Chern class
of $X$ is zero.
Suppose that the K\"ahler potential is the flat one
$K=\sum_i|\Phi^i|^2$ and the superpotential
$W=\sum a_{i_1...i_r}\Phi^{i_1}\cdots\Phi^{i_r}$
has all real coefficients, $a_{i_1...i_r}\in \R$.
Then, for the complex conjugation $\tau:\phi^i\to\overline{\phi^i}$,
$\tau \Omega$ is an A-parity.
In terms of the superfields, this is given by
$$
\Phi^i\longrightarrow \overline{\Omega_A^*\Phi^i},
$$
and the Lagrangian (\ref{actionLG})
is manifestly invariant since
$\Omega_A:\theta^{\pm}\to -\btheta^{\mp}$
maps the measure $\dd\theta^-\dd\theta^+$ to
$\dd\btheta^+\dd\btheta^-$.

Next we consider the B-parity $Q_+\leftrightarrow Q_-$.
Again by looking at the expression of the supercharges,
we find that a B-parity should map the holomorphic coordinates
to holomorphic coordinates.
Since the coefficients of $\bar\partial \overline{W}$ terms
are opposite between $Q_+$ and $Q_-$, we also find that $W$
should be mapped to minus itself, $-W$, up to a constant addition.
Thus a B-parity is given by $\tau_B\Omega$ where
$\tau_B$ {\it is a holomorphic involution of the K\"ahler manifold $X$
such that}
\begin{equation}
W(\tau_B\phi)=-W(\phi)+{\rm constant}.
\label{WB}
\end{equation}
The minus sign can also be understood by looking at the fermion mass term
$W''(\phi)\psi_+\psi_-$: the worldsheet chirality flip $\Omega$ exchanges
$\psi_+$ and $\psi_-$ and thus maps
$\psi_+\psi_-\to\psi_-\psi_+=-\psi_+\psi_-$.
To compensate this minus sign,
the superpotential itself has to flip its sign.
In terms of the superfields, the parity action is given by
$$
\Phi^i\longrightarrow \tau_B^i(\Omega_B^*\Phi).
$$
The Lagrangian expressed in the superspace (\ref{actionLG})
is manifestly invariant under this, since
$W(\tau_B\Phi)=-W(\Phi)+{\rm constant}$ and
$\Omega_B:\theta^{\pm}\to\theta^{\mp}$
flips the sign of the measure
$\dd\theta^-\dd\theta^+$.

One may also consider the variants of A- and B-parities.
For an antiholomorphic involution
$\tau_A$ such that $W(\tau_A\phi)=\e^{2i\alpha}\overline{W(\phi)}$+const,
one can define an
$A_{\alpha,0}$-parity (that does
$Q_+\leftrightarrow \e^{-i\alpha}\bQ_-$)
by combining $\tau_A\Omega$ and a vector R-rotation.
Starting with a B-parity, the 
$B_{0,\beta}$-parity (that transforms $Q_+\leftrightarrow \e^{-i\beta}Q_-$)
is obtained by an additional action of an axial R-rotation (which is
a symmetry of the model under the assumption of
$c_1(X)=0$).
For an antiholomorphic involution
$\tau_A$ such that $W(\tau_A\phi)=-\overline{W(\phi)}$+const,
the operator $(-1)^{F_R}\tau_A\Omega$ is an
$\widetilde{A}$-parity (that does $Q_+\to -\bQ_-,\bQ_-\to Q_+$).
If $\tau_B$ is a holomorphic involution such that
$W(\tau_B\phi)=W(\phi)$+const rather than (\ref{WB}),
then
$(-1)^{F_R}\tau_B\Omega$ is a
$\widetilde{B}$-parity (transforming $Q_+\to -Q_-,Q_-\to Q_+$).

\subsection{Overlap of crosscap and RR ground states}

\newcommand{\tili}{\tilde{\imath}}

Let us consider the A-parity associated with an antiholomorphic involution
$\tau$ of $X$ such that $W(\tau\phi)=\overline{W(\phi)}$+const.
We will compute the overlaps
of the crosscap states and the RR ground states.
If we use the ground state corresponding (via a B-twist) to
the $cc$ ring elements,
the overlaps do not depend on the twisted chiral parameters.
In particular, one can take the large-volume limit of $X$
where the zero-mode approximation is exact.
This is precisely as in the case of overlaps of the boundary state for
A-branes and the same set of ground states
\cite{HIV,HKKPTVVZ}.

We recall that an A-brane in a massive LG model
is wrapped on a Lagrangian submanifold $\gamma$ consisting of the collection
of gradient flow lines of ${\rm Re}(W)$ starting from a critical point.
Its image in the $W$-plane is a straight line emanating
from the critical value and extending in
the real-positive direction.
The overlaps with RR ground states are expressed as
the integrals over $\gamma$ of the ground state wavefunctions
of the zero mode quantum mechanics.
The ground states are middle-dimensional forms on $X$
annihilated by the supercharges
\beqa
&&\bQ_+=\bartial-i\partial W\wedge,
\,\,\,\,\,
Q_+=*(-\partial+i\bartial\overline{W}\wedge)*,
\nn\\
&&
Q_-=\partial+i\bartial\overline{W}\wedge,
\,\,\,\,\,
\bQ_-=*(-\bartial-i\partial W\wedge)*.
\nn
\eeqa
We choose two sets of such ground states $\{\omega_i\}$
and $\{\omega_{\tili}\}$ to be used in the overlaps
$\Pi_i^{\gamma}=\langle \Scr{B}_{\gamma} |i\rangle$ and 
$\widetilde{\Pi}_{\tili}^{\gamma}=\langle \tili |\Scr{B}_{\gamma}\rangle$.
The overlaps are expressed as
\begin{equation}
\Pi_i^{\gamma}
=\int_{\gamma}\e^{-i(W-\overline{W})}\omega_i,\quad
\widetilde{\Pi}_{\tili}^{\gamma}
=\int_{\gamma}\e^{i(W-\overline{W})}*\omega_{\tili}.
\label{piexx}
\end{equation}
The factors $\e^{\mp i(W-\overline{W})}$
are simply some constants on $\gamma$, but they
turn the integrands into  closed forms on $X$.\footnote{
The factors $\e^{\mp i(W-\overline{W})}$ are also naturally induced
from the modification of the boundary term so that
the open string ground state energy is zero. They also make
the overlaps  obey the parallelism
$\nabla \Pi=\nabla\widetilde{\Pi}=0$.}
Since the integrands are closed, we can deform the cycle $\gamma$
without changing the integrals.
We deform it to $\gamma^-$ for $\Pi^{\gamma}_i$
and to $\gamma^+$ for $\widetilde{\Pi}^{\gamma}_{\tili}$,
where $\gamma^{\mp}$ are such that
the $W$-image is rotated by the small phase $\e^{\mp i\epsilon}$
around the critical value.
Even though the boundaries are moved,
this rotation does not change the integral since
$\e^{\mp i (W-\overline{W})}$ works as a convergence factor.
At this point, one can replace $\omega_i$ (resp.~$\omega_{\tili}$)
by another representative of the 
$(\bQ_++Q_-)$ cohomology class (resp.~
$(Q_++\bQ_-)$ cohomology class).
Convenient representatives are of the forms $\e^{-i\overline{W}}\phi_i
{\mit\Omega}$
for $\omega_i$
and $\e^{-i\overline{W}}\overline{\phi_i}\overline{\mit\Omega}$
for $\omega_{\tili}$, where $\phi_i$ are chiral ring elements
and ${\mit\Omega}$ is the nowhere vanishing holomorphic $n$-form of 
$X$.
Thus, we find alternative expressions
\beq
\Pi^{\gamma}_i
=\int_{\gamma^-}\e^{-iW}\phi_i{\mit\Omega},\quad
\widetilde{\Pi}_{\bi}^{\gamma}
=\int_{\gamma^+}\e^{-i\overline{W}}\overline{\phi_i}*\overline{\mit\Omega}.
\eeq
It is useful to think of this in terms of cohomology theory.
Note that $\gamma^{\pm}$ defines an element of
the relative homology group $H_n(X,B_{\pm};\Z)$, where $B_{\pm}$ 
is a region in $X$ such that $\pm{\rm Im}(W)>R$ for a large positive $R$.
On the other hand, $\e^{-iW}\phi_i{\mit\Omega}$
and $\e^{-i\overline{W}}\bphi_i*\overline{\mit\Omega}$
form a basis of the relative cohomology groups 
$H^n(X,B_{-};\C)$ and $H^n(X,B_+;\C)$.
The overlaps are simply the natural pairings.

In the above argument, we have assumed that
the critical points are all non-degenerate and
the corresponding vacua are massive.
Suppose now that $W$ has degenerate
critical points and the theory flows to a non-trivial superconformal
field theory. Let us assume that the theory can be deformed
to a massive theory
by deforming the superpotential
${\mit\Delta}W$, which does not affect the asymptotic behaviour at infinity.
This is the case, for example, in the minimal model
$W=\Phi^{k+2}$, where the addition of a generic lower-order term splits
the degenerate critical point to $k+1$ non-degenerate critical points,
without changing the asymptotic behaviour.
In such a case, 
we can apply the above to the deformed theory, obtaining
the integral formula for
the overlap of the boundary states and RR ground states.
Since the expression obtained in this way is analytic in the perturbation
${\mit\Delta}W$, the overlaps
for the theory before deformation are obtained
by just  setting ${\mit\Delta}W\to 0$.
In this way, we see that for such a class of SCFT, we also have the same
integral formula for the overlaps of the boundary states and RR ground states.
In what follows, this line of argument will be frequently used or assumed.
In the computation involving crosscap states, the only thing to be 
careful of is that the deformation ${\mit\Delta}W$ must preserve the parity
under consideration.
As for the A-parity $\Phi\to \overline{\Omega_A^*\Phi}$ for the minimal model,
this is ensured by using a real polynomial ${\mit\Delta}W$
of lower order.

As in the case of D-branes,
the overlaps $\Pi^{\tau\Omega}_i=\langle\Scr{C}_{(-1)^F\tau\Omega}|i\rangle$
and
$\widetilde{\Pi}^{\tau\Omega}_{\bi}=\langle \bi|\Scr{C}_{\tau\Omega}\rangle$
 of the crosscap states
 and the ground states are expressed
as the integration of $\omega_i$ or $*\omega_{\bi}$
over the $\tau$-fixed locus
$X^{\tau}\subset X$
--- the {\it orientifold plane}.
As in the case of branes,
one can put the constant factor $\e^{\mp i(W-\overline{W})}$
in the integrand and deform the integration submanifold $X^{\tau}$,
in order to express the integral as
the holomorphic (or antiholomorphic) integrals over deformed
orientifold planes.
We note that the $W$-image of $X^{\tau}$ is parallel to the real axis
but not necessarily extending in the real-positive direction;
it could also be extending in the real-negative direction or
in both the real-positive and negative directions.
Thus, in order for
$\e^{\mp i (W-\overline{W})}$ to work as the convergence factor,
we deform the plane $X^{\tau}$ so that the $W$-image is rotated by the phase
$\e^{\mp i\epsilon}$ in the real-positive direction
and by $\e^{\pm i\epsilon}$ in the real-negative direction.
\begin{figure}[tb]
\centerline{\includegraphics{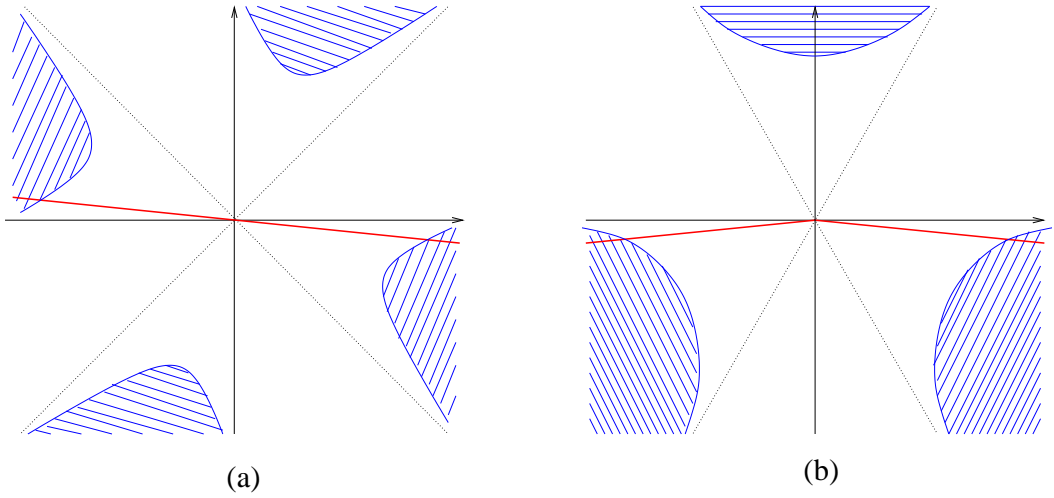}}
\caption{Deformed orientifold planes
for $\tau:\phi\to\bphi$ in the models with $X=\C$ and superpotential
(a) $W=\Phi^4$ and (b) $W=\Phi^3$. The shaded regions are
$B_-$ and the bold (broken) lines are the deformed
orientifold planes $(X^{\tau})^-$.
The other ones $B_+$ and $(X^{\tau})^+$ are obtained by reflection
with respect to the real line.}
\label{defo}
\end{figure}
We call the resulting cycle 
$X^{\tau \mp}$. (See Fig.~\ref{defo}.)
Thus the overlaps are expressed as
\beqa
&&\Pi_i^{\tau\Omega}=
\int_{X^{\tau}}
\e^{-i(W-\overline{W})}\omega_i
=\int_{X^{\tau -}}\e^{-iW}\phi_i{\mit\Omega},
\label{PiGi}\\
&&\widetilde{\Pi}_{\bi}^{\tau\Omega}
=\int_{X^{\tau}}\e^{i(W-\overline{W})}*\omega_{\bi}
=\int_{X^{\tau +}}\e^{-i\overline{W}}\overline{\phi_i}*\overline{\mit\Omega}.
\label{tPiGti}
\eeqa
The deformed orientifold planes define relative homology
classes, $[X^{\tau \mp}]\in H_n(X,B_{\mp};\Z)$.
Then, one can consider the expressions (\ref{PiGi}) and
(\ref{tPiGti}) as the pairing of
$[X^{\tau \mp}]$
and the basis of the group $H^n(X,B_{\pm};\C)$
defined by the chiral ring elements $\phi_i$.

\subsection{The twisted Witten index}

Using the expressions for the overlaps with the RR ground states,
one can find a homological formula for the Witten index.
Inserting the expressions (\ref{piexx})
into the bilinear identity (\ref{biIab}),
we find
$$
I(a,b)=\sum_{i,\bj}\int_{\gamma_a^-}\e^{-\beta(W-\overline{W})}\omega_i
\,g^{i\bj}\int_{\gamma_b^+}\e^{i\beta(W-\overline{W})}*\omega_{\bj},
$$
where $g^{i\bj}$ is the inverse of the matrix
$g_{i\bj}=\int_X\omega_i\wedge *\omega_{\bj}$.
We have similar expressions for the twisted Witten indices
$I_{\tau\Omega}$ and $I_{\tau\Omega}(a)$.

By Riemann's bilinear identity, such expressions
can be identified as certain intersection numbers (see \cite{HKKPTVVZ}).
Here, an essential role is played by Poincar\'e duality.
For each homology class $[C^+]$
in $H_n(X,B_+)$ we find a cohomology class ${\rm Pd}[C^+]$ in
$H^n(X,B_-)$ such that $\int_X{\rm Pd}[C^+]\wedge \eta_+=\int_{C^+}\eta_+$
for any $\eta_+\in H^n(X,B_+)$ and
$\int_{D^-}{\rm Pd}[C^+]=\#(D^-\cap C^+)$ for any $D^+\in
H_n(X,B_+)$.
It is roughly a delta function supported on $C^+$.
Since the group $H^n(X,B_-)$
is spanned by $\e^{-i\beta(W-\overline{W})}\omega_i$,
one can express the cohomology class ${\rm Pd}[C^+]$
as the linear combination
$\sum_ic^i\e^{-i\beta(W-\overline{W})}\omega_i$.
The coefficients $c^i$ can be found by taking the wedge product with
$\e^{i\beta(W-\overline{W})}*\omega_{\bj}\in H^n(X,B_+)$
and integrating over $X$. This shows
$c^i=\sum_{\bj}g^{i\bj}\int_{C^+}\e^{i\beta(W-\overline{W})}*\omega_{\bj}$.
Thus, we find the expression for ${\rm Pd}[C^+]$, or
$$
\sum_{i,\bj}\int_{D^-}\e^{-i\beta(W-\overline{W})}\omega_i \,g^{i\bj}
\int_{C^+}\e^{i\beta(W-\overline{W})}*\omega_{\bj}
=\int_{D^-}{\rm Pd}[C^+]=\#(D^-\cap C^+)
$$
for $D^-\in H_n(X,B_-)$.

Applying this to $D^-=\gamma_a^-$ and $C^+=\gamma_b^+$, we find
\begin{equation}
I(\gamma_a,\gamma_b)=\#(\gamma_a^-\cap\gamma_b^+).
\end{equation}
For the parity-twisted Witten index of the closed string,
we find
\begin{equation}
I_{\tau\Omega}=\#(X^{\tau -}\cap X^{\tau +}).
\end{equation}
This is the Landau--Ginzburg version of
 the index formula (\ref{indpapa}) in non-linear sigma models.
For the parity-twisted Witten index for the
$\gamma$-$\tau\gamma$ open string,
we find
\begin{equation}
I_{\tau\Omega}(\gamma)=\#(\gamma^-\cap X^{\tau +}).
\end{equation}
This is the LG counterpart of the index formula (\ref{indbapa}).
We also find $I_{\tau\Omega}(\tau\gamma)=\#(X^{\tau -}\cap \gamma^+)$,
which is consistent with the above since
$\#(\gamma^-\cap X^{\tau +})=\#(X^{\tau-}\cap (\tau\gamma)^+)$.

\subsection{The case of $W=\Phi^{k+2}$}

The LG model with superpotential (\ref{Wmin})
flows to the level $k$ 
${\mathcal N}=2$ minimal model \cite{M,VW} (see also \cite{minW}).
The system has both vector and axial R-symmetries
$$
U(1)_V:\Phi(\theta^{\pm})
\to \e^{2i\alpha\over k+2}\Phi(\e^{-i\alpha}\theta^{\pm}),
\quad
U(1)_A:\Phi(\theta^{\pm})\to
\Phi(\e^{\mp i\beta}\theta^{\pm}),
$$
which correspond to two combinations of the
$U(1)_R$ and $U(1)_L$ R-symmetries
(\ref{RLR}).
There is also a discrete symmetry
generated by
$$
\Phi(\theta^{\pm})\to\e^{\pi i\over k+2}\Phi(\pm\theta^{\pm}),
$$
which corresponds to the $\Z_{2(k+2)}$ symmetry (\ref{axsym}).
This was found by a subtle analysis of the anomaly in the gauged WZW model,
but it is evident
at the level of the classical Lagrangian in the LG model.

The chiral ring of the model is $\C[\phi]/W'(\phi)$,
namely, generated by $\phi$ and subject to the relation
$\phi^{k+1}=0$.
The ring elements correspond to the states
\beq
|j\rangle_{cc}\leftrightarrow \phi^{2j}.
\eeq
The antichiral fields correspond to the states
\beq
|j\rangle_{aa}\leftrightarrow \bphi^{2j}.
\eeq
They can also be related to the RR ground states $|j\rangle^{}_{{}_{\rm RR}}$
by  spectral flow.

In what follows, we study orientifolds (and D-branes) of this LG model, and
compare with the results obtained in the previous section.

\subsubsection{A-orientifolds}

Let us first find the A-parities of the system. As discussed above,
an A-parity is of the form $\tau_A\Omega$, where $\tau_A$ is
an antiholomorphic
map such that $(\tau_A\phi)^{k+2}=\bphi^{k+2}$ up to a possible addition
of a constant.
We find $k+2$ of them given by
\beq
\qquad\qquad\tau_A^{2m}:\phi\to\e^{2\pi i m\over k+2}\bphi,\qquad
m=0,1,\ldots,k+1.
\eeq
We also find the same number of $\widetilde{A}$-parities
$(-1)^{F_R}\tau_A^{2m+1}\Omega$,
where
\beq
\qquad\qquad\tau_A^{2m+1}:\phi\to\e^{\pi i (2m+1)\over k+2}\bphi,\qquad
m=0,1,\ldots,k+1.
\eeq
Note that $\tau_A^{\ell}$
maps the chiral primary fields $\phi^{2j}$ to antichiral
primary fields $\e^{2\pi i\ell j\over k+2}\bphi^{2j}$.
Comparing with the action of $P_A^{\ell}$, which maps
$|j\rangle_{cc}$ to
$\e^{2\pi i\ell j\over k+2}|j\rangle_{aa}$
up to an $\ell$-independent phase multiplication,
we find that these are related as
\beq
a^{2m}P_A=\tau_A^{2m}\Omega,
\quad
a^{2m+1}P_A=(-1)^{F_R}\tau_A^{2m+1}\Omega,
\eeq
possibly with a uniform shift of $m$.

In what follows we study some properties of the parity for
the involution $\tau=\tau_A^0$. Other A-parities are simply combinations
with the discrete axial symmetries.

\subsubsection*{\it Closed string Witten index}

\begin{figure}[tb]
\centerline{\includegraphics{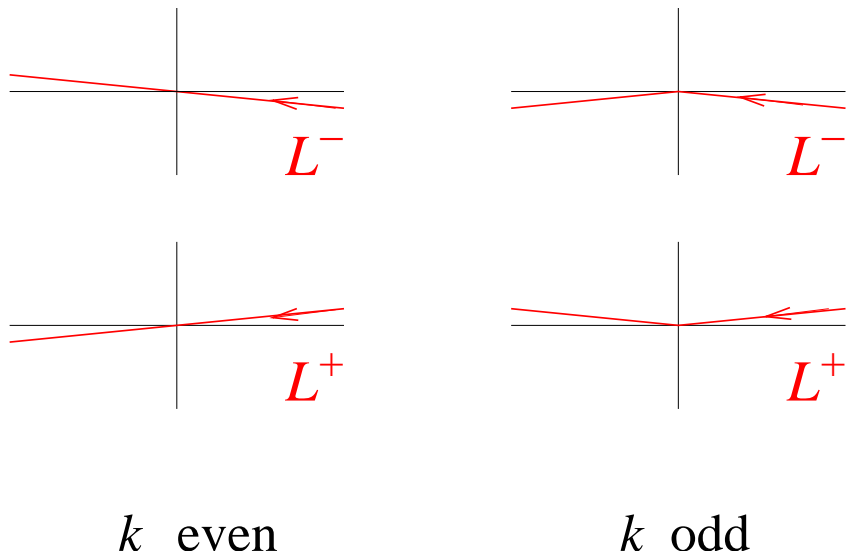}}
\caption{The deformed orientifold planes}
\label{Lpm}
\end{figure}

The orientifold plane $L$ for $\tau\Omega$ is the real line
$\phi\in \R$.
Its $W$-image is the semi-infinite line $\R_{\geq 0}$
for even $k$ and the real line itself for odd $k$.
Thus, the deformed orientifold planes $L^{\pm}$
are straight lines $\e^{\pm i\epsilon}\R$ for even $k$
while they are infinitesimally bent lines
$\e^{\pm i\epsilon}\R_{\geq 0}\cup
\e^{\mp i\epsilon}\R_{\leq 0}$ for odd $k$.
This is shown in Fig.~\ref{Lpm}, 
where we have chosen an orientation of each of
the planes.
We find from this that the parity-twisted Witten index for the closed string
is given by
\beq
I_{\tau\Omega}=\#(L^-\cap L^+)
=\left\{\begin{array}{ll}
1&\mbox{$k$ even}\\
0&\mbox{$k$ odd}.
\end{array}
\right.
\eeq
This is in agreement with the result (\ref{WindMMA}).

\subsubsection*{\it Overlap of the crosscap and the RR ground states}

Let us compute the overlap of the crosscap states and the
RR ground states using the LG model.
The differential form $\phi_{j}{\mit \Omega}$ corresponding to
the normalized ground state $|j\rangle^{}_{{}_{\rm RR}}$ is
\beq
\phi_{j}{\mit\Omega}
=c_j\phi^{2j}\dd\phi,
\eeq
where
\beq
c_j=-i\e^{-{\pi i(2j+1)\over 2(k+2)}}{\sqrt{k+2}\over
\Gamma({2j+1\over k+2})\sqrt{2\sin({\pi(2j+1)\over k+2})}}.
\label{cl}
\eeq
We have chosen this normalization so that the associated ground states
have unit norm $\langle \bj|j'\rangle=\delta_{j,j'}$.
(See Appendix~\ref{app:orv} for an explanation of this point.)
It is straightforward to compute the integrals
(\ref{PiGi}) and (\ref{tPiGti})
\beqa
&&\Pi_{j}^{\tau\Omega}
=
c_j\int_{L^-}\e^{-i\phi^{k+2}}\phi^{2j}\dd\phi,
\nn\\
&&\widetilde{\Pi}_{\bj}^{\tau\Omega}
=\overline{c_j}\int_{L^+}\e^{-i\bphi^{k+2}}\bphi^{2j}*\dd\bphi.
\nn
\eeqa
They are
\beqa
\Pi_{j}^{\tau\Omega}
&=&\left\{
\begin{array}{ll}
{\displaystyle i\e^{-{\pi i(2j+1)\over (k+2)}}{1+(-1)^{2j}\over
\sqrt{2(k+2)\sin({\pi(2j+1)\over k+2})}}}
&\mbox{$k$ even},\\[0.8cm]
{\displaystyle i\e^{-{\pi i(2j+1)\over 2(k+2)}}
{\e^{-{\pi i(2j+1)\over 2(k+2)}}
+(-1)^{2j}\e^{{\pi i(2j+1)\over 2(k+2)}}
\over
\sqrt{2(k+2)\sin({\pi(2j+1)\over k+2})}}}
&\mbox{$k$ odd},
\end{array}\right.
\\[0.7cm]
\widetilde{\Pi}_{\bj}^{\tau\Omega}
&=&\left\{
\begin{array}{ll}
{\displaystyle
{1+(-1)^{2j}\over
\sqrt{2(k+2)\sin({\pi (2j+1)\over k+2})}}}
&\mbox{$k$ even},\\[0.8cm]
{\displaystyle
\e^{\pi i(2j+1)\over 2(k+2)}{\e^{-{\pi i(2j+1)\over 2(k+2)}}
+(-1)^{2j}\e^{\pi i(2j+1)\over 2(k+2)}\over
\sqrt{2(k+2)\sin({\pi (2j+1)\over k+2})}}}
&\mbox{$k$ odd}.
\end{array}\right.
\eeqa
Let us compare these with the results
(\ref{ovPj}) and (\ref{ovFPj}) from the RCFT analysis.
It is straightforward to check that
they completely agree:
\beqa
&&
\Pi^{\tau\Omega}_j=\langle\Scr{C}_{(-1)^FP_A}|j\rangle^{}_{{}_{\rm RR}},
\nn\\
&&
\widetilde{\Pi}^{\tau\Omega}_{\bj}
={}^{}_{{}_{\rm RR}}\!\langle j|\Scr{C}_{P_A}\rangle.
\nn
\eeqa

\subsubsection*{\it Open string Witten index}

\begin{figure}[tb]
\centerline{\includegraphics{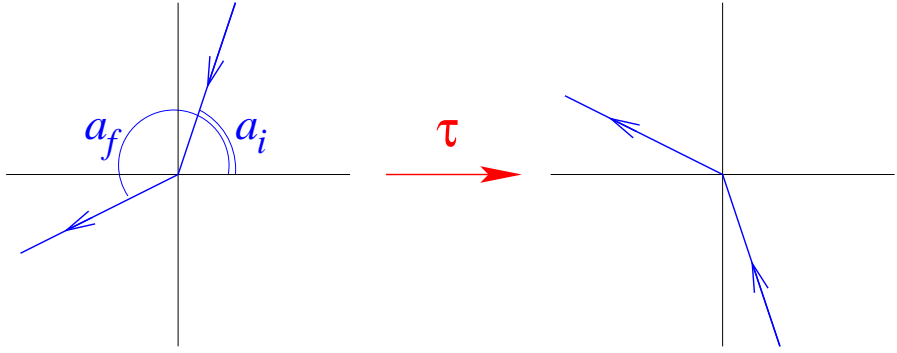}}
\caption{D-brane transform
$\tau:A_{a_f,a_i}\to A_{-a_f,-a_i}$}
\label{Abr}
\end{figure}

The A-branes of the model are studied 
in \cite{HIV}, using the deformation
to a massive model, and found to be D1-branes at the wedge-shaped
broken lines $A_{a_f,a_i}$.
The broken line $A_{a_f,a_i}$ is the inward half-line
$\{\phi\in\e^{ia_i}\R_{\geq 0}\}$ joined at the origin to the outward
half-line $\{\phi\in\e^{ia_f}\R_{\geq 0}\}$
where the angles are distinct, $a_i\ne a_f$, and are
 quantized as $a_i, a_f\in {2\pi \over k+2}\Z_{k+2}$.
See the left figure in Fig.~\ref{Abr}.
The Cardy brane $\Scr{B}_{j,n,s}$ with odd $s$ corresponds to
the broken line $A_{j,n,s}$ where
$A_{j,n,s=\pm 1}=A_{a_{\pm},a_{\mp}}=\pm A_{a_+,a_-}$
where $a_{\pm}={\pi(n\pm 2j\pm 1)\over k+2}$.
The (untwisted) open string index is computed to be \cite{HIV}
\beq
I(A_{jns},A_{j'n's'})
=\#(A_{jns}^-\cap A_{j'n's'}^+)
=(-1)^{s-s'\over 2}N_{j\,j'}^{n'-n\over 2},
\label{OwIn}
\eeq
where $N_{j_1\,j_2}^{j_3}$ is the level $k$ $SU(2)$ fusion coefficients,
which are extended outside the standard region
$0\leq j_3\leq {k\over 2}$
by
$N_{j_1\,j_2}^{j_3}=-N_{j_1\,j_2}^{-j_3-1}=N_{j_1\,j_2}^{j_3+(k+2)}$
and $N_{j_1\,j_2}^{-{1\over 2}}=0$.
Let us now compare it with the results obtained in the RCFT analysis.
We see that the above result agrees with $I(\Scr{B}_{jns},\Scr{B}_{j'n's'})$
given in (\ref{cylindex}), if we make a shift
$(n,s)\to (n-1,s-1)$ in the latter.
This implies the following identification
of the brane and the boundary states
\beq
\begin{array}{ccccc}
&&A_{jns}&&\\
&\swarrow&&\searrow&\\
\langle \Scr{B}_{j,n-1,s-1}|&&&&|\Scr{B}_{j,n,s}\rangle.
\end{array}
\label{correspondence}
\eeq
Furthermore, under this correspondence, the overlaps
of the boundary states and the RR ground states are exactly
identical in the two descriptions \cite{HIV}
(see also Appendix~\ref{app:orv}):
\beqa
&&\Pi_j^{A_{jns}}
=c_j\int_{A_{jns}^-}\e^{-i\phi^{k+2}}\phi^{2j}\dd\phi
={}^{}_{{}_{\rm RR}}\langle \Scr{B}_{j,n-1,(s-1)}|j\rangle^{}_{{}_{\rm RR}},
\nn\\
&&\widetilde{\Pi}^{A_{jns}}_{\bj}
=\overline{c_j}
\int_{A_{jns}^+}\e^{-i\bphi^{k+2}}\bphi^{2j}*\dd\bphi
={}^{}_{{}_{\rm RR}}\langle j|\Scr{B}_{j,n,(s)}\rangle^{}_{{}_{\rm RR}}.
\nn
\eeqa

Let us now look at the parity action
on the branes.
Since the parity acts on the $\phi$-plane by complex
conjugation which inverts the phases,
the A-branes are mapped under $\tau$ as
$A_{a_f,a_i}\to
A_{-a_f,-a_i}$.
(See Fig.~\ref{Abr}.)
Since $-a_{\pm}(j,n)={\pi(-n\mp 2j\mp 1)\over k+2}
=a_{\mp}(j,-n)$, this is equivalent to
\beq
\tau:A_{j,n,s=1}\to A_{j,-n,s=-1}.
\eeq
This is in agreement with the RCFT result (\ref{RulE})
under the correspondence
(\ref{correspondence}).

Let us next compute the parity-twisted Witten index for the open string
stretched from
$A_{a_f,a_i}$ to $\tau A_{a_f,a_i}$.
The index is given by the intersection number of
the deformed D-brane $A_{a_f,a_i}^-$ and the deformed orientifold plane $L^+$.
It is easy to see that
\beq
I_{\tau\Omega}(A_{a_f,a_i},A_{-a_f,-a_i})
=\#(A_{a_f,a_i}^-\cap L^+)
=\left\{\begin{array}{ll}
-1&
0< a_i \leq\pi < a_f \leq 2\pi\\
1&0 < a_f \leq \pi < a_i \leq 2\pi\\
0&\mbox{otherwise}.
\end{array}\right.
\eeq
One could reduce the computation to the computation of the brane intersection
number by using the relation $\#(A^-\cap L^+)=\pm \#(A^-\cap \tau A^+)$.
For $A_{jns=1}$, the sign here is $-$ for $0<n\leq k+2$
 and $+$ for $k+2 <n\leq 2(k+2)$.
On the other hand, we know the brane intersection number
by formula (\ref{OwIn}), $\#(A_{jn1}^-\cap \tau A_{jn1}^+)
=\#(A_{jn1}^-\cap A_{j,-n,-1}^+)
=N_{j\,j}^{n-1}$ in which the extension of
$N_{j_1j_2}^{j_3}$ outside $0\leq j_3\leq {k\over 2}$
is understood.
Thus, we find
\beq
I_{\tau\Omega}(A_{j,n,1},A_{j,-n,-1})
=\left\{
\begin{array}{ll}
N_{j\,j}^{n-1}&{1\over 2}\leq n\leq {k+2\over 2}+{1\over 2}\\[0.1cm]
-N_{j\,j}^{(k+2)-n}&{k+2\over 2}+{1\over 2}\leq n\leq (k+2)\\[0.1cm]
-N_{j\,j}^{n-(k+2)-1}&(k+2)+{1\over 2}\leq n\leq
{3(k+2)\over 2}+{1\over 2}\\[0.1cm]
N_{j\,j}^{2(k+2)-n}&{3(k+2)\over 2}+{1\over 2}\leq n\leq 2(k+2).
\end{array}\right.
\eeq
This is in complete agreement with (\ref{IpABB})
under the correspondence
(\ref{correspondence}).
As in (\ref{IpABBa}),
this can be concisely written as
\beq
I_{\tau\Omega}(A_{j,n,s},A_{j,-n,s+2})
=(-1)^{s-1\over 2}\widetilde{N}_{j,j}^{n-1},
\eeq
where $\widetilde{N}_{j_1,j_2}^{j_3}=N_{j_1,j_2}^{j_3}$
in $-{1\over 2}\leq j_3\leq {k+1\over 2}$ and is extended by
$\widetilde{N}_{j_1,j_2}^{j_3}
=\widetilde{N}_{j_1,j_2}^{-j_3-1}=-\widetilde{N}_{j_1,j_2}^{j_3+(k+2)}$
outside that region.
Yet another expression is obtained by noting that
the orientifold plane $L$ is equal  or close to one of the branes.
This leads to
\beq
I_{\tau\Omega}(A_{jns})
=(-1)^{s-1\over 2}N_{j,{1\over 2}[{k+1\over 2}]}^{n+[{k-1\over 2}]\over 2}.
\eeq

\subsubsection{B-orientifolds}

We next consider B-parities.
A B-parity should be of the form
$\tau_B\Omega$, where $\tau_B$ is a holomorphic map
such that $(\tau_B\phi)^{k+2}=-\phi^{k+2}$
plus a possible constant.
We find $k+2$ of them given by
\beq
\qquad\qquad\tau_B^{2m+1}:\phi\to\e^{\pi i (2m+1)\over k+2}\phi,\qquad
m=0,1,\ldots,k+1.
\eeq
We also find $\widetilde{B}$-parities
$(-1)^{F_R}\tau_B^{2m}\Omega$
with
\beq
\qquad\qquad\tau_B^{2m}:\phi\to\e^{2\pi im\over k+2}\phi,\qquad
m=0,1,\ldots,k+1.
\eeq
$\tau_B^{\ell}$ maps the chiral primary fields
$\phi^{2j}$ to $\e^{2\pi i\ell j\over k+2}\phi^{2j}$.
Comparing with the action of $P_B^{\ell}$ which maps
$|j\rangle_{cc}$ to
$\e^{2\pi i\ell j\over k+2}|j\rangle_{cc}$,
up to an $\ell$-independent phase multiplication,
we find that these are related as
\beq
a^{2m+1}P_B=\tau_B^{2m+1}\Omega,
\quad
a^{2m}P_B=(-1)^{F_R}\tau_B^{2m}\Omega,
\eeq
possibly with a uniform shift of $m$.

For even $k$, there is no involutive B-parity, and therefore
one cannot think about the crosscap states.
For odd $k$ there is a unique involutive B-parity,
$P^{k+2}_B=\tau_B^{k+2}\Omega$. The crosscap state
$|P_B^{k+2}\rangle$ of course consists of RR-sector states.
However, it has no overlap with RR ground states, since
$|P_B^{k+2}\rangle$ has to have zero R-charge but, for odd $k$,
there is simply no RR ground states with vanishing R-charge.
Thus, no non-trivial computation  can be done in
the LG realization.
\footnote{This is unlike the result of the study of B-branes, where
for even $k$ there is one B-brane with a non-trivial
overlap with the RR ground state $|{k\over 4}\rangle$ of zero R-charge,
which is computable in the LG realization \cite{HKKPTVVZ}.}

In the gauged WZW model, it was a non-trivial task to find
anomaly-free B-parity symmetries.
For instance, the transform ${\mathcal I}_B\Omega$ was anomalous since
it flips the sign of the fermion path-integral measure
in an odd instanton background.
In the LG realization, it can readily be found at the level of the classical
Lagrangian:
The analogue of ${\mathcal I}_B\Omega$ would be
$\tau_B^0\Omega$ and it flips the sign of the fermion mass term
$W''(\phi)\psi_+\psi_-$.
This reminds us of mirror symmetry where
the effect of worldsheet instantons in one theory is
classical in the mirror.
In fact, in a closely related $SL(2,\R)/U(1)$ gauged WZW model,
one can find a mirror transform to a LG model
whose superpotential reflects the instanton effects \cite{HK}.
The present observation suggests that this story may extend
also to the $SU(2)/U(1)$ model.
In the next section,
we will see among other things
that parity anomalies in non-linear sigma models
are reflected as the explicit breaking by
 the superpotential in the mirror LG models.

\section{Orientifolds of Linear Sigma Models and Mirror Symmetry}
\label{sec:LSM}

Linear sigma models provide a way to realize and study
non-linear sigma models on a class of target spaces from a global
view point,
and allow us to find a picture of the moduli space of theories.
In this and the next section we study parity symmetries of
linear sigma models.
We determine the conditions on the parameters for theory
to be invariant under A- and B-type parities.
The condition agrees
in the large volume limit with the one derived from the non-linear sigma model
and at the LG orbifold point to the one from the LG model.
We also determine the corresponding parity in
the mirror Landau--Ginzburg model.
In particular, we find that the information on the parity actions on
the line bundle ${\mathcal L}_{\tau^*B+B}$ (which we found to be
required in Section~\ref{subsubsec:Bfield})
 has a natural counterpart in the mirror LG model as the
type of the orientifold planes.
The general results are applied in specific examples.

\subsection{Parity symmetry of linear sigma models}

Let us consider a $(2,2)$ supersymmetric
$U(1)^k$ gauge theory with $N$ matter fields $\Phi_1,\ldots,\Phi_N$ of
charge $Q_i^a$, where $i=1,\ldots,N$ labels the matter fields
and $a=1,\ldots, k$ labels the gauge group.
The basic Lagrangian of the model is given by
\beq
L_1=\int\dd^4\theta \left[
\sum_{i=1}^N\bPhi_i\e^{Q_i\cdot V}\Phi_i-\sum_{a=1}^k
{1\over e_a^2}|\Sigma_a|^2\right]
+{\rm Re}\int\dd^2\widetilde{\theta}\sum_{a=1}^k\Bigl(
-t^a\Sigma_a\Bigr).
\eeq
$V_a$ in $Q_i\cdot V:=\sum_{a=1}^kQ_i^aV_a$ are
the vector superfields, $\Sigma_a:=\bD_+D_-V_a$ are the fieldstrengths
(twisted chiral superfields), and $e_a$ are the gauge coupling constants
(with dimension of mass). The measure
is 
$\dd^2\widetilde{\theta}=\dd\btheta^-\dd\theta^+$
for the twisted
chiral superfields
and $t^a$ are the complex combinations of the Fayet--Iliopoulos
parameters and theta angles:
\beq
t^a=r^a-i\theta^a.
\eeq
If one can find a gauge-invariant holomorphic polynomial
$W(\Phi_1,\ldots,\Phi_N)$, one may also consider adding an F-term
\beq
L_2={\rm Re}\int \dd^2\theta \,W(\Phi_1,\ldots,\Phi_N).
\eeq
With or without this additional term, the FI parameter
is renormalized as $r^a(\mu')=r^a(\mu)+\sum_{i=1}^Nb_1^a\log(\mu'/\mu)$,
where
$$
b_1^a:=\sum_{i=1}^NQ_i^a.
$$
At certain energies lower than the gauge couplings,
the system reduces to the non-linear sigma model
on the vacuum manifold if $r^a$ is in a suitable region.
If the Lagrangian is given only by $L_1$, the vacuum manifold is a 
toric manifold $X$ determined by the symplectic reduction
$\sum_{i=1}^NQ_i^a|\phi_i|^2=r^a$ mod $U(1)^k$.
With the additional term $L_2$ it is a subspace
of $X$ determined by the equation $\partial_iW=0$.
In either case, it has a second cohomology group of
rank $k$.
With respect to an integral basis $\{\omega_a\}$, the K\"ahler class
is roughly written as $\sum_{a=1}^kr^a\omega_a$ and
the first Chern class is given by $\sum_{a=1}^kb_1^a\omega_a$.
There are, however, regions of $r^a$,  where the system reduces
to  Landau--Ginzburg orbifolds
or some hybrid of them, rather than a non-linear sigma model.

We are interested in the parity symmetries of this class of
gauge systems.

\subsubsection{A-parity}

Let us consider the A-parity transformation
\beq
\begin{array}{l}
\Phi_i\longrightarrow \overline{\Omega_A^*\Phi_i},
\\[0.2cm]
V_a\longrightarrow \Omega_A^*V_a.
\end{array}
\label{Alsm}
\eeq
This is compatible with the gauge symmetry, as  can be seen from
the action on the components $(\phi(x),v_{\mu}(x))\to
(\bphi(\wtx),-(-1)^{\mu}v_{\mu}(\wtx))$, or by looking how
 the superfield gauge transformation
$\Phi\to\e^{i\Lambda}\Phi, V\to V+i(\bLambda-\Lambda)$ is affected.
One can also check that the Lagrangian $L_1$ is invariant:
The term $\int\dd^4\theta\, \bPhi\e^{V}\Phi$ is obviously invariant.
To see the rest, we note that the field strength transforms as
$$
\Sigma_a\longrightarrow -\Omega_A^*\Sigma_a.
$$
Then, the kinetic term $\int\dd^4\theta |\Sigma|^2$ is also invariant.
The twisted superpotential $-t\Sigma$ flips sign, but the sign is cancelled
since the measure $\dd^2\widetilde{\theta}$ is odd under $\Omega_A$.
This shows the invariance of $L_1$.
For the potential term $L_2$ to be invariant, however,
the superpotential $W$ has to obey the condition
$W(\bPhi_i)=\overline{W(\Phi_i)}$, that is, the coefficients of
the polynomial $W$ all have to be  real.
If the system describes a non-linear sigma model
at certain energies, the map $\phi_i\to
\bphi_i$ reduces to an antiholomorphic involution of the target
space, and the parity corresponds to the A-parity
of the non-linear sigma model associated with it.

In the quantum theory, however, we have to see if the path-integral
measure is also invariant. Here we encounter a possible anomaly of the type
considered in the minimal model (and the non-linear sigma model), since
the topology of any gauge field background is preserved under
 the parity we are considering, $v_a\to -\Omega^*v_a$.
Following the argument of Section \ref{subsub:panomaly},
we see that the fermion
measure changes as follows
\beq
{\mathcal D}_v\Psi
\longrightarrow
(-1)^{b_1\cdot c_1}
{\mathcal D}_v\Psi,
\label{measTr}
\eeq
where $b_1\cdot c_1:=\sum_{a=1}^kb_1^ac_1(V_a)$ in which
$V_a$ is the $U(1)$ bundle of the $a$-th gauge field.
Thus, if some of the $b_1^a$ are odd, the parity symmetry is anomalous.
This corresponds to the anomaly of an A-parity
in the non-linear sigma model in the case where the target space
is not spin, since $b_1^a$ provides the first Chern class of the manifold.

One may consider combining (\ref{Alsm}) with an internal action
on the fields.
Here we describe the case where this action involves a permutation of the
$\Phi_i$'s.
A permutation
$\Phi_i\to \Phi_{\sigma(i)}$
is compatible with the gauge symmetry if there is a linear
transformation $V\to V'$ such that $Q_i\cdot V'=Q_{\sigma(i)}\cdot V$.
Namely, it should be accompanied by an action on the gauge field
$V_a\to \sigma_a^{\,b}V_b$
where $\sigma_a^{\,b}$ is a matrix such that
\beq
\sum_{b=1}^kQ_i^b\sigma_b^{\, a}=Q_{\sigma(i)}^a.
\label{Qsigma}
\eeq
Thus, we consider the combined transformation
\beq
\begin{array}{l}
\Phi_i\longrightarrow \overline{\Omega_A^*\Phi_{\sigma(i)}},
\\[0.2cm]
V_a\longrightarrow \sigma_a^b\Omega_A^*V_b.
\end{array}
\label{Alsm2}
\eeq
Under this, the Lagrangian $L_1$ is invariant if the FI-theta
parameters obey $t^b\sigma_b^{\, a}=t^a$ (mod $2\pi i \Z$).
In the quantum theory, because of 
the transformation of the path-integral measure
(\ref{measTr}), the condition for this to be a symmetry is
\beq
t^b\sigma_b^{\,a}=t^a+\pi i b_1^a \qquad\mbox{mod $2\pi i\Z$}.
\label{Asymlin}
\eeq
If this is not satisfied,
or if there is no solution to this equation, the parity symmetry
is anomalous.
If the model has the potential term $L_2$, there is a further condition
that
\beq
W(\bPhi_{\sigma(i)})=\overline{W(\Phi_i)}.
\eeq
In the picture of the non-linear sigma model,
$\phi_i\to \overline{\phi_{\sigma(i)}}$ corresponds to
an antiholomorphic diffeomorphism $f$ of the target space.
The symmetry condition (\ref{Asymlin}) is nothing but the geometric
condition (\ref{Asymc}) in the sigma model, since one can identify
$c_1(X)^a=b_1^a$ and $f_a^{\, b}=-\sigma_a^{\, b}$.
The latter holds because the ${\rm Re}\Sigma_a$ correspond to
the integral basis $\omega_a$ of $H^2(X)$,
and they are transformed under the parity
as
$$
\Sigma_a\longrightarrow -\sigma_a^{\,b}\Omega_A^*\Sigma_b,
$$
while $f_a^{\, b}$ is defined as $\omega_a\to f^*\omega_a
=f_a^{\, b}\omega_b$.

One may consider yet another modification of
the parity transformation. It is to combine (\ref{Alsm}) or (\ref{Alsm2})
with one of the $U(1)^{N-k}$ torus actions, $\Phi_i\to \e^{i\theta_i}\Phi_i$.
This is a symmetry under the same condition on
$t^a$ as before but, if there is an $L_2$ term, under the modified condition
on the superpotential
$W(\e^{i\theta_i}\overline{\Phi_{\sigma(i)}})=\overline{W(\Phi_i)}$.

\subsubsection{B-parity}

Let us next consider the B-parity transformation
\beq
\begin{array}{l}
\Phi_i\longrightarrow \e^{i\theta_i}\Omega_B^*\Phi_{i},
\\[0.2cm]
V_a\longrightarrow \Omega_B^*V_a.
\end{array}
\label{Blsm}
\eeq
It acts on the component fields as $(\phi(x),v_{\mu}(x))\to
(\phi(\wtx),(-1)^{\mu}v_{\mu}(\wtx))$ and is compatible with
the gauge symmetry.
The field strength transforms as
$$
\Sigma_a\longrightarrow \overline{\Omega_B^*\Sigma_a},
$$
and therefore the action $L_1$ is invariant under the condition
that the $t^a$ are all real. The potential term $L_2$, if it is present,
is always invariant. This parity symmetry is anomaly-free.
In the sigma model picture, this corresponds to the B-parity associated
with the identity action on the target space.

As before, one may also consider combining this with an internal action on
the fields. We discuss the combination with a permutation of the fields
\beq
\begin{array}{l}
\Phi_i\longrightarrow \e^{i\theta_i}\Omega_B^*\Phi_{\sigma(i)},
\\[0.2cm]
V_a\longrightarrow \sigma_a^{\,b}\Omega_B^*V_b,
\end{array}
\label{Blsm2}
\eeq
where we need the relation (\ref{Qsigma}) for compatibility
with the gauge symmetry.
The condition of invariance of $L_1$ is
\beq
t^b\sigma_b^{\,a}=\overline{t^a}
\qquad\mbox{mod $2\pi i\Z$},
\label{Bsymlin}
\eeq
and the condition for invariance of $L_2$ is
\beq
W(\e^{i\theta_i}\Phi_{\sigma(i)})=-W(\Phi_i).
\eeq
The symmetry (\ref{Blsm2}) is an exact symmetry of the quantum theory.
In the sigma model picture, it
corresponds to the B-parity symmetry associated
with a holomorphic automorphism $f$ of the target space.
The condition (\ref{Bsymlin}) is nothing but
the geometric condition (\ref{Bsymc})
since $\sigma_a^{\, b}=f_a^{\, b}$.
The latter holds because the real part of the field strength is
transformed as
${\rm Re}\Sigma_a\to \sigma_a^{\, b}\Omega_B^*{\rm Re}\Sigma_b$
while $f_a^{\, b}$ is defined as the action on the
integral basis of $H^2$, $\omega_a\to f^*\omega_a=f_a^{\,b}\omega_b$.


\subsection{Description in the mirror LG model}

The model with Lagrangian $L_1$
has a mirror description \cite{HV}.
It is obtained by dualization of the phase of the charged matter fields,
taking into account the effect of the vortex instantons.
The dualization of $\arg(\Phi_i)$ yields a twisted chiral superfield
$Y_i$ with periodic identification $Y_i\equiv Y_i+2\pi i$.
The twisted superpotential of the dual theory
is
$$
\widetilde{W}=\sum_{a=1}^k\left(\sum_{i=1}^NQ_i^aY_i-t^a\right)\Sigma_a
+\sum_{i=1}^N\e^{-Y_i},
$$
where the $Y$-linear terms originate from the dualization
and the exponential terms come from instanton effects.
In the non-linear
sigma model limit where the gauge coupling is taken to be large,
it is appropriate to integrate out the heavy fields $\Sigma_a$,
and we are left
with the theory of fields $Y_i$ obeying the constraints
\beq
\sum_{i=1}^NQ_i^aY_i=t^a\qquad \mbox{mod $2\pi i\Z$,}
\label{constrai}
\eeq
having the twisted superpotential
\beq
\widetilde{W}=\sum_{i=1}^N\e^{-Y_i}.
\eeq
Namely, the mirror is the LG model on the algebraic torus
$Y\cong (\C^{\times})^{N-k}$ defined by
(\ref{constrai}) with the above superpotential.
We study how the A- and B-parities we considered above are described
in this mirror theory.

\subsubsection{A-parity (B-parity in LG)}
\label{subsub:AsmBlg}

Let us consider the A-parity (\ref{Alsm2}) of the original linear
sigma model.
Looking at the action on the charged matter fields $\Phi_i\to
\overline{\Omega_A^*\Phi_{\sigma(i)}}$,
we expect that their dual fields are transformed similarly,
$Y_i\to \Omega_A^*Y_{\sigma(i)}$.
This would transform the superpotential as
$\widetilde{W}\to \Omega_A^*\widetilde{W}$.
 However,  we recall an
important condition for unbroken symmetry:
since the twisted F-term measure $\dd^2\widetilde{\theta}$
flips sign under $\Omega_A$, the twisted superpotential must also flip sign
under the transformation. The only way to make the sign of
$\sum_{i=1}^N\e^{-Y_i}$ flip is to add $\pi i$ to each field $Y_i$.
Thus,  it can be concluded that the dual transformation of the fields, required
by the condition that it be a symmetry, is
\beq
Y_i\longrightarrow \Omega_A^*Y_{\sigma(i)}+\pi i.
\label{YtransA}
\eeq
In addition, we should make sure that it is compatible with
the constraint (\ref{constrai}).
Let us examine this.
The left hand side of the constraint
$\sum_{i=1}^NQ_i^aY_i=t^a$ (mod $2\pi i \Z$) is transformed to
\beqa
\sum_{i=1}^NQ_i^a(\Omega^*_AY_{\sigma(i)}+\pi i)
&=&\sum_{i=1}^NQ_{\sigma^{-1}(i)}^a(\Omega_A^*Y_i+\pi i)
\nn\\
&=&\sum_{i,b}Q_i^b(\sigma^{-1})_b^{\,a}(\Omega_A^*Y_i+\pi i)
=\sum_{b}(\sigma^{-1})_b^{\,a}(t^b+\pi i b_1^b).
\nn
\eeqa
In the last step,
 we have used the fact that the original fields obey the constraint.
Now, inserting the anomaly-free condition (\ref{Asymlin}), we infer
that $(\sigma^{-1})_b^{\,a}(t^b+\pi i b_1^b)=t^a$ mod $2\pi i\Z$.
Thus the transformed fields indeed respect the constraint.
If the condition (\ref{Asymlin}) were broken,
the transformation (\ref{YtransA}) would not be consistent with the constraint
(\ref{constrai}), or we would not be able to find
a transformation such that $\widetilde{W}\to -\Omega_A^*\widetilde{W}$.
One of the conclusions is that the parity anomaly,
which is a non-trivial quantum effect in the original (linear) sigma model,
is reflected in the dual theory as the explicit
breaking by the potential term.

We recall that one could consider the modification of A-parity using
the $U(1)^{N-k}$ torus actions.
How does it affect the parity symmetry in
the mirror side? As we have discussed, there is no freedom to change
the action (\ref{YtransA}) on the dual fields.
In fact, the change appears simply in the way it acts on the winding
sector. Recall that the LG field $Y_i$ takes values in the algebraic torus
$Y=(\C^{\times})^{N-k}$ with non-trivial topology
$\pi_1(Y)=\Z^{N-k}$, and the momentum in the sigma model on $X$
associated with $U(1)^{N-k}$ symmetry is dual
to the winding number of $Y$ in the mirror LG model.

A transformation of this type has been used in \cite{AAHV}
in the application of mirror symmetry to compute the space-time superpotential
in Type II orientifolds on non-compact Calabi--Yau threefolds.
We now consider some examples where the parity anomaly and its cancellation
play important roles.

\subsubsection{Examples and applications}

\subsubsection*{\it Example 1.}

Let us consider the case $X=\CP^n$ whose mirror is
the LG model of $n$ periodic variables $Y_1,...,Y_n$
with superpotential
$$
\widetilde{W}=\e^{-Y_1}+\cdots +\e^{-Y_n}+\e^{-t+Y_1+\cdots +Y_n},
$$
where $t$ corresponds to the complexified K\"ahler class of
$\CP^n$. Consider the involution
$$
\widetilde{\tau}: Y_i\to Y_i+\pi i.
$$
If $n$ is odd,
it flips the sign of the superpotential $\widetilde{W}\to -\widetilde{W}$
and hence is a symmetry of the system. If $n$ is even, it fails to flip
the superpotential --- the first $n$ terms of
$\widetilde{W}$ do flip but the last term
does not. Thus, $\widetilde{\tau}\Omega$ is not a symmetry of the system.
Note that the above holomorphic involution $\widetilde{\tau}$
corresponds to the
antiholomorphic involution $\tau:\Phi_i\to\bPhi_i$ in the original
$\CP^n$ sigma model.
The above trouble for even $n$ corresponds to the anomaly
of this A-parity.
Moreover one can show that there is no way of transforming the fields
$Y_i$ so that $\widetilde{W}\to -\widetilde{W}$.
This corresponds to the A-parity anomaly on  non-spin manifolds
with $b^2(X)=1$.

\subsubsection*{\it Example 2.}

Let us next consider the case $X=\CP^n\times \CP^n$ whose mirror is
the direct sum
$$
\widetilde{W}=\sum_{i=1}^n\e^{-Y_i^{(1)}}
+\e^{-t_1+Y^{(1)}_1+\cdots +Y^{(1)}_n}
+\sum_{i=1}^n\e^{-Y_i^{(2)}}+\e^{-t_2+Y^{(2)}_1+\cdots +Y^{(2)}_n}.
$$
Here $t_1$ and $t_2$ correspond to the complexified K\"ahler class
of the first and the second $\CP^n$.
Consider the involution
$$
\widetilde{\tau}:(Y_i^{(1)},Y_i^{(2)})\mapsto
(Y_i^{(2)}+\pi i,Y_i^{(1)}+\pi i).
$$
Suppose $t_1=t_2$. Then,
as in Example 1, $\widetilde{\tau}$ flips the sign of the superpotential if
and only if $n$ is odd.
The failure in the cases where $n$ is odd  is again caused by
the anomaly of the A-parity
associated with $\tau:(z_1,z_2)\mapsto (\bz_2,\bz_1)$ in
$\CP^n\times \CP^n$.
However, in this case,
one can enforce a sign flip of the superpotential under  the parity action
by taking $t_1=t_2+\pi i$.
This corresponds to the cancellation of the anomaly
by a $B$-field, which we have seen in Section~\ref{paano}.

\subsubsection*{\it Application: Mirror pair of D-branes}

In Section~\ref{dbrfp}, we have seen that one can associate a D-brane
to a parity symmetry, as long as the parity has a fixed point.
One can use this to find the mirror pairs of D-branes.
Let us consider the parity symmetry in Example 2.
The fixed-point set is
$$
Y^{\widetilde{\tau}}=\Bigl\{\,Y^{(1)}_i=Y^{(2)}_i+\pi i\,\Bigr\}\cong
(\C^{\times})^n.
$$
In the original sigma model on $X=\CP^n\times\CP^n$, the fixed-point set is
the `skew-diagonal'
$$
X^{\tau}=\Bigl\{\, z_1=\bz_2\,\Bigr\}\cong \CP^n.
$$
Thus, we see that the A-brane wrapped on $X^{\tau}$ is mirror to the
B-brane wrapped on $Y^{\widetilde{\tau}}$.
\footnote{Mirror symmetry between skew diagonal $\CP^n$
and $\{Y_i^{(1)}=Y_i^{(2)}+\pi i\}$
 was first reported to KH by C. Vafa.
At that time (October, 2000) the reason for the necessity of a
$B$-field difference
for the even $n$ case (discussed below) was not clear to them.}
To be precise, this is true only when the parity
$\tau\Omega$ (or equivalently $\widetilde{\tau}\Omega$)
is a symmetry of the theory.
For example let us consider the case with $t_1=t_2$ and $n$ even.
We know that $\tau\Omega$ is anomalous and the D-brane wrapped on
$X^{\tau}$ is expected to suffer from some pathology.
This is in fact evident on the mirror side: The superpotential
$\widetilde{W}$ is not constant on the fixed-point set
$Y^{\widetilde{\tau}}$ (that is, $Y^{\widetilde{\tau}}$ does not
lie in one of the level sets of $\widetilde{W}$), which means that
$Y^{\widetilde{\tau}}$ violates the condition of
${\mathcal N}=2_B$ supersymmetry.
On the other hand, for odd $n$ (with $t_1=t_2$),
$Y^{\widetilde{\tau}}$ lies in the level set
$$
\widetilde{W}|_{Y^{(1)}_i=Y^{(2)}_i+\pi i}=0
$$
and is indeed a good B-brane.
For  $n$ even, $Y^{\widetilde{\tau}}$ can be placed in the level set
$\widetilde{W}=0$ by taking $t_1=t_2+\pi i$.
Thus, $X^{\tau}$ becomes a consistent A-brane by adding a $B$-field of period $\pi$
in one of the $\CP^n$'s of $X=\CP^n\times\CP^n$.
This cannot be seen directly in the
sigma model without a detailed  analysis of the subtleties of the moduli space
of holomorphic discs \cite{FOOO}.
In the mirror description, this is evident at the classical level.

We note that D-branes of the above type are oriented, namely, they can have
non-trivial overlaps with the RR ground states.
This is because the brane $Y^{\widetilde{\tau}}$ in the LG description
is middle-dimensional \cite{HKKPTVVZ}.

\subsubsection{B-parity (A-parity in LG)}

Let us next consider the B-parity (\ref{Blsm2}) of the linear
sigma model.
The action on the charged matter fields $\Phi_i\to
\Omega_B^*\Phi_{\sigma(i)}$
indicates that it transforms  the dual fields as
\beq
Y_i\longrightarrow \overline{\Omega_B^*Y_{\sigma(i)}}.
\eeq
This is consistent with the constraint (\ref{constrai})
if the $t^a$ obey the reality condition
(\ref{Bsymlin}),
$t^b\sigma_b^{\,a}=\overline{t^a}$ mod $2\pi i\Z$.
This is indeed a symmetry of the dual system since
the superpotential is transformed as
$\widetilde{W}\to \overline{\Omega_B^*\widetilde{W}}$.

We recall that one could modify the B-parity using
the $U(1)^{N-k}$ torus actions.
As in the case of A-parities,
the modification will change the way the parity acts on the winding
sector in the mirror theory and induce the mixture of
the ($SO$ vs. $Sp$) types of  orientifold planes.
This is a generalization of the phenomenon that the 
orientifold of $S^1$ by a half-period shift is T-dual to
the orientifold of the dual circle $\widetilde{S^1}$
by an inversion, with two opposite types of
orientifold points \cite{DaPa,Wnovec,BH1}.

Let us examine this in more detail, in the case where the parity is
associated with the simple complex conjugation
$Y_i\to\overline{Y_i}$,
which is a symmetry of the system if the $t^a$ are all real,
$\theta^a=0$ $\forall a$.
The fixed points of this action are
$Y_i=\overline{Y_i}$ mod $2\pi i \Z$, namely
$$
Y_i\in \R+\pi ip_i,
$$
where $p_i$ are integers obeying $\sum_{i=1}^NQ_i^ap_i=0$
mod $2\Z$. The set of points with
$Y_i\in \R+\pi ip_i$ is a middle-dimensional plane $\R^{N-k}$
and will be denoted as $L_{p_1...p_N}$.
There are $2^{N-k}$ such planes.
The fixed-point set $Y^{\widetilde{\tau}}$ is the union
of these $2^{N-k}$ $L_p$'s.
The type of the orientifold plane $L_p$ (whether it is
$SO$ or $Sp$ ) depends on
the original parity symmetry on the sigma model side.
Let us first consider the basic parity of $SO$-type
associated with the identity $\Phi_i\to \Phi_i$ of $X$.
Its mirror should be such that all $2^{N-k}$ orientifold planes
are of  $SO$ type.
Then, they have the same orientations
so that the total homology class is
\beq
Y^{\widetilde{\rm id}}
=\sum_{p}L_{p_1...p_N}.
\eeq
Here the orientation of all $L_p$'s
are related by translations by $\pi i p_i$.
Let us next consider the parity of $SO$-type associated with
an involution of $X$ of the form $\tau_r:\Phi_i\to (-1)^{r_i}\Phi_i$.
This action corresponds on the mirror side to the multiplication by
$$
(-1)^{w_1r_1+\cdots+w_Nr_N}
$$
in the sector with winding number $(w_1,...,w_N)$ (where $w_i$ are
integers obeying $\sum_{i=1}^NQ_i^aw_i=0$ mod $2\Z$).
For such a parity action, the type of the orientifold plane
$L_{p_1...p_N}$ is still $SO$-type if $(-1)^{p_1r_1+\cdots +p_Nr_N}=1$,
but is flipped to $Sp$-type if
$(-1)^{p_1r_1+\cdots +p_Nr_N}=-1$.
Thus the total homology class of the orientifold plane is
\beq
Y^{\widetilde{\tau_r}}
=\sum_{p}(-1)^{p_1r_1+\cdots +p_Nr_N}L_{p_1...p_N}.
\eeq
Note that the involution $\tau_r$ is gauge-equivalent to
the involution $\tau_{r+Q\cdot \lambda}$, with $\lambda_a\in \Z$.
However, thanks to the condition $\sum_{i=1}^NQ_i^ap_i=0$ mod 2,
the gauge equivalent replacement $r\to r+Q\cdot \lambda$
does not alter the type 
$(-1)^{p_1r_1+\cdots +p_Nr_N}$ of the orientifold plane
$L_{p_1...p_N}$.

Actually, the parity associated with $Y_i\to\overline{Y_i}$
is a symmetry as long as $t^a=\overline{t^a}$ mod $2\pi i\Z$.
The theta angle (or $B$-field)  
$\theta^a$ only has to be in $\pi \Z$ and does not have to vanish.
The above story can be applied also to the case with non-zero
$\theta^a$.
An important difference appears however in the constraint on $p_i$
--- the $p_i$ must still be integers but obey
a different constraint
$$
\sum_{i=1}^NQ_i^ap_i=-{\theta^a\over \pi}\quad\mbox{mod $2\Z$}.
$$
This causes an apparent trouble. Under the gauge equivalent
replacement $r_i\to r_i+Q_i^a\lambda_a$ ($\lambda_a\in \Z$),
the type $(-1)^{p_1r_1+\cdots +p_Nr_N}$ of the orientifold plane
$L_{p_1...p_N}$
changes by a factor $\e^{i\theta^a\lambda_a}$.
However, rather than being a trouble, it is consistent with
the observation we made in the study of parity symmetry in non-linear
sigma models with non-zero $B$-field. The observation was:
{\it In addition to the geometric action $\tau:X\to X$,
 we have to specify the parity action on
the line bundle ${\mathcal L}_{\tau^*B+B}$
with first Chern class $(\tau^*B+B)/2\pi$.}
In the present case, $\tau^*B=B$ and the bundle is
${\mathcal L}_{2B}$ with first Chern class $B/\pi$.
We propose that different choices of $r_i$
within a fixed gauge equivalence class
correspond to different choices of the parity action on ${\mathcal L}_{2B}$
if they are different in the numbers $(-1)^{p_1r_1+\cdots+p_Nr_N}$.
In fact, the field $\Phi_i$ may be regarded as a section of a
line bundle ${\mathcal L}_i=\otimes_a{\mathcal L}_a^{Q^a_i}$ where
$\omega_a=c_1({\mathcal L}_a)$ are basis elements of
$H^2(X,\Z)\cong \Z^{\oplus k}$.
The bundle ${\mathcal L}_{2B}$ has first Chern class
$c_1({\mathcal L}_{2B})=\sum_{a=1}^k\omega_a\theta_a/\pi$.
Let us choose $p_i$ (from the same mod 2 integer class) so that
$\sum_{i=1}^NQ_i^ap_i=-{\theta^a/\pi}$ holds exactly (i.e. not just mod 2).
Then we find
$$
\bigotimes_{i,a}{\mathcal L}_a^{-Q_i^ap_i}
=\bigotimes_a{\mathcal L}_a^{\theta^a/\pi}\stackrel{c_1}{\longrightarrow}
\sum_{a=1}^k\omega_a{\theta^a\over \pi}={B\over \pi}.
$$
Namely, $\otimes_i {\mathcal L}_i^{-p_i}\cong {\mathcal L}_{2B}$.
Thus, $\prod_i\Phi_i^{-p_i}$ can be regarded as the section of the
bundle ${\mathcal L}_{2B}$.
The $(r_i)$, which determine the numbers $(-1)^{p_1r_1+\cdots+p_Nr_N}$,
thus specify the parity action
of the bundle ${\mathcal L}_{2B}$.
In other words, the information of
the pair $(\tau_r:X\to X,\tau_r:{\mathcal L}_{2B}\to
{\mathcal L}_{2B})$ is mapped under mirror symmetry
to the information of the
type of the orientifold planes $L_{p_1...p_N}$.

We illustrate these rules in the example of $X=\CP^1$.

\subsubsection{Example: B-parities of $\CP^1$ and their mirrors}

For $X=\CP^1$ there are two possible geometric actions of the above type.
One is the identity, and the other is the rotation by
$\pi$ along the $U(1)$-fibre, $R:\Phi_1/\Phi_2\to -\Phi_1/\Phi_2$.
Also, there are two possible values of the theta angle
$\theta=0$ and $\theta=\pi$.
The mirror is the sine-Gordon model
with superpotential
$$
W=\e^{-Y}+\e^{-t+Y}.
$$
The parity action on the dual field is
the complex conjugation $Y\to \overline{Y}$ which 
 is a symmetry if $\theta=0,\pi$. The fixed-point set consists of
$L_0=\{Y\in \R\}$ and $L_{1}=\{Y\in \R+\pi i\}$.
They are respectively $L_{00}$ and $L_{11}$ if $\theta=0$
or $L_{01}$ and $L_{10}$ if $\theta=\pi$,
in the notation of the above general discussion.
We compute the parity-twisted Witten index in both sigma model and LG model
and compare the results.

We recall that the cohomology of $X=\CP^1$ is generated by
$1\in H^0(\CP^1)$ and $H\in H^2(\CP^1)$ with $\int_{\CP^1} H=1$.
We discuss the cases of $\theta=0$ and $\theta=\pi$ separately.

\subsubsection*{$\theta=0$}

\begin{figure}[tb]
\centerline{\includegraphics{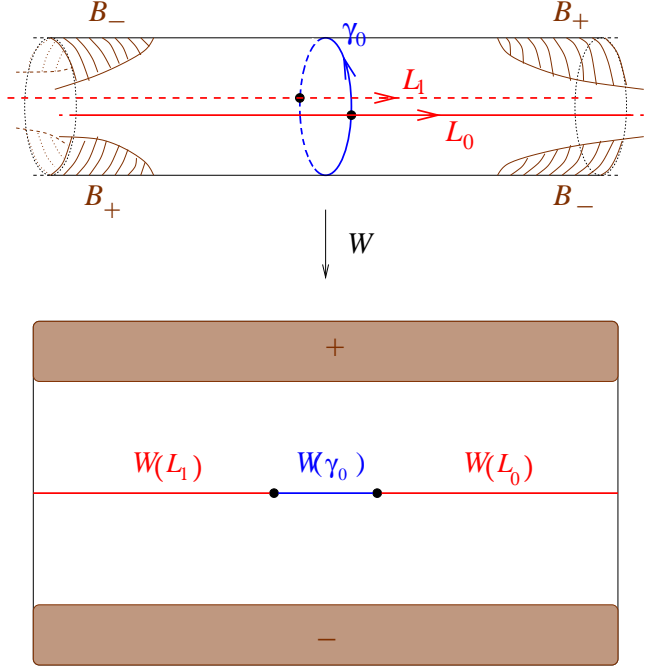}}
\caption{Orientifold planes ($L_0$ and $L_1$)
and the mirror of the D0-brane
($\gamma_0$). Their images under the superpotential $W$ are also depicted.}
\label{p1a}
\end{figure}

Let us first consider the basic parity $\Omega$ ($\tau={\rm id}$).
This maps the bundle ${\mathcal O}(n)$ to ${\mathcal O}(-n)$.
The fixed-point set is $\CP^1$ itself and hence the
normal bundle is zero.
The characteristic classes
relevant to the computation of the index are
$L(\CP^1)=0$ and ${\rm td}(\CP^1)=1+H=\e^H.$
We then find
\beqa
&&I_{\Omega}=0,
\label{IO0}
\\
&&I_{\Omega}({\mathcal O}(n),{\mathcal O}(-n))=\int_{\CP^1}\e^{-2nH}\e^H
=1-2n.
\label{IOOO0}
\eeqa
Let us next consider the parity $R\Omega$ associated with the
$\pi$-rotation. This also maps ${\mathcal O}(n)$ to ${\mathcal O}(-n)$.
The fixed-point set consists of two points, say, the north pole ${\bf N}$ and south pole
${\bf S}$.
The normal bundle is real two dimensional (complex one-dimensional)
and hence
$e(N({\bf N}))=e(N({\bf S}))=0$ and
${\rm ch}(\wedge \overline{N}_{\bf N})
={\rm ch}(\wedge \overline{N}_{\bf S})=2$.
This yields
\beqa
&&I_{R\Omega}=0,
\label{IR0}
\\
&&I_{R\Omega}({\mathcal O}(n),{\mathcal O}(-n))
=\sum_{p={\bf N}}^{\bf S}\int_p\e^{-2nH}{1\over 2}={1\over 2}+{1\over 2}=1.
\label{IROO0}
\eeqa
According to the above discussion, the orientifold planes
on the mirror side are
\beqa
&&Y^{\widetilde{\rm id}}=L_0+L_1,
\\
&&Y^{\widetilde{R}}=L_0-L_1.
\eeqa
We also know that the mirror of ${\mathcal O}(0)$ is the same as
$L_0$ and the mirror of ${\mathcal O}(n)$ is topologically
\beq
\gamma_{{\mathcal O}(n)}=L_0+n\gamma_0,
\eeq
where $\gamma_0$ (the mirror of the 0-brane)
is a circle that winds once around the $Y$-cylinder
(See Fig.~\ref{p1a}).
Under the parity action
$Y\to\overline{Y}$, it is mapped to
$L_0-n\gamma_0$ which is indeed the mirror of ${\mathcal O}(-n)$.
The Witten index is obtained by taking the intersection numbers.
We recall that the intersection of $C_1$ and $C_2$ is defined as
$\#(C_1^-\cap C_2^+)$ where $C_i^{\pm}$ is obtained from
$C_i$ by moving the asymptotes
toward the region $B_{\pm}$.
It is visible in Fig.~\ref{p1a} that
\beqa
&&\#(L_0^-\cap L_0^+)=1;\quad
\#(L_1^-\cap L_1^+)=-1;
\nn\\
&&\#(L_i^-\cap L_j^+)=0\,\,\mbox{if}\,\,i\ne j;\quad
\#(\gamma_0^-\cap L_i)=-1.
\nn
\eeqa
Using these, we find
\beqa
&&
I_{\widetilde{\Omega}}=\#((Y^{\widetilde{\rm id}})^-\cap
(Y^{\widetilde{\rm id}})^+)=1-1=0,
\\
&&
I_{\widetilde{\Omega}}(\gamma_{{\mathcal O}(n)},\gamma_{{\mathcal O}(-n)})
=\#(\gamma_{{\mathcal O}(n)}^-\cap (Y^{\widetilde{\rm id}})^+)
=1+n(-1)+n(-1)=1-2n,
\eeqa
which reproduces (\ref{IO0}), (\ref{IOOO0}), and
\beqa
&&
I_{\widetilde{R\Omega}}=\#((Y^{\widetilde{R}})^-\cap
(Y^{\widetilde{R}})^+)=1-1=0,
\\
&&
I_{\widetilde{R\Omega}}
(\gamma_{{\mathcal O}(n)},\gamma_{{\mathcal O}(-n)})
=\#(\gamma_{{\mathcal O}(n)}^-\cap (Y^{\widetilde{R}})^+)
=1+n(-1)-n(-1)=1.
\eeqa
which reproduces (\ref{IR0}), (\ref{IROO0}).

\subsubsection*{$\theta=\pi$}

\begin{figure}[tb]
\centerline{\includegraphics{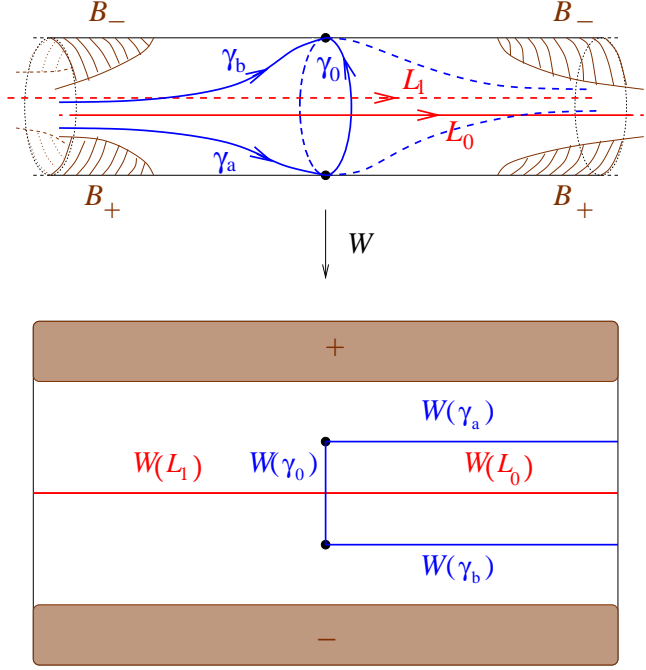}}
\caption{Orientifold planes ($L_0$ and $L_1$)
and the mirror of ${\mathcal O}(0)$
($\gamma_a$) and or ${\mathcal O}(1)$ ($\gamma_b$).
Notice that the location of $B_+$ and $B_-$ have changed from
the previous case $\theta=0$.}
\label{p1b}
\end{figure}

We now consider the case with a non-zero theta angle $\theta=\pi$.
This corresponds to turning on the $B$-field $B=\pi H$.
For all  holomorphic involutions $\tau$, 
the twist bundle ${\mathcal L}_{\tau^*B+B}$ has first Chern class
$B/\pi=H$ and hence it is ${\mathcal O}(1)$.
Thus, the parity transforms the bundles as
\beq
\tau\Omega:{\mathcal O}(n)\longrightarrow \overline{{\mathcal O}(n)}\otimes
{\mathcal L}_{\tau^*B+B}=
{\mathcal O}(1-n).
\label{buntra}
\eeq
The parities to be considered are $\tau_{r_1,r_2}$ that acts on the sections
$\Phi_1$ and $\Phi_2$ of ${\mathcal L}_{\tau^*B+B}={\mathcal O}(1)$
as $(\Phi_1,\Phi_2)\mapsto ((-1)^{r_1}\Phi_1,(-1)^{r_2}\Phi_2)$.
$\tau_{00}$ and $\tau_{11}$ project to the identity map of
$\CP^1$ while $\tau_{10}$ and $\tau_{01}$ project to the
$\pi$-rotation $R:\CP^1\to\CP^1$.
The sign function at the fixed-point set is
\beqa
&&\varepsilon_B^{\tau_{00}}=1, \quad
\varepsilon_B^{\tau_{11}}=-1,\quad\mbox{on $\CP^1$}
\nn\\
&&\varepsilon_B^{\tau_{10}}=
\left\{
\begin{array}{ll}
1&\mbox{at ${\bf N}$}\\
-1&\mbox{at ${\bf S}$}
\end{array}
\right.
\quad
\varepsilon_B^{\tau_{01}}=
\left\{
\begin{array}{ll}
-1&\mbox{at ${\bf N}$}\\
1&\mbox{at ${\bf S}$}
\end{array}
\right.
\nn
\eeqa
where we assumed that ${\bf N}$ and ${\bf S}$ are loci
of $\Phi_1=0$ and $\Phi_2=0$ respectively.
It is straightforward to compute the index.
\beqa
&&
I_{\tau_{00}\Omega}=0,
\label{IO1}
\\
&&
I_{\tau_{00}\Omega}({\mathcal O}(n),{\mathcal O}(1-n))
=\int_{\CP^1}\e^{-2nH}\varepsilon_B^{\tau_{00}}\e^{H}\e^H=2-2n.
\label{IOOO1}
\eeqa
For $\tau_{11}\Omega$ the opposite sign occurs. Also,
\beqa
&&
I_{\tau_{10}\Omega}=0,
\label{IR1}
\\
&&
I_{\tau_{10}\Omega}({\mathcal O}(n),{\mathcal O}(1-n))
=\sum_{p={\bf N}}^{\bf S}{\varepsilon_B^{\tau_{10}}(p)\over 2}
={1\over 2}-{1\over 2}=0.
\label{IROO1}
\eeqa
For $\tau_{01}\Omega$ the sign is opposite and hence the same.
By the general discussion, the orientifold planes in the mirror
are
\beqa
&&
Y^{\widetilde{\tau}_{00}}=L_0+L_1,
\\
&&
Y^{\widetilde{\tau}_{10}}=L_0-L_1,
\eeqa
and $Y^{\widetilde{\tau}_{11}}=-L_0-L_1$
and $Y^{\widetilde{\tau}_{01}}=-L_0+L_1$.
The mirror of ${\mathcal O}(0)$ and ${\mathcal O}(1)$ are
the wavefront trajectories originating in the two critical points
$Y=r/2-\pi i/2$ and $Y=r/2+\pi i/2$ respectively.
They are depicted as $\gamma_a$ and $\gamma_b$ in Fig.~\ref{p1b}
They are homologically related as $\gamma_b=\gamma_a+\gamma_0$.
The mirror of the other bundles are homologically
\beq
\gamma_{{\mathcal O}(n)}=\gamma_a+n\gamma_0.
\eeq
Under the parity action $Y\to\overline{Y}$,
$\gamma_a$ and $\gamma_b=\gamma_a+\gamma_0$ are
exchanged while $\gamma_0$ is reversed.
Thus $\gamma_{{\mathcal O}(n)}=\gamma_a+n\gamma_0$ is mapped to
$\gamma_b-n\gamma_0=\gamma_a+(1-n)\gamma_0$ which is indeed the mirror
of ${\mathcal O}(1-n)$.
To compute the intersection numbers, we note
the following
\beqa
&&\#(L_i^-\cap L_j^+)=0\quad\forall i,\,\,\forall j;
\nn\\
&&\#(\gamma_a^-\cap L_i^+)=1\quad i=0,1;\qquad
\#(\gamma_0^-\cap L_i)=-1.
\nn
\eeqa
Using these, we find
\beqa
&&
I_{\widetilde{\tau_{00}\Omega}}=\#((Y^{\widetilde{\tau}_{00}})^-\cap
(Y^{\widetilde{\tau}_{00}})^+)=0,
\\
&&
I_{\widetilde{\tau_{00}\Omega}}
(\gamma_{{\mathcal O}(n)},\gamma_{{\mathcal O}(1-n)})
=\#(\gamma_{{\mathcal O}(n)}^-\cap (Y^{\widetilde{\tau}_{00}})^+)
=1+1+n(-1)+n(-1)=2-2n,\qquad
\eeqa
which reproduces (\ref{IO1}), (\ref{IOOO1}),
and
\beqa
&&
I_{\widetilde{\tau_{10}\Omega}}=\#((Y^{\widetilde{\tau}_{10}})^-\cap
(Y^{\widetilde{\tau}_{10}})^+)=0,
\\
&&
I_{\widetilde{\tau_{10}\Omega}}
(\gamma_{{\mathcal O}(n)},\gamma_{{\mathcal O}(1-n)})
=\#(\gamma_{{\mathcal O}(n)}^-\cap (Y^{\widetilde{\tau}_{10}})^+)
=1-1+n(-1)-n(-1)=0,
\eeqa
which reproduces (\ref{IR1}), (\ref{IROO1}).

\section{Orientifolds of Compact Calabi--Yau: A First Step}
\label{sec:CY}

In this section, we initiate the discussion of orientifolds
of compact Calabi--Yau manifolds, which are of vital phenomenological
relevance.
This is a first step and
full detail must be clarified in future works.
However, the considerations using linear sigma models and mirror symmetry
already provide a basic global picture and directions to proceed.

\subsection{LSM for compact CY and parity symmetry}

Let us consider the $U(1)$ gauge theory with matter fields
$P,\Phi_1,...,\Phi_N$ of charge $-N,1,...,1$
having the term $L_2$ with gauge invariant superpotential
$$
W=PG(\Phi_1,...,\Phi_N),
$$
where $G(\Phi_i)$ is a polynomial of degree $N$.
In this system, the FI parameter $r$ does not run and
is a parameter of the theory.
At large positive $r$, the model corresponds to the non-linear sigma model
on the degree $N$ hypersurface of $\CP^{N-1}$
defined by the equation $G(z_1,...,z_N)=0$
for the homogeneous coordinates $z_i$,
which is a Calabi--Yau manifold of dimension $(N-2)$.
At large negative $r$, $p$ acquires a fixed value $\langle p\rangle$
and the model corresponds to a LG orbifold ---
the LG model with superpotential
$W=\langle p\rangle G(\Phi_1,...,\Phi_N)$ modded out by the $\Z_N$ action
generated by $\Phi_i\to \e^{2\pi i\over N}\Phi_i$.
The worldsheet theory is singular exactly at one point
\beq
t=t_*:=N\log(-N)=N\log N+\pi i N.
\eeq
The moduli space of the complexified K\"ahler class $t$ has three notable points
--- the large-volume limit ($r=+\infty$), the conifold point
($t=t_*$), and the LG orbifold point or equivalently the Gepner point
($r=-\infty$).
The moduli space is connected and the theory is singular only
at the conifold point.
The moduli space of complex structure is a complex space
of complex dimension ${2N-1\choose N-1}-N^2+\delta_{N,4}$. 

The model has a mirror description,
which is the non-linear sigma model on the
degree $N$-hypersurface in $\CP^{N-1}$
$$
\widetilde{G}(\wtz_i):=\wtz_1^N+\cdots+\wtz_N^N
+\e^{t/N}\wtz_1\cdots\wtz_N=0,
$$
modded out by the orbifold group $(\Z_N)^{N-2}$
acting as $\wtz_i\to\omega_i\wtz_i$,
where $\omega_i^N=\omega_1\cdots\omega_N=1$.
At the singular point $t=t_*$, the mirror manifold has
a conifold singularity at $\wtz_i=1$.
It has one complex structure modulus $\e^{t}$
and ${2N-1\choose N-1}-N^2+\delta_{N,4}$ K\"ahler moduli
coming from the K\"ahler class of $\CP^{N-1}$ and the resolutions of
the orbifold singularities.
The mirror may also be described as the
LG orbifold with superpotential
 $\widetilde{W}=\widetilde{G}(\widetilde{\Phi}_i)$ and orbifold group
$(\Z_N)^{N-1}$: $\widetilde{\Phi}_i\to \omega_i\widetilde{\Phi}_i$.
This mirror can be found by using the dualization
$\arg(\Phi_i)\to Y_i$ followed by the change of variables
$\e^{-Y_i}\sim \widetilde{\Phi}_i^N$ \cite{HV}.

Let us study the parity symmetries of this system.
We will consider A-parity and B-parity separately.

\subsubsection{A-parity (B-parity of mirror)}

\newcommand{\wttheta}{\widetilde{\theta}}

The transformation
\beq
P\to \overline{\Omega_A^*P},\quad
\Phi_i\to \overline{\Omega_A^*\Phi_{\sigma(i)}},
\quad
V\to \Omega_A^*V
\label{BPG}
\eeq
yields an A-parity
that is an exact symmetry of the system
if $G(z_i)$ is a polynomial obeying
$G(\overline{z_{\sigma(i)}})=\overline{G(z_i)}$.
Note that there is no anomaly since the sum of the charges is 
zero $-N+1+\cdots +1=0$ (and hence even), or equivalently,
since $M$ is Calabi--Yau (and hence spin).
There is no condition on the complexified K\"ahler modulus, but the
complex structure moduli are reduced by the constraint
$G(\overline{z_{\sigma(i)}})=\overline{G(z_i)}$.
The reduction is by one-half if $\sigma$ is an involution.

Following the duality transformation and change of variables,
we find that
the corresponding parity in the mirror side (in the LG description)
is of the form
\beq
\widetilde{\Phi}_i\longrightarrow
\e^{\pi i/N}\Omega_A^*\widetilde{\Phi}_{\sigma(i)},
\label{mirBPG}
\eeq
where the phase is chosen uniquely by the requirement
that the superpotential be flipped
$\widetilde{W}\to -\Omega_A^*\widetilde{W}$.
In the geometric description of the mirror,
the parity is simply the one associated with
the holomorphic map $\wtz_i\to \wtz_{\sigma(i)}$ of the mirror CY manifold.

If the defining polynomial obeys
$G(\e^{i\theta_i}\overline{z_{\sigma(i)}})
=\e^{-i\theta_P}\overline{G(z_i)}$,
then we have a parity symmetry given by (\ref{BPG}), combined with the
torus action $P\to \e^{i\theta_P}P$, $\Phi_i\to\e^{i\theta_i}\Phi_i$.
In this case, the parity action on the mirror LG fields is still given
by (\ref{mirBPG}), but a difference appears in the
action on the twisted sector states.
In the geometric picture, this will change the type of orientifold
planes, just as in \cite{AAHV}.

\subsubsection{B-parity (A-parity of the mirror)}

\begin{figure}[tb]
\centerline{\includegraphics{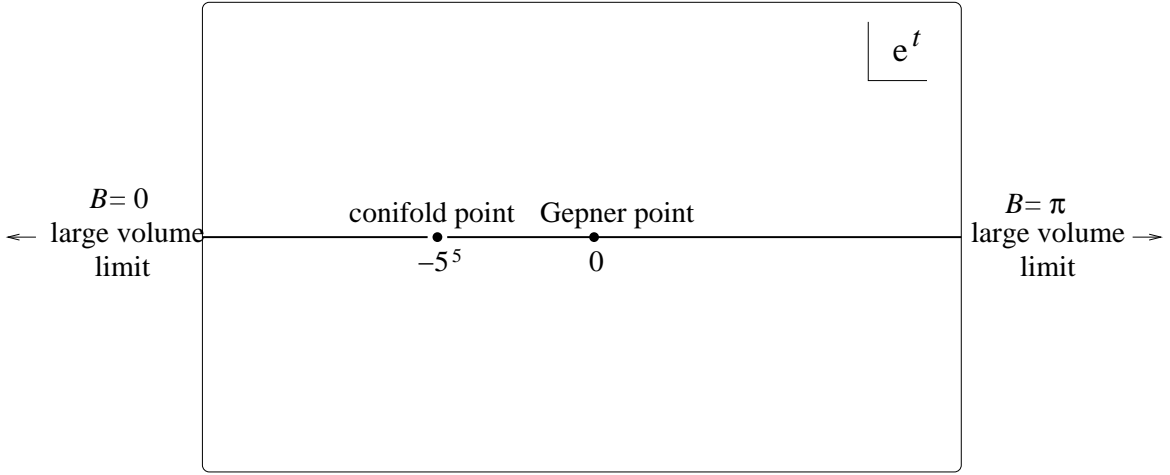}}
\caption{K\"ahler moduli space for a B-orientifold of the quintic}
\label{km}
\end{figure}

The transformation
\beq
P\to -\Omega_B^*P,\quad
\Phi_i\to \Omega_B^*\Phi_{\sigma(i)},\quad
V\to\Omega_B^*V
\eeq
yields a B-parity.
It is an exact symmetry of the system if
the defining polynomial is invariant under
the permutation $G(z_{\sigma(i)})=G(z_i)$ and also if
the FI-theta parameter $t$ obeys
$t=\overline{t}$ mod $2\pi i\Z$, namely
\beq
{\rm Im}\,t=0\quad \mbox{or}\quad \pi i.
\label{cccons}
\eeq
Thus, there is a holomorphic constraint on
the complex structure moduli, and
the complexified K\"ahler modulus
is reduced by a half --- one real dimension.
The reduced moduli space ${\rm Im}t\in \pi i\Z$
passes directly through the singular point $t_*=N\log N+\pi iN$.
Thus, the real moduli space of the worldsheet theory is separated at $t=t_*$.
It has two parts --- one includes the large-volume limit with trivial $B$-field
and the other includes the large volume limit with
a $B$-field of period $\pi$.
The LG orbifold point (Gepner point)
belongs to one of them and is separated from the other.
Note that the analytic continuation in the $t$-space
to go around the singular point $t=t_*$
\cite{phases,Coleman} is not applicable here
since the orientifold forbids us
to continuously change the $\theta$ angle.
Figure~\ref{km} depicts the restriction of the K\"ahler moduli by the B-parity
in the case of $N=5$, the quintic hypersurface in $\CP^4$
which will be discussed
in some detail below (together with the A-parities).\footnote{We thank
M.~Douglas for pointing out an error in the identification of
the value of the B-field at the two large volume regions.
The relation of the B-field and the theta angle is shifted by $N\pi$
when the $P$-field is integrated \cite{MorrisonPlesser}.}

\subsection{The case of the quintic}

\newcommand{\wttau}{\widetilde{\tau}}
\newcommand{\wtw}{\widetilde{w}}
\newcommand{\wtu}{\widetilde{u}}
\newcommand{\bw}{\overline{w}}

Let us consider the case $N=5$, quintic hypersurfaces $M$ in $\CP^4$.
We classify the involutive
parity symmetries that are present in the
Fermat-type hypersurfaces:
$$
z_1^5+z_2^5+z_3^5+z_4^5+z_5^5=0.
$$
Without orientifold, there is one complexified K\"ahler modulus
parametrized by a function of $t$,
and $101$ complex structure moduli. The latter  is the number
of monomials ${9\choose 4}=126$ minus the dimension $5^5=25$
of the group $GL(5,\C)$ of coordinate transformations.
The mirror is the resolution of the orbifold 
of
$$
\wtz_1^5+\wtz_2^5+\wtz_3^5+\wtz_4^5+\wtz_5^5
+\e^{t/5}\wtz_1\wtz_2\wtz_3\wtz_4\wtz_5=0.
$$
by the $\Z_5^3$ action
$\wtz_i\to\omega_i\wtz_i$,
$\omega_i^5=\omega_1\omega_2\omega_3\omega_4\omega_5=1$,
$\omega_i\equiv \omega \omega_i$ \cite{GreenePlesser}.
Without orientifold, there are $101$ 
complexified K\"ahler moduli and
one complex structure modulus $\e^t$.
$101$ comes from the $1+100$ harmonic two-forms where the extra
$100$ is counted at the orbifold point as follows.
Note that there are ten $\Z_5\times\Z_5$ invariant points 
(points of the form $(0,0,0,1,-1)$). At each point there are 6 localized
blow-up modes of a $\C^3/\Z_5\times\Z_5$ singularity.
For some pairs of
$\Z_5\times\Z_5$ invariant points, there is a curve of
$\C^2/\Z_5$ singularity, which carry $4$ blow-up modes.
There are ten such pairs.
Thus, in total, there are
$10\times 6+10\times 4=100$ blow-up modes.

\subsubsection{A-parities}

We start with the A-parities. In all  cases, 
the $101$ complex structure moduli are reduced to one half
($101$ real ones), but the K\"ahler moduli space remains the same.

The simplest A-parity is associated with the complex conjugation
$\tau_A:z_i\mapsto \overline{z_i}$.
(This example was also studied in \cite{BBKL}.)
The fixed-point set is the real quintic
which is a (3-dimensional) submanifold of
the quintic defined by $z_i\in \R$.
To find its topology, let us consider the map
$M\to \CP^3$ obtained by forgetting the last coordinate $z_5$.
It maps the real quintic $M^{\tau_A}$
homeomorphically onto the real locus of $\CP^3$,
namely $\RP^3$. This is because
any real number has a unique real $5$-th root,
$x_5={}^5\!\!\sqrt{-x_1^2-x_2^2-x_3^2-x_4^2}$,
and `taking the $5$-th root' defines a continuous map
$\R\to\R$.
Thus, the real quintic has the topology of a real projective space $\RP^3$.
On the mirror side, this maps to the B-parity associated with
the identity $\wttau_B:\wtz_i\mapsto \wtz_i$,
whose fixed-point set is the mirror
quintic itself (6-dimensional).
One may consider a modification of $\tau_A$ of the form
$z_i\mapsto \e^{i\theta_i}\overline{z_i}$, which is a symmetry
if $\e^{5i\theta_i}$ is independent of $i$.
But this is simply the complex conjugation with respect to the 
coordinate $z_i'=\e^{-i\theta_i/2}z_i$.
Thus, there is no essential difference.
This will be the same for the rest of the cases as well,
and hence will not be mentioned.

Next, we consider the A-parity associated with
$\tau_A^{{}^{12}}:(z_1,z_2,z_3,z_4,z_5)\mapsto
(\overline{z_2},\overline{z_1},\overline{z_3},\overline{z_4},\overline{z_5})$.
The $\tau_A^{{}^{12}}$-fixed-point set is the (3-dimensional)
submanifold defined by
$z_2=\overline{z_1}$, $z_i\in \R$ ($i=3,4,5$).
Using the $z_5$-forgetting map $M\to \CP^3$, we find that
it has the topology of $\RP^3$.
This corresponds to the B-parity of the mirror
 associated with the holomorphic involution
$\wttau_B^{{}^{12}}:(\wtz_1,\wtz_2,\wtz_3,\wtz_4,\wtz_5)
\mapsto (\wtz_2,\wtz_1,\wtz_3,\wtz_4,\wtz_5)$.
The fixed-point set is a point $\{ (1,-1,0,0,0)\}$ (0-dimensional)
and a complex hypersurface $\{(z,z,z_3,z_4,z_5))\}$ (4-dimensional).

The final A-parity is the one associated with
$\tau_A^{{}^{12,34}}:(z_1,z_2,z_3,z_4,z_5)\mapsto
(\overline{z_2},\overline{z_1},\overline{z_4},\overline{z_3},\overline{z_5})$.
The fixed-point set is the (3-dimensional) submanifold
defined by
$z_2=\overline{z_1},z_4=\overline{z_3}$, $z_5\in \R$.
Using the $z_5$-forgetting map $M\to \CP^3$, we again find that
it has a topology of $\RP^3$.
This corresponds to the B-parity of the mirror
 associated with
$\wttau_B^{{}^{12,34}}:
(\wtz_1,\wtz_2,\wtz_3,\wtz_4,\wtz_5)
\mapsto (\wtz_2,\wtz_1,\wtz_4,\wtz_3,\wtz_5)$.
The fixed-point set is a line $\{(\wtz,-\wtz,\wtw,-\wtw,0)\}$ and
a curve $\{(\wtz,\wtz,\wtw,\wtw,\wtu)\}$ (both 2-dimensional).

\subsubsection{B-parities}

We now consider B-parities.
In all the cases, the complexified K\"ahler modulus (complex) is reduced
to a real one, ${\rm Im}\,t\in \pi i\Z$.
The reduction of the complex structure moduli is holomorphic and depends on
the particular case.

The basic B-parity is associated with the identity
$\tau_B:z_i\mapsto z_i$. The fixed-point set is the quintic itself
(6-dimensional).
All the complex $101$ complex structure moduli survive.
In the mirror description, this corresponds to the A-parity associated with
the complex conjugation
$\wttau_A:\wtz_i\mapsto \overline{\wtz_i}$.
The fixed-point set is the real locus $\wtz_i\in\R$ (3-dimensional).
One may consider modifying $\tau_B$ by a phase multiplication.
For $z_i\mapsto \e^{i\theta_i}z_i$ to be a symmetry,
$\e^{5i\theta_i}$ must be $i$ independent,
and for this to be an involution, $\e^{2i\theta_i}$ must be
$i$-independent. This requires $\e^{i\theta_i}=\e^{5i\theta}/
(\e^{2i\theta_i})^2$ to be $i$-independent. Namely, $\e^{i\theta_i}$
are all the same and the transformation is a part of the $U(1)$
gauge symmetry.
The `modification' had no effect.
(It turns out that this is the same for the rest of the cases,
and hence will not be mentioned.)

Next, we consider the B-parity associated with
$\tau_B^{{}^{12}}:(z_1,z_2,z_3,z_4,z_5)
\mapsto (z_2,z_1,z_3,z_4,z_5)$.
Let us count the number of complex structure moduli.
The polynomial $G(z_i)$ must be invariant under the exchange of
$z_1$ and $z_2$. Monomials that are invariant under $z_1\leftrightarrow z_2$
are of the form $(z_1z_2)^az_3^bz_4^cz_5^d$ and there are 34 of them.
From the rest, $126-34=92$, a half ($46$ of them) is invariant.
Thus, invariant polynomials can have $46+34=80$ terms.
Coordinate transformations commuting with the exchange
$z_1\leftrightarrow z_2$ form a subgroup of $GL(5,\C)$
of dimension $17$.
Thus, the number of complex structure moduli
is $80-17=63$.
The fixed-point set is the point $\{(1,-1,0,0,0)\}$ (0-dimensional)
and the hypersurface $\{(z,z,z_3,z_4,z_5)\}$ (4-dimensional).
The mirror description of this parity is  associated with
$\wttau_A^{{}^{12}}:(\wtz_1,\wtz_2,\wtz_3,\wtz_4,\wtz_5)\mapsto
(\overline{\wtz_2},\overline{\wtz_1},\overline{\wtz_3},
\overline{\wtz_4},\overline{\wtz_5})$.
The complexified K\"ahler moduli correspond to
$\wttau_A^{{}^{12}}$-anti-invariant harmonic $(1,1)$ forms, by the condition
(\ref{Asymc}).
Let us count the number at the orbifold point.
We recall that the $101$ consists of $1$ `bulk' modulus,
$10\times 4$ moduli of the resolution of $\C^2/\Z_5$ sigularity along ten
curves, and $10\times 6$ moduli of the isolated resolution of
ten $\C^3/\Z_5\times\Z_5$ singularity.
Since $\wttau=\wttau_A^{{}^{12}}$ is antiholomorphic,
$1$ from the bulk remains (subtotal ${\bf 1}$).
Out of the ten $\Z_5$ curves, six of them (three pairs)
are exchanged by $\wttau$ while four of them are $\wttau$-invariant.
Out of $6\times 4$ from
the three pairs, a half of them are $\wttau$-anti-invariant
(${\bf 12}$).
$\wttau$ acts non-trivially on three of the four invariant curves, and all of
the $3\times 4$ remain (${\bf 12}$).
$\wttau$ acts identically on the last invariant curve $\{(0,0,*,*,*)\}$
exchanging the four exceptional curves (two pairs) of the $\C^2/\Z_5$
resolution, leaving  half of $4$ (${\bf 2}$).
Among the ten $\Z_5\times \Z_5$ points, six of them (three pairs)
are exchanged by $\wttau$ while four of them are $\tau$-invariant.
Out of the $6\times 6$ from
the three pairs, a half of them are $\tau$-anti-invariant
(${\bf 18}$).
$\wttau$ acts near three of the four invariant points as
$(z_1,z_2,z_3)\to (\bz_2,\bz_1,\bz_3)$ in $\C^3/\Z_5\times\Z_5$.
One can show by an orbifold analysis that there are $4$ anti-invariant
modulus at each such singularity, thus total of $3\times 4$
(${\bf 12}$). Near the last invariant point $(1,-1,0,0,0)$,
$\wttau$ acts as
$(z_1,z_2,z_3)\to (\bz_1,\bz_2,\bz_3)$ in $\C^3/\Z_5\times\Z_5$,
and all moduli remain (${\bf 6}$).
The total number of moduli is $1+12+12+2+18+12+6=63$,
in agreement with the number of complex moduli
of the original side.
The fixed-point set is the (3-dimensional) submanifold defined by
$\wtz_2=\overline{\wtz_1}$ and $\wtz_i\in\R$ ($i=3,4,5$).

The final B-parity is
associated with $\tau_B^{{}^{12,34}}:(z_1,z_2,z_3,z_4,z_5)
\mapsto (z_2,z_1,z_4,z_3,z_5)$.
Let us count the number of complex structure moduli.
Monomials that are invariant under $z_1\leftrightarrow z_2$,
$z_3\leftrightarrow z_4$
are of the form $(z_1z_2)^a(z_3z_4)^bz_5^c$ and there are 6 of them.
Thus, the invariant polynomials can have ${126-6\over 2}+6=66$ terms.
The group of coordinate transformation has dimension $13$.
Thus, the number of complex structure moduli
is $66-13=53$.
The fixed-point set is the line
$\{(z,-z,w,-w,0)\}$ 
and the genus 6 curve
$\{(z,z,w,w,u)\}$ (both 2-dimensional).
On the mirror side, the parity is the one associated with
$\wttau_A^{{}^{12,34}}:(\wtz_1,\wtz_2,\wtz_3,\wtz_4,\wtz_5)\mapsto
(\overline{\wtz_2},\overline{\wtz_1},\overline{\wtz_4},
\overline{\wtz_3},\overline{\wtz_5})$.
The complexified K\"ahler moduli
correspond to
$\wttau_A^{{}^{12,34}}$-anti-invariant harmonic $(1,1)$ forms.
Counting as in the case of
$\wttau_A^{{}^{12}}$-anti-invariants, the number of moduli is
$1+16+4+24+8=53$, again in agreement with the original side.
The fixed-point set is the (3-dimensional) submanifold defined by
$\wtz_2=\overline{\wtz_1}$,
$\wtz_4=\overline{\wtz_3}$, and $\wtz_i\in\R$ ($i=3,4,5$).

\begin{table}
\caption{Orientifolds of Quintic and their Mirrors}
\label{oqtable}
\begin{center}
\begin{tabular}{|l||l|}
\hline
quintic $M$&mirror quintic $\widetilde{M}$
\\
~~$(1_{\C},101_{\C})$&
~~$(101_{\C},1_{\C})$
\\
\noalign{\hrule height 0.8pt}
$\tau_A:z_i\to \overline{z_i}$
&
$\wttau_B:\wtz_i\to\wtz_i$
\\
$~~(1_{\C},101_{\R})$&$~~(101_{\R},1_{\C})$
\\
~~O6 at the real quintic ($\RP^3$)
&
~~O9 at $\widetilde{M}$
\\
\hline
$\tau_A^{{}^{12}}\!:(z_1,z_2,z_3,z_4,z_5)\to
(\overline{z_2},\overline{z_1},\overline{z_3},\overline{z_4},\overline{z_5})$
&
$\wttau_B^{{}^{12}}\!:(\wtz_1,\wtz_2,\wtz_3,\wtz_4,\wtz_5)\to
(\wtz_2,\wtz_1,\wtz_3,\wtz_4,\wtz_5)$
\\
$~~(1_{\C},101_{\R})$&$~~(101_{\R},1_{\C})$
\\
~~O6 at $\RP^3=\{(z,\bz,x_3,x_4,x_5)|x_i\in \R\}$
&
~~O3 at $\{(1,-1,0,0,0)\}$
\\
&
~~O7 at $\{(\wtz,\wtz,\wtz_3,\wtz_4,\wtz_5)\}$
\\
\hline
$\tau_A^{{}^{12,34}}\!:(z_1,z_2,z_3,z_4,z_5)\to
(\overline{z_2},\overline{z_1},\overline{z_4},\overline{z_3},\overline{z_5})$
&
$\wttau_B^{{}^{12,34}}\!:(\wtz_1,\wtz_2,\wtz_3,\wtz_4,\wtz_5)\to
(\wtz_2,\wtz_1,\wtz_4,\wtz_3,\wtz_5)$
\\
$~~(1_{\C},101_{\R})$&$~~(101_{\R},1_{\C})$
\\
~~O6 at $\RP^3=\{(z,\bz,w,\bw,x)|x\in \R\}$
&
~~O5 at $\{(\wtz,-\wtz,\wtw,-\wtw,0)\}$
\\
&
~~O5 at $\{(\wtz,\wtz,\wtw,\wtw,\wtu)\}$
\\
\noalign{\hrule height 0.8pt}
$\tau_B: z_i\to z_i$
&
$\wttau_A:\wtz_i\to\overline{\wtz_i}$
\\
$~~(1_{\R},101_{\C})$&$~~(101_{\C},1_{\R})$
\\
~~O9 at $M$
&
~~O6 at real quintic
\\
\hline
$\tau_B^{{}^{12}}\!:(z_1,z_2,z_3,z_4,z_5)\to
(z_2,z_1,z_3,z_4,z_5)$
&
$\wttau_A^{{}^{12}}\!:(\wtz_1,\wtz_2,\wtz_3,\wtz_4,\wtz_5)\to
(\overline{\wtz_2},\overline{\wtz_1},\overline{\wtz_3},\overline{\wtz_4},
\overline{\wtz_5})$
\\
$~~(1_{\R},63_{\C})$&$~~(63_{\C},1_{\R})$
\\
~~O3 at $\{(1,-1,0,0,0)\}$
&
~~O6 at $\{(\wtz,\overline{\wtz},\wtx_3,\wtx_4,\wtx_5)|\wtx_i\in\R\}$
\\
~~O7 at $\{(z,z,z_3,z_4,z_5)\}$
&
\\
\hline
$\tau_B^{{}^{12,34}}\!:(z_1,z_2,z_3,z_4,z_5)\to
(z_2,z_1,z_4,z_3,z_5)$
&
$\wttau_A^{{}^{12,34}}\!:(\wtz_1,\wtz_2,\wtz_3,\wtz_4,\wtz_5)\to
(\overline{\wtz_2},\overline{\wtz_1},\overline{\wtz_4},\overline{\wtz_3},
\overline{\wtz_5})$
\\
$~~(1_{\R},53_{\C})$&$~~(53_{\C},1_{\R})$
\\
~~O5 at $\{(z,-z,w,-w,0)\}$ (line)
&
~~O6 at $\{(\wtz,\overline{\wtz},
\wtw,\overline{\wtw},\wtx)|\wtx\in\R\}$
\\
~~O5 at $\{(z,z,w,w,z_5)\}$ (genus 6)
&
\\
\hline
\end{tabular}
\end{center}
\end{table}
\subsubsection{Summary and remarks}

The results are summarized in Table \ref{oqtable}.
The left column and the right column are mirror of each other.
The number $(n,m)$ in each column
means that there are $n$ complexified K\"ahler moduli and
$m$ complex structure moduli. The subscript ${}_{\R}$
or ${}_{\C}$ shows that the moduli is real or complex.
We have in mind to embed these orientifolds  in Type II superstring theory
on ${\rm CY}^3\times \R^{3+1}$ in which the parity acts trivially
on the $3+1$ dimensional Minkowski-coordinates.
Thus, the fixed-point sets of dimension
$p$ are called orientifold $(p+3)$-planes, or
simply O$(p+3)$.

We have not determined the types of the orientifold planes.
In the large volume limit, each orientifold plane
can be of several types --- roughly two types, $SO$ or $Sp$.
We have not determined which combinations of the types are possible
and how they are related by mirror symmetry. This will be determined,
for example, by going to the Gepner point at which an exact construction
of the crosscap states is now available as an application of the construction
in Section~\ref{sec:Omin} (see \cite{BlWi,ABPSS} for existing results
on Type I strings on Calabi--Yaus).
Also, for a consistent compactification in string theory, we need to cancel
the tadpole generated by the O-planes by, say, including D-branes.
This too can be done once we know the types of the orientifold planes.
This will be carried out in a future publication.

\subsection{Spacetime picture}

So far, we have been studying the theory on the string worldsheet
with a focus on the parity symmetries that commute with halves of
the $(2,2)$ worldsheet supersymmetry.
In the full string theory, more emphasis is put on the spacetime physics,
especially when we consider Type II orientifolds of the form
$M^6\times \R^{3+1}/\tau$, where
$\tau$ is an involution of $M^6$ and acts trivially on the
Minkowski coordinates.
There are several important issues, one of which
is tadpole cancellation mentioned above.
Here we comment on three others ---
supersymmetry,
massless fields, and superpotential
(all in the sense of
the $3+1$ dimensional spacetime).
In what follows,
$M$ is a general simply connected
Calabi--Yau 3-fold with an involution $\tau$.
We note that the number of moduli
of complexified K\"ahler class and complex structures of $M$
(before orientifolding) are
\beqa
&&\dim_{\C}H^{1,1}(M)=:h^{1,1},
\nn\\
&&\dim_{\C}H^1(M,T_M)=\dim_{\C}H^{2,1}(M)=:h^{2,1}.
\nn
\eeqa
It is useful to bear in mind the example of the 
quintic and the six involutions in Table \ref{oqtable}.

\subsubsection{Spacetime supersymmetry}

We have seen that $\tau\Omega$ preserves an ${\mathcal N}=2$
worldsheet supersymmetry if $\tau:M\to M$
is an antiholomorphic involution (A-parity)
a holomorphic involution (B-parity).
However, not all of them preserve a spacetime supersymmetry,
that is, ${\mathcal N}=1$ supersymmetry in $3+1$ dimensions.
There is no extra condition on B-parities, but
the orientifold
by an A-parity preserves a spacetime supersymmetry only if
$\tau$ maps the holomorphic three-form ${\mit\Omega}$
to its complex conjugate with
a possible constant phase,
$\tau^*{\mit\Omega}=\e^{i\theta}\overline{\mit \Omega}$.
For the quintic, all the three A-parities in fact satisfy this constraint.
Let us consider the Fermat type quintic.
Using the inhomogeneous coordinates
at the $z_5\ne 0$ patch, $\zeta_i=z_i/z_5$
obeying $\sum_{i=1}^4(\zeta_i)^5+1=0$, the holomorphic three-form
is expressed as
${\mit\Omega}=\dd\zeta_1\wedge\dd\zeta_2\wedge\dd\zeta_3/(\zeta_4)^4
=-\dd\zeta_1\wedge\dd\zeta_2\wedge\dd\zeta_4/(\zeta_3)^4$.
Then it is easy to see that
$$
\tau_A^*{\mit\Omega}=\overline{\mit\Omega},
\quad
\tau_A^{{}^{12}*}{\mit\Omega}=-\overline{\mit\Omega},
\quad
\tau_A^{{}^{12,34}*}{\mit\Omega}=\overline{\mit\Omega}.
$$
For the corresponding mirror involutions, we find
$$
\wttau_B^*\widetilde{\mit\Omega}=\widetilde{\mit\Omega},
\quad
\wttau_B^{{}^{12}*}\widetilde{\mit\Omega}=-\widetilde{\mit\Omega},
\quad
\wttau_B^{{}^{12,34}*}\widetilde{\mit\Omega}=\widetilde{\mit\Omega}.
$$
Note that the signs on the right hand sides
have an invariant meaning for B-parities. It is $+1$
for orientifolds with O9/O5-planes
and it is $-1$ for those with O7/O3-planes \cite{AAHV}.

\subsubsection{Light fields}

Moduli of the worldsheet theory give rise to
massless scalar fields in the $3+1$ dimensional spacetime.
We have seen that the moduli compatible with
parity symmetry can be real ---
orientifolds by a B-parity (resp. A-parity) reduce
the complexified K\"ahler moduli (resp. complex structure moduli)
by a real constraint.
For example, see Fig.~\ref{km}, which depicts the K\"ahler moduli
constrained by a real condition.
Thus, they give rise to real massless scalar fields in $3+1$ dimensions.
However, in the full string theory, these real fields are paired with
real fields from RR gauge potentials, and they form 
complex fields, or chiral superfields together with their fermionic
superpartners.

To see the detail, let us recall how the NSNS and RR fields
transform under the parity symmetry.
The NSNS fields, dilaton-gravity-$B$-field $(\phi,g,B)$,
are always transformed as
$$
\phi\to \tau^*\phi,\quad
g\to\tau^*g,\quad
B\to-\tau^*B.
$$
Transformation of the RR fields depends on the type of the involution.
Let $\tau={\mathcal I}_p$ be the inversion of $(9-p)$-coordinates.
\footnote{As is well-known, in Type II superstring theory,
the involutive parity symmetry
giving rise to the BPS O$p$-plane is of the form
${\mathcal I}_p\Omega$ for $p=9,8,5,4,1,0$ and 
$(-1)^{\widehat{F}_L}{\mathcal I}_p\Omega$ for
$p=8,7,6,3,2$ where $(-1)^{\widehat{F}_L}$ is $-1$ on the left-moving
Ramond sector. $p$ is even for Type IIA
and odd for Type IIB.}
Type IIB RR fields are transformed by the parity as
$$
p=9,5,1:\left\{
\begin{array}{l}
A_0\to -\tau^*A_0\\
A_2\to \tau^*A_2\\
A_4^+\to-\tau^*A_4^+,
\end{array}\right.
\qquad
p=7,3:\left\{
\begin{array}{l}
A_0\to \tau^*A_0\\
A_2\to -\tau^*A_2\\
A_4^+\to\tau^*A_4^+,
\end{array}\right.
$$
and for Type IIA,
$$
p=8,4,0:\left\{
\begin{array}{l}
A_1\to \tau^*A_1\\
A_3\to -\tau^*A_3,
\end{array}\right.
\qquad
p=6,2:\left\{
\begin{array}{l}
A_1\to -\tau^*A_1\\
A_3\to \tau^*A_3.
\end{array}\right.
$$
It is also useful to recall the massless spectrum
 before orientifolding
where the spacetime theory has ${\mathcal N}=2$ supersymmetry,
given in Table~\ref{msslss2}
(see for example
\cite{Shamit}).
\begin{table}[htb]
\caption{Massless Fields in ${\mathcal N}=2$ Compactifications}
\label{msslss2}
\begin{center}
\begin{tabular}{|l@{\quad\vrule width0.8pt\quad}c|c|}
\hline
&hypermultiplets&vector multiplets
\\
\noalign{\hrule height 0.8pt}
~~IIA
&$h^{2,1}+1$&$h^{1,1}$
\\
\hline
~~IIB&$h^{1,1}+1$&$h^{2,1}$
\\
\hline
\end{tabular}
\end{center}
\end{table}
In Type IIA, there are $h^{1,1}$ vector multiplets
from the complexified K\"ahler moduli and $A_3$ reduced on $H^{1,1}(M)$,
$h^{2,1}$ hypermultiplets from the complex structure moduli
and $A_3$ reduced on $H^{2,1}(M)\oplus H^{1,2}(M)$, and  
one hypermultiplet from $(\phi,B)$ and $A_3$ reduced on
$H^{3,0}(M)\oplus H^{0,3}(M)$, and a gravity multiplet
from $(g,A_1)$.
In Type IIB, there are $h^{2,1}$ vector multiplets
from the complex structure moduli and $A_4^+$ reduced on
$H^{2,1}(M)\oplus H^{1,2}(M)$,
$h^{1,1}$ hypermultiplets from
the complexified K\"ahler moduli and
$(A_2,A_4^+)$ reduced on $(H^{1,1}(M),H^{2,2}(M))$,
one hypermultiplet from $(\phi, B)$ and $(A_0,A_2)$,
and a gravity multiplet from $g$ and $A_4^+$
reduced on $H^{3,0}(M)\oplus H^{0,3}(M)$.
In terms of the ${\mathcal N}=1$
supersymmetry, an ${\mathcal N}=2$ vector multiplet splits into
a vector and a chiral multiplets
while a hypermultiplet splits into two chiral multiplets.
We would like to see which of them survive and which of them are projected out
by the orientifolds.

The involution $\tau$ induces an
involution on the space of Harmonic forms by $\eta\to\tau^*\eta$.
We denote by $H^{\bullet}_{+}$ and $H^{\bullet}_-$
the subspaces of $H^{\bullet}(M)$
consisting of $\tau^*$-invariant and $\tau^*$-anti-invariant forms.
We also use $H^{p,q}_{\pm}$ when applicable, and denote its dimension
by $h^{p,q}_{\pm}$.

\subsubsection*{\it Type IIA orientifolds}

Let us first consider Type IIA orientifold
associated with an antiholomorphic involution $\tau$ (such that
$\tau^*{\mit\Omega}={\rm const}\times\overline{\mit\Omega}$).
Since $\tau$
flips 3 real coordinates of $M$, it is of the type $p=6$, and therefore
the RR fields are transformed as $A_1\to -\tau^*A_1$
and $A_3\to \tau^*A_3$.
The moduli space of complex structure
is reduced to a half by the orientifold and thus has real dimension
$h^{2,1}$.
The moduli space of complexified the K\"ahler class is reduced by
the constraint (\ref{Asymc}). The reduced moduli space has complex dimension
$h^{1,1}_-$, the dimension of the space $H^{1,1}_-$ of
$\tau$-anti-invariant harmonic $(1,1)$-forms.

The surviving fields out of
the $h^{1,1}$ (${\mathcal N}=2$) vector multiplets
are $h^{1,1}_-$ (${\mathcal N}=1$) chiral multiplets
from the complexified K\"ahler moduli
and $h^{1,1}_+$ (${\mathcal N}=1$) vector multiplets
from $A_3$ reduced on $\tau$-invariant harmonic 2-forms.
Let us next see which of the $h^{2,1}$ hypermultiplet fields survives.
We find $h^{2,1}$ real scalars from the complex structure moduli.
We also find $h^{2,1}$ real scalars from $A_3$ reduced on
the $\tau^*=1$ subspace of $H^{2,1}(M)\oplus H^{1,2}(M)$.
Note that $\tau$, being antiholomorphic,
exchanges $H^{p,q}(M)$ and $H^{q,p}(M)$ and hence $\tau^*=1$
(or $\tau^*=-1$) on
half of $H^{p,q}(M)\oplus H^{q,p}(M)$ if $p\ne q$.
They combine into complex scalars in $h^{2,1}$ chiral multiplets.
From the hypermultiplet including the dilaton,
we find one chiral multiplet whose lowest component consists of
the dilaton and $A_3$ reduced on
the $\tau^*=1$ subspace of $H^{3,0}(M)\oplus H^{0,3}(M)$.
The ${\mathcal N}=2$ gravity multiplet is projected to
a ${\mathcal N}=1$ gravity multiplet.

\subsubsection*{\it Type IIB orientifolds with O9 or O5-planes}

Let us next consider Type IIB orientifold
associated with a holomorphic involution $\tau$
of the type $p=9$ or $p=5$ (such as
$\tau_B$ and $\tau_B^{{}^{12,34}}$
in the quintic case).
The RR fields are transformed as
$A_0\to -\tau^*A_0$, $A_2\to \tau^*A_2$, $A_4^+\to-\tau^*A_4^+$.
The holomorphic 3-form ${\mit\Omega}$ is preserved by $\tau$,
$\tau^*{\mit\Omega}={\mit\Omega}$.
The moduli space of the complexified K\"ahler class is reduced
to a half and has real dimension $h^{1,1}$.
The complex structure deformations are generated by the $\tau=1$ subspace of
$H^1(M,T_M)$. Since the isomorphism $H^1(M,T_M)\cong H^{2,1}(M)$
is given by contraction with ${\mit\Omega}$ and since ${\mit\Omega}$
is $\tau^*$-invariant, we see that the deformation space is isomorphic
to $H^{2,1}_+$.
Thus, the complex structure moduli are reduced to $h^{2,1}_+$.

Out of the $h^{2,1}$ ${\mathcal N}=2$ vector multiplet,
we find $h^{2,1}_+$ chiral multiplets from the complex structure moduli
and $h^{2,1}_-$ vector multiplets from
$A_4^+$ reduced on $H^{2,1}_-\oplus H^{1,2}_-$.
Let us next see which of the $h^{1,1}$ hypermultiplet fields
are unprojected.
First, we find $h^{1,1}$ real scalars from the reduced
K\"ahler moduli.
We also find real scalars from $A_2$ reduced on
$\tau$-invariant harmonic 2-forms and from $A_4^+$ reduced on
$\tau$-anti-invariant harmonic 4-forms.
Since $\tau$ preserves the orientation and the metric of $M$,
$\tau^*$ commutes with the Hodge $*$-operator which in particular sends
$H^{1,1}(M)$ to $H^{2,2}(M)$.
We thus find $H^{1,1}_+$ is isomorphic to $H^{2,2}_+$ by the $*$-operation.
This means that
$$
\dim (H^{1,1}_+\oplus H^{2,2}_-)
=\dim (H^{1,1}_-\oplus H^{2,2}_+)=h^{1,1}.
$$
Thus, the RR potentials $A_2$ and $A_4^+$ reduce to $h^{1,1}$
real scalars.
They combine with the real massless scalars from the K\"ahler moduli
and form $h^{1,1}$ chiral multiplet fields.
Next, the single hypermultiplet including the dilaton
is projected to a single chiral multiplet consisting of
the dilaton and the dual of $A_2$, plus fermions.
Finally, the ${\mathcal N}=2$ gravity multiplet projects to
an ${\mathcal N}=1$ gravity multiplet ($A_4^+$ reduced on
$H^{3,0}(M)\oplus H^{0,3}(M)$ is projected out since
$A_4^+\to -\tau^*A_4^+$ and $\tau^*{\mit\Omega}={\mit\Omega}$).

\subsubsection*{\it Type IIB orientifolds with O7 and/or O3-planes}

Finally, let us consider
Type IIB orientifolds associated with a holomorphic involution $\tau$
of the type $p=7$ and/or $p=3$ (such as $\tau_B^{{}^{12}}$
in the quintic case).
The RR fields are transformed as
$A_0\to \tau^*A_0$, $A_2\to -\tau^*A_2$, $A_4^+\to \tau^*A_4^+$,
and $\tau^*$ flips the sign of
the holomorphic 3-form ${\mit\Omega}$.
The moduli space of K\"ahler class has real dimension $h^{1,1}$.
The complex structure deformation are generated by the $\tau=1$ subspace of
$H^1(M,T_M)$, as in the previous case but
that is isomorphic to $H^{2,1}_-$ in this case since
$\tau^*{\mit\Omega}=-{\mit\Omega}$.
Thus, there are $h^{2,1}_-$ complex structure moduli.

The counting of unprojected fields proceeds as before.
We find $h^{2,1}_-$ chiral multiplets (complex structure moduli),
$h^{2,1}_+$ vector multiplets ($A_4^+$ reduced on $H^{2,1}_+\oplus H^{1,2}_+$),
$h^{1,1}$ chiral multiplets (real K\"ahler moduli
plus real fields from
$(A_2,A_4^+)$ reduced on $(H^{1,1}_-,H^{2,2}_+)$),
one chiral multiplet ($\phi,A_0$),
and a gravity multiplet
($A_4^+$ reduced on
$H^{3,0}(M)\oplus H^{0,3}(M)$ is projected out since
$A_4^+\to \tau^*A_4^+$ and $\tau^*{\mit\Omega}=-{\mit\Omega}$).

\begin{table}
\caption{Light Fields from Closed Strings}
\label{lf}
\begin{center}
\begin{tabular}{|ll@{\quad\vrule width0.8pt\quad}c|c|}
\hline
&$\begin{array}{l}
\\
\\
\end{array}$
&chiral multiplets&vector multiplets
\\
\noalign{\hrule height 0.8pt}
$\begin{array}{l}
\\
\\
\end{array}$
&IIAO(6)
&$h^{1,1}_-+h^{2,1}+1$&$h^{1,1}_+$
\\
\hline
$\begin{array}{l}
\\
\\
\end{array}$
&IIBO(9,5)~~&$h^{2,1}_++h^{1,1}+1$&$h^{2,1}_-$
\\
\hline
$\begin{array}{l}
\\
\\
\end{array}$
&IIBO(7,3)&$h^{2,1}_-+h^{1,1}+1$&$h^{2,1}_+$
\\
\hline
\end{tabular}
\end{center}
\end{table}
\begin{table}[htb]
\caption{The number for the six orientifolds of the quintic}
\label{lfq}
\begin{center}
\begin{tabular}{|l@{\quad\vrule width0.8pt\quad}c|c|}
\hline
&chiral multiplets&vector multiplets
\\
\noalign{\hrule height 0.8pt}
IIAO by $\tau_A, \tau_A^{{}^{12}},
\tau_A^{{}^{12,34}}$
&$103$&$0$
\\
\hline
IIBO by $\tau_B$
&$103$&$0$
\\
\hline
IIBO by $\tau_B^{{}^{12}}$
&$65$&$38$
\\
\hline
IIBO by $\tau_B^{{}^{12,34}}$
&$55$&$48$
\\
\hline
\end{tabular}
\end{center}
\end{table}

To summarize, we list in Table~\ref{lf} the number of light ${\mathcal N}=1$
supermultiplets. Table~\ref{lfq} shows the numbers
for the six orientifolds of the quintic.
In general, D-branes should also be included in the setup for 
tadpole cancellation.
Thus, there are other fields than in Table~\ref{lf},
associated with open string modes.
Some of them correspond to the location of the D-branes
and others are Yang-Mills gauge fields and charged matter
fields on the branes or brane intersections.

\subsubsection{Spacetime superpotential}

An important part of the low energy effective theory
is the superpotential. It is a holomorphic
function of the fields from the above `light' chiral multiplets
as well as other chiral multiplet fields such as those
associated with D-branes.
We must first compute the superpotential,
find the minima of the potential, and expand around a chosen vacuum.
Only after that one can discuss the actual mass spectrum.

Certain origins of the superpotential terms are known (see e.g.
\cite{AAHV}).
For Type IIA orientifolds, superpotential can be generated by
holomorphic disks ending on A-branes \cite{KKLM,KKLM2,AgaVaf,AgaKleVaf}
as well as
``holomorphic $\RP^2$'s'' --- holomorphic maps of $\CP^1$
to $X$ which are equivariant with respect to
the anti-podal map of $\CP^1$ and
the involution $\tau$ of $X$.
For Type IIB orientifolds,
we have the so called flux superpotential
$W=\int_X\Omega\wedge G$ where $G$ is a linear combination of
RR and NSNS 3-form field strengths with dilaton-axion as the coefficients
\cite{GVW}. 
(The flux superpotential is computed or applied in the context of
(mostly toroidal) orientifolds in \cite{FluxSup,KKLT}.
The geometry underlying ${\cal N}= 1$ superpotential
is discussed in \cite{LMW}.)
For D5-branes wrapped on 2-cycles,
the superpotential for the 3-form flux generated by D5
is equivalent \cite{AAHV,AgaVaf}
to the holomorphic Chern-Simons action
for the open topological B-model \cite{WCS,BCOV},
which in turn is equivalent \cite{KKLM}
to the superpotential associated with
the obstruction to the deformation
of holomorphic curves \cite{BDLR,Katz}.

One may wonder whether the deformation theory of the orientifold itself
is obstructed as well. Namely, whether
the deformation of a Calabi--Yau with holomorphic
involution is obstructed.
We claim that it is not. This can be shown as follows.\footnote{We thank
M. Kapranov for this argument and Y.-G. Oh for his help.}
Let us fix the underlying differentiable manifold $M$. A
`Calabi--Yau with holomorphic involution' is just a pair
$(J,\tau)$ where $J$ is a complex structure of $M$ whose canonical bundle
is holomorphically trivial and $\tau$ is an involutive diffeomorphism
of $M$ that commutes with $J$. We are interested in the deformation space of
the pair $(J,\tau)$, especially whether it is smooth or not.
(We identify pairs that are related by diffeomorphisms.)
First, we note that there is no deformation of $\tau$ itself
--- any involution close to $\tau$ is diffeomorphic to
$\tau$ itself. Thus one can fix $\tau$ and consider deformations only of
$J$ commuting with $\tau$.
Let us for now forget about $\tau$ and
consider the full space of deformations of the Calabi--Yau.
This deformation theory is not obstructed.
Namely, the Teichm\"uller space
(the space of Calabi--Yau $J$
divided by the group of diffeomorphisms homotopic to the identity) is smooth.
On the Teichm\"uller space there is a $\Z_2$ action generated by
$[J]\mapsto [\tau_*J\tau_*^{-1}]$.
The $\tau$-invariant $J$'s we are interested in are nothing but the
fixed points of this action.
Since the fixed-point locus of
a $\Z_2$ action on a smooth manifold is always smooth,
our deformation space is smooth.
This proves that the deformation theory is not obstructed.

This does not mean, however, that there is no superpotential
of purely orientifold origin.
For example if there are two O5-planes $C_+$ and $C_-$
of opposite RR charges, we do have a flux superpotential
which is $\int_{Y}\Omega$ where $Y$ is a three dimensional submanifold of $M$
bounded by $C_+-C_-$, as occured in the example studied in \cite{AAHV}.

\section*{Acknowledgement}

We would like to thank
B.~Acharya, M.~Aganagic, S.~Akbulut,
 C.-H. Cho, K.~Hosomichi,
M.~Kapranov, W.~Lerche, J.~Maldacena, Y.-G.~Oh, K.~Ono,
E.~Poppitz,
C.~Vafa for useful discussions and/or collaboration on related works.
KH thanks R. Myers for a motivating
conversation at Montr\'eal, May 2000.
IB and KH thank KITP, Santa Barbara,
KH thanks Rutgers University, Aspen Center for Physics
and University of Wisconsin,
 where parts of this project are carried out,
for their kind hospitality.
KH was supported in part by NSF-DMS 9709694,
NSF-DMS 0074329 and NSF PHY 0070928.

\appendix{Index $\Tr (-1)^FP$ in Non-linear Sigma Models}
\label{app:index}

Let $(X,g)$ be a Riemannian manifold of real dimension $n$.
We consider the supersymmetric non-linear sigma model
in $1+1$ dimensions whose classical
action on Minkowski space reads
\beqa
\Scr{L}&=&{1\over 2}g_{IJ}(\partial_0\phi^I\partial_0\phi^J-
\partial_1\phi^I\partial_1\phi^J)
+{i\over 2}g_{IJ}\psi_-^I(D_0+D_1)\psi_-^J
+{i\over 2}g_{IJ}\psi_+^I(D_0-D_1)\psi_+^J
\nn\\
&&-{1\over 4}R_{IJKL}\psi_+^I\psi_-^J\psi_+^K\psi_-^L.
\nn
\eeqa
Supersymmetry is generated by
$\delta\phi^I=i\epsilon_+\psi_-^I+i\epsilon_-\psi_+^I$,
$\delta\psi_{\pm}^I
=-\epsilon_{\mp}(\partial_0\pm\partial_1)\phi^I$.
Let $\tau:X\to X$ be an isometric involution.
The classical action is invariant under the parity symmetry
$\tau\Omega$
\beqa
&&\phi^I(x)\to \tau^I(\phi(\wtx)),
\nn\\
&&\psi^I_{\pm}(x)\to \tau^I_{*J}\psi_{\mp}^J(\wtx),
\nn
\eeqa
where $\wtx=(x^0,-x^1)$ for $x=(x^0,x^1)$, and we assume that
it is anomaly-free.
We shall compute the twisted supersymmetric index $\Tr(-1)^F\tau\Omega$
for both closed and open strings.

The basic strategy is to represent the index as the partition function
on a flat surface and localize the path-integral
on the fixed point of the supersymmetry.
For example, the ordinary Witten index $\Tr(-1)^F$ on the RR sector is
path-integral on the 2-torus with periodic boundary condition in both
directions. The boundary condition is fully supersymmetric and the fixed points
are the constant maps.
In a constant background, the action consists only of the four-fermi terms.
The 1-loop integral on non-constant modes is $1$ as a consequence of
boson-fermion cancellation, and we are left with the zero mode integral,
with the $\exp$(four-fermi terms) as the weight.
This leads to integration over $X$ of
the Pfaffian of the Riemannian curvature.
The latter is the Euler class of the tangent bundle $T(X)$ of $X$,
and hence
\beq
\Tr\!\!\mathop{}_{{\mathcal H}_{\rm RR}}\!\!(-1)^F
=\int_Xe(T(X))=\chi(X).
\eeq
This method was introduced in \cite{AG,FWin}.

Now we compute the twisted index $\Tr(-1)^F\tau\Omega$
on the RR-sector (closed string).
It is represented as the path-integral on the Klein bottle
$(x_1,x_2)\equiv (x_1+L_1,x_2)\equiv
(-x_1,x_2+L_2)$
with the periodic boundary condition along $x_1$,
but with the twisted boundary condition along $x_2$:
\beqa
&&\phi^I(x_1,x_2)=\tau^I(\phi(-x_1,x_2+L_2)),
\label{bcb1}\\
&&
\psi_{\pm}^I(x_1,x_2)=\tau_{*J}^I \psi_{\mp}^J(-x_1,x_2+L_2).
\label{bcf1}
\eeqa
This periodicity preserves the diagonal part of the
$(1,1)$ supersymmetry and one can still use the localization
method.
The fixed points of the supersymmetry are constant maps
into the set $X^{\tau}$ of $\tau$-fixed points.
We first note that the integral of modes that are not constant along
$x_1$ yields $1$ due to boson-fermion cancellation.
Thus, we can focus on modes that depend only on $x_2$.
This computation has been done beautifully in \cite{SS}.
Here is the outline.
We first separate the coordinates into tangent directions
$\phi^{\mu}$ to $X^{\tau}$ and normal directions $\phi^i$.
Let $n_1$ be the real dimension of $X^{\tau}$, so that
$\mu=1,...,n_1$ and $i=1,...,n-n_1$.
By (\ref{bcb1}), $\phi^{\mu}$ is periodic and $\phi^i$ is antiperiodic
along $x_2$.
By (\ref{bcf1}), the periodicity of the fermions are
\beqa
\psi_+^{\mu}+\psi_-^{\mu}:&&\mbox{periodic}
\nn\\
\psi_+^{\mu}-\psi_-^{\mu}:&&\mbox{anti-periodic}
\nn\\
\psi_+^{i}+\psi_-^{i}:&&\mbox{anti-periodic}
\nn\\
\psi_+^{i}-\psi_-^{i}:&&\mbox{periodic}
\nn
\eeqa
We also note that $R_{IJKL}\psi_+^I\psi_-^J\psi_+^K\psi_-^L$
is proportional to $R_{IJKL}(\psi_++\psi_-)^I(\psi_++\psi_-)^J
(\psi_+-\psi_-)^K(\psi_+-\psi_-)^L$.
It follows that the zero mode action is the curvature of the normal bundle
to $X^{\tau}$.
The one-loop integral of the non-zero modes yields
\beqa
\phi^{\mu}:&&
{\det}^{-{1\over 2}}_P(\partial^2+iR_{T}\partial)
=(2\pi \beta)^{-{n_1\over 2}}
\prod_{\lambda_{T}} {\lambda_{T}\beta/2\over
\sinh(\lambda_{T}\beta/2)}
\nn\\
\psi_+^{\mu}+\psi_-^{\mu}:&&
{\rm Pf}_P(i\partial)=1
\nn\\
\psi_+^{\mu}-\psi_-^{\mu}:&&
{\rm Pf}_A(i\partial+R_{T})=\prod_{\lambda_{T}}
2\cosh(\lambda_{T}\beta/2)
\nn\\
\phi^i:&&
{\det}_A^{-{1\over 2}}(\partial^2+iR_{N}\partial)
=\prod_{\lambda_{N}}{1\over 4\cosh(\lambda_{N}\beta/2)}
\nn\\
\psi_+^i+\psi_-^i:&&
{\rm Pf}_A(i\partial)=\prod_{\lambda_{N}}2
\nn\\
\psi_+^i-\psi_-^i:&&
{\rm Pf}_P(i\partial +R_{N})
=\prod_{\lambda_{N}}
{\sinh(\lambda_{N}\beta/2)\over \lambda_{N}\beta/2}
\nn
\eeqa
Here $\partial$ is the derivative with respect to $x_2$.
 $R_{T}$ is the curvature of the tangent bundle of $X^{\tau}$
and $(i\lambda_{T},-i\lambda_{T})$
are its eigenvalues, and similarly for the curvature of the normal bundle
$R_{N}$. (Here we assume that both $n$ and $n_1$ are even,
for simplicity.)
$\beta$ is the circumference $L_2$ in the $x_2$ direction,
on which the final expression should not depend.
In this notation, the zero mode action contributes by a factor
$$
\prod_{\lambda_{N}}\lambda_{N}\beta.
$$
Collecting all together, the zero mode integral is
 $$
\int_{X^{\tau}}(\pi\beta)^{-{n_1\over 2}}\prod_{\lambda_{T}}
{\lambda_{T}\beta/2\over\tanh(\lambda_{T}\beta/2)}
\prod_{\lambda_{N}}{\tanh(\lambda_{N}\beta/2)\over
\lambda_{N}\beta/2}
\prod_{\lambda_{N}}\lambda_{N}\beta/2.
$$
This indeed does not depend on $\beta$ because $n_1=\dim X^{\tau}$
and $\lambda_{T}$, $\lambda_{N}$ are 2-forms.
Setting $\beta=1/\pi$,
we find the following index formula
\beq
\Tr\!\!\mathop{}_{{\mathcal H}_{\rm RR}}\!\!(-1)^F\tau\Omega
=\int_{X^{\tau}}{L(T(X^{\tau}))\over
L(N(X^{\tau}))}e(N(X^{\tau})),
\eeq
 $L(V)$ is the Hirzebruch L-genus
defined by $\prod_{\lambda_V}(\lambda_V/2\pi)/\tanh(\lambda_V/2\pi)$.

We next consider the twisted index for an open string.
The string has one end on a D-brane wrapped on
$W\subset X$ and supporting a complex vector bundle $E$,
and the other end on the image brane $(\tau W,\tau E)$.
The boundary condition preserves the same supersymmetry as
$\tau\Omega$ preserves, and one can consider twisted Witten index
$\Tr (-1)^F\tau\Omega$.
This is represented as the path-integral on the M\"obius strip
$(x_1,x_2)=(L_1-x_1,x_2+L_2)$, $0\leq x_1\leq L_1$,
with the boundary condition at one end
\beqa
&&\partial_2\phi^I(0,x_2),\,
(\psi_+^I+\psi_-^I)(0,x_2):\quad \mbox{tangent to $W$},
\label{bc21}
\\
&&\partial_1\phi^I(0,x_2),\,
(\psi_+^I-\psi_-^I)(0,x_2):\quad
\mbox{normal to $W$},
\label{bc22}
\eeqa
and similar condition at the other end $(L_1,x_2)$.
The coupling to the gauge field $A$ for the bundle
$E$ is through the Chan-Paton factor
\beq
\tr_{\overline{E}}P\exp\left\{
-i\int_{x_1=0}
\Bigl(\,-{}^t\!A_M\partial_2\phi^M
-i\,{}^t\!F_{MN}(\psi_++\psi_-)^M(\psi_++\psi_-)^N\Bigr)
\dd x_2\right\},
\label{cpf}
\eeq
(similarly at $x_1=L_1$)
but not through the change in the boundary condition.
In the above formula, $M$, $N$ are coordinate indices on the brane $W$.
Also, we have $\overline{E}$ with the dual gauge field
$-{}^t\!A$ because the left boundary of the string worldsheet
 is oriented toward negative
time direction.
Periodicity along $x_2$ is the same as (\ref{bcb1})-(\ref{bcf1})
except that $-x_1$ is replaced by $L_1-x_1$.
Because of supersymmetry, path-integral localizes on
the fixed points, which are constant maps to the intersection
of the brane and the orientifold plane,
$W\cap X^{\tau}$.
By the deformation invariance of the index, we deform $W$ continuously
so that $W$ and $X^{\tau}$ intersects normally.
This means that $\tau$ sends $W$ to itself in a neighborhood of
$X^{\tau}$. One can then choose the coordinates
$(x^a,x^{\mu},x^{\alpha},x^i)$ where $(x^a,x^{\mu})$ are coordinates
on $W$ and the involution acts as
$$
\tau:(x^a,x^{\mu},x^{\alpha},x^i)\mapsto (-x^a,x^{\mu},x^{\alpha},-x^i)
$$
so that $(x^{\mu},x^{\alpha})$ are coordinates on $X^{\tau}$.
We name the dimensions of the respective directions as
$$
a=1,...,n_1;\quad
\mu=1,...,n_2;\quad
\alpha=1,...,n_3;\quad
i=1,...,n_4.
$$
For simplicity we assume that all four dimensions are even.
Since $\tau W$ is the same as $W$ in a neighborhood of $X^{\tau}$,
the boundary conditions are the same on the two end points of the string.
In the intersection $W\cap X^{\tau}$,
$\tau E$ can be topologically identified with the complex conjugate
of $E$. Thus, the Chan-Paton factor at $x_1=L_1$ is the same as
(\ref{cpf}).
We can thus ignore the $x_1$ dependence of the fields ---
the integral of $x_1$-dependent modes will give $1$ by boson-fermion
cancellation.
The boundary condition (\ref{bc21})-(\ref{bc22}) then forces the constraints
$\psi_+^a=\psi_-^a=:\psi^a$,
$\psi_+^{\mu}=\psi_-^{\mu}=:\psi^{\mu}$,
$\phi^{\alpha}=0$,
$\psi_+^{\alpha}=-\psi_-^{\alpha}=:\psi^{\alpha}$,
$\phi^i=0$,
$\psi_+^i=-\psi_-^i=:\psi^i$.
By (\ref{bcb1}) and (\ref{bcf1}), the remaining fields obey
the following periodicity
\beqa
\phi^a:&&\mbox{anti-periodic}
\nn\\
\psi^a:&&\mbox{anti-periodic}
\nn\\
\phi^{\mu}:&&\mbox{periodic}
\nn\\
\psi^{\mu}:&&\mbox{periodic}
\nn\\
\psi^{\alpha}:&&\mbox{anti-periodic}
\nn\\
\psi^i:&&\mbox{periodic}
\nn
\eeqa
The one-loop integral of the non-zero modes yields
\beqa
\phi^a:&&{\det}_A^{-{1\over 2}}(\partial^2+iR_1\partial)
=\prod_{\lambda_1}{1\over 4\cosh(\lambda_1\beta/2)}
\nn\\
\psi^a:&&{\rm Pf}_A(i\partial)=\prod_{\lambda_1}2
\nn\\
\phi^{\mu}:&&{\det}_P^{-{1\over 2}}(\partial^2+iR_2\partial)
=(2\pi \beta)^{-{n_2\over 2}}\prod_{\lambda_2}{\lambda_2\beta/2
\over \sinh(\lambda_2\beta/2)}
\nn\\
\psi^{\mu}:&&{\rm Pf}_P(i\partial)=1
\nn\\
\psi^{\alpha}:&&{\rm Pf}_A(i\partial +R_3)
=\prod_{\lambda_3}2\cosh(\lambda_3\beta/2)
\nn\\
\psi^i:&&{\rm Pf}_P(i\partial +R_4)=\prod_{\lambda_4}{\sinh(\lambda_4\beta/2)
\over \lambda_4\beta/2}
\nn
\eeqa
The zero mode bulk action yields the factor
$$
\prod_{\lambda_4}\lambda_4\beta,
$$
and the boundary action provides
$$
\tr_{\overline{E}}\e^{2\beta \,{}^t\!F},
$$
where the exponent has the factor $2\beta$ rather than $\beta$ because
the circumference of the closed boundary circle is twice as large as
$\beta=L_2$.
Collecting all together, we find the following expression for
the zero mode integral
\beqa
&&\int_{W\cap X^{\tau}}\tr_{\overline{E}}\e^{2\beta \,{}^t\!F}\times
\nn\\
&&
2^{-{n_1\over 2}+{n_3\over 2}}(2\pi\beta)^{-{n_2\over 2}}
\prod_{\lambda_1}{1\over \cosh({\lambda_1\beta\over 2})}
\prod_{\lambda_2}{{\lambda_2\beta\over 2}\over\sinh({\lambda_2\beta\over 2})}
\prod_{\lambda_3}\cosh({\lambda_3\beta\over 2})
\prod_{\lambda_4}{\sinh({\lambda_4\beta\over 2})\over
{\lambda_4\beta\over 2}}
\prod_{\lambda_4}\lambda_4\beta.
\nn
\eeqa
This is indeed independent of
the value of $\beta$ on dimensional ground,
$n_2=\dim W\cap X^{\tau}$.
At this stage, we use the elementary relations
$$
{\sinh({x\over 2})
\over {x\over 2}}
=\sqrt{{\sinh(x)\over x}{\tanh({x\over 2})\over
{x\over 2}}},
\qquad
\cosh\Bigl({x\over 2}\Bigr)
=\sqrt{{\sinh(x)\over x}{{x\over 2}\over \tanh({x\over 2})}}.
$$
to re-express the integral as
\beqa
&&2^{-{n_1\over 2}+{n_3\over 2}}(2\pi\beta)^{-{n_2\over 2}}
\int_{W\cap X^{\tau}}\tr_{\overline{E}}\e^{2\beta \,{}^t\!F}
\sqrt{
\prod_{\lambda_1}{{\lambda_1\beta}\over\sinh(\lambda_1\beta)}
{\tanh({\lambda_1\beta\over 2})\over {\lambda_1\beta\over 2}}
\prod_{\lambda_2}{\lambda_2\beta\over\sinh(\lambda_2\beta)}
{{\lambda_2\beta\over 2}\over \tanh({\lambda_2\beta\over 2})}}
\times\qquad
\nn\\
&&
\qquad\qquad\qquad\qquad\times\sqrt{
\prod_{\lambda_3}{\sinh(\lambda_3\beta)\over\lambda_3\beta}
{{\lambda_3\beta\over 2}\over\tanh({\lambda_3\beta\over 2})}
\prod_{\lambda_4}{\sinh(\lambda_4\beta)\over \lambda_4\beta}
{\tanh({\lambda_4\beta\over 2})\over{\lambda_4\beta\over 2}}
}
\prod_{\lambda_4}\lambda_4\beta
\nn
\eeqa
Setting $\beta=1/4\pi$ we find that the index is expressed as
\beqa
&&\Tr\!\!\mathop{}_{{\mathcal H}_{(W,E),(\tau W,\tau E)}}\!\!
(-1)^F\tau\Omega
\nn\\
&&\qquad
=2^{-{n_1\over 2}+{n_3\over 2}+{n_2\over 2}-{n_4\over 2}}
\int_{W\cap X^{\tau}}
{\rm ch}(\overline{E})
\sqrt{\widehat{A}(T(W))\over\widehat{A}(N(W))}
\sqrt{L({1\over 4}T(X^{\tau}))\over
L({1\over 4}N(X^{\tau}))}e(N(W)\cap N(X^{\tau})).
\nn\\
\label{appindf}
\eeqa
Here $\widehat{A}(V)$ is the A-roof genus 
$\prod_{\lambda_V}(\lambda_V/4\pi)/\sinh(\lambda_V/4\pi)$
and $L({1\over 4}V)$ is the modified L-genus
$\prod_{\lambda_V}(\lambda_V/8\pi)/
\tanh(\lambda_V/8\pi)$. We note here that
the power of $2$ in the formula is
\beq
-{n_1\over 2}+{n_3\over 2}+{n_2\over 2}-{n_4\over 2}
=n_2+n_3-{n_1+n_2+n_3+n_4\over 2}=\dim_{\R}X^{\tau}-{1\over 2}\dim_{\R}X.
\eeq
It way happen that $\tau$ has fixed-point submanifolds of various dimensions.
It is clear form the above derivation that, in such a case,
the index is the sum
over components of intersection where the summand is the above
in each component of intersection.

One may consider the application of the result to superstring theory.
In particular, one can read off from the index the WZ coupling
of the D-branes and O-planes to the bulk RR fields
\cite{GHM,Zheng,MM,FrWi,DjM,MSS,CRoos,Stef}.
The index in a special case has been computed in such a context
\cite{chri}.

\subsubsection*{\it The case $W=X$}

In the case $W=X$, the formula (\ref{appindf}) 
with a modification $\hat{A}(X)\to {\rm td}(X)$ simplifies considerably.
It is straightforward to see that
$$
\int_{X^{\tau}}2^{\dim_r\!X^{\tau}-{1\over 2}\dim_r\!X}
{\rm ch}(\overline{E})\sqrt{{\rm td}(X)
{L({1\over 4}T(X^{\tau}))\over
L({1\over 4}N(X^{\tau}))}}
=\int_{X^{\tau}}{\rm ch}(\overline{E})
\prod_{\lambda_t}{{\lambda_t\over 2\pi}\over
1-\e^{-{\lambda_t\over 4\pi}}}
\prod_{\lambda_n}{1\over 1+\e^{-{\lambda_n\over 4\pi}}}.
$$
In the integrand, only the term of the same degree as $\dim X^{\tau}$
contribute.
This allows us to replace it by
$$
\int_{X^{\tau}}{\rm ch}(2\overline{E})
\prod_{\lambda_t}{{\lambda_t\over 2\pi}\over
1-\e^{-{\lambda_t\over 2\pi}}}
\prod_{\lambda_n}{1\over 1+\e^{-{\lambda_n\over 2\pi}}}.
$$
which is nothing but
$$
\int_{X^{\tau}}{\rm ch}(2\overline{E})\,
{\rm td}(X^{\tau}){1\over {\rm ch}(\wedge\overline{N\,\,}_{\!\!\!X^{\tau}})}.
$$

\appendix{Computation of the weights in the coset construction}
\label{app:weights}

\def\cH{{\cal H}}
\def\CN{{\cal N}}

In this appendix, we compute the weights of all fields of
the minimal model $SU(2)_k\times U(1)_2/U(1)_{k+2}$
exactly, in particular not only modulo integers. From 
this we can then infer exact expressions for $\sqrt{T}$.

As is standard in the coset construction, one starts by embedding
the denominator theory $U(1)_{k+2}$ into the numerator
$SU(2)_k\times U(1)_2$. This is achieved by the identification
\beq
J^{H} (z) = J^3(z) - J^{(2)}(z),
\eeq
which is the same as in (\ref{JH}). As before,
$J^3(z)$ is the  current associated with the Cartan-subalgebra
of $su(2)$ and  $J^{H}$ and $J^{(2)}=\frac{1}{2} J^f$ 
are the level $k+2$ and $2$
$U(1)$ currents. Note however that here we are talking about the
model $after$ GSO projection.

The Hilbert space of the numerator theory can then be decomposed in
the following way
\beq
{\cal H}_j \otimes {\cal H}_s = \bigoplus_{n\in \Z_{2k+4}, 2j+n+s \, even}
{\cal H}_{(j,n,s)} \otimes {\cal H}_{-n},
\eeq
where ${\cal H}_j$, ${\cal H}_s$, ${\cal H}_{j,n,s}$ and ${\cal H}_{-n}$
denote  representation spaces of the $SU(2)_k$, $U(1)_2$, the GSO
projected minimal model
and $U(1)_{k+2}$ respectively. Compared to the discussion of the
unprojected theory, $\hat{V}_{j}^{G,k} = {\cal H}_j$, but the Hilbert
spaces for the two $U(1)$'s are different.

To determine the weights of the
primary fields of the coset theory, one has to understand in detail
how the ground states $\ket{j,n,s} \otimes \ket{-n}$ of
${\cal H}_{j,n,s} \otimes {\cal H}_{-n}$ are realized within the
representation space ${\cal H}_j \otimes {\cal H}_s$. 

To find these ground states we fix $j,s$ and choose $n$ such 
that $2j+n+s$ is even. Within the subspace 
\beq \label{Hlsm} \cH_{js}^{(n)} \ = \ (\cH_j \otimes \cH_s)^{(n)} \ := \ 
    \{ \psi \in \cH_j \otimes \cH_s \ |\ 
   \e^{\frac{2 \pi i}{2k+4} J_0^{H}} \, \psi \ = \ 
   \e^{\frac{2\pi i n}{2k+4}} \, \psi \ \} 
\eeq 
we then search for eigenstates $\psi^{(n)}_{js}$ of 
$L^{SU(2)_k}_0 + L^{U(1)_2}_0$ 
with minimal eigenvalue.    

Some of these states are easily identified. These are the states 
which are realized in terms of ground states of the numerator
theory,  
\beq
\psi^{(n)}_{j\, s} \ = \ |j,n,s\rangle \otimes |-n\rangle \ = \ 
|j, \nu=-n+s\rangle \otimes 
                                      |s\rangle\ \ 
\eeq
where $n$ is restricted by $|n-s|\leq 2j$. In this way we have realized 
all fields from the so-called standard range of $\CN=2$ minimal models, 
\beqa \label{srange} 
 2j \ \leq\  k\ , \ \ \quad &&|n-s| \ \leq \ 2j, \  \ 2j + n + s 
 \ \mbox{\rm even} \ \ \\
&& n\in \{ -k-1,\dots, k+2 \}, \ \ \ s\in \{-1,0,1,2 \}
\eeqa
For these fields, the following formula for their conformal weights 
holds exactly (not just up to an integer),  
\beq \label{srweight}
h_{(j,n,s)} \ = \ \frac{j(j+1)}{k+2} - \frac{ n^2}{4(k+2)} + \frac{s^2}{8}
   \ \ . 
\eeq
But the $2j+1$ states we have found do not exhaust the ground states 
of the denominator theory. Additional states can be constructed with
the help of 
$$
(J^+)^\mu_{-1} |j,2j\rangle \ \ , \ \ (J^-)^\mu_{-1} |j,-2j\rangle  \ \in 
\cH_j  \ \ \mbox{ for } \ \ \mu = 1,2,\dots \ \ ,
$$
where $J^+(z)$ and $ J^-(z)$ are the raising and lowering operators of the
$su(2)$ algebra.
These states carry the charge $\nu = 2j + 2\mu$ or $\nu = -2j-2\mu$, 
respectively. When combined with appropriate states from $\cH_s$ 
(not necessarily the ground state) they furnish all the ground 
states for the denominator theory of the coset.

There is a second class of fields whose weight is easy to write
down, namely those which can be reflected to the standard range
by field identification. More precisely, these are primaries
labelled $(j,n,s)$, for which $(k/2-j, n\hat{+} (k+2), s\hat{+}2)$
is in the standard range. The notation $\hat{+}$ denotes the addition
modulo $2k+4$ for the label $n$ and modulo $4$ for the label $s$, in
such a way that the result of the addition is in the range
$-k-1, \dots, k+2$ for the labels $n$ and $-1,0,1,2$ for $s$. 
The hatted sum can be rewritten as
$n\hat{+} (k+2)= n -\frac{n}{|n|} (k+2)$, where we made use of
the choice that $n$ itself is in the range $-k-1, \dots, k+2$.
We can now apply the standard
range formula for the weights to the reflected field, with the
result that
\beq\label{reflweight}
h_{(j,n,s)} =  \ \frac{j(j+1)}{k+2} - \frac{n^2}{4(k+2)} + \frac{s^2}{8} +
\frac{|n|-|s|-2j}{2} \quad {\rm for} \ \ (\frac{k}{2}-j, n\hat{+} (k+2), s\hat{+}2) \in S.R.
   \ \ . 
\eeq
There is a list of primaries which cannot be reflected to the standard
range, namely $(j,-2j,2), (j,2j+2, 0), (j,2j+1,-1)$ and $(j,-2j-1,1)$.
The primaries $(j,-2j,2)$ with $j \geq 1$
have a coset realization with minimal weight as
$\ket{j,2j-2} \otimes \ket{-2} \in \cH_j \otimes \cH_s$. 
Note that the $U(1)_2$-label is not
in the standard domain for $s$. We can then use the expression (\ref{srweight})
for the weight and obtain
\beq
h_{(j,-2j,2)} = \frac{j}{(k+2)} + \frac{1}{2} \quad \quad j\geq 1.
\eeq
In the special case $j=0$ corresponding to the primary $(0,0,2)$ one
picks the realization $J^+_{-1} \ket{0,0}\otimes \ket{2}$, resulting in
the weight
\beq
h_{(0,0,2)} = \frac{3}{2}.
\eeq
Note that the primaries with higher $j$'s could have similarly
been represented
as $J^+_{-1}\ket{j,2j} \otimes\ket{2}$, but these states do not have
minimal weight and hence are not ground states.

The primaries $(j,2j+2,0)$, $j\neq k/2$, have a coset representation as 
$J^-_{-1}\ket{j,-2j} \otimes \ket{0}$, which is of minimal weight.
This results in a weight given by (\ref{srweight}) shifted
by one, or explicitly
\beq
h_{(j,2j+2,0)} = -\frac{j+1}{(k+2)} +1
\eeq
For the special case $j=k/2$ we consider the state 
$\ket{k/2,-k+2} \otimes\ket{4}$,
leading to the result that the standard range result has to be shifted by $2$,
or
\beq
h_{(\frac{k}{2},k+2,0)} = \frac{3}{2}
\eeq
The set of states $(j,-2j,2)$ is mapped to $(j,2j+2,0)$ under field
identification and our computation of the weights is compatible with that.
For later use, we emphasize that the standard range formula (\ref{srweight})
still holds for $(j,-2j,2)$, whereas the reflected formula (\ref{reflweight})
applies to the field identified primaries $(j,2j+2,0)$. Therefore, these
primaries can for our purposes be effectively  included in the discussion of
the standard range fields and no further case distinction is required.

In the Ramond sector, we realize the exceptional cases $(j,2j+1,-1)$
as $ J^-_{-1}\ket{j,-2j} \otimes \ket{-1}$ leading to a shift of the standard
range weight by $1$
\beq
h_{(j,2j+1,-1)} = \frac{c}{24} +1.
\eeq
Similarly, we have $\ket{j,-2j-1,1} = J^+_{-1}\ket{j,2j} \otimes \ket{1}$ and
a weight of
\beq
h_{(j,-2j-1,1)} = \frac{c}{24} +1.
\eeq
In the same way as in the NS sector,
 $(j,2j+1,-1)$ get mapped to $(j,-2j-1,1)$ under field identification and
our computation of the weights is compatible with that. However, neither
(\ref{srweight}) nor (\ref{reflweight}) holds in the R-sector, so that
we need to consider the fields that cannot be reflected to the standard
range separately.

In terms of the $T$ matrices, we can conclude from the above discussion
that
\beq
T^{\frac{1}{2}}_{(j,n,s)} = \sigma_{j,n,s}
\ T^{\frac{1}{2}}_j \ T^{-\frac{1}{2}}_n \ T^{\frac{1}{2}}_s
\eeq
where
$$
\sigma_{j,n,s} :=
\left\{\begin{array}{ll}
1 & (j,n,s) \in S.R. \\
1 & (j,-2j,2), \ l\geq 2 \\
1 & (\frac{k}{2},k+2,0)\\
(-1)^{\frac{|n| -|s|-2j}{2}} & 
(\frac{k}{2}-j, n\hat{+}(k+2), s\hat{+}2) \in S.R.\\
(-1)^{\frac{|n| -|s|-2j}{2}} & (j,2j+2,0), \ j \neq \frac{k}{2} \\
-1 & (j,\pm (2j+1), \mp 1)\\
-1 & (0,0,2)
\end{array}\right.
$$
One can derive the following general relation between the $\sigma$
\beq\label{sigmarelation}
\sigma_{\frac{k}{2}-j,n\hat{+}(k+2),s\hat{+}2} = 
(-1)^{\frac{|n|-|s|-2j}{2}} \sigma_{j,n,s}
\eeq
This holds independent of whether $(j,n,s)$ is or can be brought
to the standard range. It is however important that the label $n$ is
chosen in the range $-k-1, \dots, k+2$ and $s$ in $-1,0,1,2$.
Note also that the sign factor 
$(-1)^{\frac{|n|-|s|-2j}{2}}$ is invariant under field identification
$j\to \frac{k}{2}-j, n\to n\hat{+}(k+2), s\to s\hat{+}2$.

\appendix{P-matrix for the minimal model}

In this appendix, we write down an explicit expressions
of the P-matrix of the ${\mathcal N}=2$ minimal model.
We start with the expressions for
the P-matrix and Y-tensor of the constituent theories.

\subsubsection*{\underline{$SU(2)_k$}}

The P-matrix of the level k $SU(2)$ WZW model is
$$
P_{j j'}={2\over\sqrt{k+2}}
\sin\left[{\pi(2j+1)(2j'+1)\over 2(k+2)}\right]\delta_{2j+2j'+k}^{(2)}.
$$
Some part of Y-tensor is given by
$$
Y_{j 0}^{j'}=(-1)^{2j+j'}N_{j\,j}^{j'},\quad
Y_{j {k\over 2}}^{j'}=N_{j\,({k\over 2}-j)}^{j'}.
$$
In particular, $Y_{j\,0}^0=(-1)^{2j}$ and
$Y_{j\,{k\over 2}}^{k\over 2}=1$ for any $j\in\Pk$.

\subsubsection*{\underline{$U(1)_k$}}

The P-matrix and Y-tensor of the level $k$ $U(1)$ is
\beqa
&P_{n n'}={1\over \sqrt{k}}\e^{-{\pi i \widehat{n}\widehat{n'}\over 2k}}
\delta_{n+n'+k}^{(2)},
\nn\\[0.2cm]
&Y_{n n'}^{n''}
=\delta_{n'+n''}^{(2)}\left(
\delta_{n+{\widehat{n'}-\widehat{n''}\over 2}}^{(2k)}
+(-1)^{n'+k}
\delta_{n+{\widehat{n'}-\widehat{n''}\over 2}+k}^{(2k)}\right),
\nn
\eeqa
where $\widehat{n}$ is the unique member of $n+2k\Z$ in the standard range
$\{-k+1,...,k-1,k\}$. In the following, we will omit the $\hat{}$, but it
is understood that all labels are chosen in this way.

\subsubsection*{\it Minimal model}

We first note that the Q-matrix of the minimal model
can be expressed in terms of the Q-matrices of the constituent
theories in the following way
\beq
Q_{(j,n,s)(j',n',s')} = Q_{jj'} Q_{nn'}^* Q_{ss'} + 
Q_{j(\frac{k}{2}-j')} Q_{n(n'\hat{+}(k+2))}^* Q_{s(s'\hat{+}2)}
\label{factorizedformofQ}
\eeq
The P-matrix is then obtained as
\beqa
P_{(j,n,s)(j'n's')} &=& T_{(j,n,s)}^{\frac{1}{2}} Q_{(j,n,s)(j',n',s')}
T_{(j',n',s')}^{\frac{1}{2}} \\ \no
&=& \sigma_{j,n,s} \sigma_{j'n's'} P_{jj'} P_{nn'}^* P_{ss'} \\ \no
&&~~~+ \sigma_{j,n,s} \sigma_{\frac{k}{2}-j', n'\hat{+}(k+2), s'\hat{+} 2}
P_{j,\frac{k}{2}-j'} P_{n,n'\hat{+}(k+2)}^* P_{s,s'\hat{+}2} \\ \no
&=&\sigma_{j,n,s} \sigma_{j'n's'} \left( P_{jj'} P_{nn'}^* P_{ss'}
+(-1)^{\frac{|n'|-|s'|-2j'}{2}}
P_{j,\frac{k}{2}-j'} P_{n,n'\hat{+}(k+2)}^* P_{s,s'\hat{+}2} \right)
\eeqa
In the last step, we have used (\ref{sigmarelation}).
One can further evaluate this formula as
\beqa\no
P_{n,n'\hat{+}(k+2)}^* &=& \frac{1}{\sqrt{k+2}} \delta^{(2)}_{n+n'}
\e^{\frac{\pi i n(n'\hat{+}(k+2))}{2(k+2)}} = 
\frac{1}{\sqrt{k+2}} \delta^{(2)}_{n+n'}
\e^{\frac{\pi i nn'}{k+2}} \e^{\frac{\pi i n}{2}} (-1)^{\frac{|n'|+n'}{2}}\\ \no
P_{s,s'\hat{+}2} &=& \frac{1}{\sqrt{2}} \delta^{(2)}_{s+s'}
\e^{-\frac{\pi i s(s'\hat{+}2)}{4}} =
\frac{1}{\sqrt{2}} \delta^{(2)}_{s+s'} \e^{-\frac{\pi i ss'}{4}}
\e^{\frac{\pi i s}{2}} 
(-1)^{\frac{|s'|-s'}{2}}.
\eeqa
Note that these expressions are only valid if $n,n',s,s'$ are chosen
in the range $-k-1, \dots, k+2$ and $-1,0,1,2$.

The explicit expression for the P-matrix is
\beqa\no
P_{(j,n,s)(j'n's')}\!\!&=&\!\!
 \sigma_{j,n,s} \sigma_{j'n's'} \frac{\sqrt{2}}{k+2}
\delta^{(2)}_{s+s'} \ \e^{\frac{\pi i nn'}{2(k+2)}} \ \e^{-\frac{\pi i ss'}{4}}
\Bigl( \sin\left[\pi\mbox{$\frac{(2j+1)(2j'+1)}{2(k+2)}$}\right]
 \delta^{(2)}_{2j+2j'+k}
\ \delta^{(2)}_{n+n'+k} \\ 
&& + (-1)^{\frac{2j'+n'+s'}{2}} \e^{\frac{\pi i (n+s)}{2}}
\sin\left[\pi \mbox{$\frac{(2j+1)(k-2j'+1)}{2(k+2)}$}\right]
 \delta^{(2)}_{2j+2j'}
\ \delta^{(2)}_{n+n'} \Bigr)
\eeqa
In the calculations involving B-type parities, it is often 
convenient to use the following form of the P-matrix, where one inserts
the $U(1)$ data explicitly, and then uses general formulas for the
$SU(2)$ part:
\beqa
P_{(j,n,s)(j'n's')} &=& \sigma_{j,n,s} \sigma_{j'n's'}
\frac{1}{\sqrt{2(k+2)}} \e^{\frac{\pi i nn'}{2(k+2)}} \e^{-\frac{\pi iss'}{4}}
\delta^{(2)}_{s+s'}
\nn\\
&& \left( P_{jj'} \delta^{(2)}_{n+n'+k} + 
(-1)^{\frac{2j'+n'+s'}{2}} \e^{\frac{\pi i (n+s)}{2}} P_{j,\frac{k}{2}-j'}
\delta^{(2)}_{n+n'} \right)
\label{psu2u1}
\eeqa

\appendix{Formulae for the crosscap states}
\label{app:cc}

\subsection{A-type}

We compute the explicit coefficients of the A-type crosscap
states 
\beq
\ket{\Scr{C}_{\bar{n},\bar{s}}} = \sum_{(j,n,s)\in \Mk} 
\frac{P_{(0,\bar{n},\bar{s})(j,n,s)}}{\sqrt{S_{(0,0,0)(j,n,s)}}} \ 
\cket{(j,n,s)}
\eeq
using the expressions  for the P-matrix given in the previous section:
\beqa
\ket{\Scr{C}_{\bar{n},\bar{s}}} \!\!&=&\!\!
\sqrt{\frac{2}{k+2}} \sum_{(j,n,s)\in \Mk} 
\frac{\sigma_{j,n,s}
\sigma_{0,\bar{n},\bar{s}}}{\sqrt{\sin\pi\frac{2j+1}{k+2}}} 
\ \delta^{(2)}_{\bar{s}+s} \
\e^{\pi i \frac{\bar{n}n}{2(k+2)}} \ \e^{-\pi i \frac{\bar{s}s}{4}}
\no \\
&&\Biggl( \sin\left[\pi \mbox{$\frac{2j+1}{2(k+2)}$}\right]
\delta^{(2)}_{2j+k}\delta^{(2)}_{\bar{n}+n+k} 
 + (-1)^{\frac{2j+n+s}{2}+\frac{\bar{n}+\bar{s}}{2}}
\cos\left[\pi\mbox{$\frac{2j+1}{2(k+2)}$}\right]
\delta^{(2)}_{2j} \delta^{(2)}_{\bar{n}+n} \Biggr) \cket{(j,n,s)}
\nn\\
\eeqa

\subsubsection*{\it M\"obius strip}

The M\"obius strips are expressed using the $Y$-tensor,
\beqa
\langle \Scr{B}_{j,n,s}| q_t^H |\Scr{C}_{\bar{n},\bar{s}}\rangle
&=& \sum_{(j',n',s') \in \Mk}
Y_{(j,-n,-s)(0,\bar{n},\bar{s})}^{\ \ (j',n',s')}
\widehat{\chi}_{j',-n',-s'}(\tau)
\nn\\
&=& {1\over 2}\sum_{(j',n',s'):\, {\rm even} }
\sqrt{T_{0\bar{n}\bar{s}}\over T_{j'n's'}}
\widetilde{Y}_{(j,-n,-s)(0,\bar{n},\bar{s})}^{\ \ (j',n',s')}
\widehat{\chi}_{j',-n',-s'}(\tau) 
\nn\\
&=& \sum_{(j',n',s'):\, {\rm even}} 
\sqrt{T_{0\bar{n}\bar{s}}\over T_{j'n's'}}\widetilde{Y}_{j0}^{j'} 
\overline{\tilde{Y}}_{-n,\bar{n}}^{n'} \tilde{Y}_{-s,\bar{s}}^{s'}
\widehat{\chi}_{j',-n',-s'}(\tau)
\nn\\
&=&
\sum_{(j',n',s'):\, {\rm even}} 
{\sigma_{0\bar{n}\bar{s}}\over \sigma_{j'n's'}}Y_{j0}^{j'} 
\overline{Y}_{-n,\bar{n}}^{n'} Y_{-s,\bar{s}}^{s'}
\widehat{\chi}_{j',-n',-s'}(\tau)
\nn\\
&=&\sum_{(j',n',s'):\, {\rm even}} 
{\sigma_{0\bar{n}\bar{s}}\over \sigma_{j'n's'}}
\delta_{n'+\bar{n}}^{(2)}\delta_{s'+\bar{s}}^{(2)}
\left(
\delta_{n,{\widehat{\bar{n}}-\widehat{n'}\over 2}}^{(2k+4)}
+(-1)^{\bar{n}+k}
\delta_{n,{\widehat{\bar{n}}-\widehat{n'}\over 2}+k+2}^{(2k+4)}
\right)
\nn\\
&&\qquad
\times\left(
\delta_{s,{\widehat{\bar{s}}-\widehat{s'}\over 2}}^{(4)}
+(-1)^{\bar{s}}
\delta_{s,{\widehat{\bar{s}}-\widehat{s'}\over 2}+2}^{(4)}
\right)
(-1)^{2j+j'}N_{jj}^{j'}
\widehat{\chi}_{j',-n',-s'}(\tau)
\nn
\eeqa
Simplifying the delta functions for the $s$-indices by
$$
\delta_{s,{\widehat{\bar{s}}-\widehat{s'}\over 2}}^{(4)}
+(-1)^{\bar{s}}
\delta_{s,{\widehat{\bar{s}}-\widehat{s'}\over 2}+2}^{(4)}
=(-1)^{\bar{s}\cdot {s-{\widehat{\bar{s}}-\widehat{s'}\over 2}\over 2}}
\delta^{(4)}_{s',\bar{s}-2s},
$$
we obtain the formula
\beqa
\langle \Scr{B}_{j,n,s}| q_t^H |\Scr{C}_{\bar{n},\bar{s}}\rangle
&=&
\sum_{(j',n',s'):\, {\rm even}}
\delta_{n'+\bar{n}}^{(2)}
\left(
\delta_{n,{\widehat{\bar{n}}-\widehat{n'}\over 2}}^{(2k+4)}
+(-1)^{\bar{n}+k}
\delta_{n,{\widehat{\bar{n}}-\widehat{n'}\over 2}+k+2}^{(2k+4)}
\right)\delta^{(4)}_{s',\bar{s}-2s}
\nn\\
&&\qquad\qquad\quad
\times
(-1)^{2j+j'+\bar{s}\cdot {s-{\widehat{\bar{s}}-\widehat{s'}\over 2}\over 2}}
N_{jj}^{j'}
{\sigma_{0\bar{n}\bar{s}}\over
\sigma_{j'n's'}}\widehat{\chi}_{j',-n',-s'}(\tau).
\quad
\label{thecorrectone}
\eeqa
Taking the complex conjugation, we also find
\beqa
\langle \Scr{C}_{\bar{n},\bar{s}}| q_t^H |\Scr{B}_{j,n,s}\rangle
&=&
\sum_{(j',n',s'):\, {\rm even}}
\delta_{n'+\bar{n}}^{(2)}
\left(
\delta_{n,{\widehat{\bar{n}}-\widehat{n'}\over 2}}^{(2k+4)}
+(-1)^{\bar{n}+k}
\delta_{n,{\widehat{\bar{n}}-\widehat{n'}\over 2}+k+2}^{(2k+4)}
\right)\delta^{(4)}_{s',\bar{s}-2s}
\nn\\
&&\qquad\qquad\quad
\times
(-1)^{2j+j'+\bar{s}\cdot {s-{\widehat{\bar{s}}-\widehat{s'}\over 2}\over 2}}
N_{jj}^{j'}
{\sigma_{0\bar{n}\bar{s}}\over
\sigma_{j'n's'}}\widehat{\chi}_{j',n',s'}(\tau).
\quad
\label{thecorrectone2}
\eeqa
One may further simplify the $\delta$-function for $n$-indices.
This leads to the following expression
$$
\langle \Scr{C}_{\bar{n},\bar{s}}| q_t^H |\Scr{B}_{j,n,s}\rangle
=
\sum_{j'\in \Pk}
N_{jj}^{j'}\delta_{2n+n'-\bar{n}}^{(2k+4)}\delta^{(4)}_{2s+s'-\bar{s}}
\epsilon_{\bar{n},\bar{s}}^{j,n,s}(j',n',s')
\widehat{\chi}_{j',n',s'}(\tau),
$$
where
$$
\epsilon_{\bar{n},\bar{s}}^{j,n,s}(j',n',s')
:=(-1)^{2j+j'
+\bar{s}\cdot {s-{\widehat{\bar{s}}-\widehat{s'}\over 2}\over 2}
+(\bar{n}+k){n-{\widehat{\bar{n}}-\widehat{n'}\over 2}\over k+2}
}
{\sigma_{0\bar{n}\bar{s}}\over
\sigma_{j'n's'}}.
$$

\subsection{B-type}

The set of simple currents $(0,n,s)$ splits up into two orbits
under the orbifold group $\Z_{k+2} \times \Z_2$. The first orbit is the one
of $(0,0,0)$, which contains only currents $(0,n,s)$ with $n,s$ even, the
other orbit is the one of $(0,1,1)$, which contains only currents with
$n,s$ odd. Accordingly, there are two types of crosscap states. We
first construct A-type crosscaps in the orbifold and then apply the
mirror map. The crosscaps of the orbifold corresponding to the even orbit
are 
\beq
\ket{\Scr{C}_{P^{\theta_{rq}}_{(0,0,0)}}} = 
\frac{1}{\sqrt{2(k+2)}} \ \sum_{n,s}
\delta^{(2)}_n \ \delta^{(2)}_s \ \e^{-\pi i \theta_{rq}(n,s)}
\ket{\Scr{C}_{n,s}},
\eeq
where $\theta_{rq}(n,s)= -rn/(k+2) +qs/2$, as explained in the main text.
The resulting crosscap states on the orbifold are
\beq
\ket{\Scr{C}_{P^{\theta_{rq}}_{(0,0,0)}}}= 
2 \sum_j \delta^{(2)}_{2j} \ (-1)^{j}
\ (-1)^{\frac{\widehat{2r}}{2}+q} \ 
\frac{\cos \pi \frac{2j+1}{2(k+2)}}
{\sqrt{\sin \pi \frac{2j+1}{k+2}}} \ \sigma_{j,-2r,-2q} 
\ \cket{j,-2r,-2q}
\eeq
Applying the mirror map, one obtains the B-type states
\beq
\ket{\Scr{C}_{P^{\theta_{rq}}_{(0,0,0)}}^B}
= 2 \sum_j \delta^{(2)}_{2j} \ (-1)^{j}
\ (-1)^{\frac{\widehat{2r}}{2}+q} \ \frac{\cos \pi \frac{2j+1}{2(k+2)}}
{\sqrt{\sin \pi \frac{2j+1}{k+2}}} \ \sigma_{j,-2r,-2q} 
\ \cket{j,2r,2q}_B
\label{ilkaB1}
\eeq
The crosscaps of the orbifold corresponding to the odd currents are
\beq
\ket{\Scr{C}_{P^{\theta_{rq}}_{(0,1,1)}}} 
= \frac{1}{\sqrt{2(k+2)}} \ \sum_{n,s}
\delta^{(2)}_n \ \delta^{(2)}_s \ 
\e^{-\pi i (\theta_{rq}(n,s) -\hat{Q}_{(0,1,1)}(0,n,s))} \ 
\ket{\Scr{C}_{n+1,s+1}}
\eeq
We first compute
$$
\e^{\pi i \hat{Q}_{(0,1,1)}(0,n,s)} = 
\e^{\pi i (\frac{n}{2(k+2)}- \frac{s}{4})} \ \e^{\pi i \frac{|n|}{2}} \
(-1)^{(k+1)\frac{n+1-(n \hat{+} 1)}{2k+4}} \ \sigma_{0,n+1,s+1}
$$
In the calculation, one has to replace various hatted sums by
ordinary sums. One uses
\beqa
e^{\pi i \frac{(n\hat{+}1)n'}{2(k+2)}} \delta^{(2)}_{1+k+n+n'}
&=& \e^{\pi i \frac{(n+1)n'}{2(k+2)}}
(-1)^{(k+1)\frac{n+1-(n \hat{+} 1)}{2k+4}}\delta^{(2)}_{1+k+n+n'} \\ \no
e^{\pi i \frac{(n\hat{+}1)n'}{2(k+2)}} (-1)^{\frac{n\hat{+}1}{2}}
\delta^{(2)}_{1+n+n'}
&=& \e^{\pi i \frac{(n+1)n'}{2(k+2)}} (-1)^{\frac{n+1}{2}}
(-1)^{(k+1)\frac{n+1-(n \hat{+} 1)}{2k+4}}\delta^{(2)}_{1+n+n'} \\ \no
e^{-\frac{(s\hat{+}1)s'}{4}} \delta^{(2)}_s \delta^{(2)}_{s+s'}
&=& \e^{-\frac{(s+1)s'}{4}} (-1)^{\frac{s}{2}} \ 
\delta^{(2)}_s \delta^{(2)}_{s+s'}
\eeqa
Inserting this, one obtains the following crosscap states for the orbifold
theory
\beqa
\ket{\Scr{C}_{P^{\theta_{rq}}_{(0,1,1)}}}&=& 2 \e^{i \omega_{(0,1,1)}}
\sum_{j} \frac{\sigma_{j,-2r-1,-2q-1}}{\sqrt{\sin \pi \frac{2j+1}{k+2}}}
\e^{-\pi i \frac{\widehat{2r+1}}{k+2}} \ \e^{\pi i \frac{\widehat{2q+1}}{4}} \
\ \delta^{(2)}_{2j} \  \ \no \\ 
&& (-1)^{j +1} (-1)^{\frac{\widehat{2r+1}+2q+1}{2}}
\cos \pi \frac{2j+1}{2(k+2)}
\cket{j,-2r-1,-2q-1}
\eeqa
An application of the mirror map leads to the following odd B-type 
crosscap states
\beqa
\ket{\Scr{C}_{P_{(0,1,1)}^{\theta_{rq}}}^B}&=& 2 \e^{i \omega_{(0,1,1)}}
\sum_{j} 
\frac{\sigma_{j,-2r-1,-2q-1}}{\sqrt{\sin \pi \frac{2j+1}{k+2}}}
\e^{-\pi i \frac{\widehat{2r+1}}{2(k+2)}} \ 
\e^{\pi i \frac{\widehat{2q+1}}{4}} \
\ \delta^{(2)}_{2j} \  \ \no \\
&& (-1)^{j +1} (-1)^{\frac{\widehat{2r+1}+2q+1}{2}}
\cos \pi \frac{2j+1}{2(k+2)}
\cket{j,2r+1,2q+1}_B
\label{ilkaB2}
\eeqa
One can now choose $\omega_{(0,1,1)}$ in such a way that the crosscap
becomes real
$$
e^{i\omega_{(0,1,1)}} = 
\e^{\pi i \frac{\widehat{2r+1}}{2(k+2)}} \ \e^{-\pi i \frac{\widehat{2q+1}}{4}}
$$
At this point, it is convenient to relabel the states: We keep
the label $(r,q)\in \Z_{k+2}\times \Z_2$ and introduce a new
orbit-label $p \in {0,1}$. $p=0$ refers to the orbit of $(0,0,0)$
and $p=1$ to the orbit of $(0,1,1)$. We can then give a closed
formula for the crosscap states
\beq
\ket{\Scr{C}_{rqp}} = (2(k+2))^{\frac{1}{4}}
\sum_{j} \sigma_{j,-2r-p,-2q-p} 
\ \frac{P_{\frac{k}{2}j}}{\sqrt{S_{0j}}}
(-1)^{\frac{\widehat{2r+p}-p}{2} +q} 
\cket{j,2r+p,2q+p}_B
\label{appCrqp}
\eeq
Here, $S$ and $P$ are the modular matrices of $SU(2)$.

\subsubsection*{\it M\"obius strip}

We present the computation of the M\"obius strip.
\beqa
\lefteqn{\langle \Scr{C}_{rqp}|q_t^H|\Scr{B}^B_{[j,s]}
\rangle_{g_{4r+2p,2p}}}
\nn\\
&=&
(-1)^{sq}
\sqrt{2k+4} \sum_j \frac{S_{jj''} P_{\frac{k}{2}j''}}{S_{0j''}}
(-1)^{\frac{\widehat{2r+p}-p}{2}+q}
\sigma_{j'',-2r-p,-2q-p}
\hat\chi_{j'',2r+p,2q+p}(\tau).\quad
\eeqa
A modular transformation using the $P$-matrix as given in (\ref{psu2u1}) yields
\beqa
\lefteqn{
\langle \Scr{C}_{rqp}|q_t^H|\Scr{B}^B_{[j,s]} \rangle_{g_{4r+2p,2p}}}
\nn\\ &=&
(-1)^{sq}\sum_{(j'n's')\in \Mk} \delta^{(2)}_{s'+p}
\Bigl( Y_{jj'}^{\frac{k}{2}}
(-1)^{\frac{\widehat{2r+p}-p}{2}+q} \delta^{(2)}_{n'+k+p}
+ (-1)^{\frac{2j'+n'+s'}{2}+p}Y_{j \frac{k}{2}-j'}^{\frac{k}{2}}
\delta^{(2)}_{n'+p}
\Bigr) 
\nn\\
&& ~~~~~~~~~~~~~~~~
\times 
\e^{\frac{\pi i(\widehat{2r+p})n'}{2(k+2)}-\frac{\pi i (\widehat{2q+p})s'}{4}}
\sigma_{j',n',s'} \hat\chi_{j'n's'}(\tau)
\nn\\[0.2cm]
&=& 
(-1)^{sq}\sum_{(j'n's')\in \Mk} \delta^{(2)}_{s'+p}
\Bigl( N_{jj}^{\frac{k}{2}-j'}
(-1)^{\frac{\widehat{2r+p}-p}{2}+q} 
\delta^{(2)}_{n'+p+k}
+ (-1)^{\frac{2j'+n'+s'}{2}+p} N_{jj}^{j'} \delta^{(2)}_{n'+p}
\Bigr)
\nn\\
&& ~~~~~~~~~~~~~~~~~
\times 
\e^{\frac{\pi i (\widehat{2r+p})n'}{2(k+2)}-\frac{\pi i (\widehat{2q+p})s'}{4}}
\sigma_{j',n',s'} \hat\chi_{j'n's'}(\tau) 
\nn\\[0.2cm]
&=&
(-1)^{sq}\sum_{(j'n's'):\, {\rm even}} \delta^{(2)}_{s'+p} \delta^{(2)}_{n'+p}
 (-1)^{\frac{2j'+n'+s'}{2}+p} N_{jj}^{j'}
e^{\frac{\pi i (\widehat{2r+p})n'}{2(k+2)}-\frac{\pi i (\widehat{2q+p})s'}{4}}
\sigma_{j',n',s'} \hat\chi_{j'n's'}(\tau)
\nn\\
\label{CBBapp}
\eeqa
Here we have used some relations between $SU(2)$ the $Y$-matrix and the
fusion rules: $Y_{jj'}^{\frac{k}{2}} = N_{j,\frac{k}{2}-j}^{j'} = 
N_{jj}^{\frac{k}{2}-j'}, Y_{j,\frac{k}{2}-j'}^{j'} = N_{jj}^{j'}$.
To determine the M\"obius strip involving the short orbit boundary states 
note that
there is no overlap of the crosscap state with the extra RR-part
of the boundary state.
The formula presented in the main text is obtained from 
the second line of the above equation by
inserting $N_{\frac{k}{4}\frac{k}{4}}^{j'}
=N_{\frac{k}{4}\frac{k}{4}}^{\frac{k}{2}-j'}
=1$ for all $j'=0, \dots, \frac{k}{2}$.

\appendix{Supercurrent Conditions}\label{app:SC}

In this appendix, we derive the supercurrent conditions obeyed by
the A-type and B-type crosscaps of the superparafermion RCFT.
The starting point is the conditions on the boundary states.
The Cardy states $|\Scr{B}_{j,n,s=\pm 1}\rangle$ obey the A-type
supercurrent condition
\beq
\overline{\tG}_{-r}-iG_r=\tG_r-i\bG_{-r}=0,
\label{Acon}
\eeq
where $r\in \Z$ (resp.~$r\in\Z+\half$)
when they act on the RR-part (resp.~NSNS-part) of the boundary state.
This shows that the combinations of the Ishibashi states
$$
\bket{j,n,s}-\bket{j,n,s+2}
$$
obey the same condition (\ref{Acon}) with
$r\in \Z+\half$ if $s=0$ and $r\in\Z$ if $s=-1$.
Then, the following combination of the crosscap Ishibashi states
$$
\sqrt{T_{j,n,s}}\cket{j,n,s}-\sqrt{T_{j,n,s+2}}\cket{j,n,s+2}=
\e^{\pi i L_0}\Bigl(\bket{j,n,s}-\bket{j,n,s+2}\Bigr)
$$
obey the condition
$
\e^{\pi i L_0}\Bigl(\overline{\tG}_{-r}-iG_r\Bigr)\e^{-\pi i L_0}
=\e^{\pi i L_0}\Bigl(\tG_r-i\bG_{-r}\Bigr)\e^{-\pi i L_0}=0.
$
This is nothing but the supercurrent condition on
the $A_{0,0}$-parity for $s=-1$ and $A_{{\pi\over 2},{\pi\over
2}}$-parity for $s=0$.
It is also easy to see that the other combination
$$
\sqrt{T_{j,n,s}}\cket{j,n,s}+\sqrt{T_{j,n,s+2}}\cket{j,n,s+2}
$$
obey the $A_{\pi,0}$-condition  for $s=-1$ and
the $A_{{\pi\over 2},-{\pi\over 2}}$-condition for $s=0$.

We now consider the PSS crosscaps
$$
|\Scr{C}_{j,n,s}\rangle
=\sqrt{T_{j,n,s}}\sum_{j,n,s}
{Q_{j,n,s,}^{\,\,\,\,j',n',s'}\over\sqrt{S_{0,0,0}^{\,\,\,j',n',s'}}}
\sqrt{T_{j',n',s'}}\cket{j',n',s'}.
$$
Using the ``factorized form'' (\ref{factorizedformofQ})
of $Q$-matrix and the following
property of
$Q_{s,s'}$
$$
{Q_{0,0}\,\,Q_{0,2}\choose
Q_{2,0}\,\,Q_{2,2}}\propto
{\,\,1\,\,\,-\!i\,\choose \!-i\,\,\,\,1\!},\quad\,
{Q_{-1,-1}\,\,Q_{-1,1}\choose
Q_{1,-1}\,\,Q_{1,1}}\propto
{\!\!-i\,\,\,\,1\!\choose \,1\,\,\,-\!i\,},\qquad
$$
it is easy to see that the $s'$ dependence of
$Q_{j,n,s}^{\,\,\,j',n',s'}\mp Q_{j,n,s+2}^{\,\,\,j',n',s'}$
is proportional to $1$ for $s'$ and $\mp 1$ for $s'+2$.
This shows that
$$
{1\over \sqrt{T_{j,n,s}}}|\Scr{C}_{j,n,s}\rangle
\mp {1\over \sqrt{T_{j,n,s+2}}}|\Scr{C}_{j,n,s+2}\rangle
$$
is spanned by
$\sqrt{T_{j',n',s}}\cket{j',n',s}\mp \sqrt{T_{j',n',s+2}}\cket{j',n',s+2}$,
which obey the supercurrent conditions of the types given above.
This shows the first table in Section~\ref{subsub:right}.

Next, let us apply the mirror map to the combinations
of the crosscap Ishibashi states considered above.
This shows that
$$
\sqrt{T_{j,n,s}}\cket{j,n,s}_B\mp \sqrt{T_{j,n,s+2}}\cket{j,n,s+2}_B
$$
obey for the $(-)$-sign
the supercurrent conditions of the type
$B_{0,0}$ ($s$ odd) and $B_{{\pi\over 2},{\pi\over 2}}$ ($s$ even),
while they obey for the $(+)$-sign
the conditions of the type
$B_{0,\pi}$ ($s$ odd) and $B_{{\pi\over 2},-{\pi\over 2}}$
($s$ even).
Note that the B-type crosscaps that appear in
(\ref{Crqp})
((\ref{ilkaB1}) and (\ref{ilkaB2}) or their combined form (\ref{appCrqp}))
has the following structure
$$
|\Scr{C}^B_{r,q,p}\rangle
=\sum_{j\,{\rm integer}}
c_{j,r,p}{(-1)^q\over \sqrt{T^{(2)}_{2q+p}}}
\sqrt{T_{j,2r+p,2q+p}}\cket{j,2r+p,2q+p}_B,
$$
where $c_{j,r,p}$ is a number depending on $(j,r,p)$ only.
This motivates us to find the combinations (\ref{BCexp})
which obey the supercurrent conditions
of the types in the
second table of Section~\ref{subsub:right}.

\appendix{Normalization of RR Ground States}
\label{app:orv}

In this appendix, we review the Landau--Ginzburg computation of
the overlaps of RR ground states and A-branes in
${\mathcal N}=2$ minimal model \cite{HIV,HKKPTVVZ}.
This explains the choice of the normalization constant (\ref{cl}) for
the ground state wavefunctions.

For the ground state wave functions, we use those of the form
\beq
\omega_{l\over 2}=c_{l\over 2}
\e^{-i\overline{W}}\phi^{l}\dd\phi+(\bQ_++Q_-)(\cdots),
\quad l=0,1,\ldots,k,
\eeq
for some constant $c_{l\over 2}$.
Their inner-products are computed as
\begin{eqnarray}
g_{{l\over 2},\overline{l'\over 2}}
&=&
\int \omega_{l\over 2}\wedge *\overline{\omega_{l'\over 2}}
\nn\\
&=&c_{l\over 2}\overline{c_{l'\over 2}}
\int \e^{-i(\phi^{k+2}+\bphi^{k+2})}\phi^l\bphi^{l'}\dd\phi\wedge
*\dd\bphi
\nn\\
&=&
2c_{l\over 2}\overline{c_{l'\over 2}}\int \e^{-2ir^{k+2}\cos((k+2)\theta)}
\e^{i(l-l')\theta}r^{l+l'+1}\dd r\dd\theta,
\nn
\end{eqnarray}
where we have used the polar coordinates $\phi=r\e^{i\theta}$.
We see that they are orthogonal, $g_{{l\over 2},\overline{l'\over 2}}=0$
if $l\ne l'$.
Expanding the exponential and performing the $\theta$ integral,
we find
\begin{eqnarray}
g_{{l\over 2},\overline{l\over 2}}
&=&
4\pi |c_{l\over 2}|^2\int_0^{\infty}\sum_{m=0}^{\infty}
{(-1)^m\over (m!)^2}(r^{k+2})^{2m}r^{2l+1}\dd r
\nn\\
&=&
4\pi |c_{l\over 2}|^2 \int_0^{\infty} J_0(2r^{k+2})r^{2l+1}\dd r
\nn\\
&=&
|c_{l\over 2}|^2{2\over k+2}\left[\Gamma\left({l+1\over k+2}\right)\right]^2
\sin\left({\pi(l+1)\over k+2}\right),
\nn
\end{eqnarray}
where $J_0(x)=\sum_{m=0}^{\infty}(-1)^m(x/2)^{2m}/(m!)^2$ is the
Bessel function, and we have used the integral formula
$\int_0^{\infty}x^{\mu-1}J_0(ax)\dd x
=2^{\mu-1}[\Gamma(\mu/2)]^2\sin(\pi\mu/2)/
(\pi a^{\mu})$.
Thus, they form an orthonormal basis if $c_{l\over 2}$ are chosen as
\begin{equation}
c_{l\over 2}=\e^{i\gamma_l}{\sqrt{k+2}\over
\Gamma({l+1\over k+2})\sqrt{2\sin({\pi (l+1)\over k+2})}},
\label{clnorm}
\end{equation}
where $\e^{i\gamma_l}$ is some phase.

Now, let us perform the overlap integrals.
We consider the brane $A_{jns=1}=A_{a_+,a_-}$
where $a_{\pm}={\pi(n\pm 2j\pm 1)\over k+2}$.
It is a broken line coming from the
infinity in the direction $z_i=\e^{\pi i(n-2j-1)\over k+2}$,
cornering at $\phi=0$, and then going out to infinity
in the direction $z_f=\e^{\pi i(n+2j+1)\over k+2}$.
The overlap is
\begin{eqnarray}
\Pi_{l\over 2}^{A_{jn1}}
&=&
c_{l\over 2}\int_{A_{jn1}^-}\e^{-i\phi^{k+2}}\phi^l\dd \phi
\nn\\
&=&
c_{l\over 2}
(z_f^{l+1}-z_i^{l+1})\int_0^{+\infty-i\cdot 0}\e^{-it^{k+2}}t^l\dd t
\nn\\
&=&c_{l\over 2}{1\over k+2}\e^{\pi i(n-1/2)(l+1)\over k+2}
2i\sin\left[{\pi(2j+1)(l+1)\over k+2}\right]\Gamma\left({l+1\over
k+2}\right)
\nn
\end{eqnarray}
and
\begin{eqnarray}
\widetilde{\Pi}_{\overline{l\over 2}}^{A_{jn1}}
&=&\overline{c_{l\over 2}}
\int_{A_{jn1}^+}\e^{-i\overline{\phi}^{k+2}}
\overline{\phi}^l*\dd \overline{\phi}
\nn\\
&=&i\,\overline{c_{l\over 2}}
\overline{\int_{A_{jn1}^+}\e^{i\phi^{k+2}}\phi^l\dd \phi}
\nn\\
&=&
i\,\overline{c_{l\over 2}}{1\over k+2}
\e^{-\pi i (n+1/2)(l+1)\over k+2}
(-2i)\sin\left[{\pi(2j+1)(l+1)\over k+2}\right]\Gamma\left({l+1\over
k+2}\right)
\nn
\end{eqnarray}
Using the normalization formula (\ref{clnorm})
with the phase choice $\e^{i\gamma_l}=-i\e^{-\pi i(l+1)\over 2(k+2)}$,
we find that the above reproduce the known formula for the
overlaps:
\begin{equation}
\Pi^{A_{jn1}}_{l\over 2}
={}^{}_{{}_{\rm RR}}\langle \Scr{B}_{j,n-1,(0)}|\,
\mbox{${l\over 2}$}\,\rangle^{}_{{}_{\rm RR}},
\,\,\,\,\,
\widetilde{\Pi}^{A_{jn1}}_{\overline{l\over 2}}
={}^{}_{{}_{\rm RR}}\langle \,\mbox{${l\over 2}$}\, 
|\Scr{B}_{j,n,(1)}\rangle^{}_{{}_{\rm RR}}.
\end{equation}

\end{document}